\documentclass[twocolumn]{aastex62}
\usepackage{booktabs}
\usepackage{chngcntr}

\received{January 1, 2019}
\revised{?, 2019}
\accepted{\today}
\submitjournal{ApJ}

\shorttitle{SEDs of AGN dust II: the data}
\shortauthors{Gonz\'alez-Mart\'in et al.}

\begin{document}

\title{Exploring the mid-infrared SEDs of six AGN dusty torus models II: the data}

\correspondingauthor{Omaira Gonz\'alez-Mart\'in, faculty}
\email{o.gonzalez@irya.unam.mx}

\author{Omaira Gonz\'alez-Mart\'in}
\affil{Instituto de Radioastronom\'ia y Astrof\'isica (IRyA-UNAM), 3-72 (Xangari), 8701, Morelia, Mexico}
\author{Josefa Masegosa}
\affil{Instituto de Astrof\'isica de Andaluc\'ia, CSIC, Glorieta de la Astronom\'ia s/n E-18008, Granada, Spain}
\author{Ismael Garc\'ia-Bernete}
\affil{Instituto de F\'isica de Cantabria (CSIC-UC), Avenida de los Castros, 39005 Santander, Spain}
\author{Cristina Ramos Almeida}
\affil{Instituto de Astrof\'isica de Canarias (IAC), C/V\'ia L\'actea, s/n, E-38205 La Laguna, Spain}
\affil{Departamento de Astrof\'isica, Universidad de La Laguna (ULL), E-38205 La Laguna, Spain}
\author{Jos\'e Miguel Rodr\'iguez-Espinosa}
\affil{Instituto de Astrof\'isica de Canarias (IAC), C/V\'ia L\'actea, s/n, E-38205 La Laguna, Spain}
\affil{Departamento de Astrof\'isica, Universidad de La Laguna (ULL), E-38205 La Laguna, Spain}
\author{Isabel M\'arquez}
\affil{Instituto de Astrof\'isica de Andaluc\'ia,  CSIC, Glorieta de la Astronom\'ia s/n E-18008, Granada, Spain}
\author{Donaji Esparza-Arredondo}
\affil{Instituto de Radioastronom\'ia y Astrof\'isica (IRyA-UNAM), 3-72 (Xangari), 8701, Morelia, Mexico}
\author{Natalia Osorio-Clavijo}
\affil{Instituto de Radioastronom\'ia y Astrof\'isica (IRyA-UNAM), 3-72 (Xangari), 8701, Morelia, Mexico}
\author{Mariela Mart\'inez-Paredes}
\affil{Instituto de Radioastronom\'ia y Astrof\'isica (IRyA-UNAM), 3-72 (Xangari), 8701, Morelia, Mexico}
\affil{Korea Astronomy and Space Science Institute 776, Daedeokdae-ro, Yuseong-gu, Daejeon, Republic of Korea (34055)}
\author{C\'esar Victoria-Ceballos}
\affil{Instituto de Radioastronom\'ia y Astrof\'isica (IRyA-UNAM), 3-72 (Xangari), 8701, Morelia, Mexico}
\author{Alice Pasetto}
\affil{Instituto de Radioastronom\'ia y Astrof\'isica (IRyA-UNAM), 3-72 (Xangari), 8701, Morelia, Mexico}
\author{Deborah Dultzin}
\affil{Instituto de Astronom\'ia (IA-UNAM), Apartado Postal 70-264, 04510, Mexico DF, Mexico}

\begin{abstract}

This is the second in a series of papers devoted to explore a set of six dusty models of active galactic nuclei (AGN) with available spectral energy distributions (SEDs). These models are the smooth torus by \citet{Fritz06}, the clumpy torus by \citet{Nenkova08B}, the clumpy torus by \citet{Hoenig10B}, the two phase torus by \citet{Siebenmorgen15}, the two phase torus by \citet{Stalevski16}, and the wind model by \citet{Hoenig17}. The first paper explores discrimination among models and the parameter restriction using synthetic spectra (Gonz\'alez-Mart\'in et al. 2019A). Here we perform spectral fitting of a sample of 110 AGN drawn from the Swift/BAT survey with \emph{Spitzer}/IRS spectroscopic data. The aim is to explore which is the model that describes better the data and the resulting parameters. The clumpy wind-disk model by \citet{Hoenig17} provides good fits for $\sim$50\% of the sample, and the clumpy torus model by \citet{Nenkova08B} is good at describing $\sim$30\% of the objects. The wind-disk model by \citet{Hoenig17} is better for reproducing the mid-infrared spectra of Type-1 Seyferts \citep[with 60\% of the Type-1 Seyferts well reproduced by this model compared to the 10\% well represented by the clumpy torus model by][]{Nenkova08B} while Type-2 Seyferts are equally fitted by both models (roughly 40\% of the Type-2 Seyferts). Large residuals are found irrespective of the model used, indicating that the AGN dust continuum emission is more complex than predicted by the models or that the parameter space is not well sampled. We found that all the resulting parameters for our AGN sample are roughly constrained to 10-20\% of the parameter space. Contrary to what is generally assumed, the derived outer radius of the torus is smaller (reaching up to a factor of $\rm{\sim 5}$ times smaller for 10\,pc tori) for the smooth torus by \citet{Fritz06} and the two phase torus by \citet{Stalevski16} than the one derived from the clumpy torus by \citep{Nenkova08B}. Covering factors and line-of-sight viewing angles strongly depend on the model used. The total dust mass is the most robust derived quantity, giving equivalent results for four of these models. 

\end{abstract}



\section{Introduction} \label{sec:intro}

The nuclear obscurer of active galactic nuclei (AGN), dubbed the torus \citep[][]{Antonucci85}, produces a broad infrared spectral energy distribution (SED), whose power and shape depend on the fraction of the source absorbed, and the geometry of the absorber, respectively \citep[see][for a review on the topic]{Netzer15,Ramos-Almeida17}. This emitting region is expected to be concentrated within the inner $\sim$5 pc of the AGN \citep[e.g.][]{Ramos-Almeida09,Alonso-Herrero11,Burtscher13,Lopez-Gonzaga16} which makes almost impossible to image it with current single mirror telescopes due to a combination of insufficient spatial resolution and foreground contamination \citep[][]{Pasetto19}. Therefore, trying to reproduce the infrared SED of nearby AGN with torus models is one of the methods to constrain the properties of the nuclear obscurer.

We can divide torus models into three generic types according to the distribution of dust: continuous or smooth \citep[e.g.][]{Pier92,vanBemmel03,Fritz06}, clumpy, \citep[e.g.][]{Dullemond05, Nenkova08A,Nenkova08B, Hoenig10A,Hoenig10B,Hoenig17}, and composite (a combination of clumpy and continuous) \citep{Stalevski12,Stalevski16,Siebenmorgen15}. However, this is not the only way to classify them. For instance, this can be done by looking at the morphological distribution of dust: torus-like \citep[e.g.][]{Fritz06,Nenkova08A,Nenkova08B,Hoenig10A,Hoenig10B,Stalevski12,Siebenmorgen15,Stalevski16} or wind-like \citep[e.g.][]{Siebenmorgen15,Hoenig17} morphologies. Another way to classify torus models is according to the chemical composition and size of dust in: graphite grains \citep[e.g.][]{Fritz06}, standard ISM composition \citep[e.g.][]{Nenkova08A,Nenkova08B,Hoenig10A,Hoenig10B,Stalevski12,Stalevski16,Hoenig17}, large-grains ISM composition \citep[e.g.][]{vanBemmel03,Hoenig10A,Hoenig10B, Hoenig17}, or silicates and amorphous carbon \citep[e.g.][]{Siebenmorgen15}.

\renewcommand{\baselinestretch}{0.6}
\begin{table*}
\scriptsize
\begin{center}
\begin{tabular}{ l l c c c c | l l c c c c}
\hline \hline
    &   Obj. Name            &    Dist.             & Scale  &   AGN & $\rm{log(L_{X})}$ &     &   Obj. Name            &    Dist.             & Scale  &   AGN & $\rm{log(L_{X})}$  \\ 
  &             &    (Mpc)             & (kpc)  &   class &         &   &             &    (Mpc)             & (kpc)  & class  &      \\ 
    &   (1)   &  (2)    &   (3)   &    (4)    & (5)   &  &  (1)   &  (2)    &   (3)   &   (4)    & (5)\\
  \hline
    1 & NGC235A                   &  95.2 &   1.7 & Sy1.9      & 43.77 &    56 & NGC4151                   &   9.9 &   0.2 & Sy1.5      & 43.17 \\
    2 & Mrk348                    &  21.5 &   0.4 & B. AGN & 43.86 &    57 & NGC4235                   &  24.2 &   0.4 & Sy1.2      & 42.74 \\
    3 & Mrk352                    &  63.7 &   1.1 & Sy1.2      & 43.19  &    58 & M106                      &   7.3 &   0.1 & Sy1.9      & 41.06\\
    4 & NGC454E                   &  51.9 &   0.9 & Sy2        & 42.79 &   59 & Mrk50                     & 100.4 &   1.8 & Sy1        & 43.43\\
    5 & NGC526A                   &  81.8 &   1.4 & Sy2        & 43.78  &    60 & NGC4388                   &  19.3 &   0.3 & Sy2        & 43.64 \\
    6 & ESO297-018                & 108.0 &   1.9 & Sy2        & 43.99 &   61 & NGC4395                   &   4.2 &   0.1 & Sy2        & 40.86 \\
    7 & NGC788                    &  58.3 &   1.0 & Sy2        & 43.51  &   62 & NGC4507                   &  50.5 &   0.9 & Sy1.9      & 43.76 \\
    8 & Mrk590                    &  91.2 &   1.6 & Sy1.5      & 43.23  &    63 & ESO506-G027               & 107.2 &   1.9 & Sy2        & 44.11 \\
    9 & IC1816                    &  72.6 &   1.3 & Sy1.8      & 43.14  &   64 & NGC4686                   &  71.7 &   1.3 & Sy2        & 43.20 \\
   10 & NGC973                    &  60.0 &   1.0 & Sy2        & 43.46 &    65 & NGC4941                   &  14.2 &   0.2 & Sy2        & 41.78 \\
   11 & NGC1052                   &  20.6 &   0.4 & B. AGN & 42.24 &    66 & NGC4939                   &  36.1 &   0.6 & Sy2        & 42.81 \\
   12 & ESO417-G006               &  69.8 &   1.2 & Sy2        & 43.27 &   67 & ESO323-077                &  64.3 &   1.1 & Sy1.5      & 43.19 \\
   13 & NGC1275                   &  69.0 &   1.2 & B. AGN & 43.76 &    68 & NGC4992                   & 107.7 &   1.9 & Sy2        & 43.89 \\
   14 & ESO548-G081               &  62.0 &   1.1 & Sy1.9      & 43.29 &    69 & IISZ010                   & 146.8 &   2.6 & Sy1.5      & 43.52 \\
   15 & 2MASXJ0350-5018    & 156.3 &   2.7 & Sy2        & 43.83  &    70 & MCG-03-34-064             &  85.6 &   1.5 & Sy1.9      & 43.28 \\
   16 & ESO549-G049               & 112.6 &   2.0 & Sy1.9      & 43.52 &    71 & CenA                      &   3.8 &   0.1 & B. AGN & 42.98 \\
   17 & 3C120                     & 141.4 &   2.5 & . AGN & 44.38 &    72 & MCG-06-30-015             &  33.2 &   0.6 & Sy1.9      & 42.89 \\
   18 & MCG-02-12-050             & 155.7 &   2.7 & Sy1.2      & 43.74 &   73 & NGC5252                   &  83.6 &   1.5 & Sy2        & 44.09 \\
   19 & MCG-01-13-025             &  68.1 &   1.2 & Sy1.5      & 43.29 &    74 & IC4329A                   &  68.8 &   1.2 & Sy1.5      & 44.18 \\
   20 & CGCG420-015               & 125.9 &   2.2 & Sy2        & 43.72 &    75 & UM614                     & 140.0 &   2.4 & Sy1.5      & 43.60 \\
   21 & 2MASXJ0505-2351    & 150.1 &   2.6 & Sy2        & 44.22 &    76 & Mrk279                    & 130.4 &   2.3 & Sy1.5      & 43.87 \\
   22 & CGCG468-002NED01          &  75.0 &   1.3 & Sy1.9      & 43.26 &    77 & CircinusGalaxy            &   4.2 &   0.1 & Sy2        & 42.07 \\
   23 & Ark120                    & 138.2 &   2.4 & Sy1        & 44.25 &    78 & NGC5506                   &  23.8 &   0.4 & Sy1.9      & 43.31 \\
   24 & PICTORA                   & 150.1 &   2.6 & Sy2        & 44.03 &    79 & NGC5548                   & 107.6 &   1.9 & Sy1.5      & 43.76 \\
   25 & NGC2110                   &  35.6 &   0.6 & Sy2        & 43.65  &    80 & ESO511-G030               &  64.1 &   1.1 & Sy1        & 43.65 \\
   26 & 2MASXJ0558-3820    & 145.1 &   2.5 & Sy1.2      & 43.86 &    81 & Mrk477                    & 161.6 &   2.8 & Sy1.9      & 43.66  \\
   27 & Mrk3                      &  61.4 &   1.1 & Sy1.9      & 43.79 &    82 & IC4518A                   &  69.6 &   1.2 & Sy2        & 43.19 \\
   28 & ESO426-G002               &  96.1 &   1.7 & Sy2        & 43.44 &   83 & Mrk841                    & 156.0 &   2.7 & Sy1.2      & 44.01 \\
   29 & UGC03478                  &  45.9 &   0.8 & Sy1.2      & 42.49 &    84 & Mrk1392                   & 154.8 &   2.7 & Sy1.5      & 43.75 \\
   30 & UGC03601                  &  73.3 &   1.3 & Sy1.9      & 43.08 &    85 & Mrk290                    & 126.7 &   2.2 & Sy1.5      & 43.68 \\
   31 & Mrk78                     & 159.1 &   2.8 & Sy2        & 43.50 &   86 & ESO138-G001               &  39.1 &   0.7 & Sy2        & 42.55 \\
   32 & Mrk10                     &  83.7 &   1.5 & Sy1.5      & 43.47 &    87 & Mrk501                    & 119.0 &   2.1 & B. AGN & 44.27 \\
   33 & IC0486                    & 116.3 &   2.0 & Sy1.9      & 43.75 &    88 & NGC6300                   &  12.3 &   0.2 & Sy2        & 42.46 \\
   34 & Mrk1210                   &  57.8 &   1.0 & Sy1.9      & 43.37 &   89 & Fairall49                 &  85.7 &   1.5 & Sy1.9      & 43.11 \\
   35 & Mrk622                    &  99.5 &   1.7 & Sy2        & 43.10 &   90 & ESO103-035                &  56.9 &   1.0 & Sy1.9      & 43.63 \\
   36 & Mrk18                     &  47.5 &   0.8 & Sy1.9      & 42.61 &   91 & Fairall51                 &  45.9 &   0.8 & Sy1.5      & 43.22 \\
   37 & MCG-01-24-012             &  84.1 &   1.5 & Sy2        & 43.61 &    92 & ESO141-G055               & 158.9 &   2.8 & Sy1.2      & 44.25 \\
   38 & MCG+04-22-042             & 138.5 &   2.4 & Sy1.2      & 43.93  &    93 & NGC6814                   &  11.8 &   0.2 & Sy1.5      & 42.59 \\
   39 & Mrk110                    & 141.0 &   2.5 & Sy1.5      & 44.25 &    94 & Mrk509                    & 147.3 &   2.6 & Sy1.2      & 44.44 \\
   40 & Mrk705                    & 124.8 &   2.2 & Sy1.2      & 43.54  &    95 & IC5063                    &  37.9 &   0.7 & Sy2        & 43.29 \\
   41 & MCG+10-14-025             & 168.6 &   2.9 & Sy1.9      & 43.41 &    96 & NGC7130                   &  69.2 &   1.2 & Sy1.9      & 43.01 \\
   42 & MCG-05-23-016             &  36.3 &   0.6 & Sy1.9      & 43.53 &    97 & Mrk520                    & 108.0 &   1.9 & Sy2        & 43.71 \\
   43 & NGC3081                   &  25.3 &   0.4 & Sy2        & 43.07 &    98 & NGC7172                   &  33.9 &   0.6 & Sy2        & 43.43 \\
   44 & ESO374-G044               & 121.9 &   2.1 & Sy2        & 43.64 &    99 & NGC7212NED02              & 114.2 &   2.0 & Sy2        & 43.30 \\
   45 & NGC3227                   &  18.8 &   0.3 & Sy1.5      & 42.58 &  100 & NGC7213                   &  22.0 &   0.4 & B. AGN & 42.46 \\
   46 & NGC3281                   &  45.7 &   0.8 & Sy2        & 43.32 &   101 & NGC7314                   &  16.7 &   0.3 & Sy1.9      & 42.47 \\
   47 & NGC3393                   &  53.6 &   0.9 & Sy2        & 42.98  &   102 & Mrk915                    & 103.3 &   1.8 & Sy1.9      & 43.63 \\
   48 & Mrk417                    & 140.3 &   2.4 & Sy2        & 43.91 &   103 & MCG+01-57-016             & 107.0 &   1.9 & Sy1.5      & 43.42 \\
   49 & Mrk421                    &  85.2 &   1.5 & B. AGN & 44.46 &   104 & UGC12282                  &  71.2 &   1.2 & Sy2        & 43.15 \\
   50 & NGC3783                   &  47.8 &   0.8 & Sy1.2      & 43.56  &   105 & NGC7603                   & 126.4 &   2.2 & Sy1        & 44.02 \\
   51 & NGC3786                   &  50.9 &   0.9 & Sy1.9      & 42.41 &   106 & UGC00488                  & 143.5 &   2.5 & Sy1        & 43.61 \\
   52 & UGC06728                  &  27.9 &   0.5 & Sy1.2      & 42.40 &   107 & Mrk1066                   &  51.7 &   0.9 & Sy2        & 42.51 \\
   53 & 2MASXJ1145-1827    & 141.1 &   2.5 & Sy1.2      & 44.08 &   108 & NGC3147                   &  39.6 &   0.7 & Sy2        & 42.20 \\
   54 & Ark347                    &  96.1 &   1.7 & Sy2        & 43.52 &   109 & ESO439-G009               & 105.8 &   1.8 & Sy2        & 43.27 \\
   55 & UGC07064                  & 107.1 &   1.9 & Sy1.9      & 43.28 &   110 & Mrk273                    & 161.8 &   2.8 & Sy2        & 43.24 \\
\hline \hline
\end{tabular}
\caption{Observational details of the AGN sample. Col. 1 gives the object name, Col. 2 the distance in Mpc, Col. 3 the spatial scale obtained with the short-low \emph{Spitzer}/IRS spectral module (note that the long-low IRS/\emph{Spitzer} module gives roughly 3 times lower resolution), Col. 4 the optical class reported by  \citet{Oh18}, and Col. 5 the 2-10 keV intrinsic X-ray luminosity reported by \citet{Oh18}. ``B. AGN" refers to beamed AGN (jet closely oriented toward the line of sight to the observer), according to the nomenclature reported by \citet{Oh18}. }
\label{tab:sample}
\end{center}
\end{table*}

We study in a series of two papers six of these radiative transfer codes with available SEDs. Each code uses a different set of dust chemical composition, global morphology, and/or internal distribution: 

\begin{enumerate}
\item {\bf Smooth toroidal model by \citet{Fritz06}} \citep[see also][]{Feltre12}. Radiative transfer code used to produce the SED of a simple torus geometry consisting in a flared disc that can be represented as two concentric spheres having the polar cones removed.

\item {\bf Clumpy toroidal model by \citet{Nenkova08B}} \citep[see also][]{Nenkova08A}. They developed a formalism that properly accounts for the concentration of dust in clumps or clouds, referred to as clumpy, to describe the nature of the AGN torus. 

\item {\bf Clumpy toroidal model by \citet{Hoenig10B}} \citep[see also][]{Hoenig06,Hoenig10A}. Radiative transfer model of 3D clumpy dust tori using optically thick dust clouds and a low torus volume filling factor. 

\item {\bf Two phase (clumpy + smooth) toroidal model by \citet{Siebenmorgen15}}. They assumed that the dust near the AGN is distributed in a torus-like geometry with the inclusion of polar dust. The dust is distributed as a clumpy medium or an homogeneous disk, or a combination of the two.

\item {\bf Two phase (clumpy + smooth) toroidal model by \citet{Stalevski16}} \citep[see also][]{Stalevski12}. They model the dust with a torus geometry and a two-phase medium, consisting in a large number of high-density clumps embedded in a smooth dusty component of low density. 

\item {\bf Clumpy disk and outflowing model by \citet{Hoenig17}}. This model consists in clumpy disk-like models \citep[following that described by][]{Hoenig10B} plus a polar, clumpy outflow. 

\end{enumerate}

\noindent For consistency with Paper I, these models are referred hereafter as [Fritz06], [Nenkova08], [Hoenig10], [Sieben15], [Stalev16], and [Hoenig17], respectively. Note that the main differences between [Nenkova08] versus [Hoenig10] and [Sieben15] versus [Stalev16] are the chemical composition of the dust and the radiative transfer equation solution. It is also worth to remark that the viewing angle is measured in all the cases from the pole to the equator of the system except for [Fritz06], which is measured in the opposite direction. See Gonz\'alez-Mart\'in et al. 2019A (hereinafter Paper I) for a complete description of these SED libraries.

The model that has been more extensively compared with the data is the clumpy model described by \citet{Nenkova08B} \citep[e.g.][]{Ramos-Almeida09,Mor12,Alonso-Herrero11,Gonzalez-Martin15,Martinez-Paredes15,Fuller16,Gonzalez-Martin17,Martinez-Paredes17,Garcia-Bernete19}. A systematic confrontation of these models against infrared SEDs is still lacking, but in those cases where it has been done, the resulting torus or cloud properties differ significantly \citep{Hoenig06,Schartmann08}. 

In paper I we produced a set of synthetic spectra from current instruments GTC/CanariCam and \emph{Spitzer}/IRS and future \emph{JWST}/MIRI and \emph{JWST}/ NIRSpec instruments using this set of six SED libraries. We found that, for a reasonable source brightness ($\rm{F_{12\mu m}>100 mJy}$), we can actually distinguish among models except for parent models. We also found that the torus parameters can be constrained within 15\% error, irrespective of the instrument used, for all but [Hoenig17]. The questions we try to answer in this second paper are: (1) are the models good enough to describe the mid-infrared SED of AGN? (2) which is the preferred model by the data? and (3) what can we say about the parameters involved and derived quantities, as the outer radius of the torus, covering factor, and total dust mass, depending on the model used?

With these aims we confronted this set of SED libraries against a set of mid-infrared \emph{Spitzer}/IRS spectra of 110 AGN selected from the \emph{Swift}/BAT survey. The paper is organized as follows. Section \ref{sec:sample} gives a brief summary of the AGN sample used along this paper. Section \ref{sec:spectralfit} describes the spectral fitting procedure. The main results on the spectral fitting are included in Section \ref{sec:results} and discussed in Section \ref{sec:discussion}. The paper is summarized in Section \ref{sec:summary}. 

\section{The sample and the data} \label{sec:sample}

We built an AGN sample with available low spectral resolution mid-infrared IRS/\emph{Spitzer} spectra to confront with models. As shown in Paper I, low spectral resolution IRS/\emph{Spitzer} spectra provide enough coverage and sensitivity to constrain most of the parameters of the models at least for sources with intermediate brightness (i.e. $\rm{F_{12\mu m}\sim}$100 mJy), although host galaxy dilution should be taken into account. 

To avoid biases against obscured objects (i.e. highly obscured AGN undetected by, e.g., optical wavelengths) we based our sample in the 105-months \emph{Swift}/BAT survey \citep{Oh18}. Out of the 1632 sources detected by the survey, 447 are identified as AGN with a redshift lower than 0.04 (i.e. distance below 170 Mpc, using $\rm{H_{0}=70 km/s/Mpc}$). We restricted the sample to the nearby Universe to ensure relatively high spatial resolution using IRS/\emph{Spitzer} spectra (see spatial scales computed for the short-low IRS/\emph{Spitzer} module\footnote{The short-low IRS/\emph{Spitzer} module gives 3.6 arcsec spatial resolution. The long-low IRS/\emph{Spitzer} module gives roughly 3 times lower resolution.} in Table\,\ref{tab:sample}, Col.\,3). We then used CASSIS\citep{Lebouteiller11} to look for the data and downloaded the optical extraction provided. We avoided HR (high resolution) \emph{Spitzer} data because they show poorer cosmetic continuum emission in the form of jumps when putting together each small wavelength module that conforms these spectra. LR (low resolution) \emph{Spitzer} spectra are better joined because they are made up only by four of these wavelength modules. Among the 447 AGN, we found observations for 110 AGN. Table \ref{tab:sample} includes the observational details of our sample. 

According to their AGN classification, the sample contains 60 type-1 Seyferts (five Sy1, 13 Sy1.2, 17 Sy1.5, one Sy1.8, and 24 Sy1.9), 41 type-2 Seyferts, and eight beamed AGN. Beamed AGN are those with a jet pointing close to the line of sight to the observer. They cover a wide range of distances (4-168 Mpc) and therefore, a spatial scale for the unresolved IRS/\emph{Spitzer} spectra of $\rm{\sim}$70 pc up to $\rm{\sim}$3 kpc. This spatial resolution reinforces the need for the spectral decomposition performed in this paper to decontaminate AGN dust from circumnuclear contributors. We also excluded all the observations extracted as extended within CASSIS because they are clearly dominated by off-nuclear processes (that have a major impact on the determination of the torus parameters, see Paper I). Our sample expands almost four orders of magnitude in X-ray luminosities ($\rm{log(L_{X}) = 40.9-44.5}$), with $\rm{12\,\mu m}$ fluxes from $\rm{\sim}$10 mJy up to $\rm{\sim}$30 Jy. Note that this selection includes intermediate luminosity (ILAGN, $\rm{log(L_{bol}=42-45}$) and low-luminosity AGN (LLAGN, $\rm{log(L_{bol}<42}$). However, it excludes high luminosity AGN (HLAGN or QSOs, $\rm{log(L_{bol}>45}$) due to the cut in distance. The AGN with strong silicate emission features, including a set of QSOs, will be the subject of a subsequent investigation (Martinez-Paredes in prep.). 

We converted the spectra into XSPEC \citep{Arnaud96} format using {\sc flx2xsp} task within HEASOFT. These files are easily read by XSPEC to perform statistical tests when fitting to models\footnote{Note that models were also converted into XSPEC format, see Paper I.}.

\begin{figure*}[!ht]
\begin{flushleft}
\includegraphics[width=0.76\columnwidth]{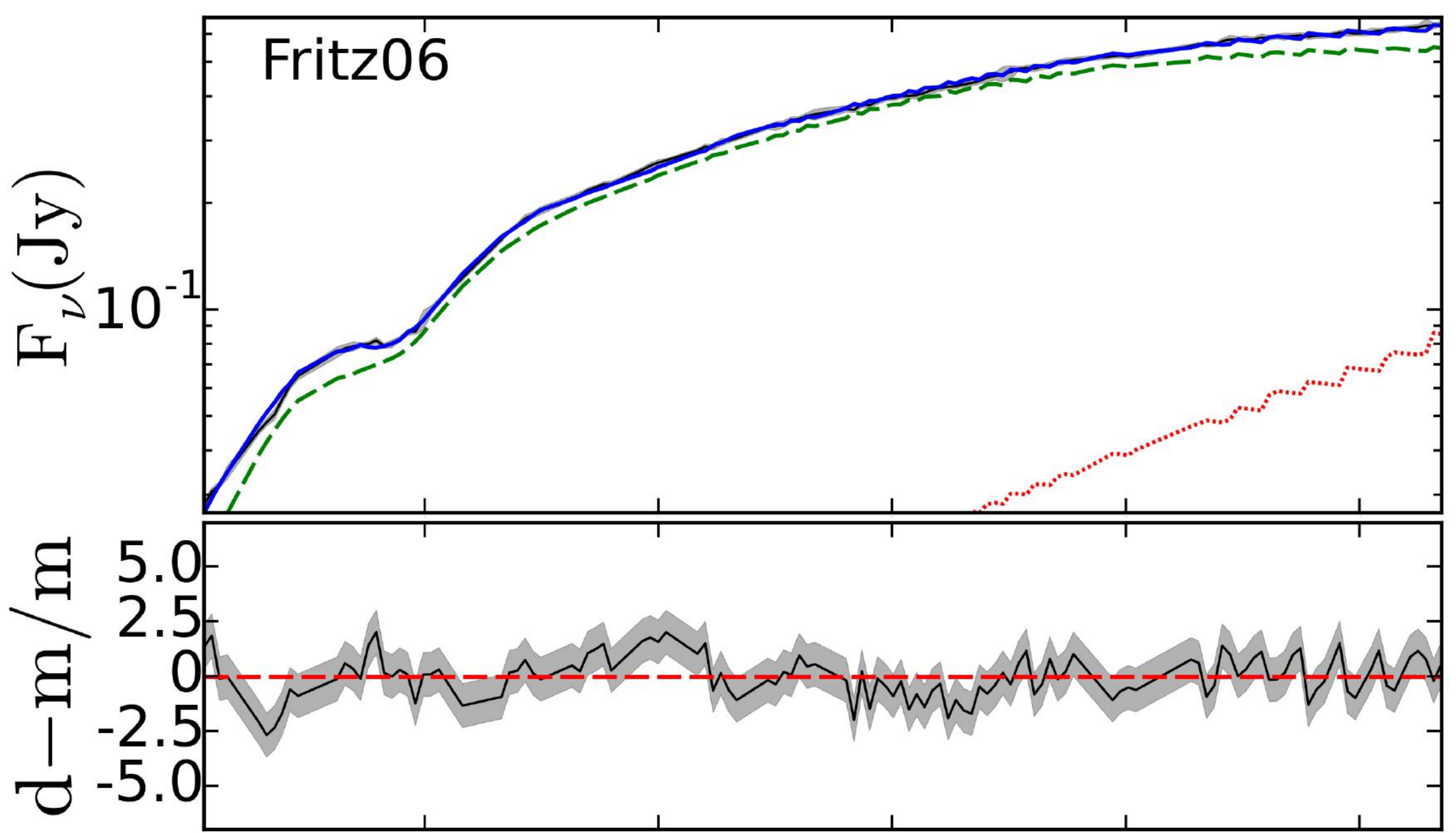}
\includegraphics[width=0.66\columnwidth]{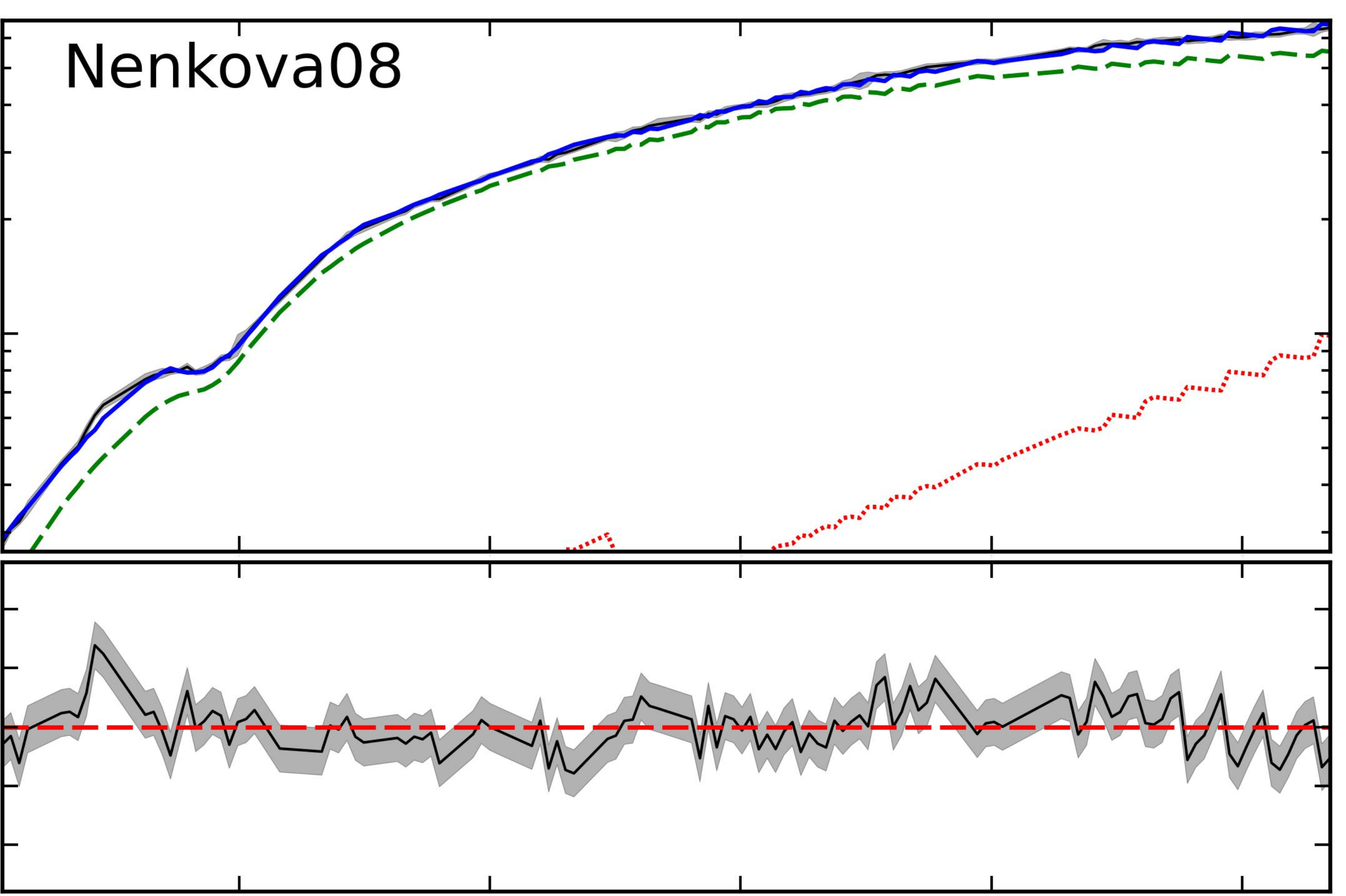}
\includegraphics[width=0.66\columnwidth]{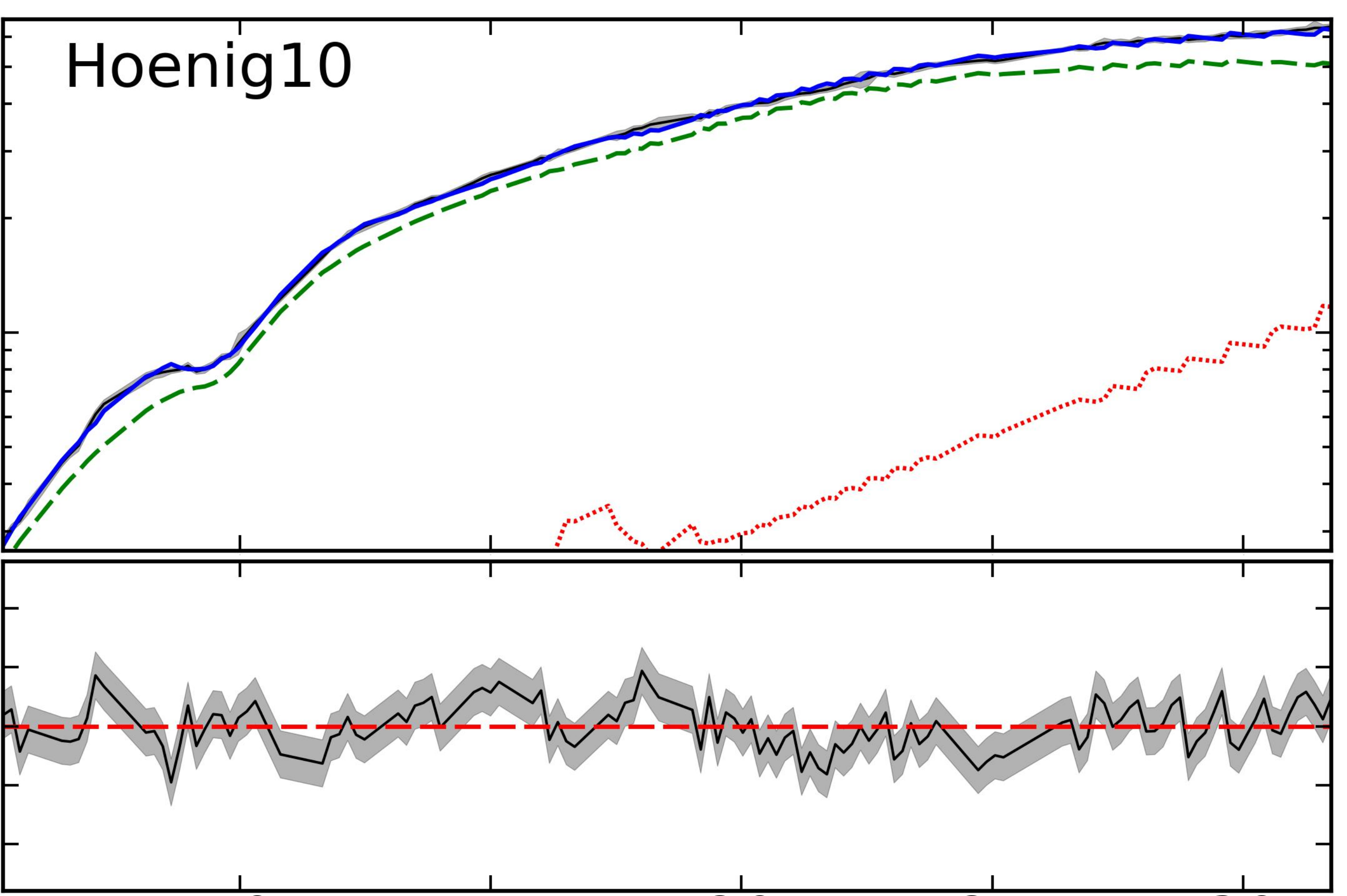}
\includegraphics[width=0.76\columnwidth]{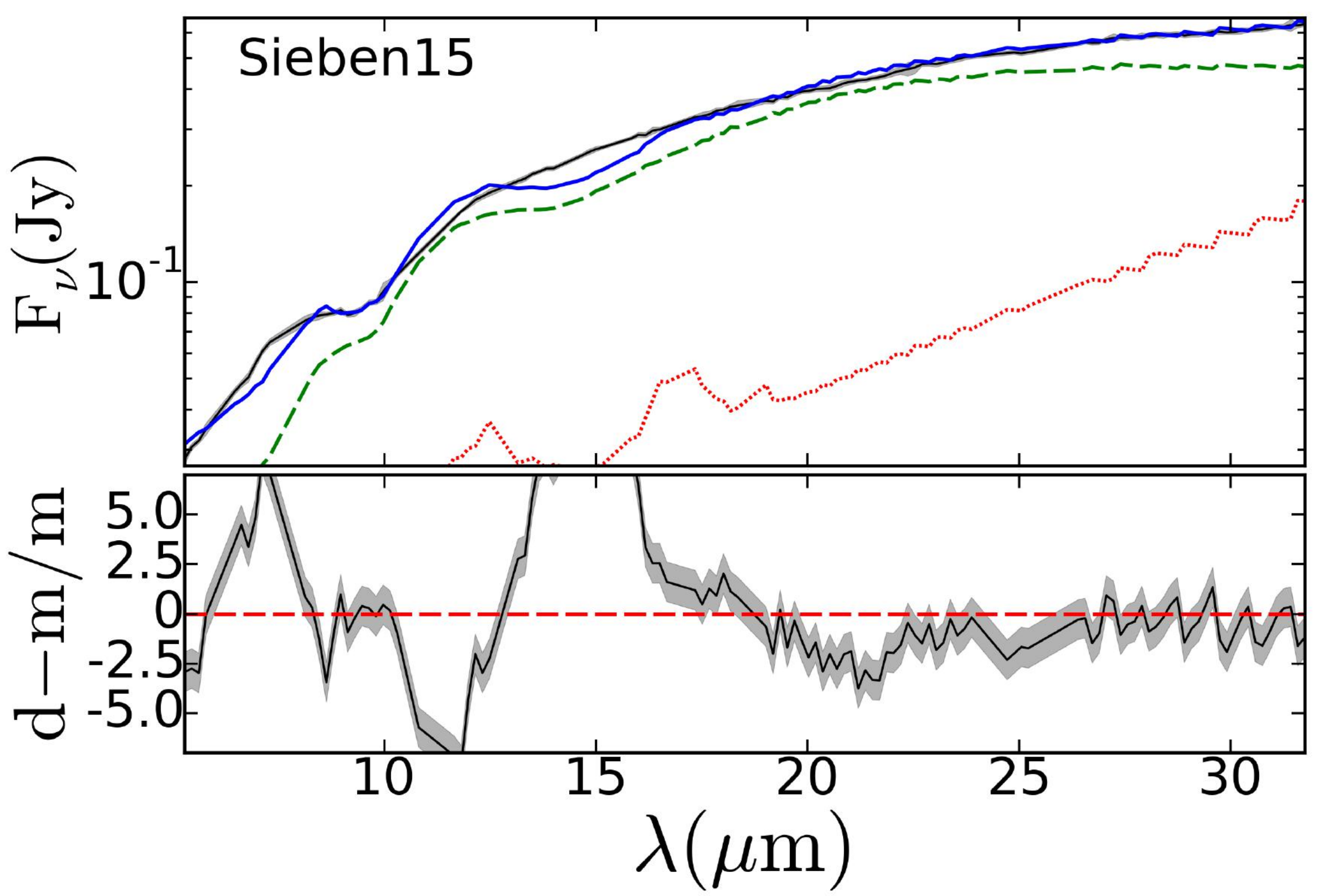}
\includegraphics[width=0.66\columnwidth]{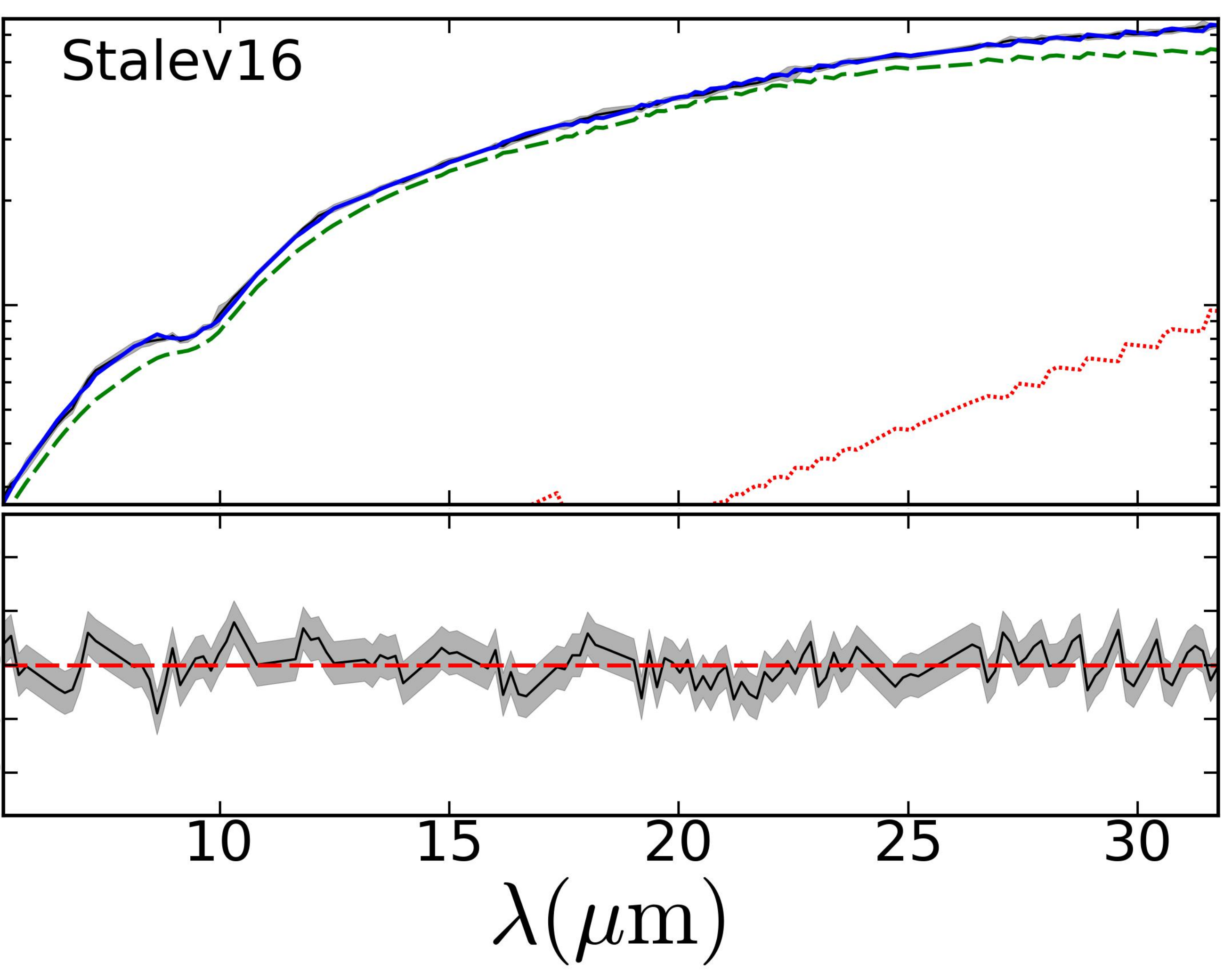}
\includegraphics[width=0.66\columnwidth]{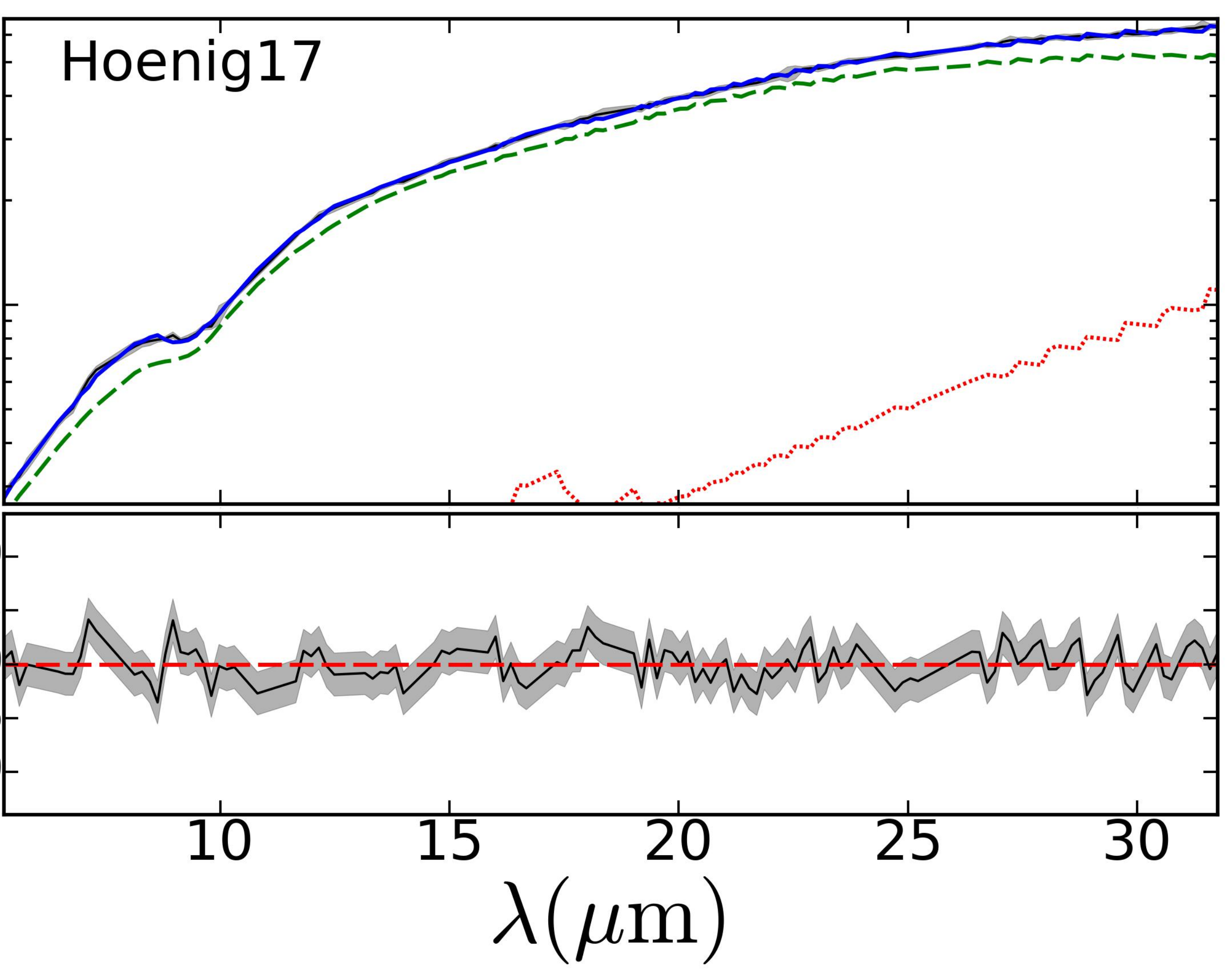}
\end{flushleft}
\begin{center}
\caption{Best fit (top panel) and residuals (bottom panel) per parameter resulting when fitting the type-2 Seyfert galaxy IC\,5063 with each model. Green dashed, red dotted, and blue continuous lines show AGN dust, the ISM components, and the sum of all the components, respectively. Note that, although not seen in the plot due to its low contribution, stellar component is included to explain wavelengths below $\rm{<7\mu m}$ for [Fritz06], [Nenkova08], and [Sieben15] (see Table\,\ref{tab:samplefit}).}
\label{fig:dataspecfit}
\end{center}
\end{figure*}

\begin{figure*}[!ht]
\begin{flushleft}
\includegraphics[width=0.75\columnwidth]{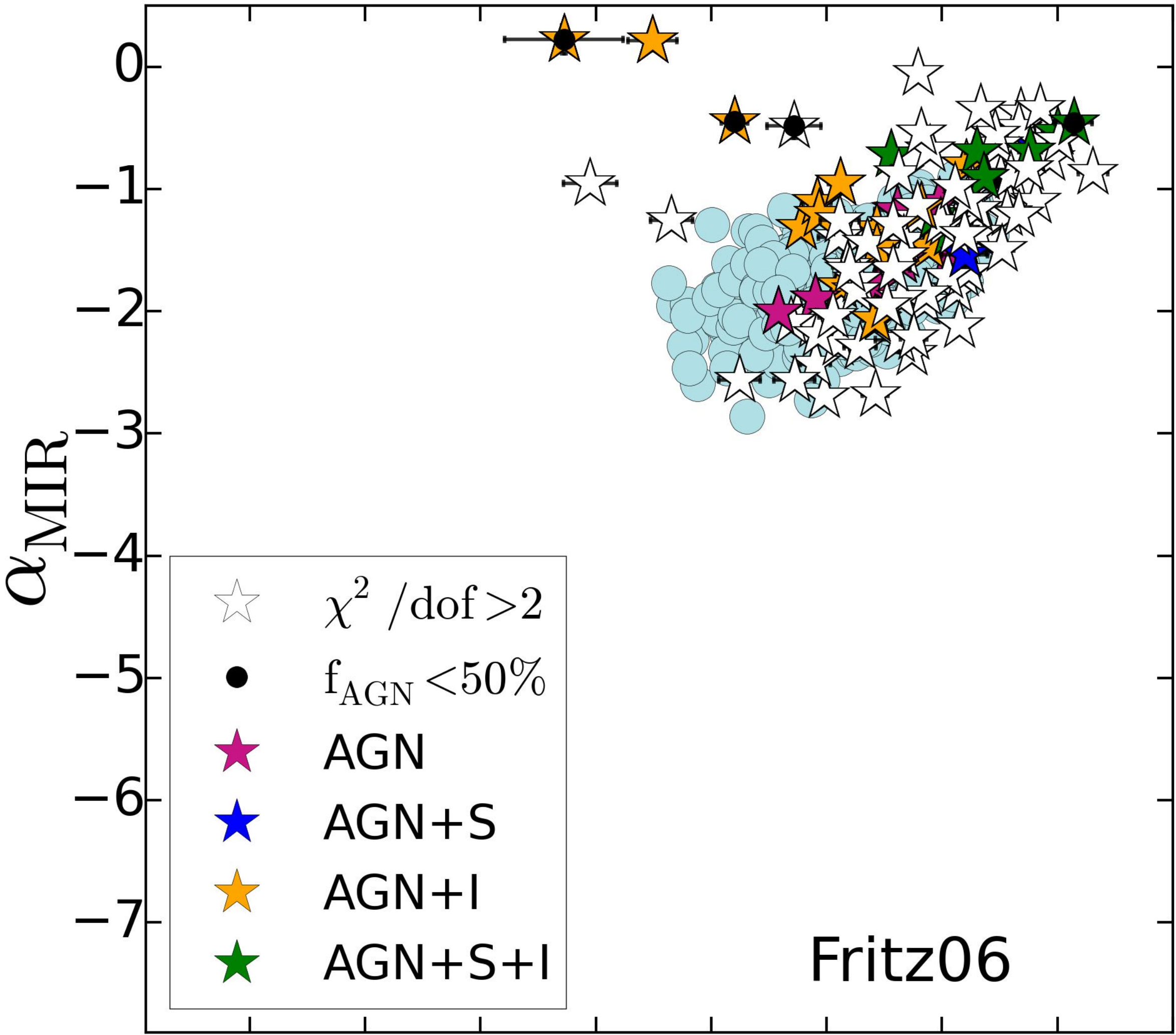}
\includegraphics[width=0.66\columnwidth]{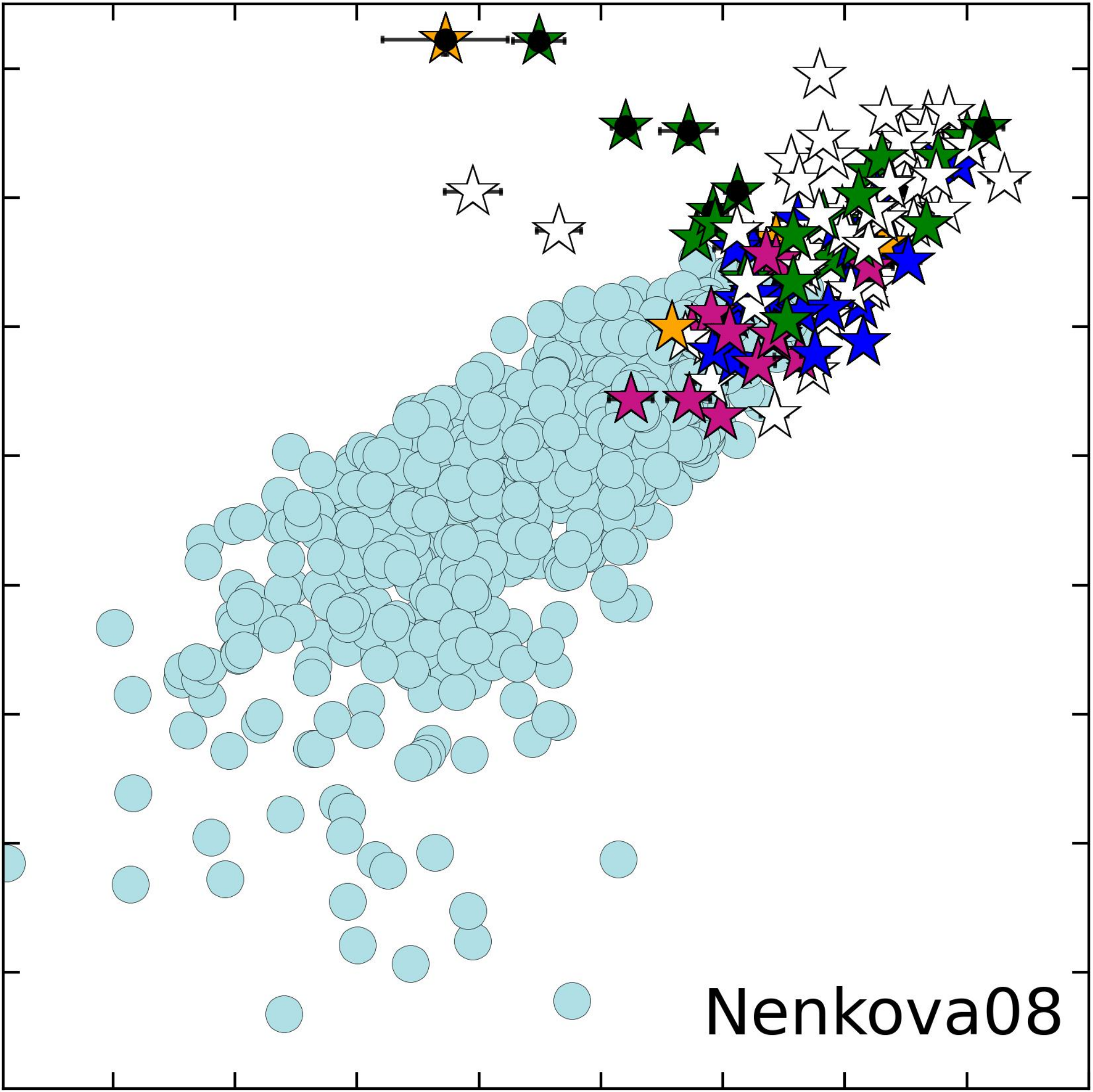}
\includegraphics[width=0.66\columnwidth]{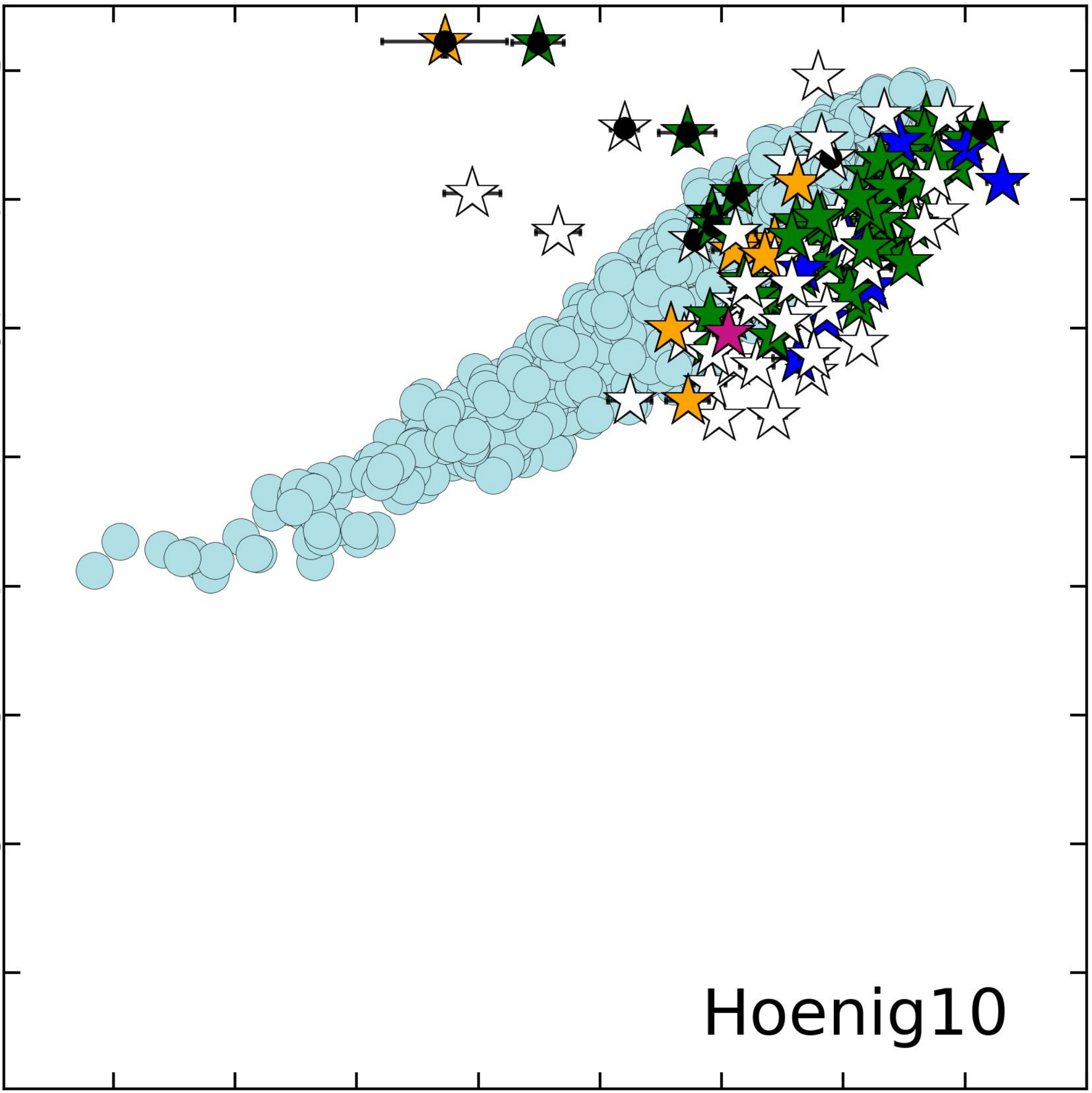}
\includegraphics[width=0.75\columnwidth]{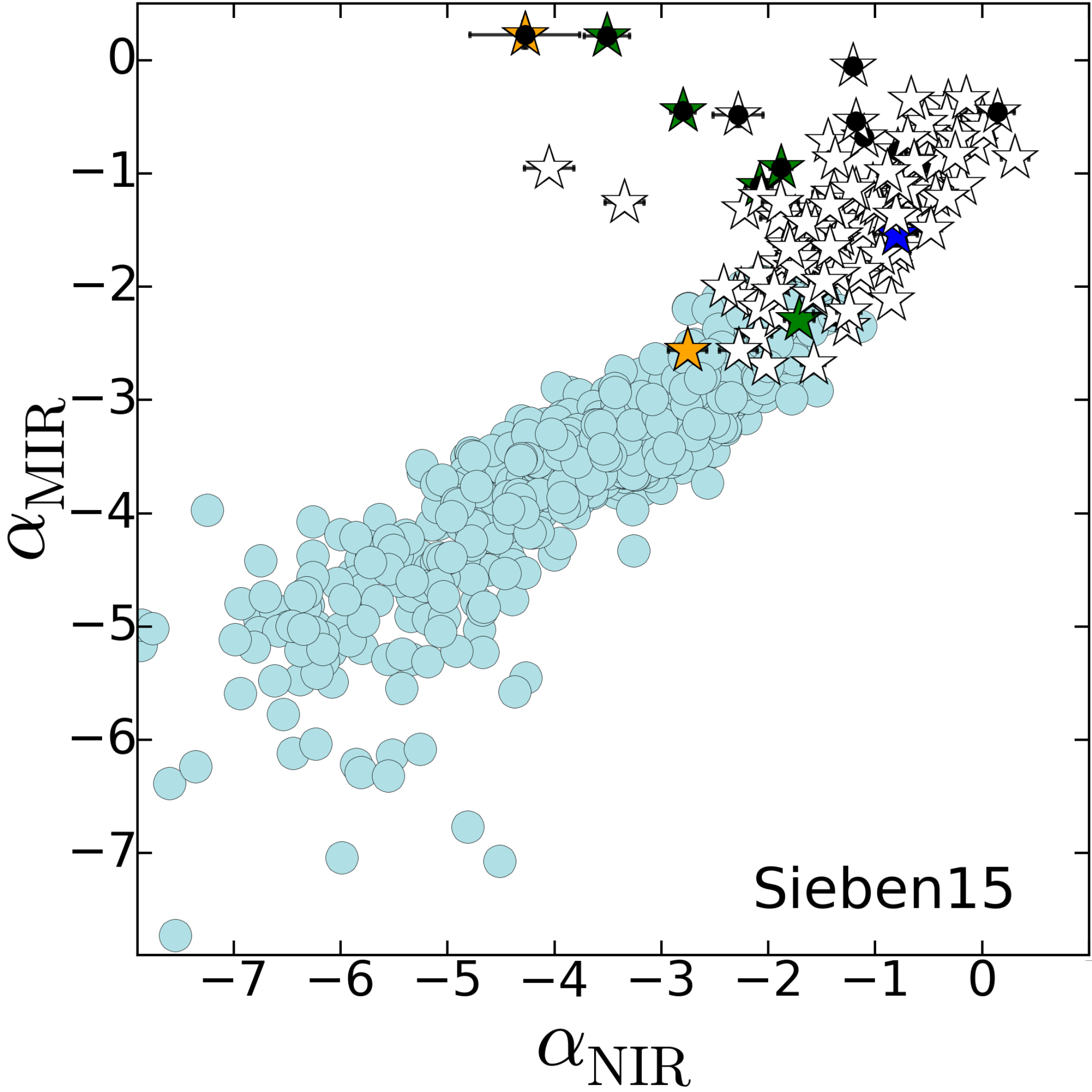}
\includegraphics[width=0.66\columnwidth]{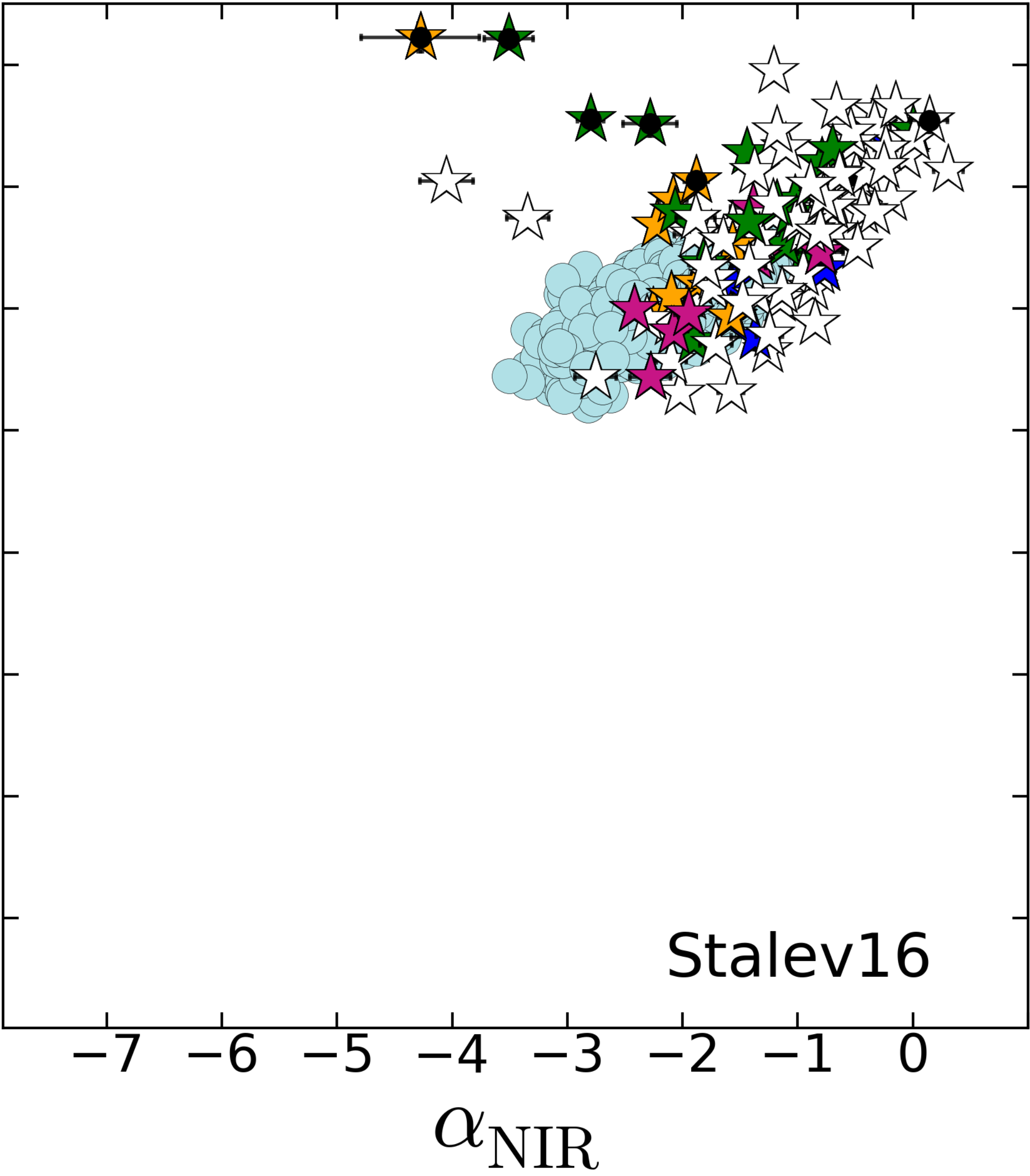}
\includegraphics[width=0.66\columnwidth]{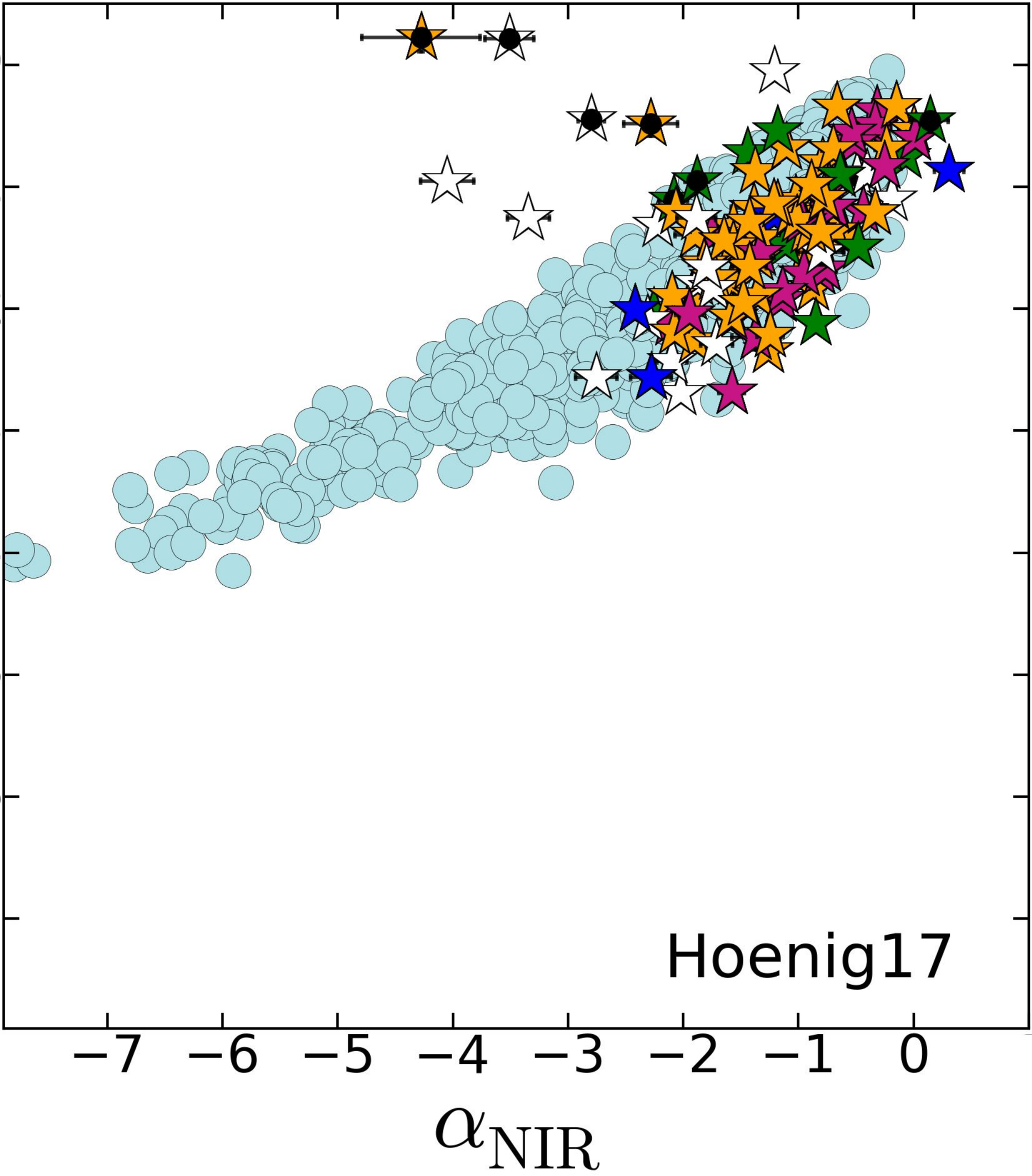}
\end{flushleft}
\begin{center}
\caption{Spectral slope computed as the flux ratio between 14 and 7.5$\rm{\mu m}$ ($\alpha_{MIR}$) versus the spectral slope computed as the flux ratio between the 7.5 and 5.5$\rm{\mu m}$ ($\alpha_{NIR}$). Synthetic spectral results (Paper I) are shown with cyan circles and objects with stars. Objects where we were able to fit the spectrum (i.e. $\rm{\chi^2/dof < 2}$) to pure AGN dust model, AGN+stellar, AGN+ISM, and AGN+stellar+ISM are shown with purple, blue, orange, and green stars. Objects not well fitted to any of the combinations are shown with white stars. Objects with low contribution of AGN dust are marked with a black dot. }
\label{fig:genfit1}
\end{center}
\end{figure*}

\begin{figure*}[!ht]
\begin{flushleft}
\includegraphics[width=0.75\columnwidth]{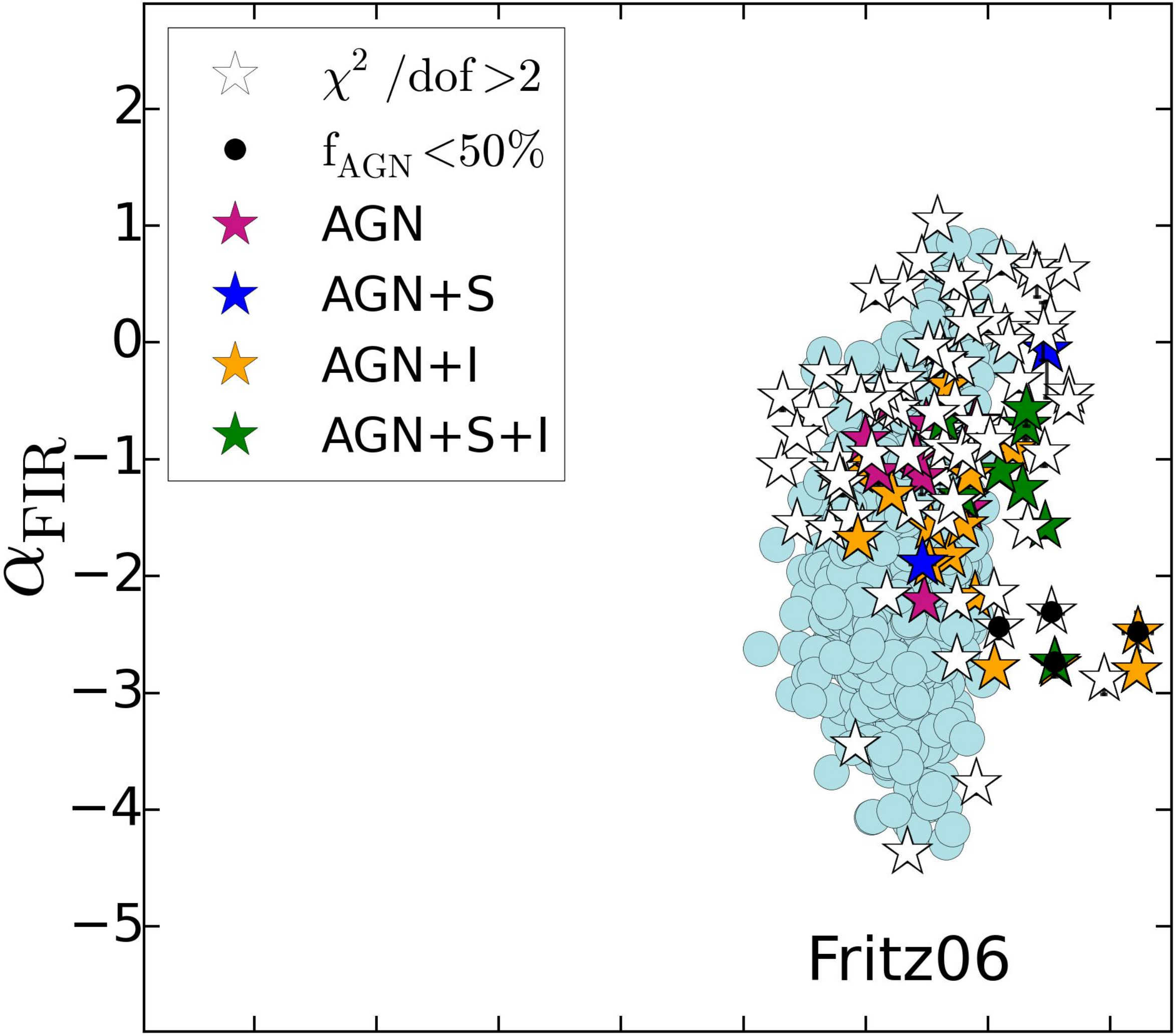}
\includegraphics[width=0.66\columnwidth]{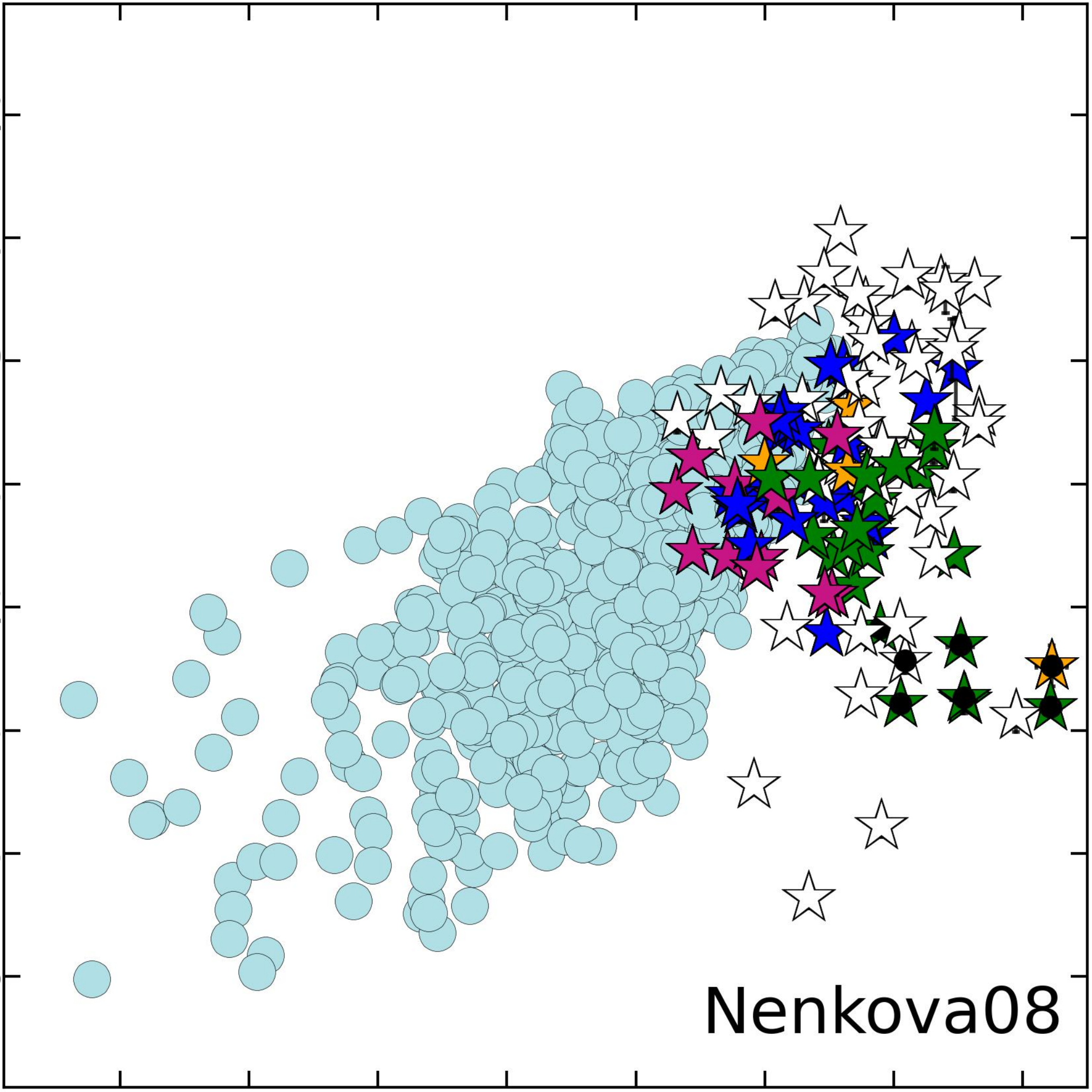}
\includegraphics[width=0.66\columnwidth]{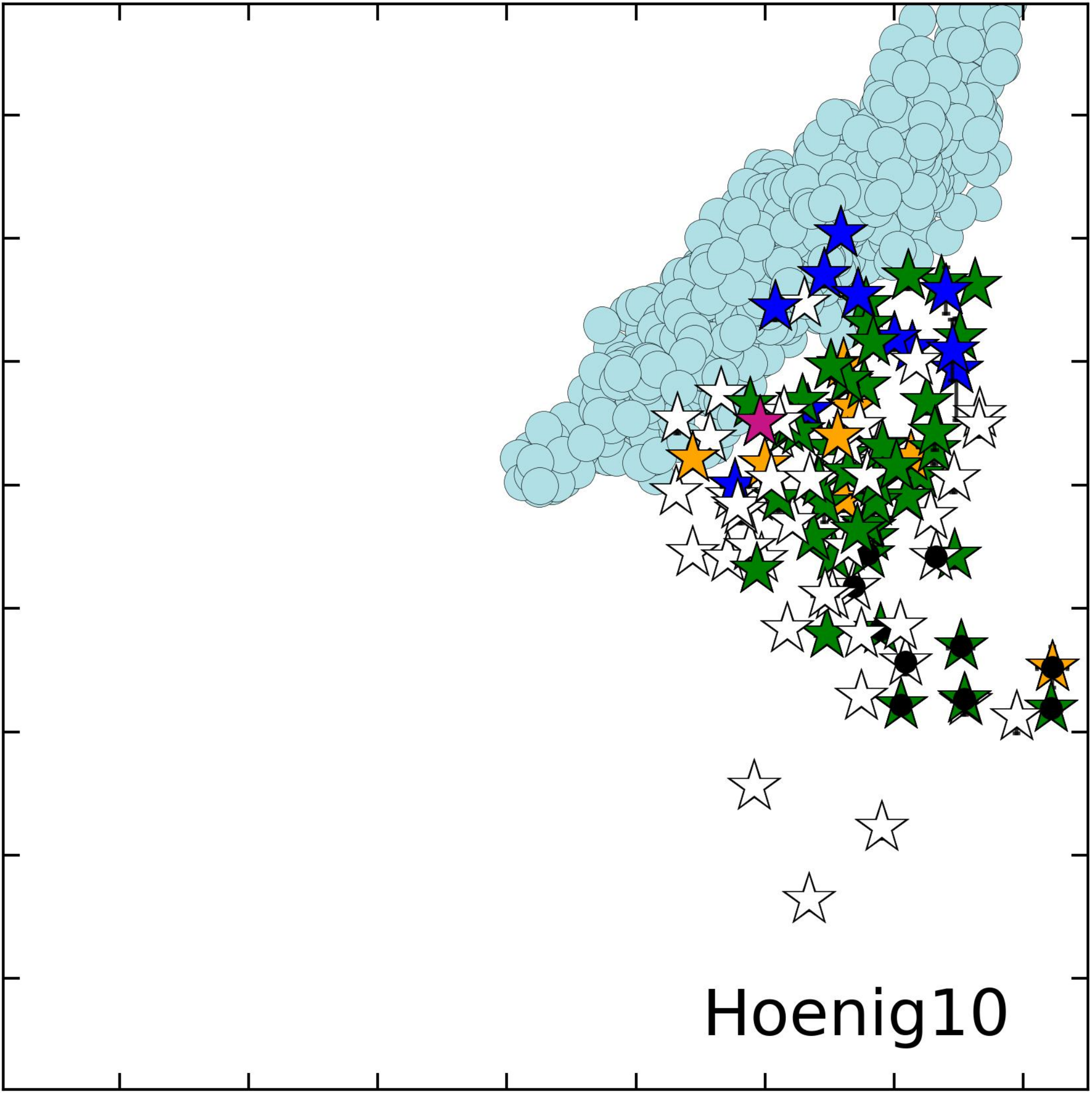}
\includegraphics[width=0.75\columnwidth]{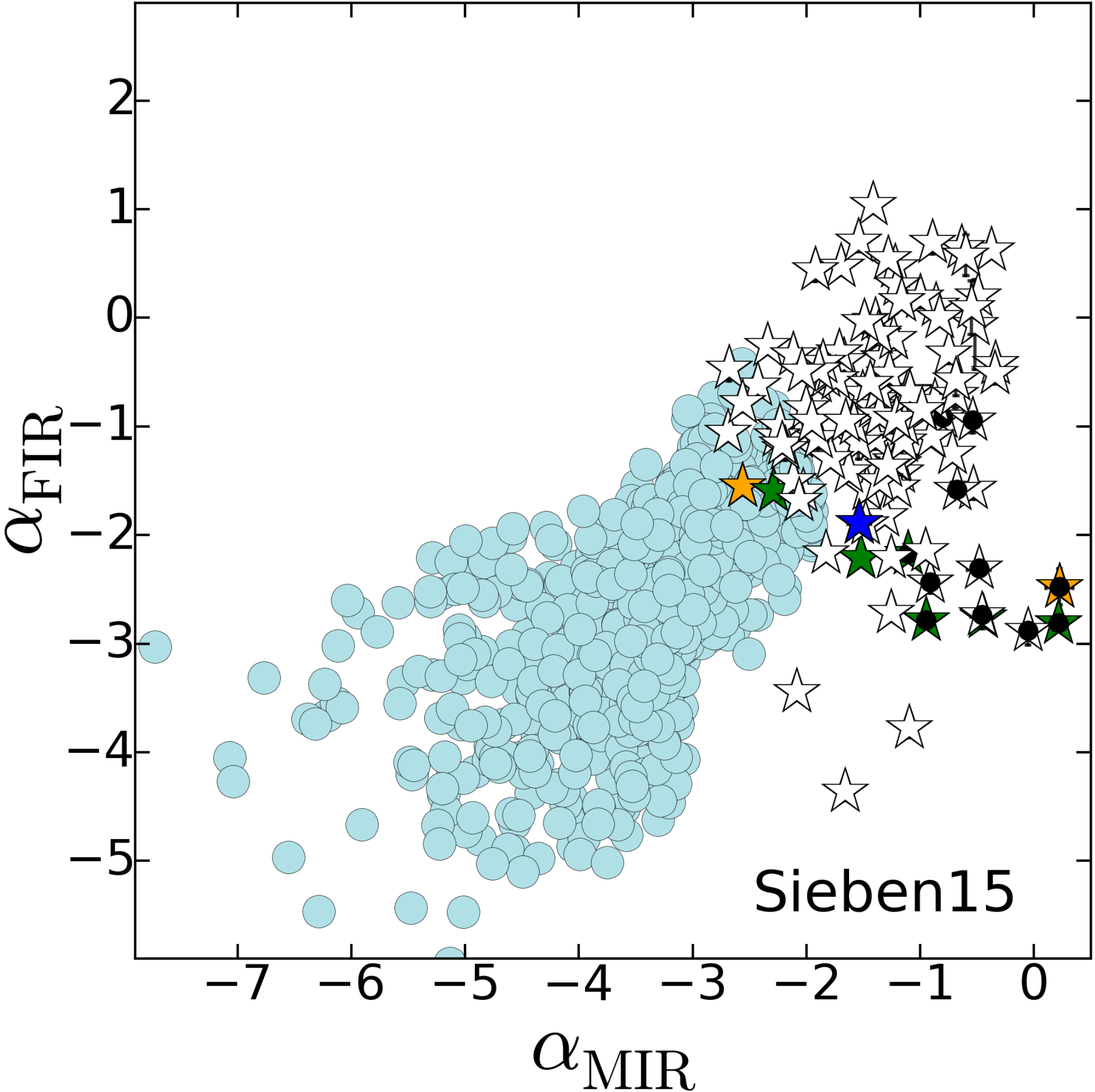}
\includegraphics[width=0.66\columnwidth]{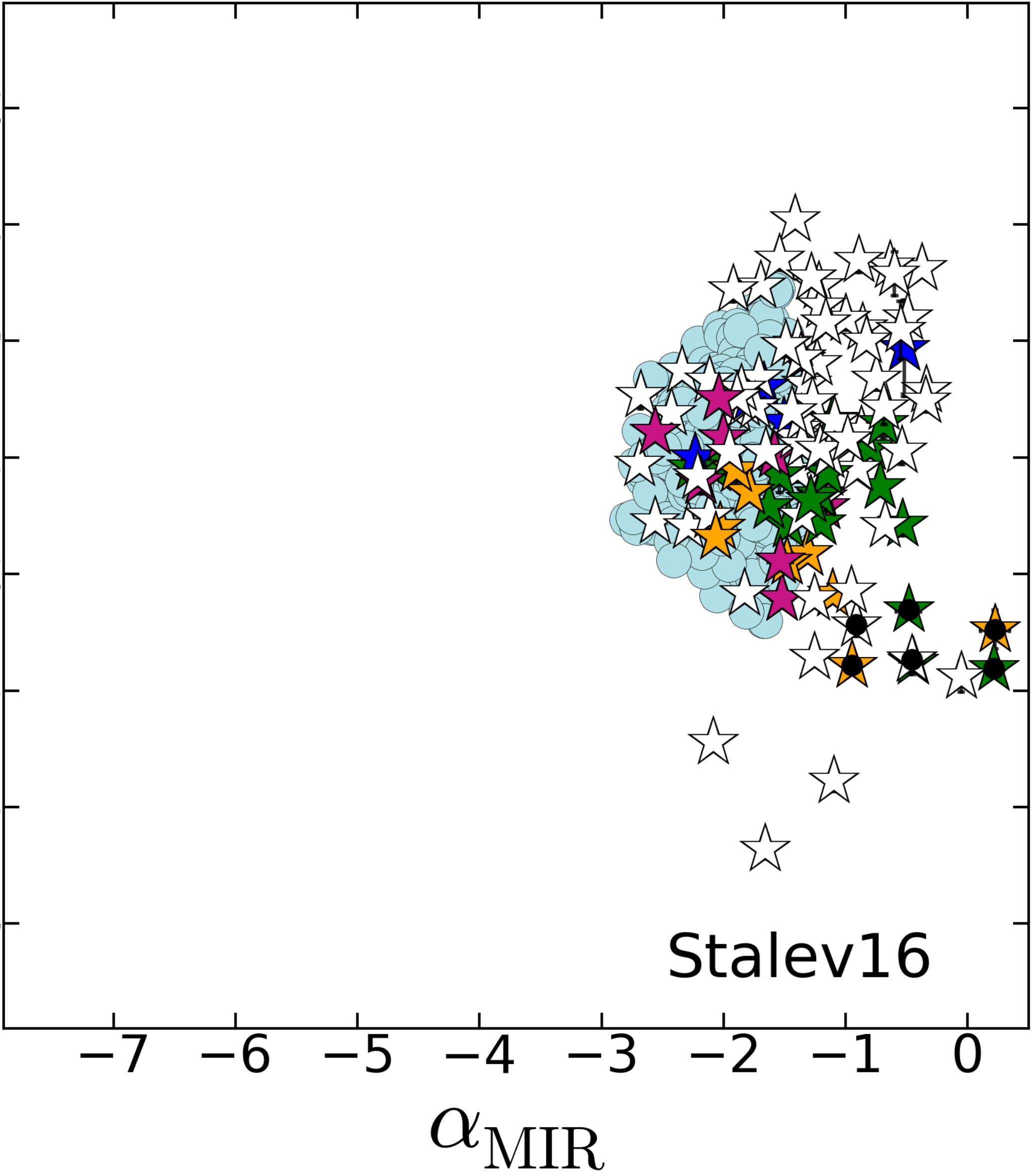}
\includegraphics[width=0.66\columnwidth]{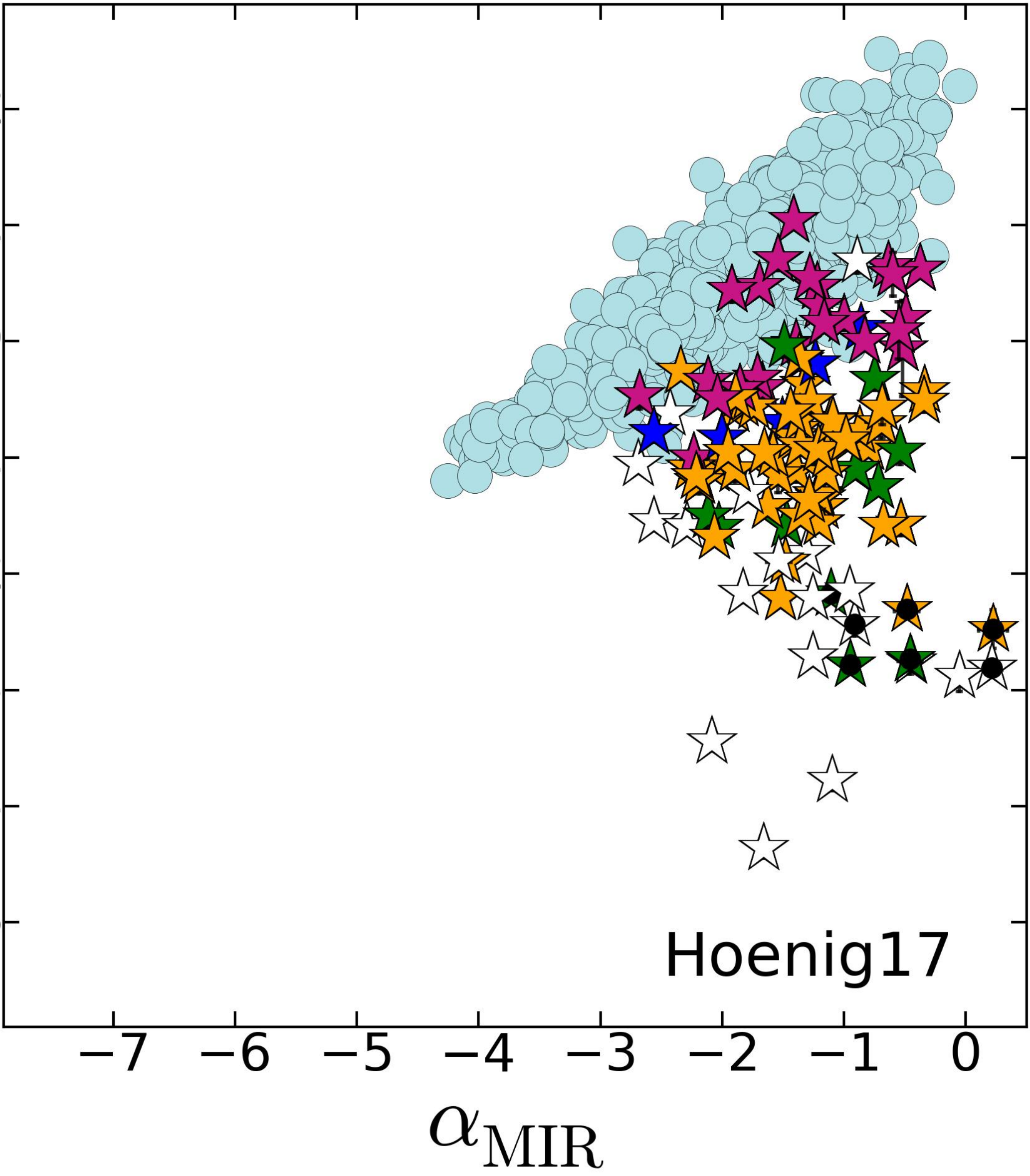}
\end{flushleft}
\begin{center}
\caption{Spectral slope computed as the flux ratio between 30 and 25$\rm{\mu m}$ ($\alpha_{FIR}$) versus the spectral slope computed as the flux ratio between the 14 and 7.5$\rm{\mu m}$ ($\alpha_{MIR}$). Symbols as in Fig.\,\ref{fig:genfit1}.}
\label{fig:genfit2}
\end{center}
\end{figure*}

\begin{figure*}[!ht]
\begin{flushleft}
\includegraphics[width=0.78\columnwidth]{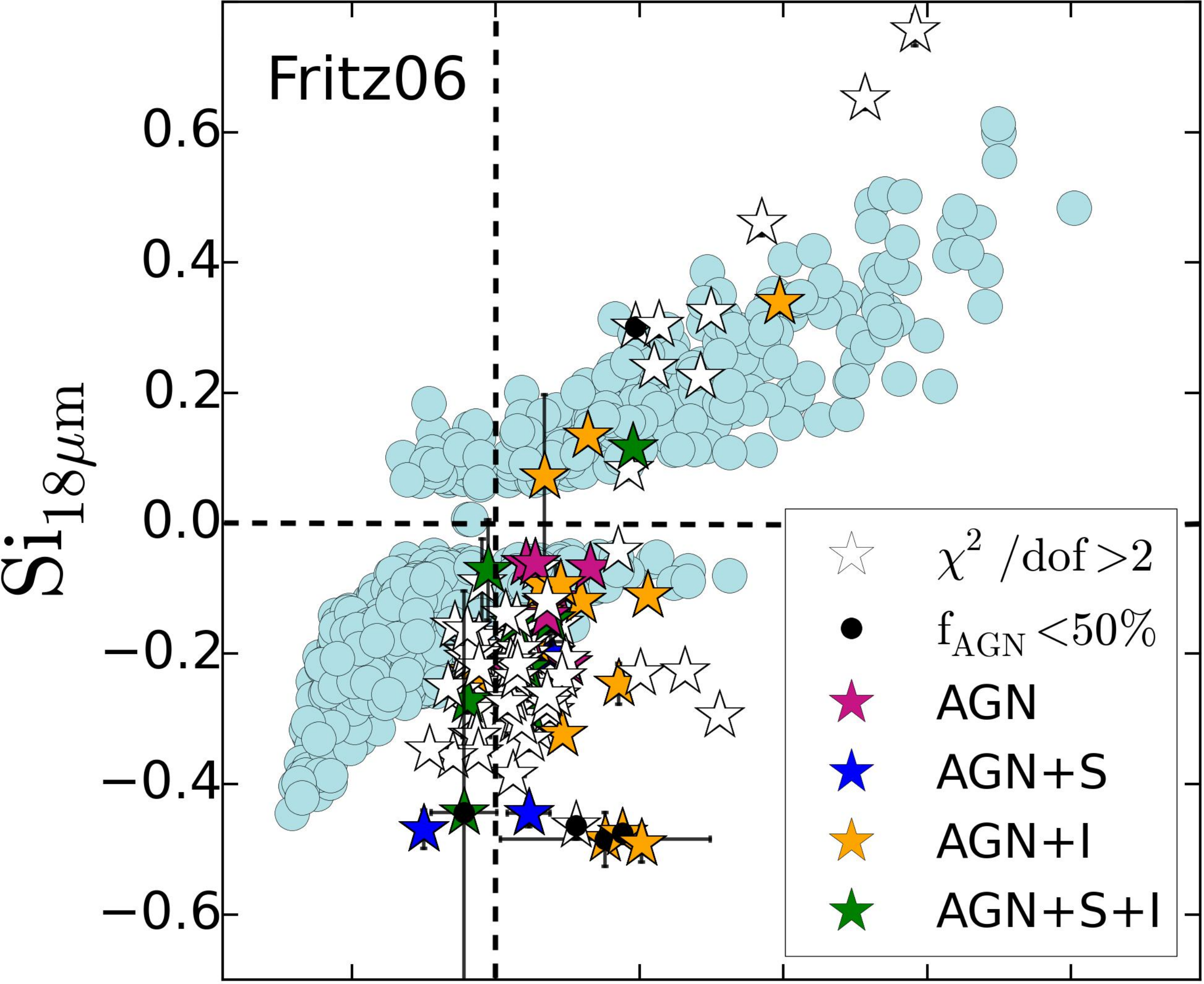}
\includegraphics[width=0.64\columnwidth]{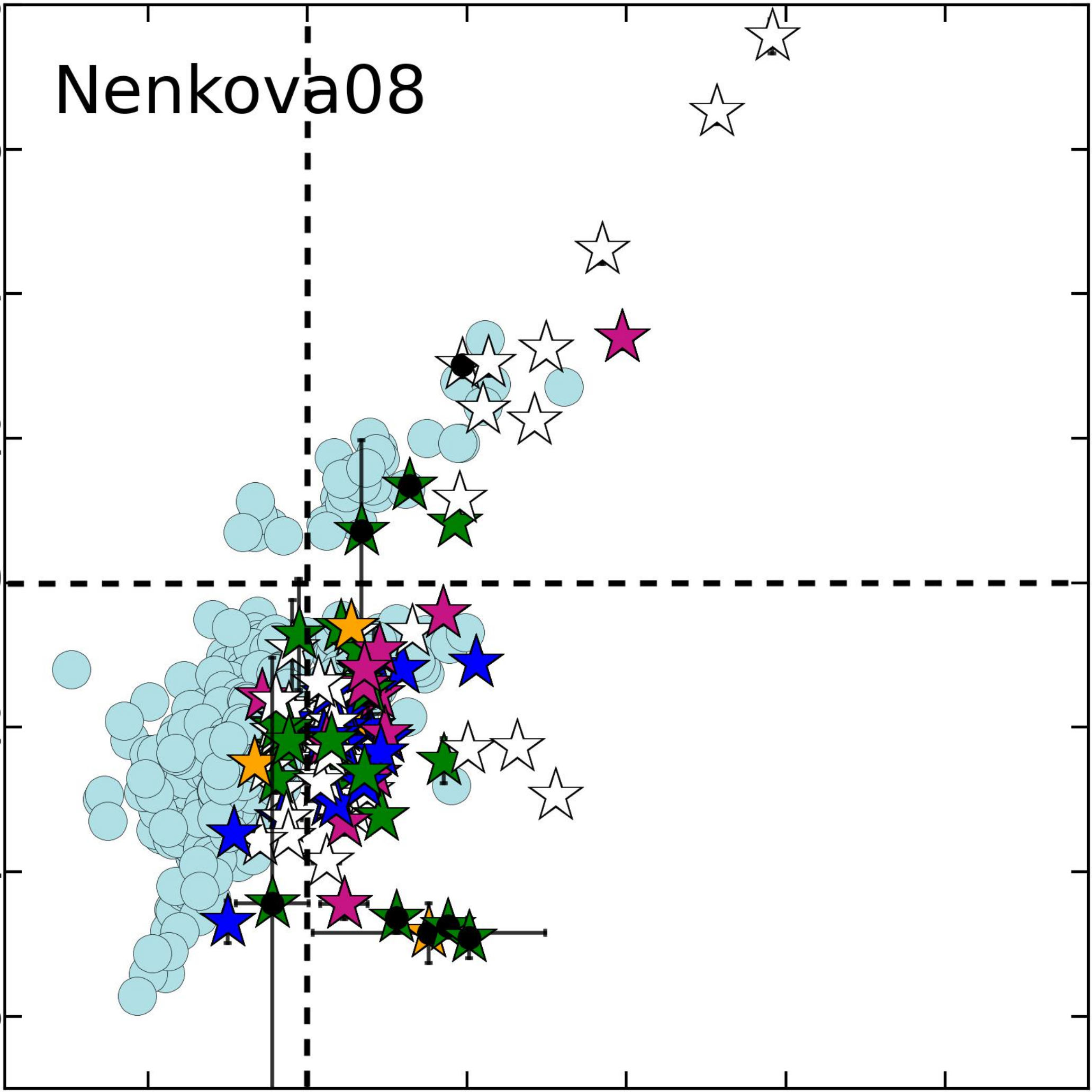}
\includegraphics[width=0.64\columnwidth]{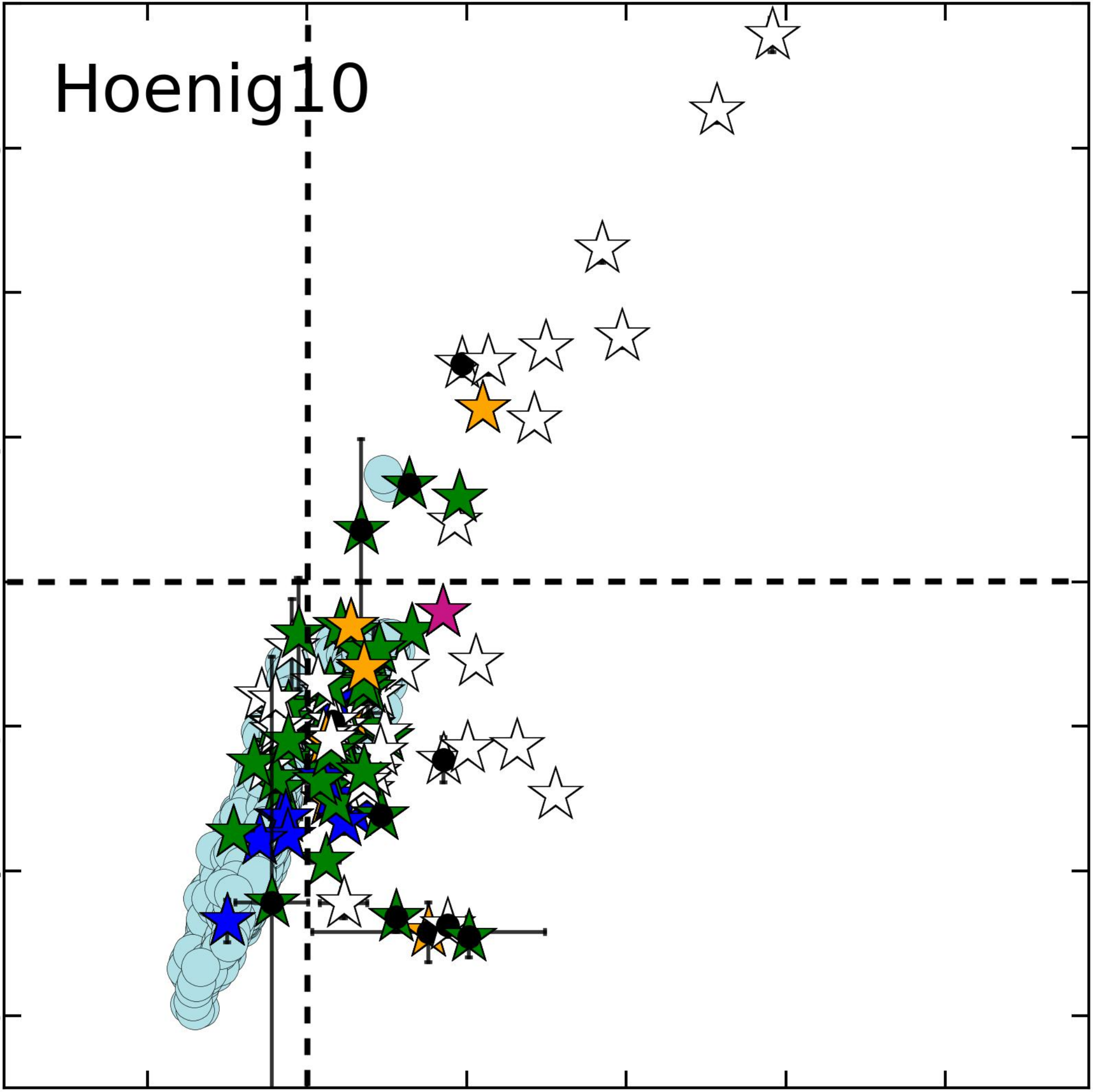}
\includegraphics[width=0.78\columnwidth]{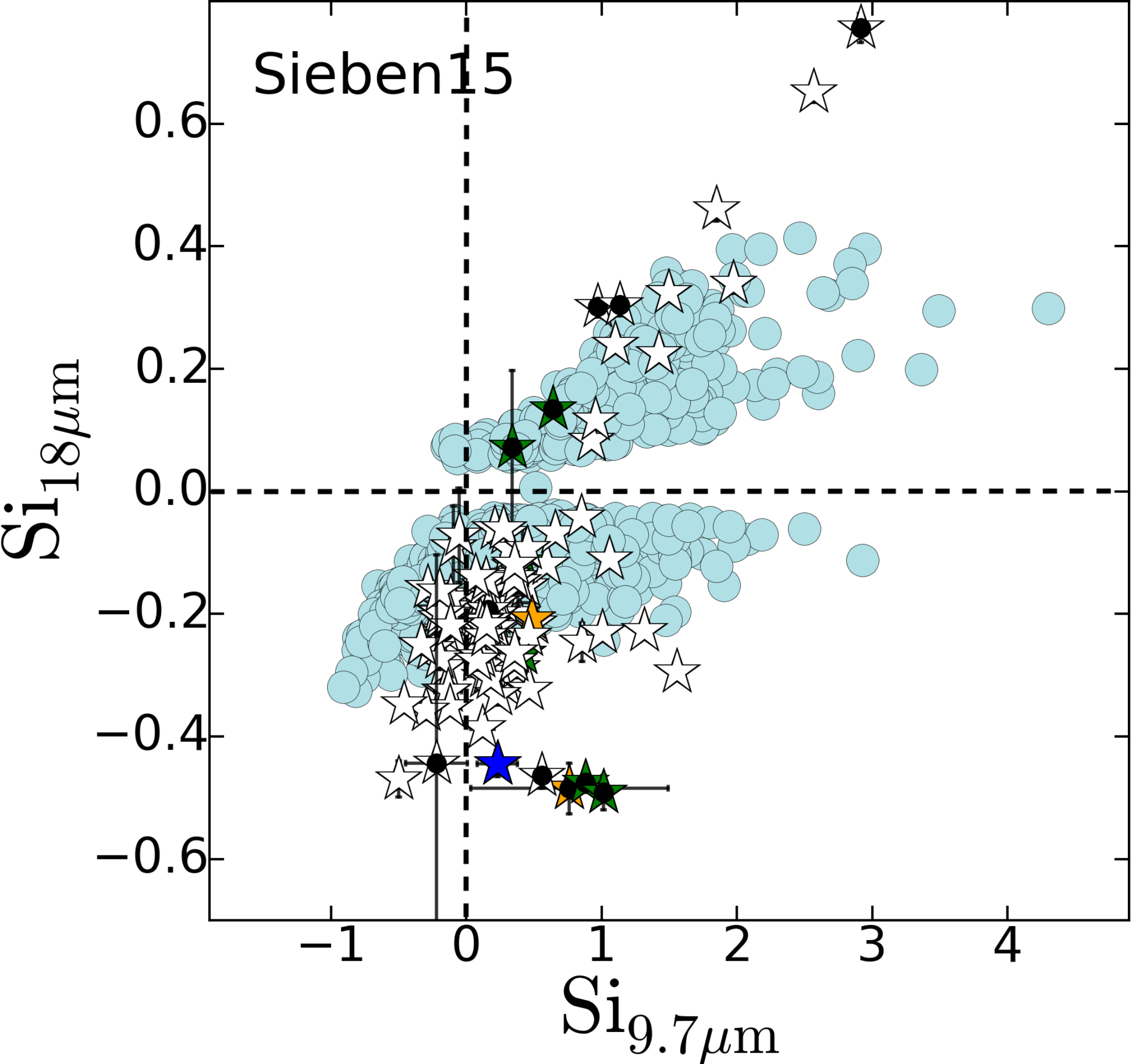}
\includegraphics[width=0.64\columnwidth]{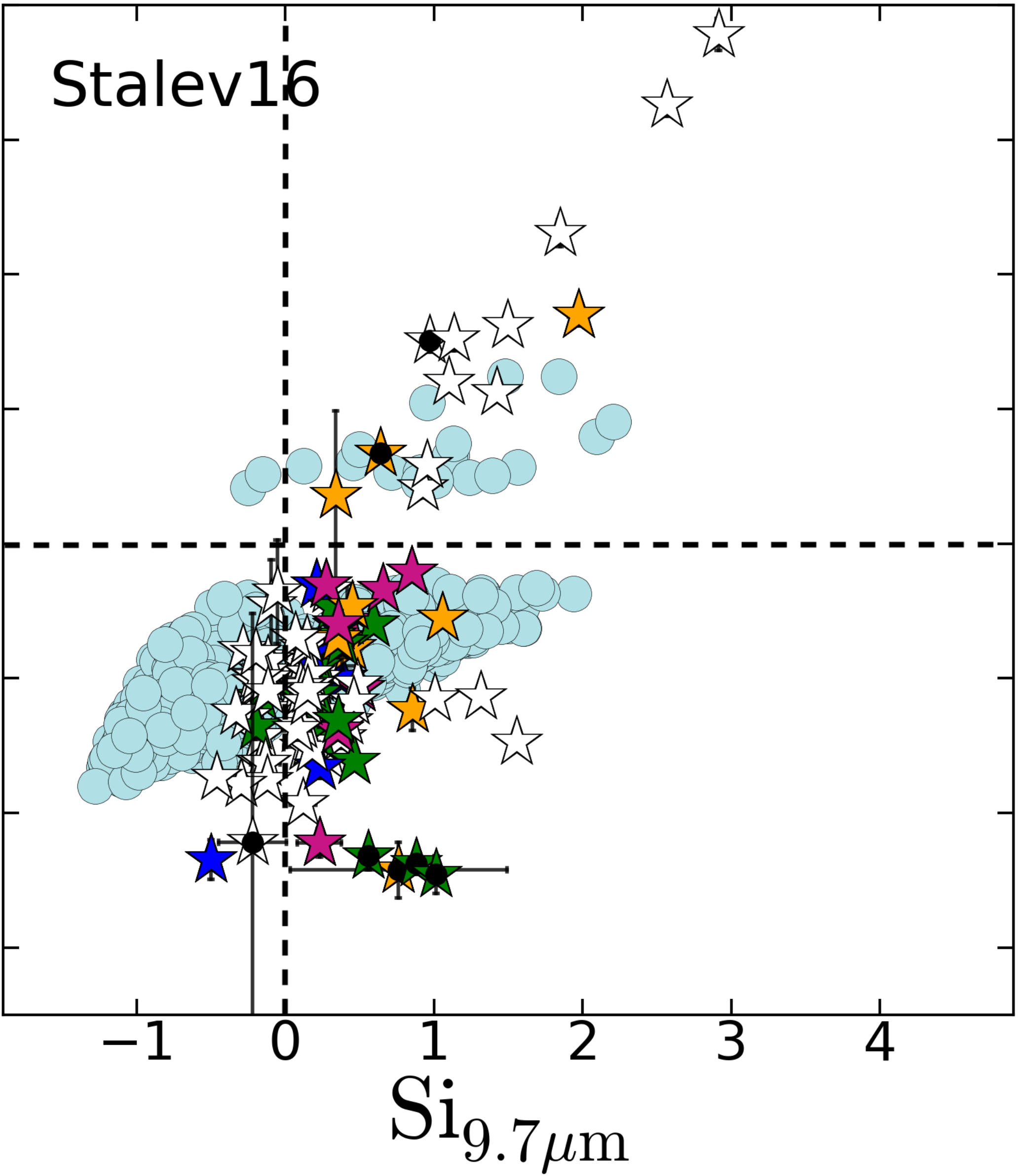}
\includegraphics[width=0.64\columnwidth]{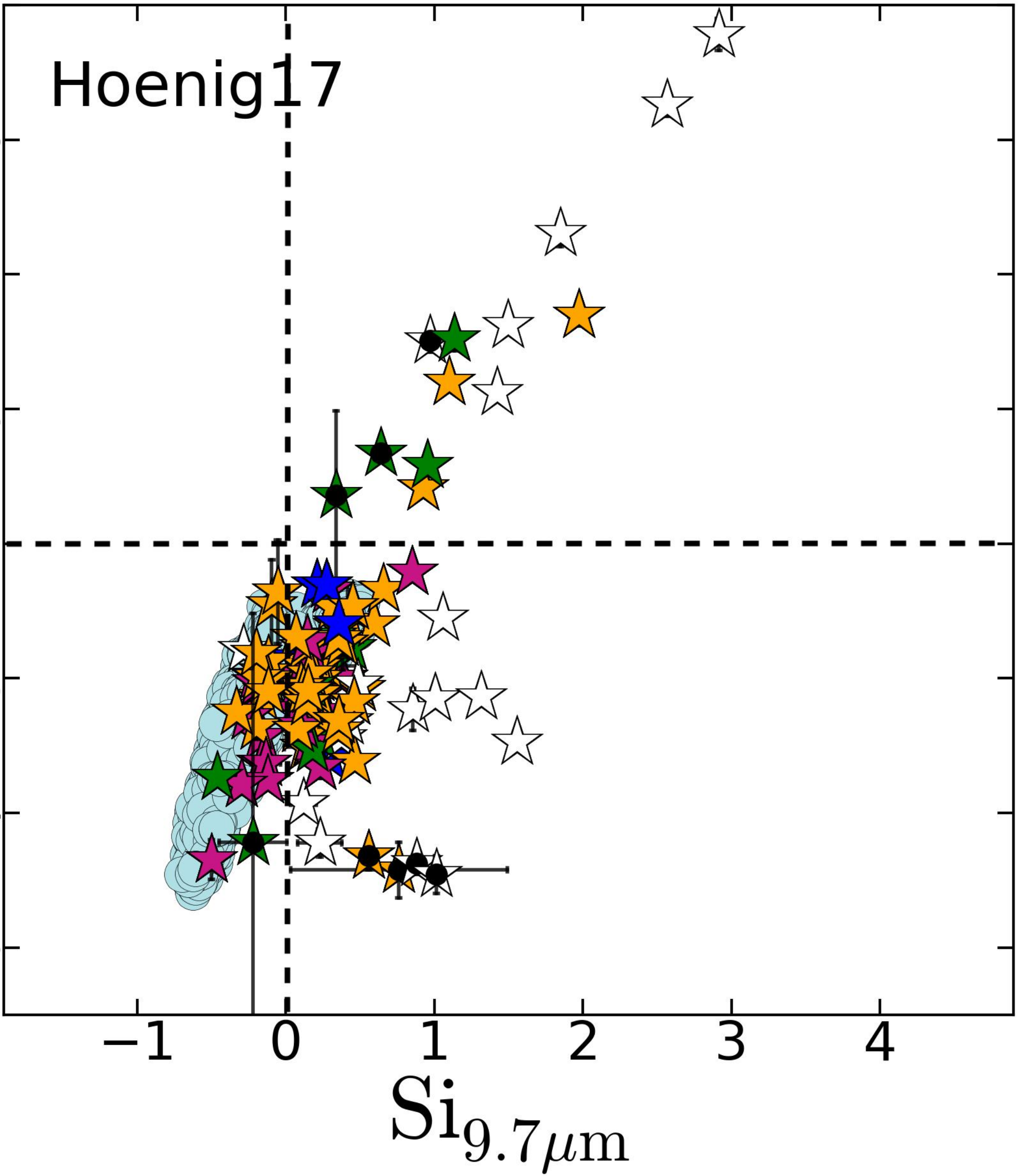}
\end{flushleft}
\begin{center}
\caption{18\,$\rm{\mu m}$ Silicate feature strength versus 9.7\,$\rm{\mu m}$ Silicate feature strength. Symbols as in Fig.\,\ref{fig:genfit1}.}
\label{fig:genfit3}
\end{center}
\end{figure*}

\section{Spectral fitting procedure} \label{sec:spectralfit}

\subsection{Model independent fit}\label{sec:nomodelprocedure} 

In Paper I we compare the spectral shape of the different models by quantifying some features in our \emph{Spitzer}/IRS synthetic spectra. These features are:

\begin{itemize} 
\item Spectral slopes: We computed three spectral slopes of the form $\rm{\alpha = - log(F_{\nu}(\lambda_{2})/ F_{\nu}(\lambda_{1}))}$/ $\rm{log(\lambda_{2}/\lambda_{1})}$, with $\rm{\lambda_2 > \lambda_1}$. Note that, under this nomenclature, negative (positive) values mean that the flux increases (decreases) with wavelength. We called $\rm{\alpha_{NIR}}$, $\rm{\alpha_{MIR}}$, and $\rm{\alpha_{FIR}}$ to the slopes evaluated at [$\rm{\lambda_1,\lambda_2}$] equal to [5.5,7.5], [7.5,14], and [25,30]\,$\rm{\mu m}$, respectively\footnote{Note that we called NIR, MIR and FIR slopes to those closer to the near-infrared, in the middle of the mid-infrared, and close to the far-infrared, respectively, although the three are in the mid-infrared wavelength range.}. These wavelengths were motivated from the residual analyses of the synthetic spectra and to compare with previous analysis \citep{Hernan-Caballero15,Garcia-Gonzalez17,Hoenig17}.

\item Silicate features strength: We also computed the silicate feature strength using the formula $\rm{Si_{\lambda}=- ln(F_{\nu}(\lambda)/F_{\nu}(continuum))}$, for the two silicate features located at $\rm{\sim 9.7\mu m}$ and $\rm{\sim 18\mu m}$. Silicate features in emission (absorption) show negative (positive) $\rm{Si_{\lambda}}$. 
\end{itemize}

We computed the same quantities for the 110 objects analyzed in this paper. Errors are computed as one standard deviation from 100 realizations with random variations of the spectral bins fluxes within the flux error. The resulting values for the 9.7$\rm{\mu m}$ silicate feature and $\rm{\alpha_{MIR}}$ are fully consistent with those derived by \citet{Garcia-Gonzalez17}.

\subsection{Dusty models}\label{sec:modelprocedure}

We fitted the \emph{Spitzer}/IRS spectra of our sample of 110 AGN using the six models discussed throughout this paper. We excluded the narrow wavelength ranges where the brightest emission lines are expected in order to isolate the continuum emission. We also added foreground extinction by dust grains to the dusty models using the {\sc zdust} component \citep{Pei92}, already included as a multiplicative component within XSPEC. This component is suitable to describe foreground extinction in the infrared, optical, and UV wavebands. We used the Milky Way extinction curve and let free to vary the color excess $\rm{E_{(B-V)}}$. We remark that extinction does not have a strong impact on the results; i.e. without foreground extinction we found identical results.

We added the stellar and/or ISM components to the best fitting AGN dust model for each source to account for circumnuclear components. The ISM component is taken from \citet{Smith07}, which are averaged Starburst templates in the $\rm{\sim}$5-160\,$\rm{\mu m}$ wavelength range for different 6.2, 7.7, 11.3, and 17\,$\rm{\mu m}$ PAH feature strengths (see their Fig.\,13). The stellar component corresponds to a stellar population of $\rm{10^{10}}$ years and solar metallicity from the stellar libraries provided by \citet{Bruzual03} (see Paper I for more details). Overall, the baseline models used to fit the data are:
\begin{eqnarray}
M1 = zdust \times  \{  Dust\ model \} \\
M2 = zdust \times  \{  Dust\ model \} + Stellar \\
M3 = zdust \times  \{  Dust\ model \} + ISM  \\
M4 = zdust \times  \{  Dust\ model \} + ISM + Stellar 
\end{eqnarray}

\noindent For all of them the initial parameters are set to the mean value. We compute the $\rm{\chi^2}$ statistics throughout to guarantee that we found and absolute minimum. We then used f-statistics to test whether the inclusion of the stellar, ISM, or the stellar+ISM components significantly improves the simpler model when f-test probability is below $\rm{10^{-4}}$. We also obtained the $\rm{\chi^2}$ statistics for the parameter space in 50 equally spaced bins with the SPEC command {\sc steppar}, which yields to the probability distribution function (PDF).

\begin{figure}[!t]
\begin{center}
\includegraphics[width=1.\columnwidth]{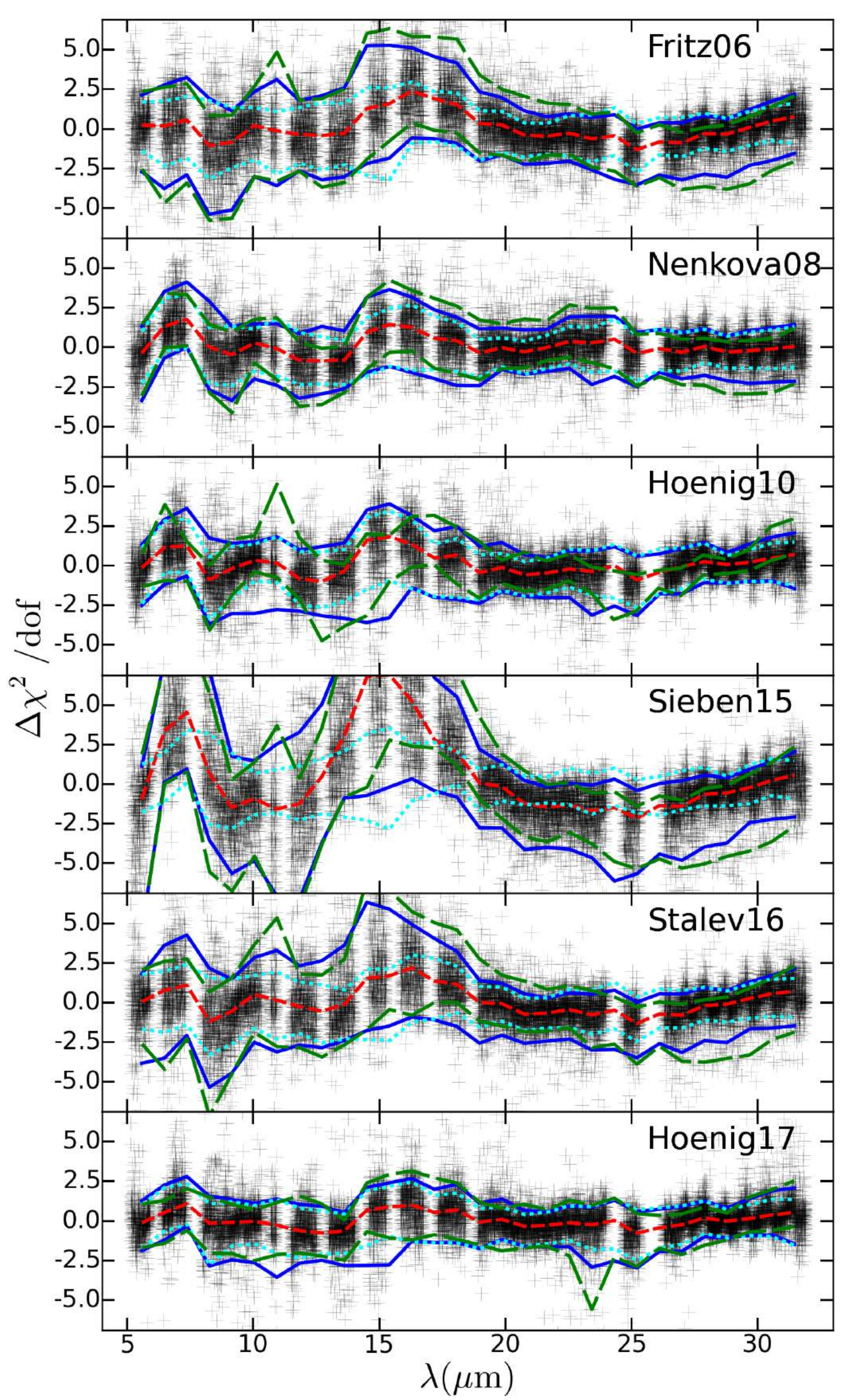}
\caption{Residuals (expressed in terms of $\rm{\Delta \chi^2/dof}$) on the spectral fit for the AGN sample with \emph{Spizer}/IRS data versus wavelength. The red short-dashed line shows the median value versus wavelength. The blue continuous, green long-dashed, and cyan dotted lines represent the 5 and 95\% percentiles of the distribution for the entire sample, AGN-dominated (less than 50\% of stellar component compared to the torus component at 5\,$\rm{\mu m}$ and less than 50\% of ISM component compared to the torus component at 30\,$\rm{\mu m}$), and spectral fits with $\rm{\chi^2/dof <2}$, respectively.}
\label{fig:data:residuals}
\end{center}
\end{figure}

\begin{figure}[!t]
\begin{center}
\includegraphics[width=1.\columnwidth]{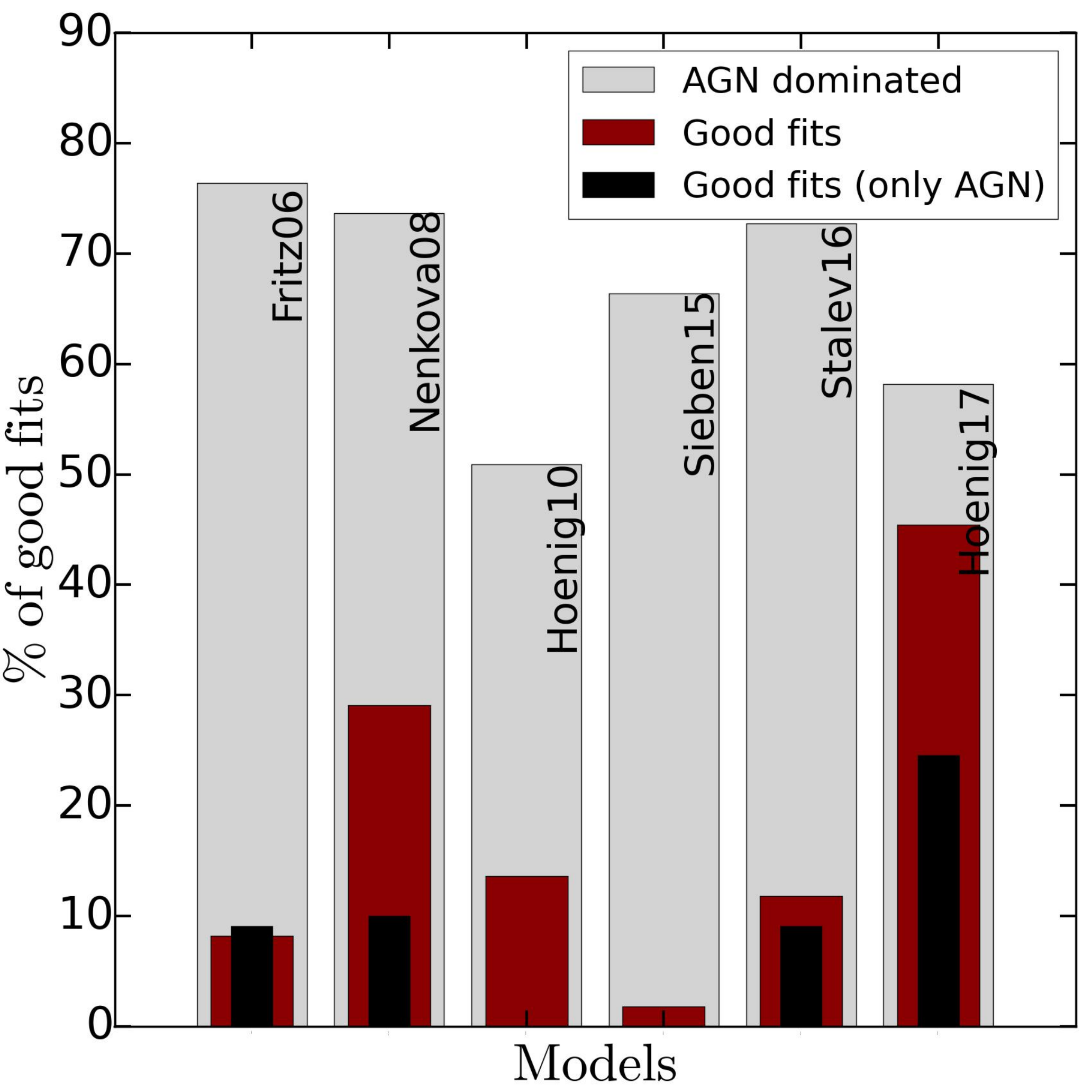}
\caption{Percentage of objects with the mid-infrared flux dominated by the AGN component (AGN-dominated, gray filled bars), the percentage of those objects that are well fitted ($\rm{\chi^2/dof < 2}$) with the AGN models plus circumnuclear contributors (brown filled bars), and those objects that are well fitted with only AGN models (black filled bars). }
\label{fig:data:goodfitstat}
\end{center}
\end{figure}

\begin{figure*}[!ht]
\begin{center}
\includegraphics[width=1.0\columnwidth]{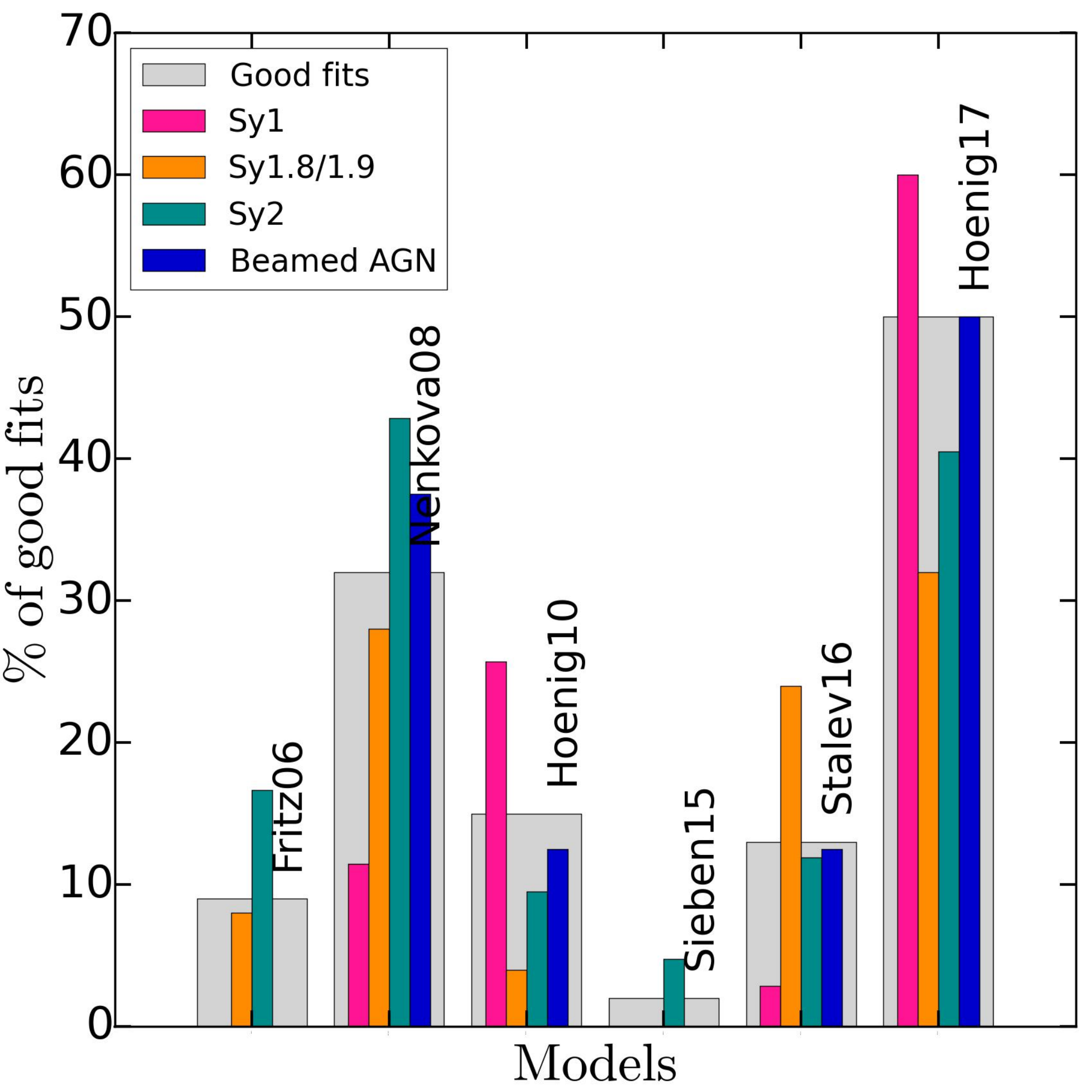}
\includegraphics[width=1.0\columnwidth]{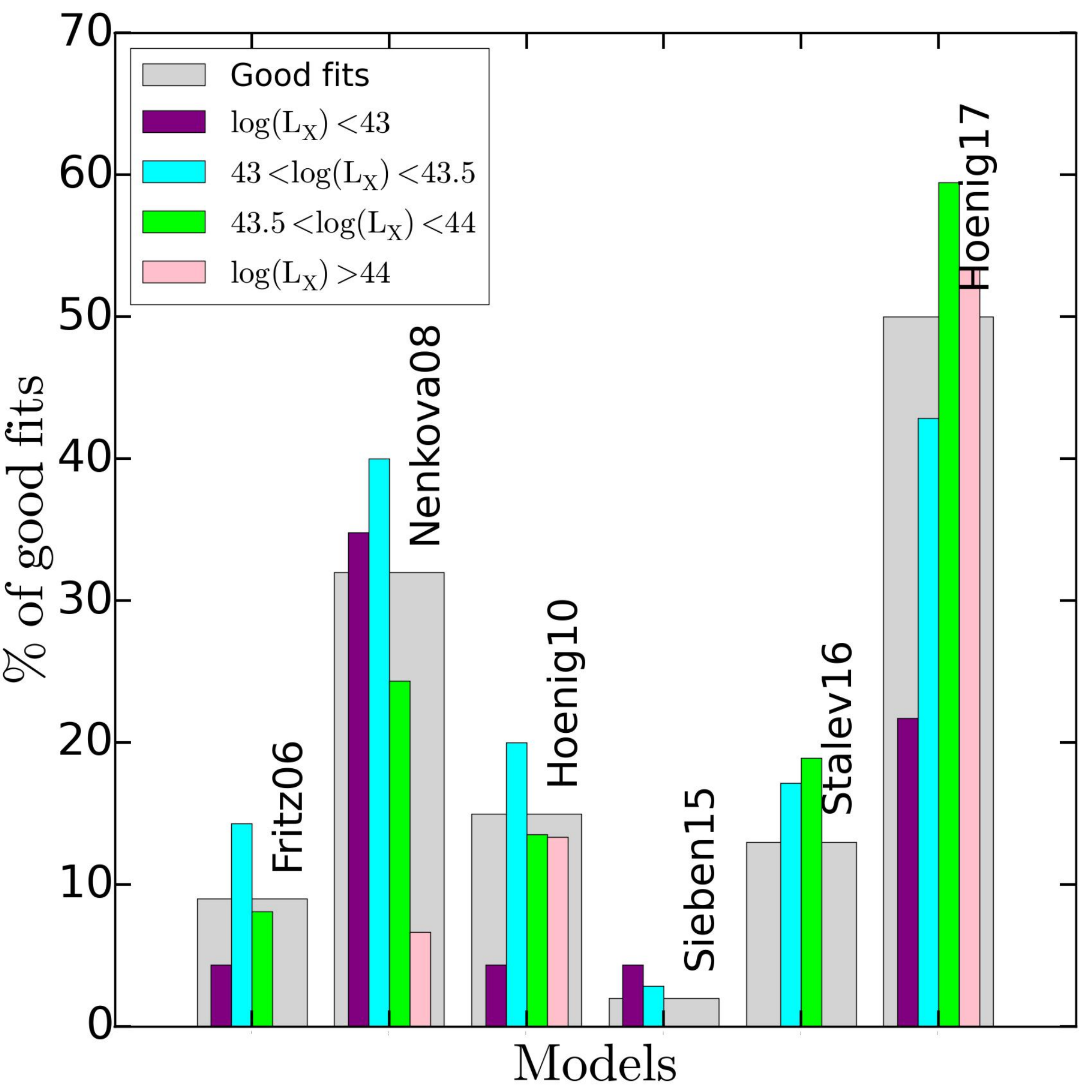}
\caption{Broad gray bars show the total the percentage of objects with good fits. Narrow bars show percentage of objects with good fits per optical type (left panel) and luminosity range (right panel).}
\label{fig:data:goodfitstat2}
\end{center}
\end{figure*}

\begin{figure*}[!ht]
\begin{center}
\includegraphics[width=2.1\columnwidth]{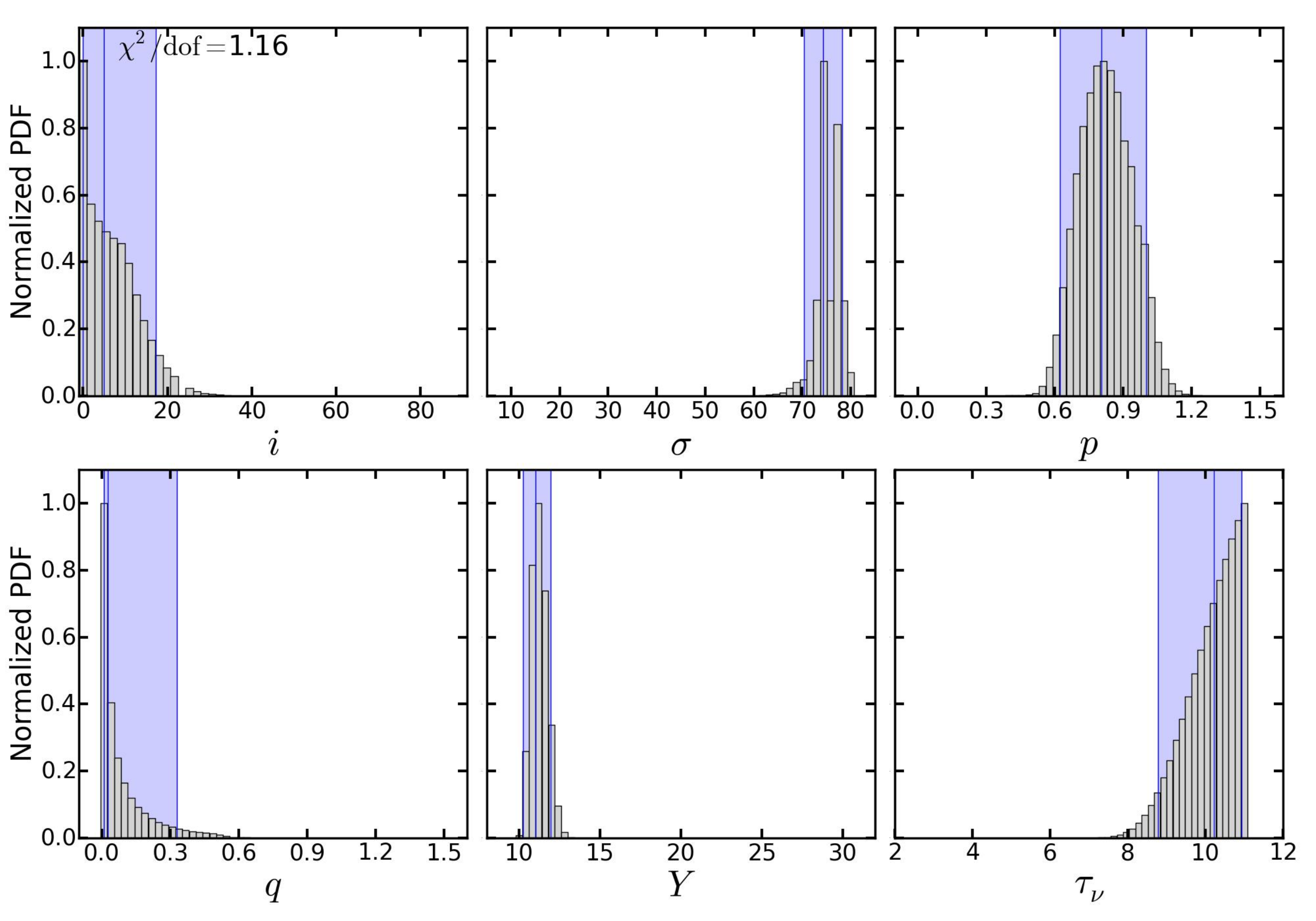}
\caption{Probability distribution function (PDF, gray filled histogram) per parameter resulting when fitting IC5063 to [Stalev16]. Blue vertical line and blue shadowed area show the weighted median and 10-90\% percentiles of the distribution per parameter.}
\label{fig:dataPDFIC5063}
\end{center}
\end{figure*}

\begin{table}
\scriptsize
\begin{center}
\begin{tabular}{ l c c c c c c}
\hline \hline
											&	F06		&	N08 	&	H10		&	S15 	&	S16		&	H17		\\  \\ \hline
$\rm{\alpha_{NIR}}$							&	77.3		&	48.2		&	92.3		&	43.6		&	51.8		&	96.4		\\
$\rm{\alpha_{MIR}}$							&	73.6		&	41.8		&	97.3		&	18.2		&	44.5		&	98.2		\\
$\rm{\alpha_{FIR}}$							&	96.4		&	88.2		&	65.5		&	73.6		&	82.7		&	64.5		\\
$\rm{Si_{9.7\mu m}}$							&	97.3		&	84.5		&	64.5		&	98.2		&	96.4		&	72.7		\\
$\rm{Si_{18\mu m}}$							&	96.4		&	93.6		&	85.5		&	88.2		&	85.5		&	83.6		\\
$\rm{\alpha_{NIR}}$ vs	$\rm{\alpha_{MIR}}$	&	42.7		&	11.8		&	27.3		&	6.4		&	21.8		&	51.8		\\
$\rm{\alpha_{MIR}}$ vs	$\rm{\alpha_{FIR}}$	&	41.8		&	14.5		 &	4.5		&	6.4		&	30.0		&	9.1		\\
$\rm{Si_{9.7\mu m}}$ vs $\rm{Si_{18\mu m}}$	&	17.3		&	27.3		&	28.2		&	24.5		&	30.9		&	35.5		\\
\hline \hline
\end{tabular}
\caption{Percentage of objects with the spectral parameters recovered by the synthetic SEDs. F06: [Fritz06]; N08: [Nenkova08]; H10: [Hoenig10]; S15: [Sieben15]; S16: [Stalev16]; and H17: [Hoenig17].}
\label{tab:shape}
\end{center}
\end{table}

\section{Results} \label{sec:results}

Table \ref{tab:samplefit} (Appendix \ref{appendix:tableresults}) shows the best-fit model results per object (F06: [Fritz06]; N08: [Nenkova08]; H10: [Hoenig10]; S15: [Sieben15]; S16: [Stalev16]; and H17: [Hoenig17]), including the percentage contribution to the 5-30\,$\rm{\mu m}$ waveband per component (A: AGN; S: Stellar; and I: ISM), the reduced $\rm{\chi^2}$ ($\rm{\chi^2/dof}$, where dof is the degree of freedom), color excess for the foreground extinction, and the final parameters per model. Fig.\,\ref{fig:dataspecfit} shows the results on the spectral fitting to the six dusty models for IC\,5063 as an example. 

Following our analysis on the synthetic spectra (Paper I), we selected objects showing high percentage of AGN (AGN-dominated hereinafter) as those with less than 50\% of stellar component compared to the torus component at 5\,$\rm{\mu m}$ and less than 50\% of ISM component compared to the torus component at 30\,$\rm{\mu m}$. They are marked with empty circles next to the model name in Table \ref{tab:samplefit}. We then considered as the best fit that showing the minimum $\rm{\chi^2/dof}$, and comparably good fits those with $\rm{\chi^2/dof< min(\chi^2/dof)+0.5}$. They are marked with filled circles next to the model name in Table \ref{tab:samplefit}. 

\subsection{Adequacy of the models}

We investigate the goodness of the fits to the data by comparing spectral slopes and silicate feature strengths for data and models. Figs.\,\ref{fig:genfit1}, \ref{fig:genfit2}, and \ref{fig:genfit3} show the results for $\rm{\alpha_{NIR}}$ versus $\rm{\alpha_{MIR}}$, $\rm{\alpha_{MIR}}$ versus $\rm{\alpha_{FIR}}$, and $\rm{Si_{18\mu m}}$ versus $\rm{Si_{9.7\mu m}}$. These plots allow us to compare the spectral shape from models (circles) with our sample results (stars). The percentage of objects that are reproduced by the models are recovered in Table\,\ref{tab:shape}.

Objects requiring only the AGN dust model to fit its spectrum nicely fall in the same area compared to the synthetic spectra. However, many objects are in this area but they still require additional contributors or they cannot be fitted to the model at all. The reason is that they cannot predict at the same time all these features.

[Hoenig17] is the best at describing the near-infrared and mid-infrared spectral slopes, with $\rm{\sim 52\%}$ of the objects within the grid of models (see Fig.\,\ref{fig:genfit1}). [Fritz06] is also good at describing these slopes with $\rm{\sim 43\%}$ of the sample within the grid of models. The other models are only able to describe $\rm{\sim 6-27\%}$ of the objects, failing at describing at the same time flat near- and mid-infrared slopes. The best models at describing at the same time mid- and far-infrared slopes are [Fritz06] and [Stalev16] (see Fig.\,\ref{fig:genfit2}), with 30-40\% of the AGN spectra with compatible slopes by these models compared to the $\rm{\sim}$5-15\% of the sample with the rest of models.

These discrepancies between the observations and models are compensated by the inclusion of ISM and/or stellar components. For instance, [Hoenig17] manages to find a good fit including an ISM component to make the far-infrared slopes steeper (clearly seen by the large number of AGN+ISM best fits in Fig.\,\ref{fig:genfit2}). Furthermore, the lack of steepness of the near- and far-infrared slopes in [Hoenig10] is compensated by adding a stellar and ISM component in many objects (Figs.\,\ref{fig:genfit1} and \ref{fig:genfit2}).

Interestingly, all the models produce a large fraction of SEDs that are not describing any observed spectrum, suggesting that the parameter space is not realistic. The best models on that aspect are [Fritz06] and [Stalev16], although they also miss a significant proportion of the parameter space that we observe for our targets. We further explore this result by tracking different ranges of the model parameters to evaluate if a single parameter can produce this unrealistic parameter space. We found that the flat spectral index for the distribution of dust ($\rm{a>-1}$) explains most of the SEDs producing simultaneously the steep near- and mid-infrared slopes for [Hoenig10] and [Hoenig17]. Small viewing angles with respect to the disk/torus equator also produce SEDs with steep near- and mid-infrared slopes. However, small viewing angles can also produce realistic values for these SED slopes and not all the unrealistic SED slopes are produced by SEDs with small viewing angles. Therefore, it seems that a single parameter cannot explain this effect.

In general, all the models fail at describing the strength of both silicate features at the same time, with 17-35\% of the sample well described by the models (see Table\,\ref{tab:shape}). Large silicate absorption features are not described by any of the models (see Fig.\,\ref{fig:genfit3}). These features might be associated with deeply dust-enshrouded objets like ULIRGs. However, some of them were not well fitted including the ISM component. This might indicate that our ISM templates need to be further extended in order to find good fits for these objects. Most objects reproduce the 9.7$\rm{\mu m}$ but fail to predict the 18$\rm{\mu m}$ silicate emission feature strength when the 9.7$\rm{\mu m}$ silicate feature is in absorption (also discussed in Martinez-Paredes et al. in prep.). Extremely large 18$\rm{\mu m}$ silicate emission features ($\rm{Si_{18\mu m}<-0.4}$) are associated to low AGN dust contribution. In those cases the silicate feature is associated to dust heated by strong star-forming processes rather than the AGN. Again, the range of strengths of the silicate features of the models is larger compared to the data for all of them except [Hoenig10] and [Hoenig17]. In particular, all of them include SEDs with strong silicate emission features ($\rm{Si_{9.7\mu m}<-0.5}$) which are not observed in the data. This is a well reported issue for smooth models and in fact it is one of the reasons why clumpy models gained certain credibility against smooth models \citep{Dullemond05,Feltre12}. However, the clumpy torus model by [Nenkova08] also predicts extreme silicate emission features which are not observed in our AGN sample. We do not expect any kind of bias against AGN with strong silicate emission features in our sample (that might be the case for absorption features contaminated by circumnuclear contributors). The clumpy torus model by [Hoenig10], although similar to [Nenkova08], better reproduces the range of silicate feature strengths. This yield to the conclusion that the differences might arise from parameter spaces or details on the radiative transfer solution. Indeed [Hoenig10] shows narrower parameter ranges than [Nenkova08] for the opacity of the clouds and the number of clouds along the equatorial plane (see Table 1 in Paper I). Moreover, in contrast to [Nenkova08], [Hoenig10] does not consider a limit on the outer radius of the torus (fixed to $\rm{Y=150}$). We also attempted here to find if a single model parameter could explain the unrealistic strengths of the silicate emission features. Although, in general, we could not find a single parameter for each model, it is interesting to notice that a small angular width of the torus ($\rm{\sigma<30^{\circ}}$) produces many of these SEDs with unrealistic strengths of the silicate emission feature for [Stalev16].

The goodness of the fit per model can also be evaluated in Fig.\,\ref{fig:data:residuals}, which shows the residuals for all the spectra. The largest discrepancies between data and models are found for [Sieben15]. Even for objects with $\rm{\chi^2/dof<2}$ (cyan dotted line), the residuals are significantly larger than those obtained when fitting the synthetic spectra to the same model used to produce them (see Paper I). This might indicate that the complexity of the spectra is not recovered by any of the models discussed here. AGN-dominated spectra (green long-dashed lines in Fig.\,\ref{fig:data:residuals}) show discrepancies near the 9.7 and 18\,$\rm{\mu m}$ silicate features (with residuals peaking at 11 and 15-17\,$\rm{\mu m}$), in the slopes below $\rm{\sim}$7\,$\rm{\mu}$m, and in the slopes above 25\,$\rm{\mu m}$. Among them, apart from [Sieben15], slightly larger residuals are shown when fitting to [Fritz06] and [Stalev16]. Indeed, both models show quite similar residuals, most probably because they come from the same radiative transfer code (SKIRT) with the same dust geometry. Objects showing a large contribution of stellar or ISM components (blue continuous) tend to show deep silicate absorption features compared to AGN-dominated spectra (green long-dashed lines). 

Fig.\,\ref{fig:data:goodfitstat} shows the percentage of AGN-dominated spectra (grey filled bars). 60\% of objects where the nucleus is isolated at a resolution better than 500 pc are AGN-dominated. The AGN-dominated spectra are not directly linked to the spatial scales achieved by \emph{Spitzer}/IRS because it is also dependent on the AGN bolometric luminosity and the particular environment of the source. The AGN isolation needs to be examined object by object. Moreover, even this result depends on the model used; the largest number of objects that are AGN-dominated is found for [Fritz06] ($\rm{\sim}$75\%) and the lowest if found for [Hoenig10] ($\rm{\sim}$50\%). This is probably due to the fact that two of the main differences between models are the slopes at short and long wavelengths, which is compensated by the inclusion of different fractions of ISM and stellar components when fitting the data. Fig.\,\ref{fig:data:goodfitstat} also shows the percentage of (AGN-dominated) objects that can be well fitted to each model (brown filled bars) and the same results but only using AGN dust model (i.e. without including circumnuclear components, see black filled bars in Fig.\,\ref{fig:data:goodfitstat}). The largest number of objects well fitted to a model is recovered when using [Hoenig17] ($\rm{\sim}$45\%) and the lowest number is found for [Sieben15] ($\rm{<}$5\%). [Nenkova08] also represents well the spectra for almost $\rm{\sim}$30\% of the spectra. The number of objects well fitted to [Fritz06], [Hoenig10], and [Stalev16] drops to $\rm{\sim}$10\%.  Note that these results change if we fit the spectra using only AGN dust models (black filled bars in Fig.\,\ref{fig:data:goodfitstat}). In this case, the percentage of objects with good fits decreases to 25\% for [Hoenig17], and [Fritz06], [Nenkova08], and [Stalev16] show 10\% of good fits.

We also investigate if the goodness of the fit depends on the optical type (Fig.\,\ref{fig:data:goodfitstat2}, left panel) and the X-ray luminosity (Fig.\,\ref{fig:data:goodfitstat2}, right panel). We split the sample into 31 Sy1s (including Sy1, Sy1.2, and Sy1.5), 25 Sy1.8/Sy1.9, 41 Sy2, and 8 beamed AGN \citep[][]{Oh18}. Moreover, we divided the sample into four segments of X-ray luminosities ($\rm{\Delta log(L_X)=0.5}$) that contain 23, 35, 37, and 15 objects from the lowest ($\rm{42.5<log(L_X)<43}$) to the highest ($\rm{44<log(L_X)<44.5}$) luminosity bin. Around 60\% of the Sy1 and 4 out of the 8 beamed AGN are well fitted to [Hoenig17]. [Nenkova08] failed to reproduce Sy1, obtaining 10\% of good fits. Note that this might be improved by a proper account of the AGN disk component. Furthermore, roughly 40\% of the Sy2 and beamed AGN are well fitted to [Nenkova08] or [Hoenig17]. A lower percentage is obtained for Sy1.8/1.9 spectra (roughly 30\% using [Nenkova08] or [Hoenig17]. Not far from these numbers and despite the low percentage of good fits, [Stalev16] seems to work for some Sy1.8/1.9 (20\% of them). An important result is that, among the two models showing the largest number of good fits (i.e. [Nenkova08] and [Hoenig17]), [Hoenig17] is better suited for high-luminous AGN and [Nenkova08] for low-luminosity AGN. [Stalev16] seems to work only for intermediate luminosities.

\begin{figure*}[!ht]
\begin{center}
\includegraphics[width=0.66\columnwidth]{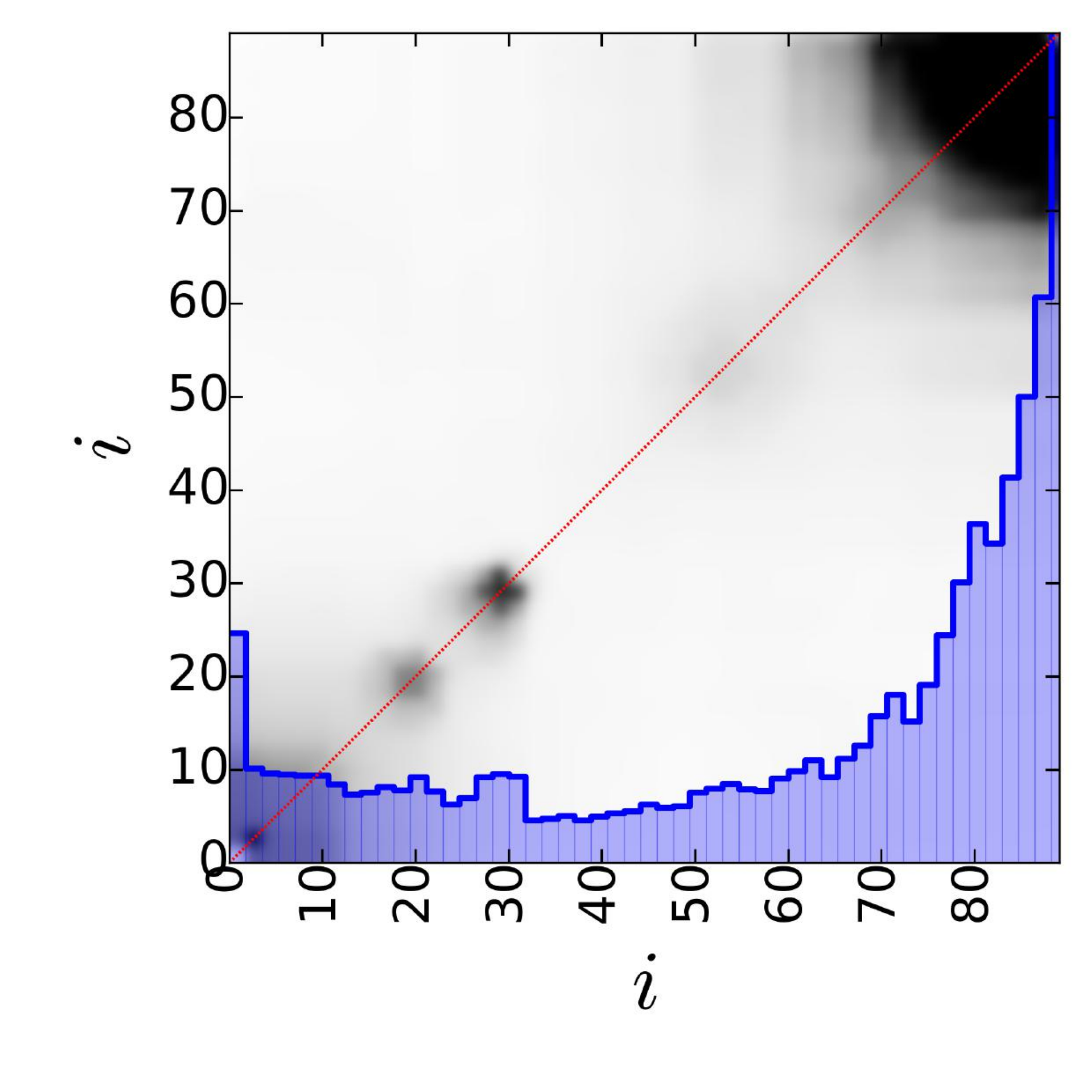}
\includegraphics[width=0.66\columnwidth]{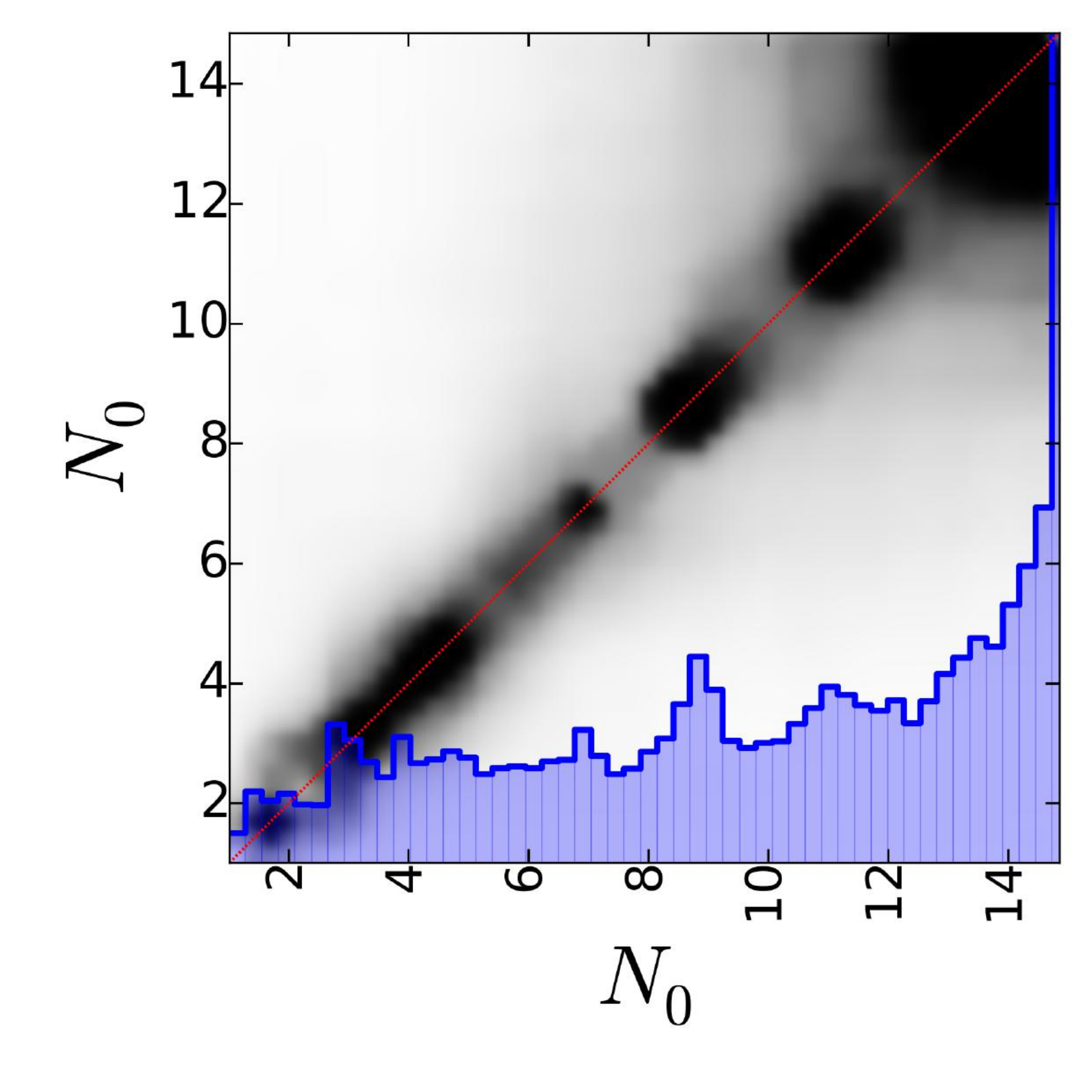}
\includegraphics[width=0.66\columnwidth]{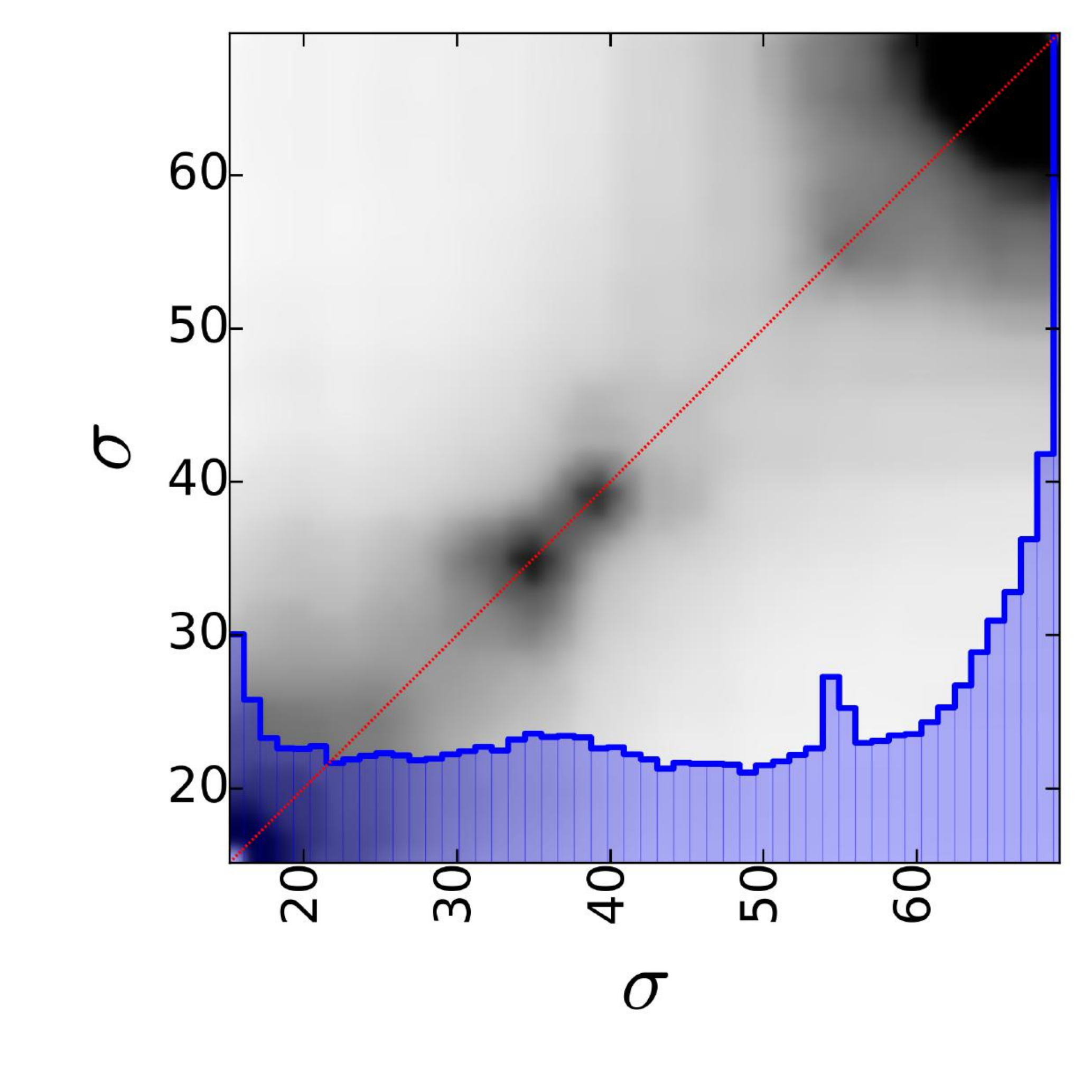}
\includegraphics[width=0.66\columnwidth]{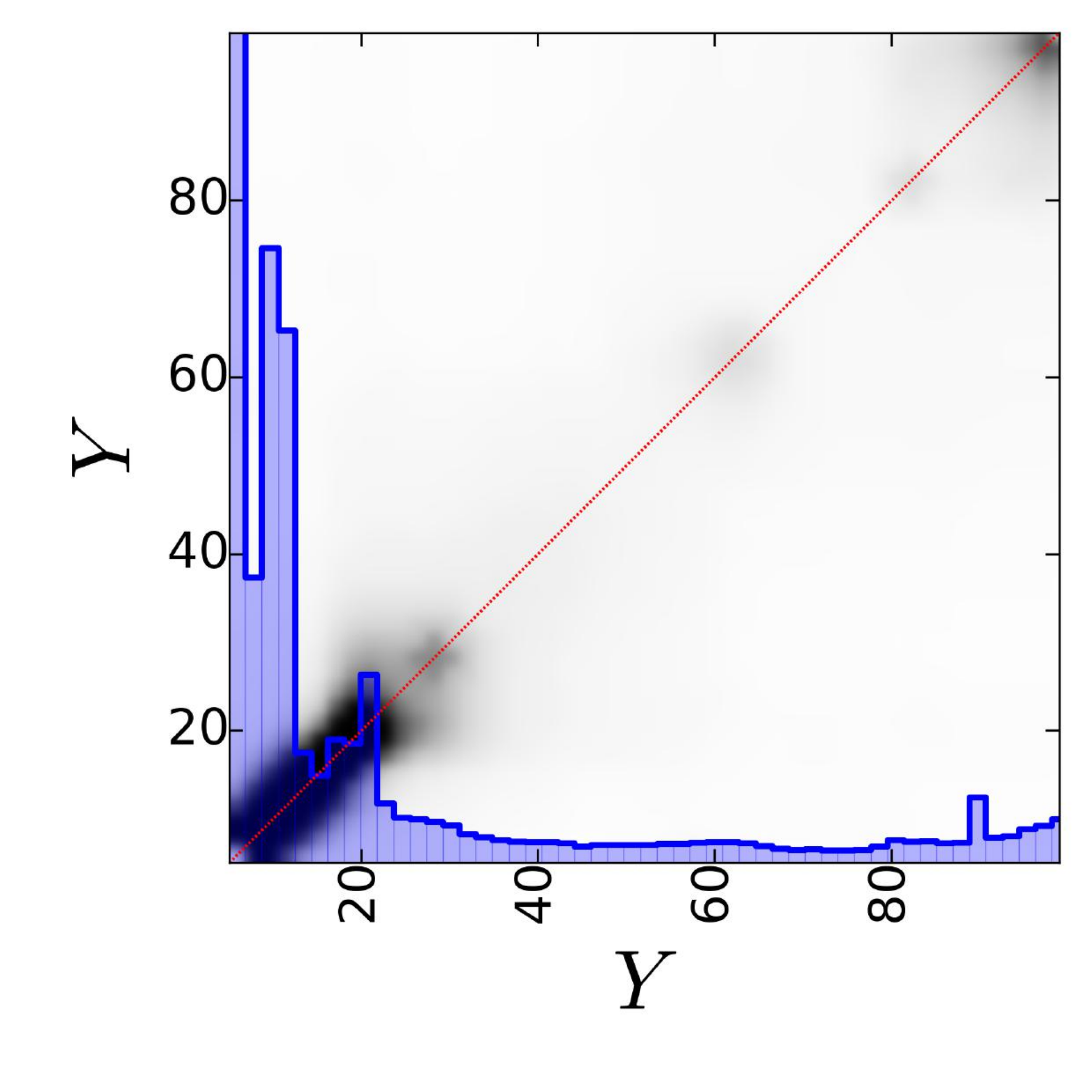}
\includegraphics[width=0.66\columnwidth]{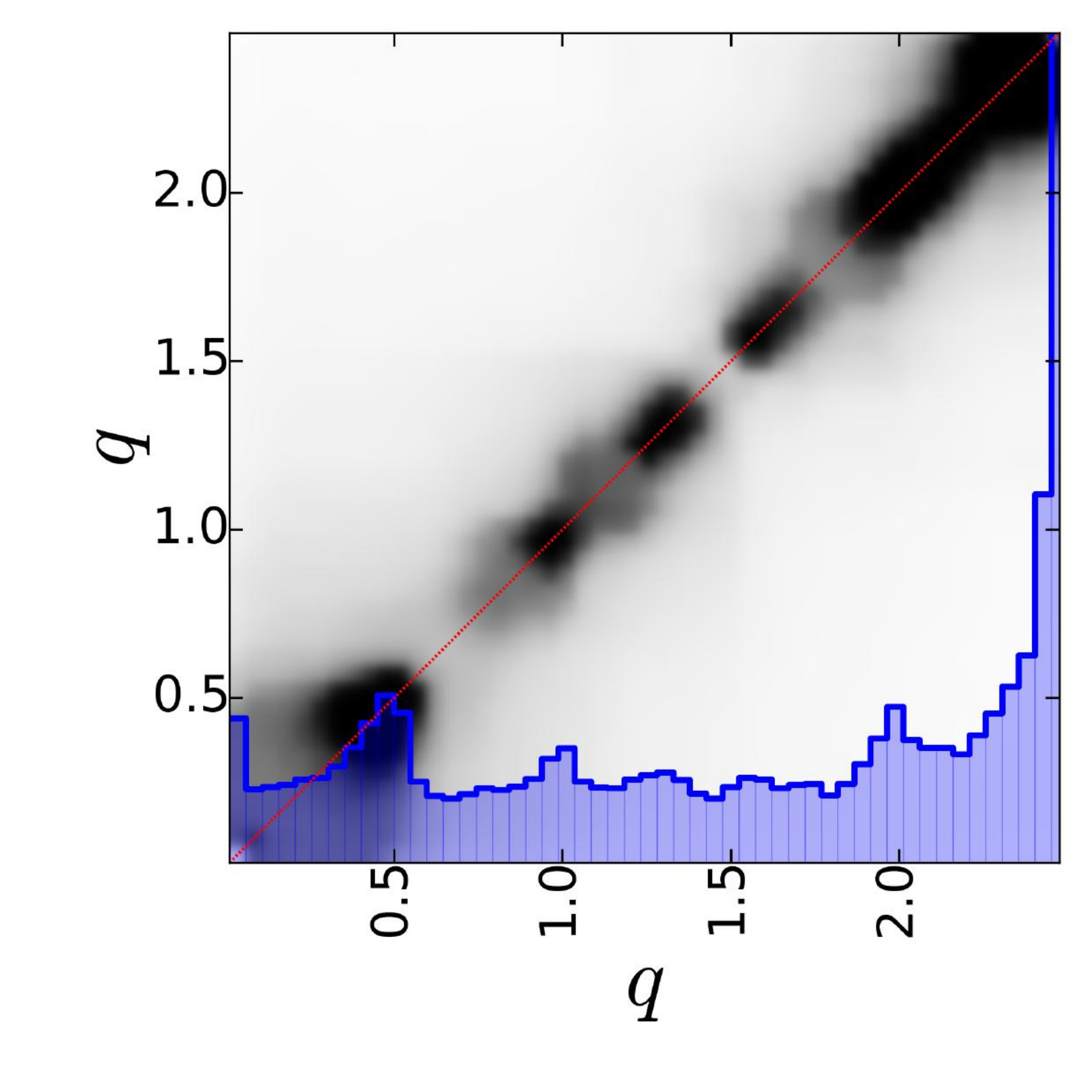}
\includegraphics[width=0.66\columnwidth]{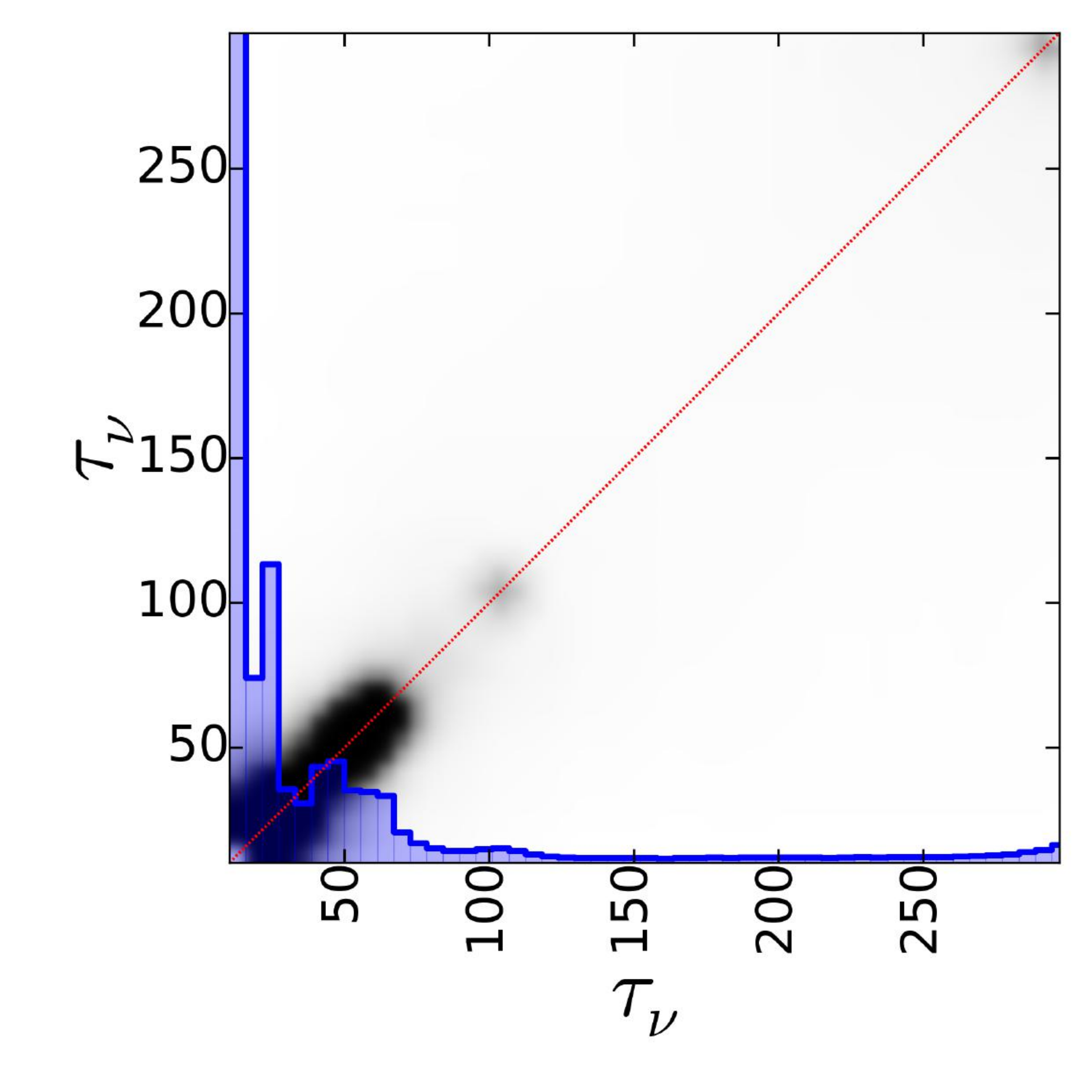}
\caption{Parameter versus parameter estimated from the sample drawn from the probability distribution function (PDF, see text) for the model [Nenkova08]. The blue filled histogram shows the total AGN sample distribution per parameter. The minimum color level shows probabilities of 1\%. The red line shows the one-to-one relation.}
\label{fig:dataspecfitPDFN08}
\end{center}
\end{figure*}

\renewcommand{\baselinestretch}{0.6}
\begin{table}
\scriptsize
\begin{center}
\begin{tabular}{l c c c l l }
\hline \hline
				&				&		&	&							&						\\
Model  		&  Par &  Error    &  & \multicolumn{2}{c}{Comments}  \\ \cline{5-6}
				&				&		&	&							&						\\
				&				&	(\%)	&    & Error             				& Range       			\\
				&				&		&    & dependency	 			& adequacy  			\\
(1)       			&    (2)           	&      (3)	&    & (4)               				&   (5)            			\\ \hline 
[F06]			&		$i$		&	9	&	&	low limit					&pegged at $i=20$						\\
				&	$\sigma$	&	8	&	&	low limit 					&pegged at $\sigma=20$	\\
				&	$\gamma$	&	11	&	&							&pegged at $\gamma=0$	\\
				&	               	&	 	&	&							&pegged at $\gamma=6$*	\\
				&	$\beta$		&	15	&	&	high limit					&pegged at $\beta=-1$	\\
				&	        		&	 	&	&	                				&pegged at $\beta=0$	\\
				&		Y		&	4	&	&							&pegged at $Y=10$		\\
				&$\tau_{\nu}$	&	10	&	&							&pegged at $\tau_{\nu}=10$	\\  
\hline 
[N08]			&	$i$			&	17	&	&	low and high limit			&pegged at $i=90$		\\
				&	$N_0$		&	15	&	&	high limit					&pegged at $N_{0}=15$   	\\
				&	$\sigma$	&	19	&	&	low and high limit			&pegged at $\sigma=70$	\\
				&		Y		&	13	&	&							&only $Y<50$			\\
				&		 		&	  	&	&							&pegged at $Y=10$		\\
				&		q		&	15	&	&	high limit					&pegged at $q=2.5$		\\
				&$\tau_{\nu}$	&	10	&	&							& only $\tau_{\nu}<100$ 	\\
				&                    	&	 	&	&							& pegged at $\tau_{\nu}=10$ \\ 
\hline 
[H10]			&		$i$		&	15	&	&			low limit			&pegged at $i=0$						\\
				&				&		&	&							&pegged at $i=90$*						\\
				&	$N_0$		&	13	&	&			high limit			&pegged at $N_{0}=10$   	\\
				&	$\theta$		&	17	&	&			high limit			&pegged at $\theta=60$   \\
				&				&		&	&							& only $\theta>40$**	      \\
				&     $a$			&	6	&	&							&pegged at $a=0$   		\\
				&$\tau_{\nu}$ 	&	12	&	&		 low and high limit       &pegged at $\tau_{\nu}=80$ \\  
\hline 
[S15]			&		$i$		&	10	&	&							&pegged at $i=90$		\\
				&	$R_{in}$		&	5	&	&							&pegged at $R_{in}=3$	\\
				&	$\eta$		&	6	&	&							&pegged at $\eta=0$*	\\
				&$\tau_{cl}$        &	9	&	&		high limit				&pegged at $\tau_{cl}=100$\\
				&$\tau_{disk}$    &	11	&	&							&						\\  
\hline 
[S16]			&	$i$			&	12	&	&							&pegged at $i=90$		\\
				&	$\sigma$	&	7	&	&							&pegged at $\sigma=10$  \\
				&				&	 	&	&							&pegged at $\sigma=80$  \\
				&	$p$			&	10	&	&							&pegged at $p=0$		\\
				&	 			&	 	&	&							&pegged at $p=1.5$		\\
				&	$q$			&	18	&	&							&pegged at $q=0$		\\
				&	 			&	 	&	&							&pegged at $q=1.5$		\\
				&	$Y$			&	7	&	&							&pegged at $Y=10$		\\
				&	        		&	 	&	&							& only $Y<15$**			\\
				&	$\tau_{\nu}$	&	11	&	&							&pegged at $\tau_{\nu}=3$	\\   
\hline 
[H17]			&		$i$		&	16	&	&							&pegged at $i=0$		\\
 				&		 		&	 	&	&							&pegged at $i=90$		\\
				&$N_0$			&	20	&	&		high limit				&pegged at $N_{0}=10$ 	\\
				&	$a$			&	11	&	&		low limit				&pegged at $a=-3$ 		\\
				&$\sigma_{\theta}$&	21	&	&		high limit				&pegged at $\sigma_{\theta}=7.5$\\
				&				&		&	&							&pegged at $\sigma_{\theta}=15$\\
				&${\theta}$		&	20	&	&		high limit				&pegged at $\theta=45$	\\
				&	$a_w$		&	15	&	&		high limit				&pegged at $a_w=-0.5$	\\
				&	$h$			&	18	&	&	low and high limit 		&pegged at $h=0.1$		\\
				&	$f_{wd}$	&	18	&	&							&pegged at $f_w=0.75$	\\
\hline \hline
\end{tabular}
\caption{Summary of parameter results (F06: [Fritz06]; N08: [Nenkova08]; H10: [Hoenig10]; S15: [Sieben15]; S16: [Stalev16]; and H17: [Hoenig17]). (Col.\,1): model name; (Col.\,2): parameter name; (Col.\,3): average percentage of error according to the standard deviation of the distributions compared to the parameter range; (Col.\,4): comments on the dependency of the error on the parameter range; and (Col\,.5): comments on the adequacy of the parameter range to the sample. We refer to the parameter results that are clustered toward the low and/or high limit as ``pegged" following XSPEC syntax. In Col.\,5 we mark with single asterisks the parameters that do not longer cluster to limits of parameter space when using only good fit results and with double asterisks those restrictions on the parameter that appear only when using only good fits  (i.e. $\rm{\chi^2/dof<2}$). }
\label{tab:parameterresults}
\end{center}
\end{table}

\subsection{Goodness of the parameter determination}

We also investigate how well we constrain the parameters of the fit for those spectra that are AGN-dominated with $\rm{\chi^2/dof<2}$. Following the procedure used for the synthetic spectra (see Paper I), we consider that a parameter is restricted when its error bar is less than 15\% of the parameter space. 
Irrespective of the model, roughly 80\% of the parameters are restricted when a good fit is found. No particular differences are found for groups of either optical type or X-ray luminosity. Note, however, that many parameters are clustered to the high or low limits. This might indicate that better results could be achieved if some parameters cover a broader parameter space (we further illustrate this below). 

The analysis above assumes that the errors show a Poisson distribution so the standard deviation is a good representation of its error. However, this is not the case in general, with asymmetric parameter posterior distributions (see Fig.\,\ref{fig:dataPDFIC5063} for an example of the resulting PDF for IC\,5063 using [Stalev16] model). From now on we use the PDFs to study the goodness of the parameters. Note that we also operate with the PDFs to compute derived quantities (e.g. dust mass, see below).

We use individual PDFs to build a parameter versus parameter plot by adding the normalized PDFs for all the objects in the sample. Fig.\,\ref{fig:dataspecfitPDFN08} shows the parameter versus parameter plots for [Nenkova08] (all the models included in Appendix\,\ref{appendix:dataspecfitPDF}). This plot is computed as the addition of all the PDFs of the individual objects. Any horizontal or vertical cut shows the PDF found for a particular parameter value. The dispersion of the plot accounts for the accuracy on the determination of the parameter because it reflects the broadening of the PDFs for the full sample. A good parameter estimate will bring a good correlation along with the diagonal axis of the plot. Broad PDFs will yield to a non-linear relation in this parameter versus parameter plot. How the broadening of the distribution changes along the parameter space also gives information on the accuracy of the parameter determination. Finally, we can also explore with this plot which range of parameters is preferred by the sample (studying the distribution per parameter overlaid as blue-filled histograms) and if the parameter space is not enough to cover the spectral shapes shown by the data, i.e. when objects show parameters that tend to be clustered to the limits of the parameter space.

Fig.\,\ref{fig:dataspecfitPDFN08} confirms some of our results on the constraints of the parameters for [Nenkova08] model using synthetic spectra (Paper I): Parameters $Y$, $q$, and $\tau_{\nu}$ are much better constrained than $i$, $N_H$, and $\sigma$. However, these plots show further information. Firstly, some parameters are less constrained toward the lower and/or upper limits of the parameter range (e.g. $i$). Furthermore, objects tend to cluster toward the lower and upper boundaries for some parameters. For instance, above 30\% of objects reaches the upper limit for $N_0$. Finally, our sample prefers a narrow range of values compared to the parameter range (e.g. $\rm{\tau_{\nu}<100}$ and $\rm{Y<40}$). Different results might come from a sub-set of objects well fitted with a particular model. In order to investigate that, we repeated the analysis using only results with $\rm{\chi^{2}/dof<2}$ and AGN-dominated spectra. We recover similar results wen using only good fits except for the clustering of some parameters at the lower and upper boundaries. For instance the density distribution slope $\gamma$ in [Fritz06] reaches the high limit only when using all the results. Moreover, further restriction of the parameters are found. For instance, $Y$ in [Stalev16] is restricted to values $Y<15$ when using good fits. No differences are found when using the sub-set of AGN-dominated spectra.

Table\,\ref{tab:parameterresults} shows a summary of the results found using these parameter versus parameter plots for all the models. Note that all the parameters show a correlation coefficient $r>0.8$, indicating that they are constrained. In general, the parameters are restricted to the 10-20\% of its parameter space (consistent with our synthetic spectral analysis, see Paper I). We detect clear broadening of this percentage to the low and/or high limit for most of the parameters and many of them tend to be clustered at the low and high limit. This might further suggest that a strong revision of the SED libraries could result on a better match of the mid-infrared spectra. Table\,\ref{tab:parameterresults} summarizes some information that modelers may use to produce SED libraries adequate for the diversity of AGN dust.

An extension of these plots, in which we confront different parameters for a particular model, is useful to investigate if any parameter of the model is linked to another, therefore studying internal degeneracies among parameters. In practice, we repeat the analysis for each parameter with all the other parameters of the model looking for linear relations among them. This resulted in 10, 15, and 21 plots ($\rm{N(N-1)/2}$, with N the number of parameters) for models with five (i.e. [Hoenig10] and [Sieben15]), six (i.e. [Fritz06], [Nenkova08], and [Stalev16]), and eight (i.e. [Hoenig17]) parameters, respectively. We did not find any linear relation among parameters of the same model, at least for the present sample. Note, however, that this relationship might be coupling several parameters at the same time or could not be linear. The study of such a complex coupling of parameters is out of the scope of this research.

We attempted a final extension of these plots, looking for linear relations between a particular parameter and all the parameters of the other models, could be used to find equivalencies between parameters of different models. This resulted in 537 parameter versus parameter plots. However, we did not find any linear relation between parameters. This implies that even external parameters as the viewing angle toward the observer, depend on the model assumption. Therefore, we confirmed the result from synthetic spectra (Paper I); the parameter results strongly depend on the chosen model. 

We also investigated if different optical types or luminosity ranges lie in different parameter ranges by plotting the histograms of these sub-sets of objects. However, we did not find any significant relation among the parameters.

\begin{figure*}[!ht]
\begin{flushleft}
\includegraphics[width=0.68\columnwidth]{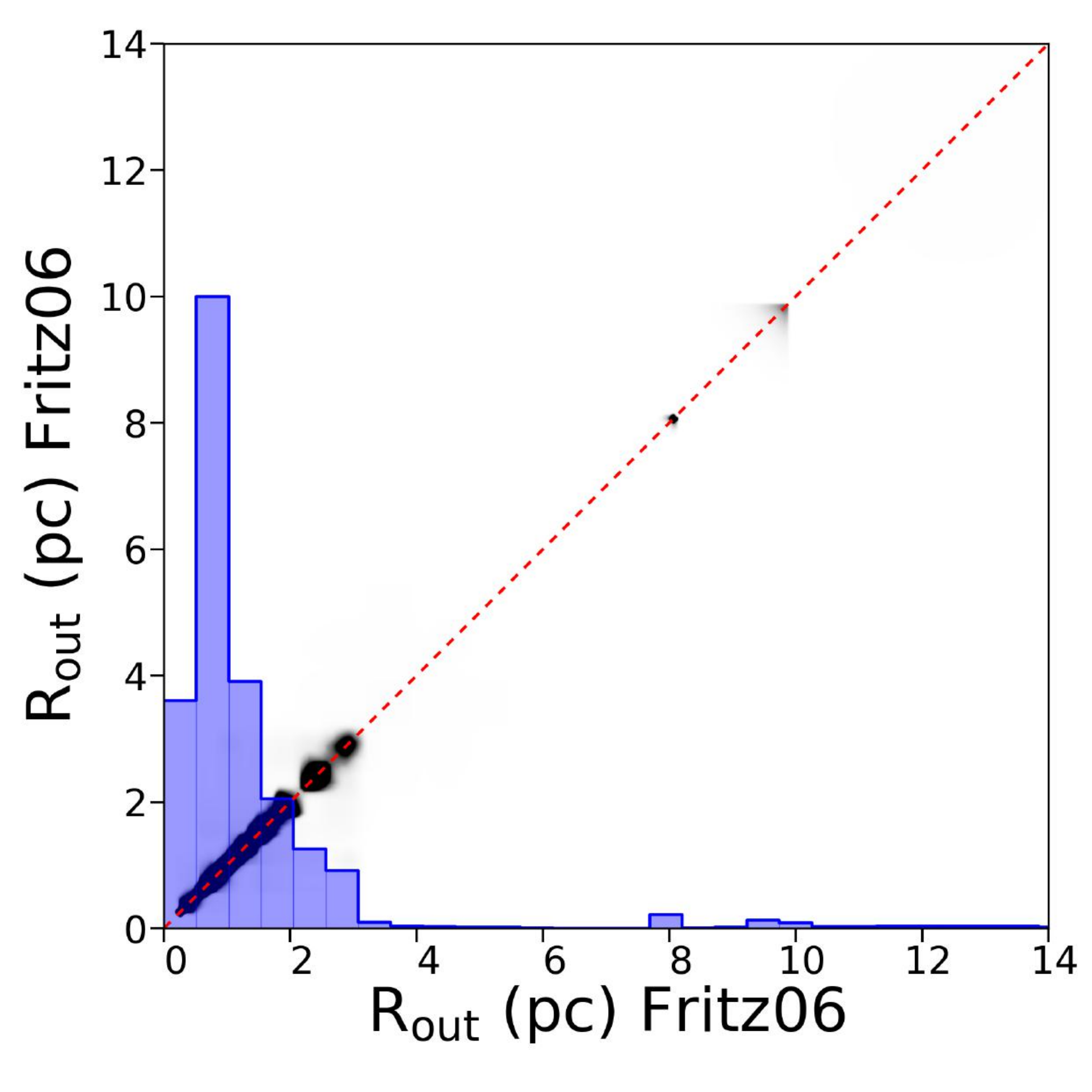}
\includegraphics[width=0.68\columnwidth]{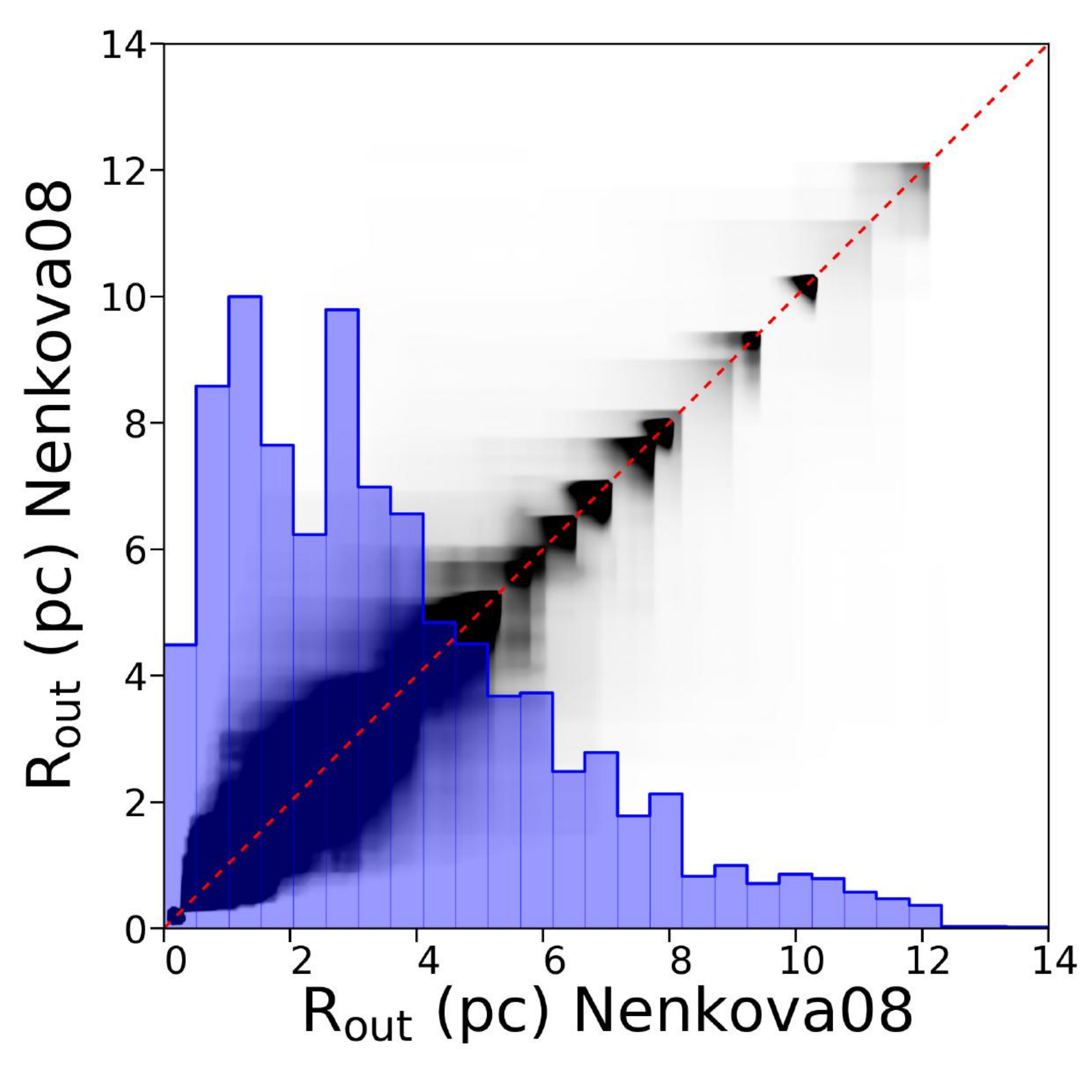}
\includegraphics[width=0.68\columnwidth]{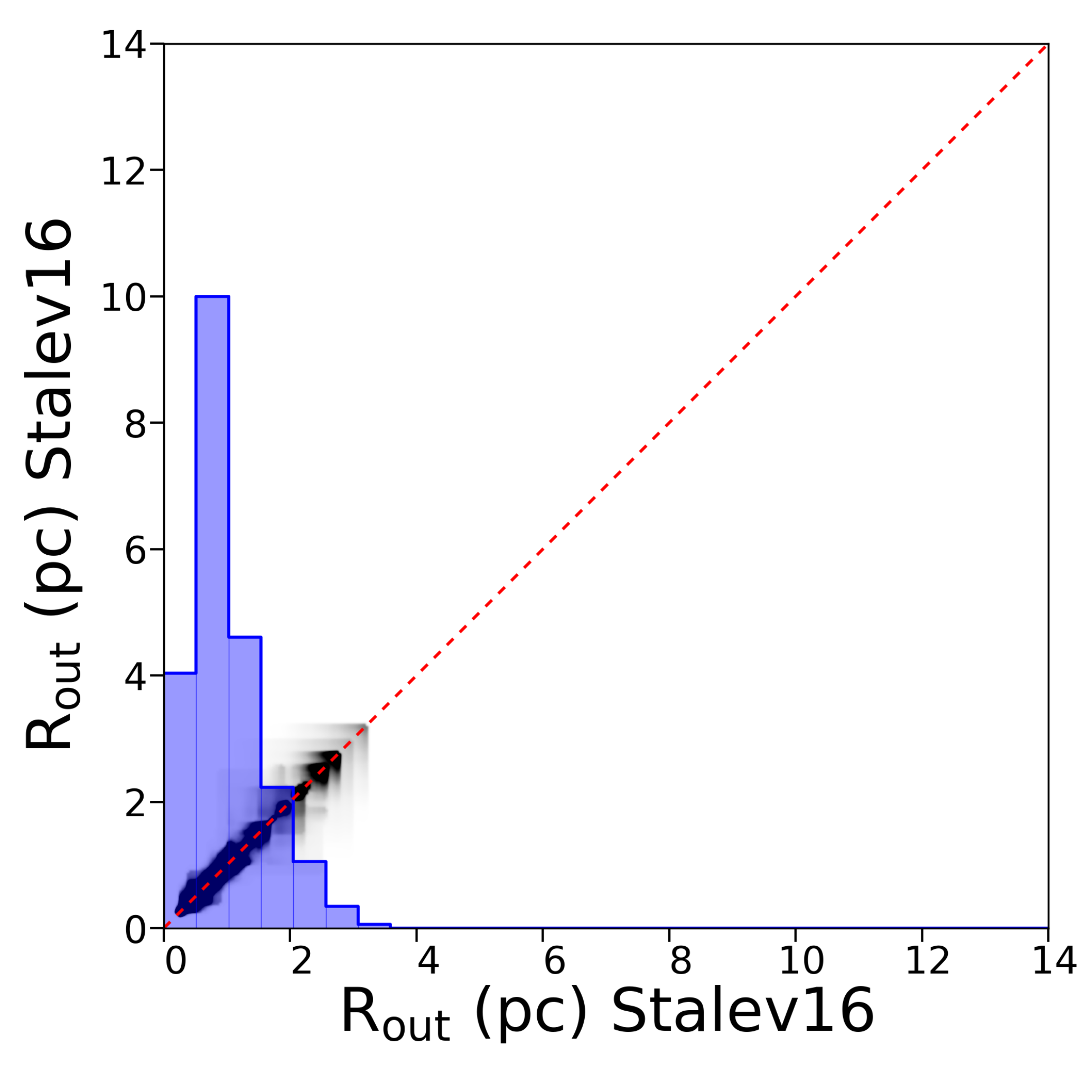}
\includegraphics[width=0.68\columnwidth]{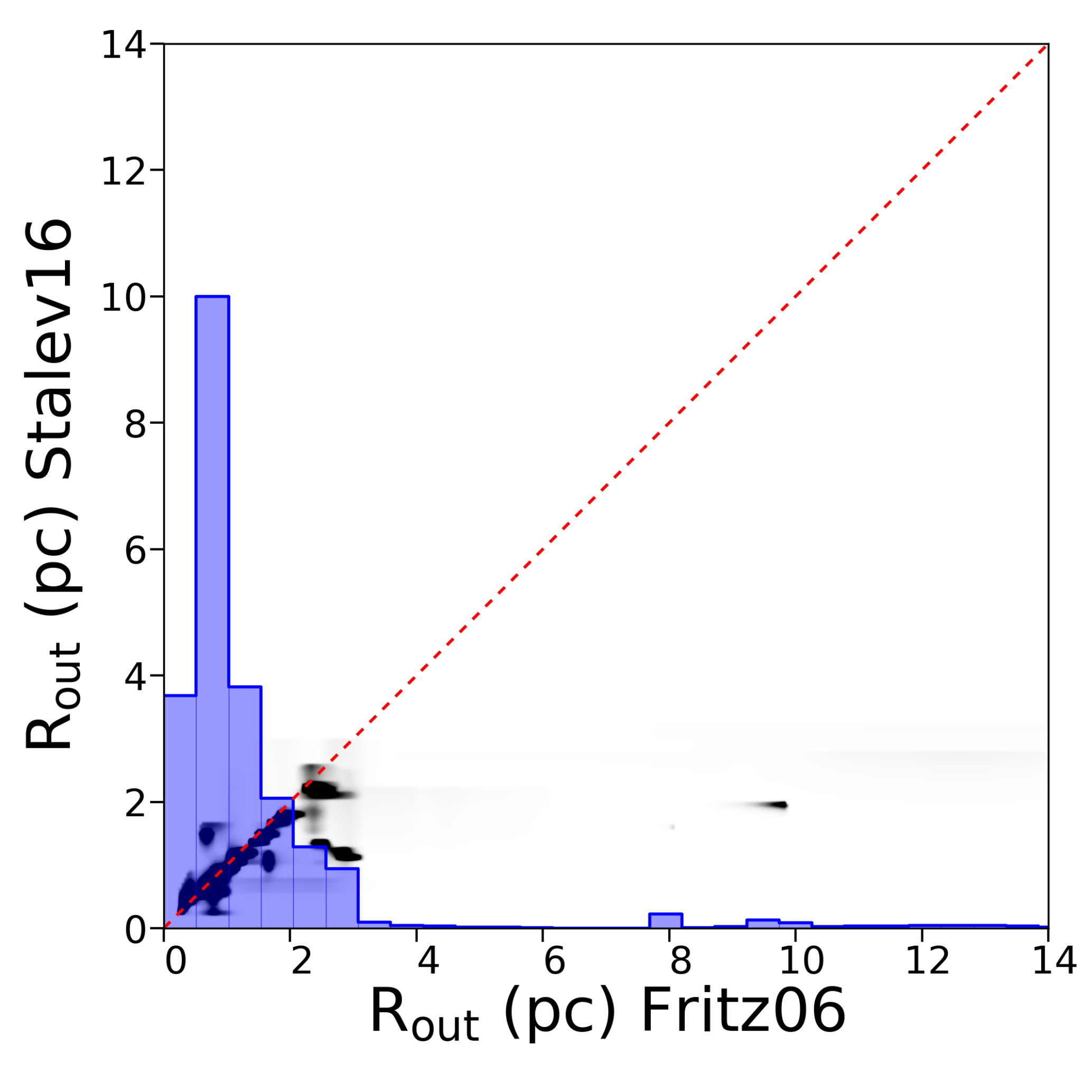}
\includegraphics[width=0.68\columnwidth]{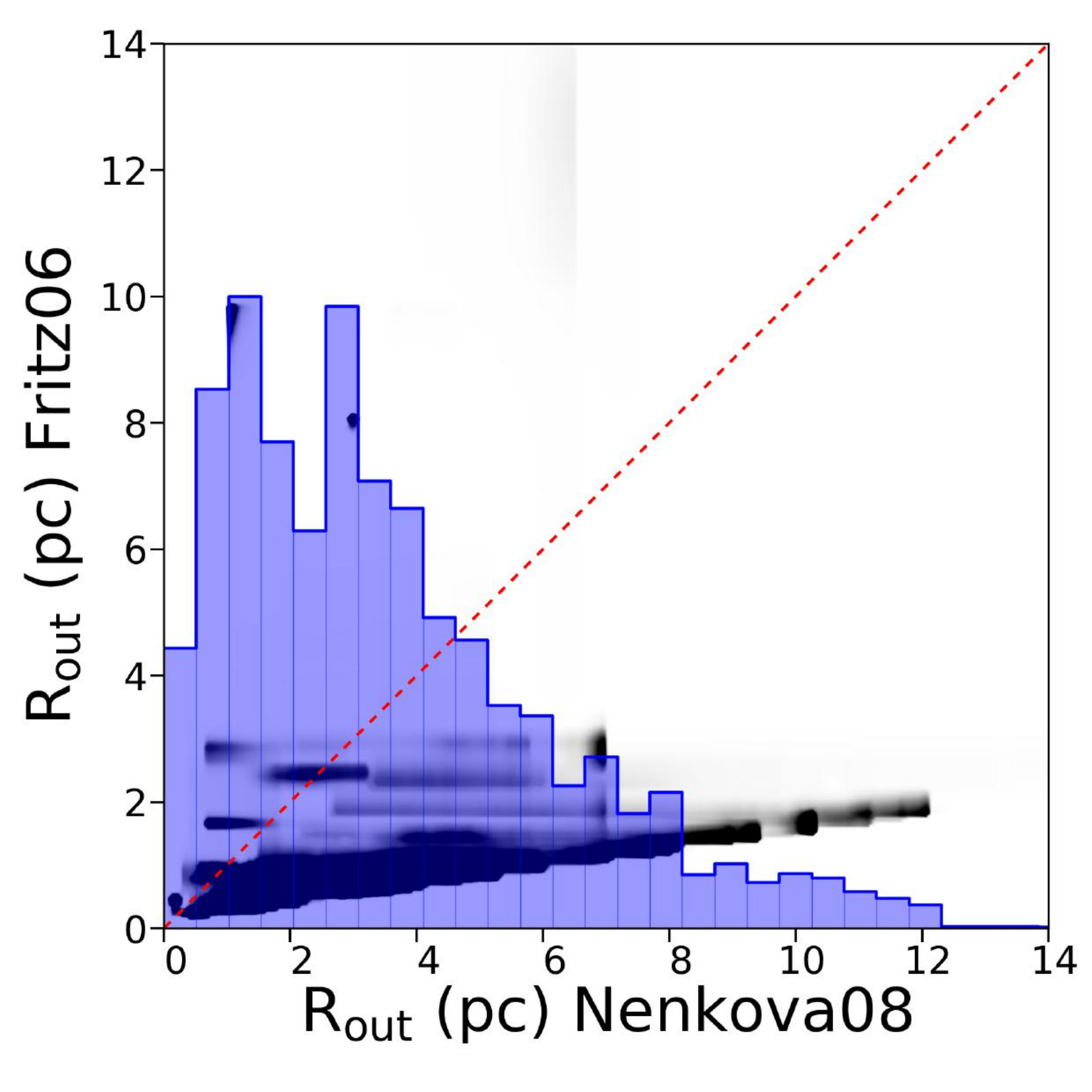}
\includegraphics[width=0.68\columnwidth]{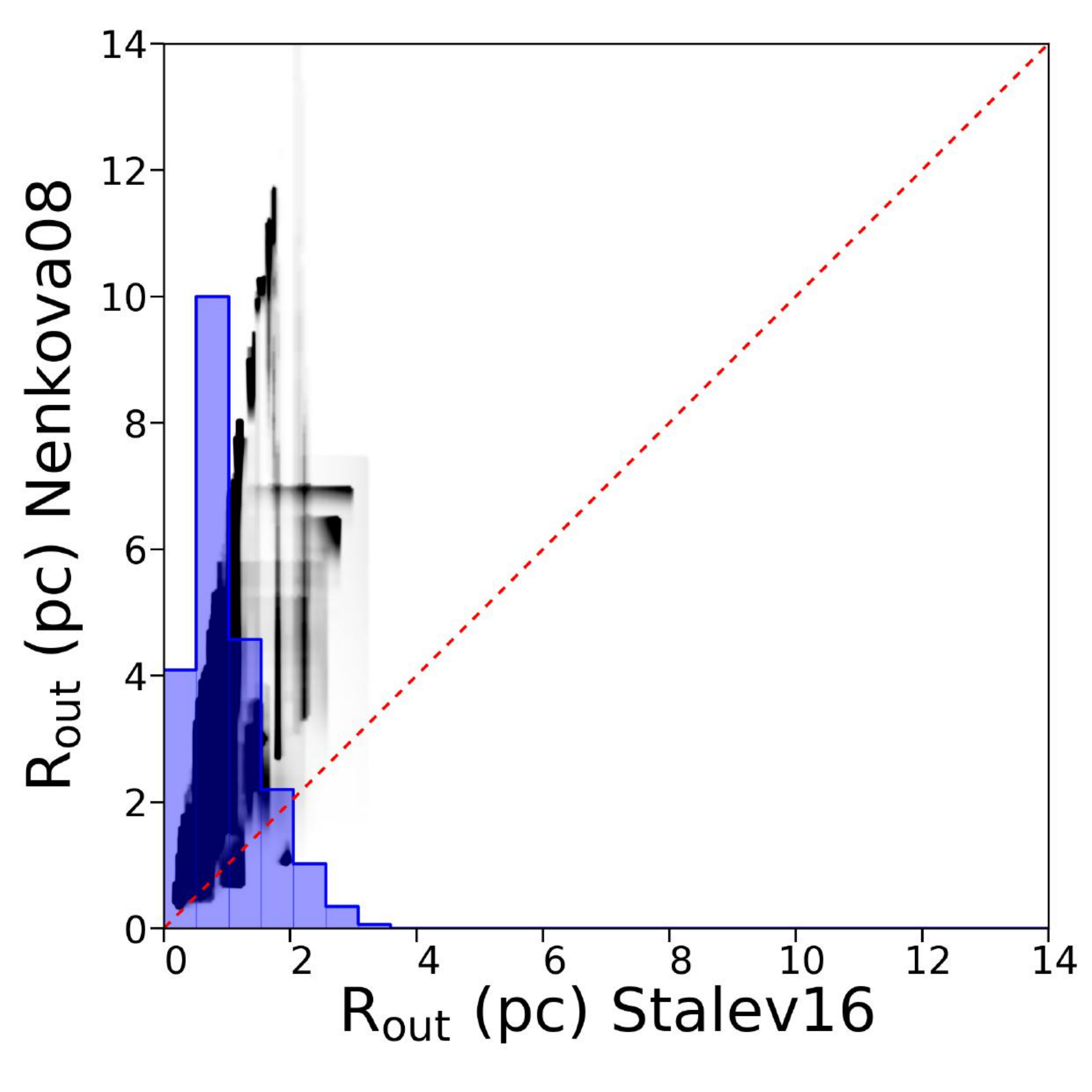}
\caption{(top): Sum of the probability distribution function (PDF) for the outer radius of the dusty structure for our sample using [Fritz06] (left), [Nenkova08] (middle), and [Stalev16] (right). (bottom): Outer radius of the torus obtained for [Stalev16] versus [Fritz06] (left), [Fritz06] versus [Nenkova008] (middle), and [Nenkova08] versus [Stalev16] (right). The blue-filled histograms show the outer radius of the torus distribution per model (maximum of the distributions arbitrarily scaled to $\rm{R_{out}=10pc}$). The dotted line shows the one-to-one relationship expected for the best accuracy determination between parameters.}
\label{fig:routselffit}
\end{flushleft}
\end{figure*}

\begin{figure*}[!ht]
\begin{flushleft}
\includegraphics[width=0.68\columnwidth]{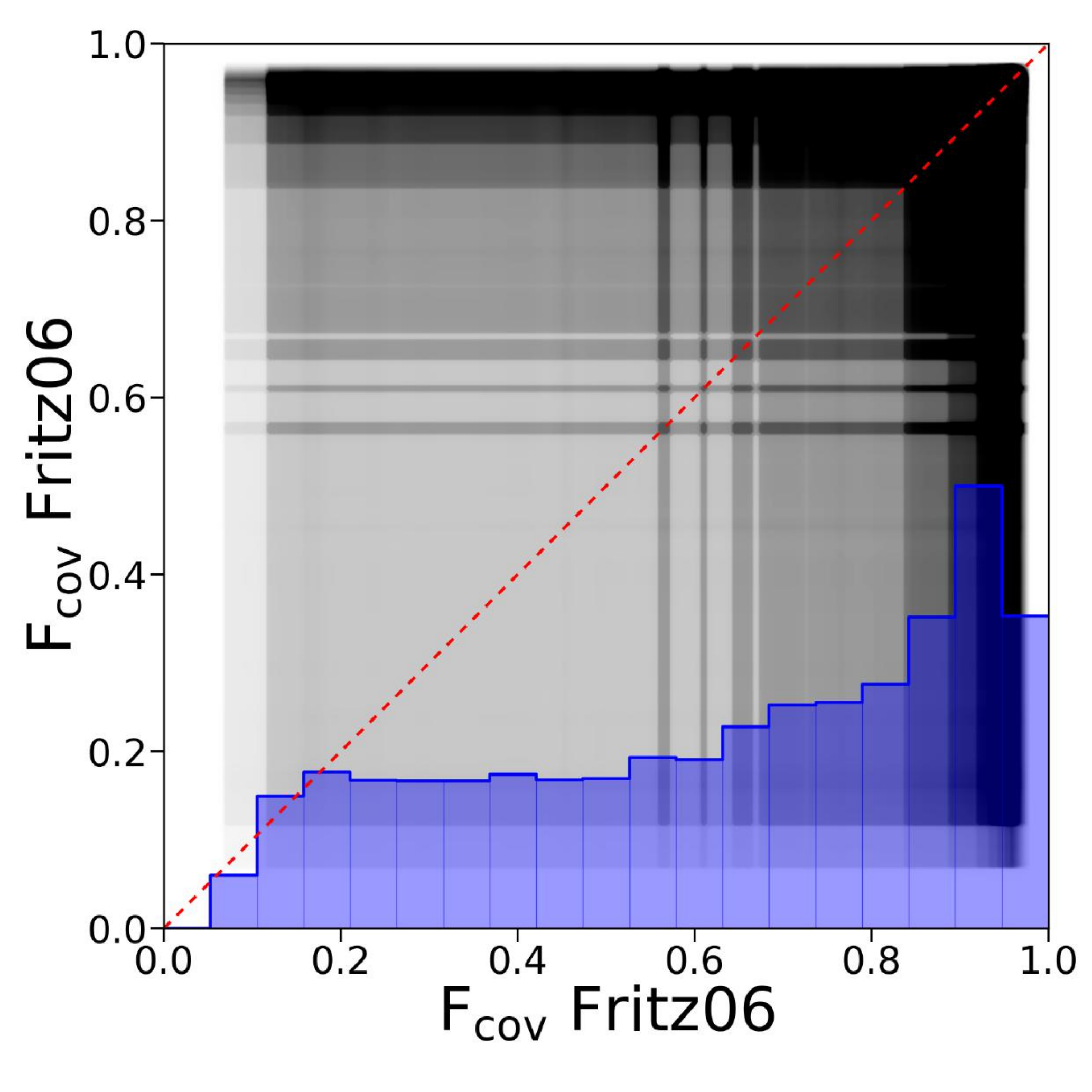}
\includegraphics[width=0.68\columnwidth]{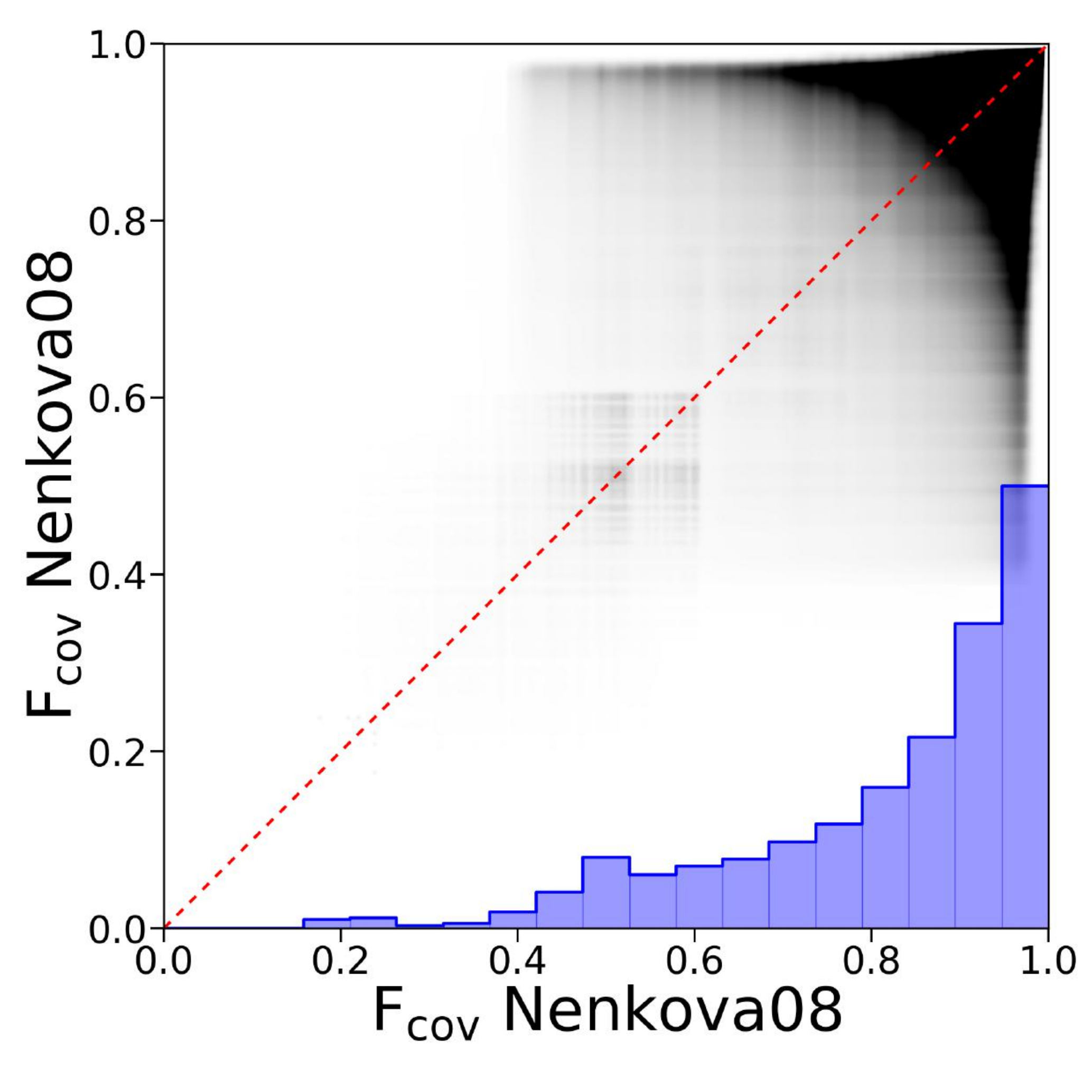}
\includegraphics[width=0.68\columnwidth]{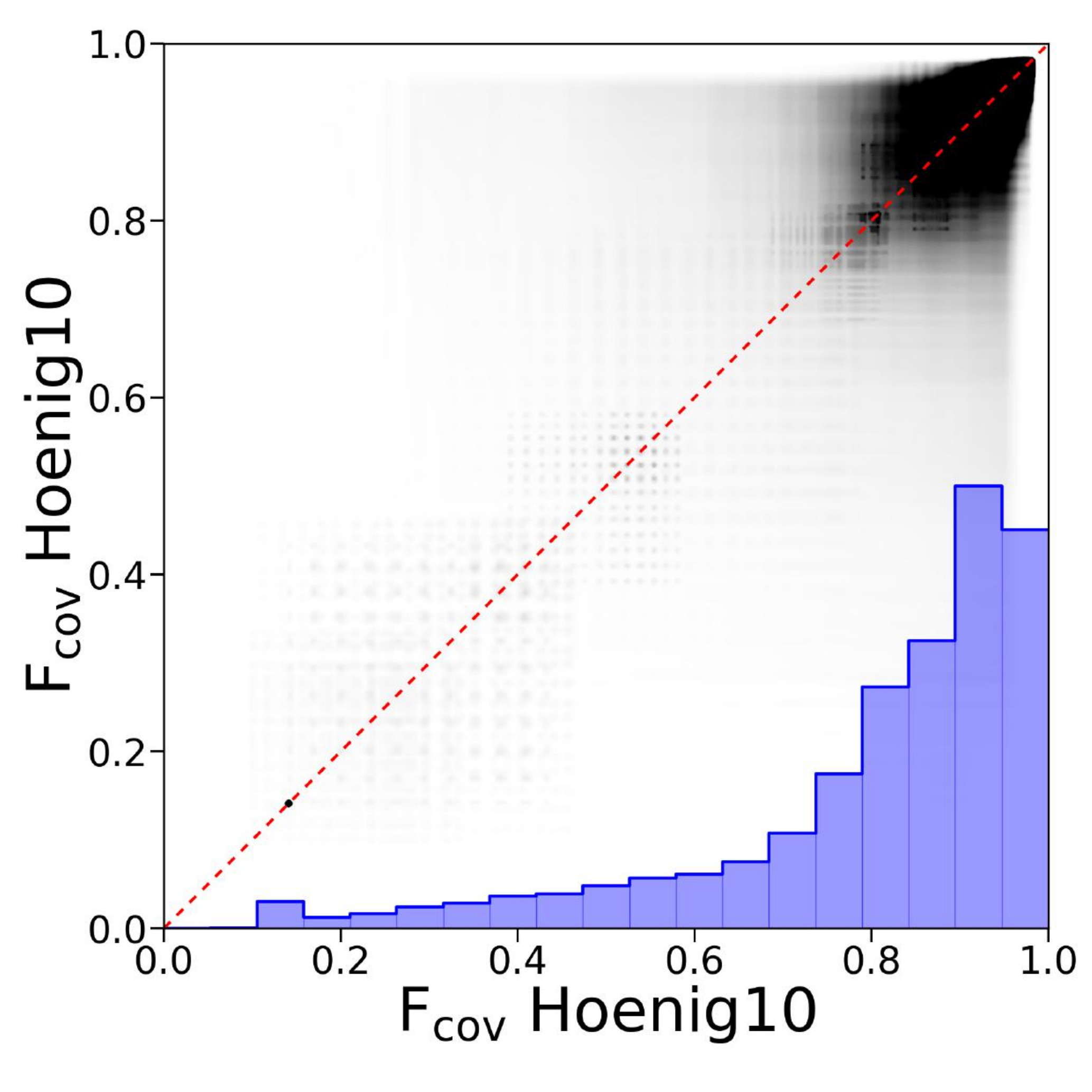}
\includegraphics[width=0.68\columnwidth]{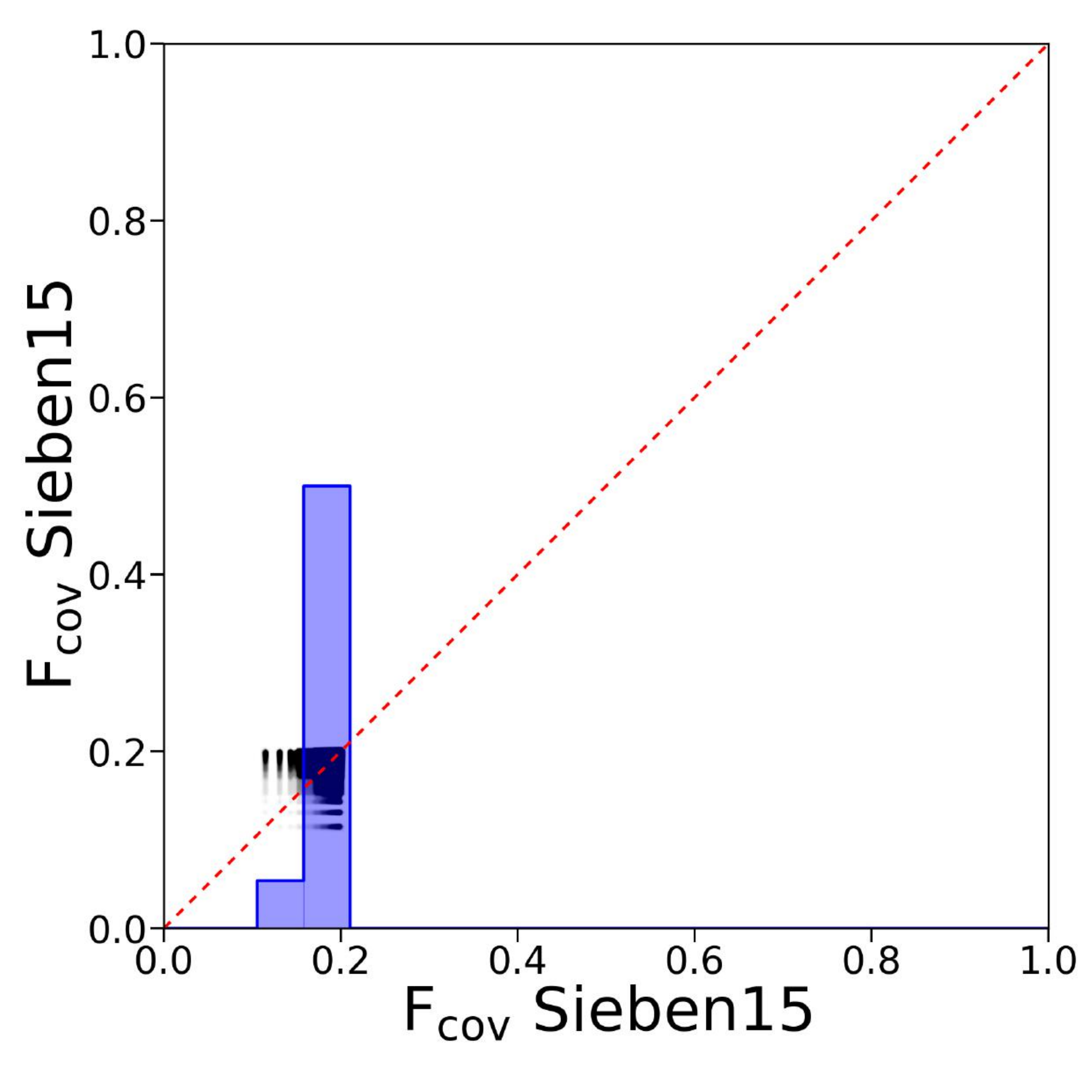}
\includegraphics[width=0.68\columnwidth]{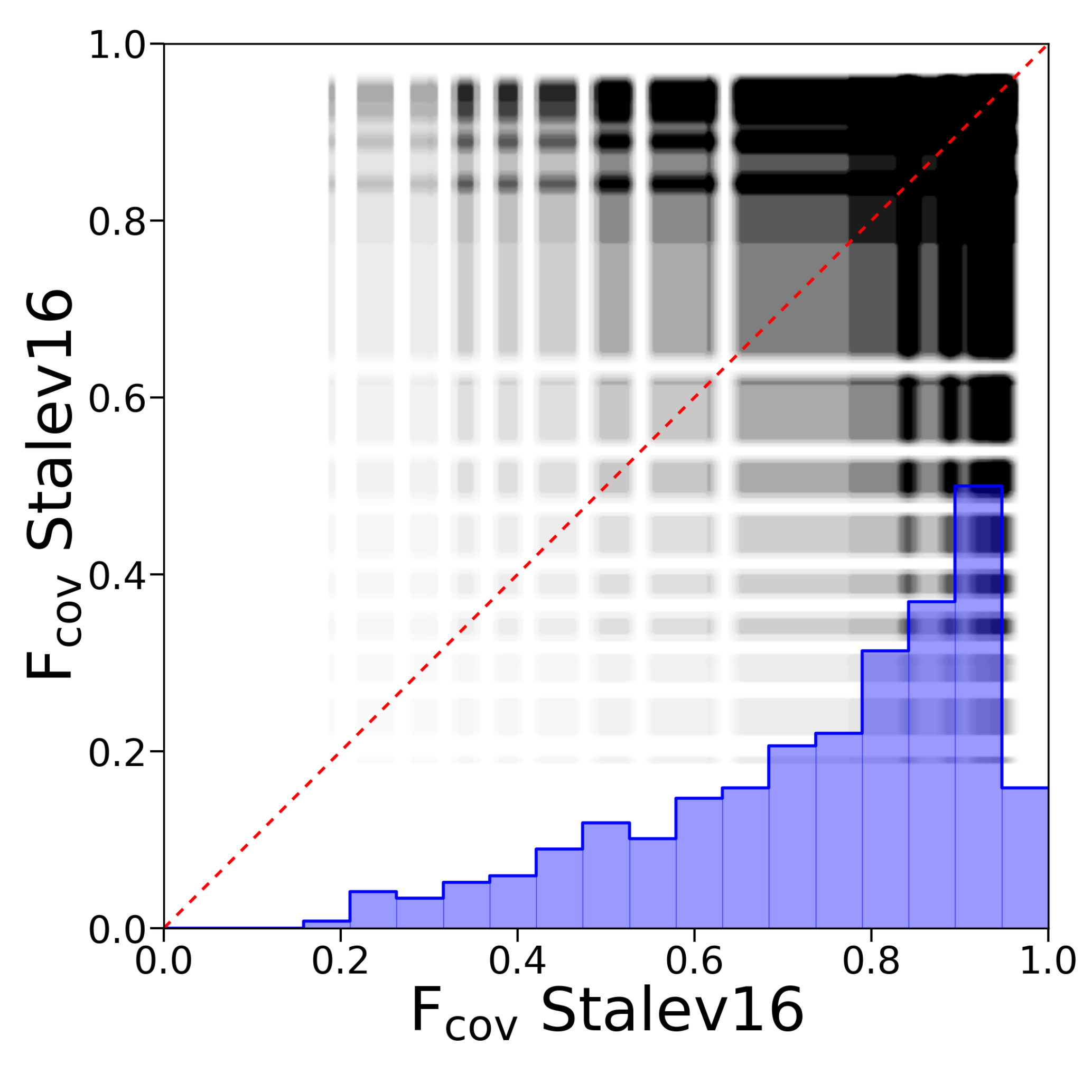}
\includegraphics[width=0.68\columnwidth]{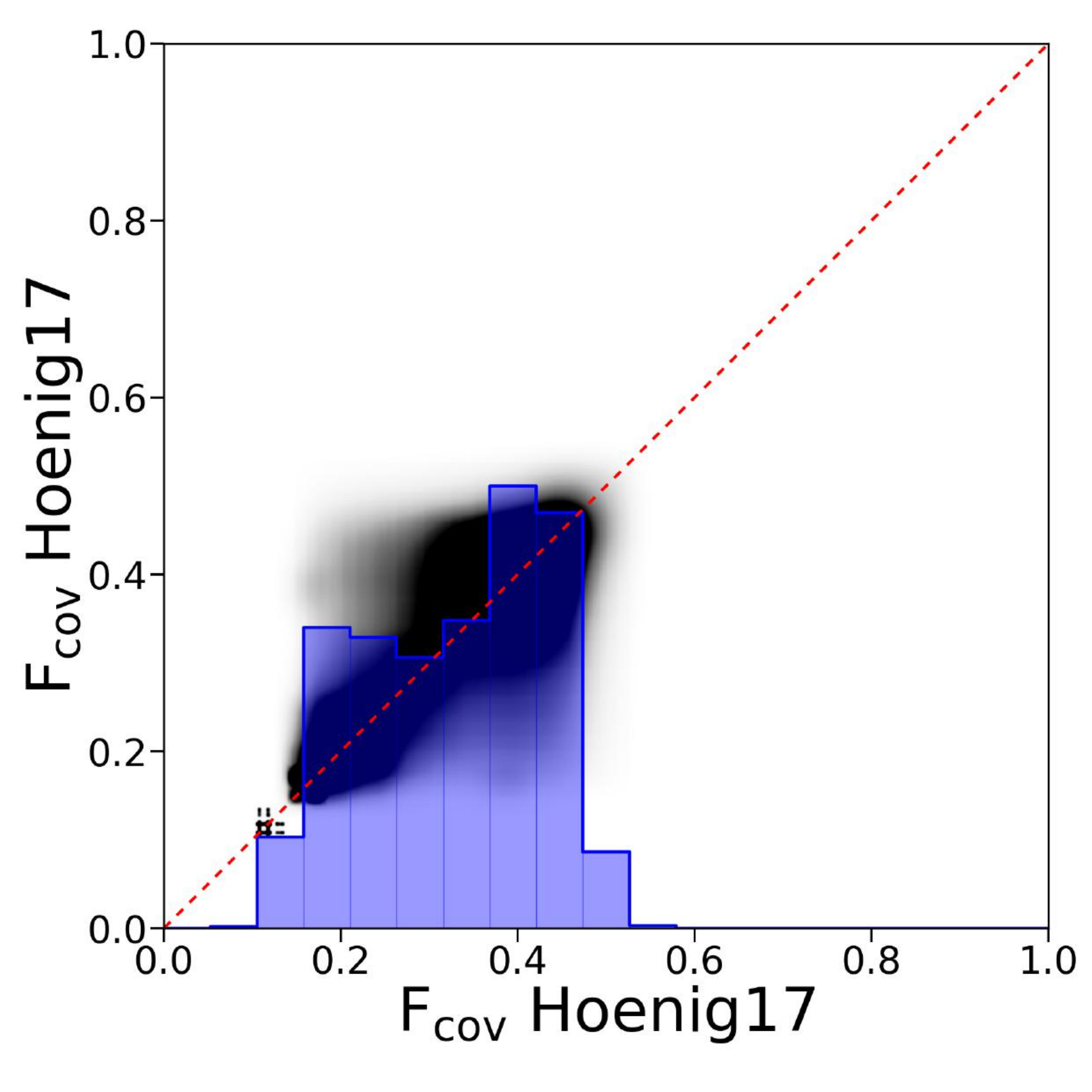}
\caption{Total PDF distribution of the covering factor for the full sample. The blue-filled histograms show the covering factor distribution per model (maximum of the distribution scaled to the half of the y axis of the plot). The dotted line shows the one-to-one relationship expected for the best accuracy determination between parameters.}
\label{fig:fcovselffit}
\end{flushleft}
\end{figure*}

\begin{figure*}[!ht]
\begin{flushleft}
\includegraphics[width=0.68\columnwidth]{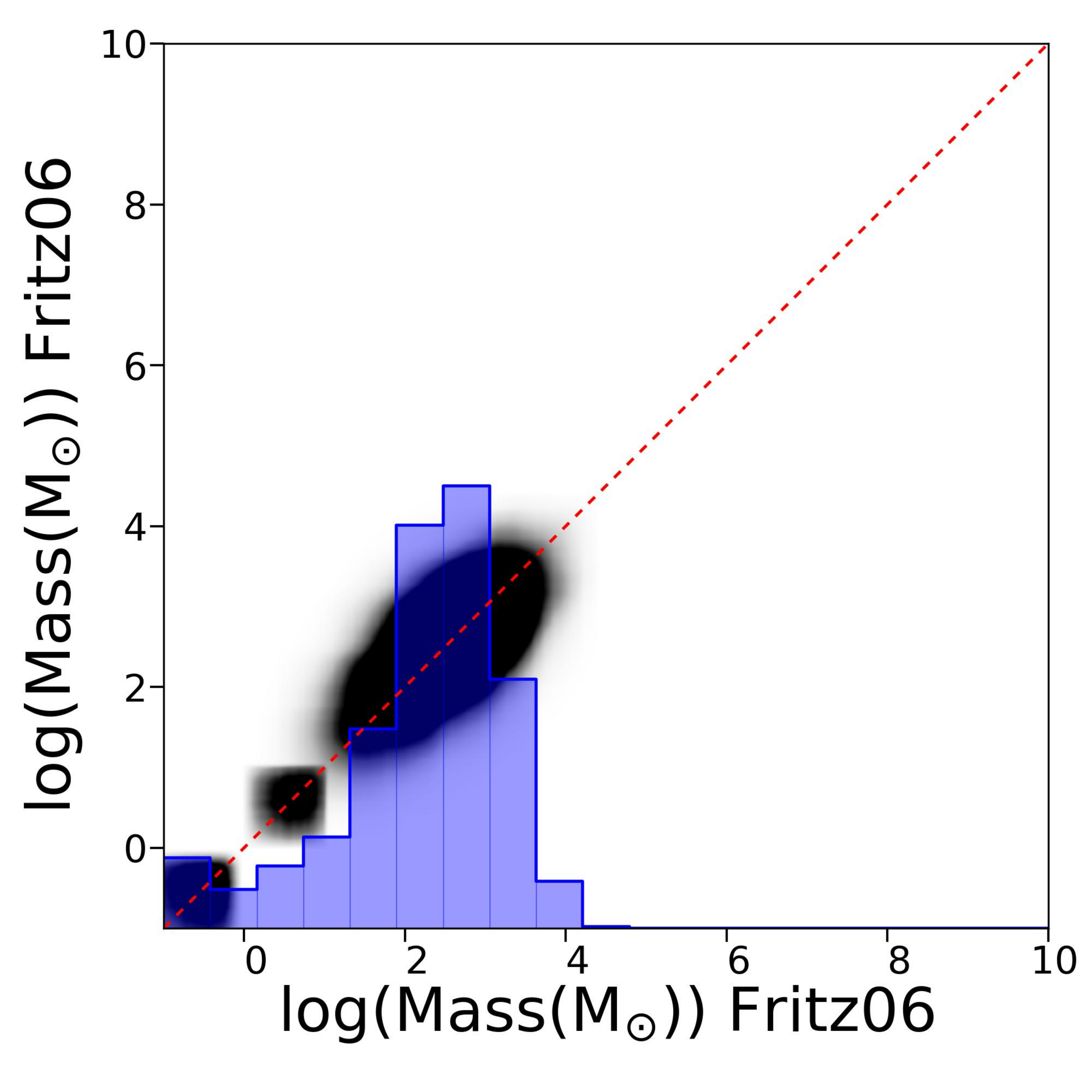}
\includegraphics[width=0.68\columnwidth]{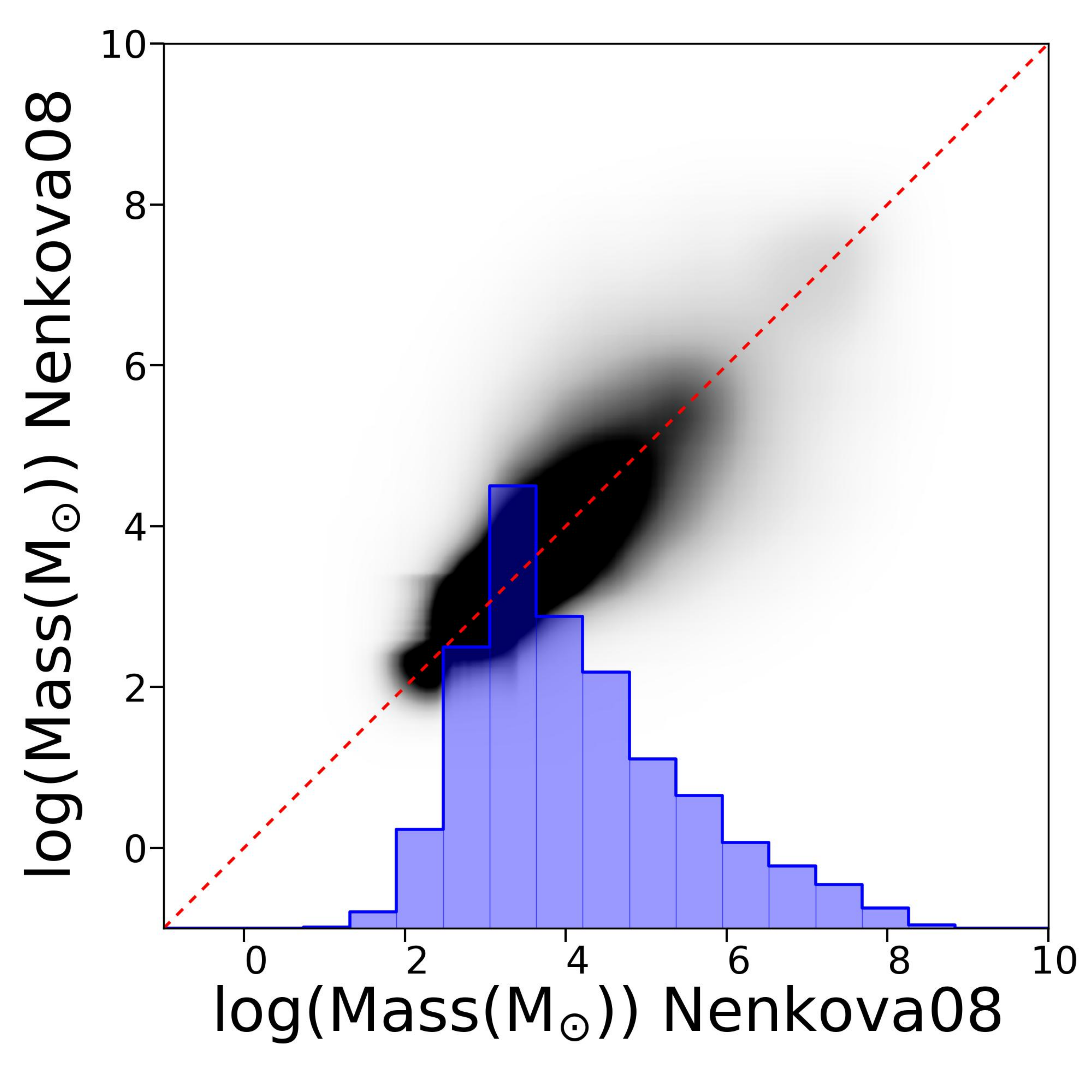}
\includegraphics[width=0.68\columnwidth]{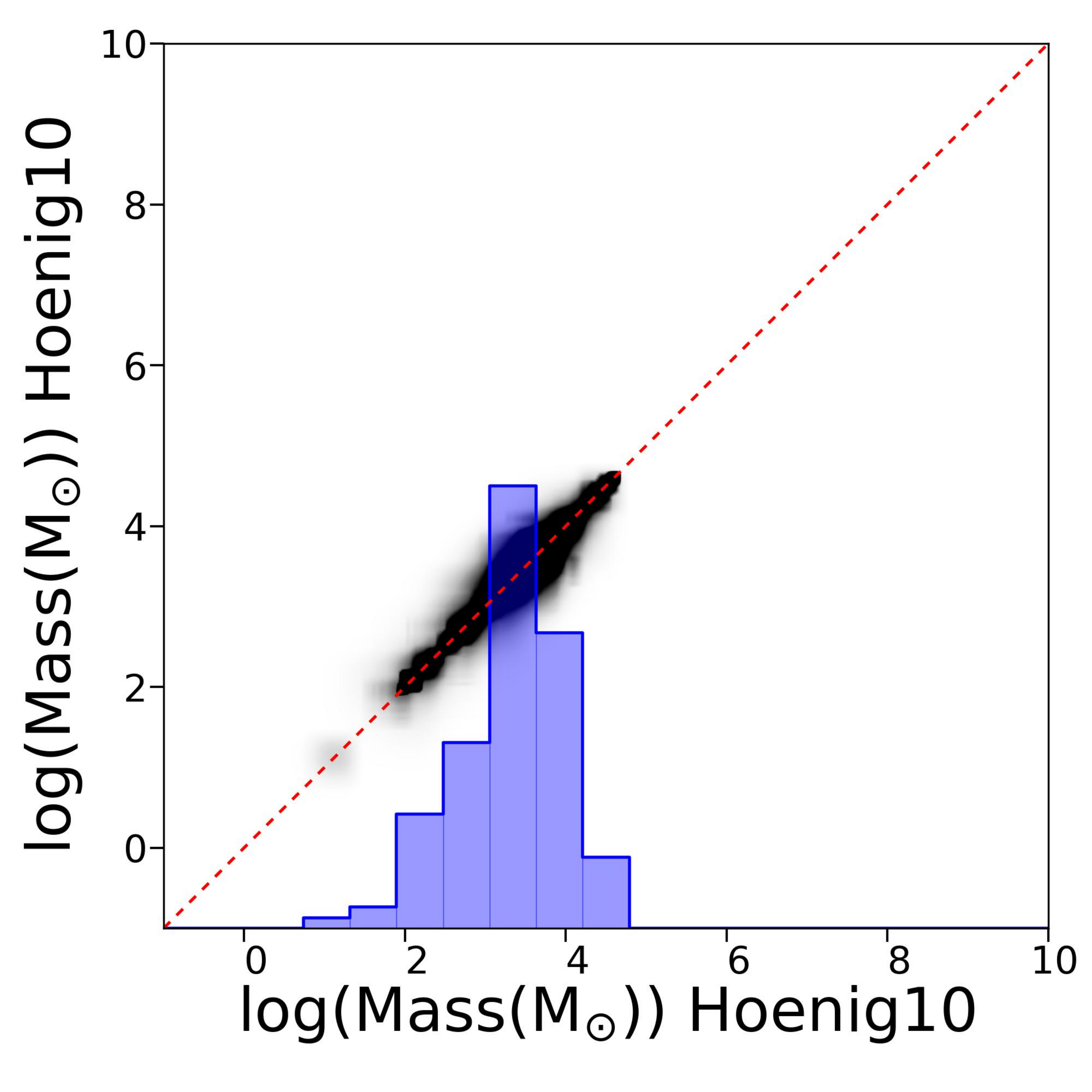}
\includegraphics[width=0.68\columnwidth]{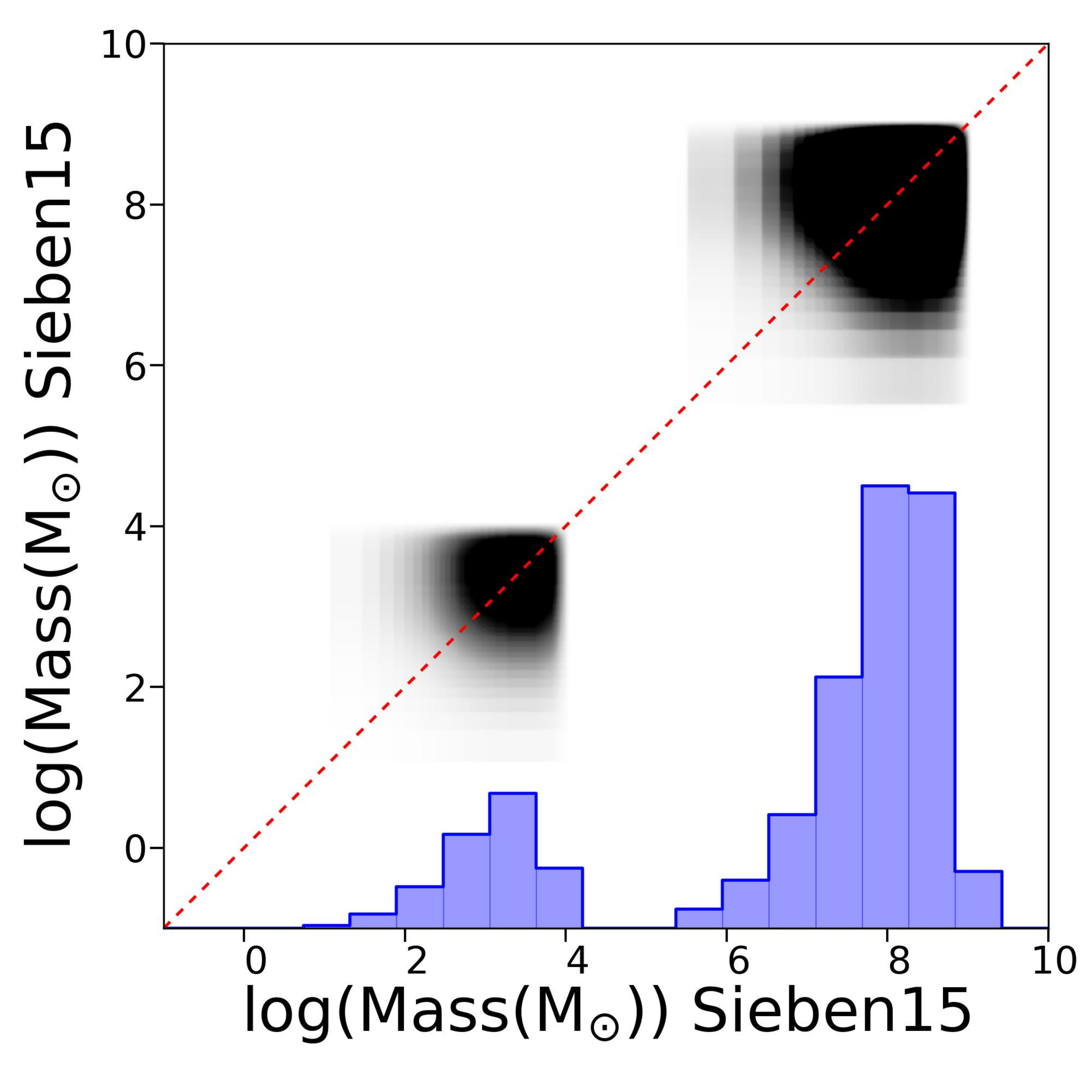}
\includegraphics[width=0.68\columnwidth]{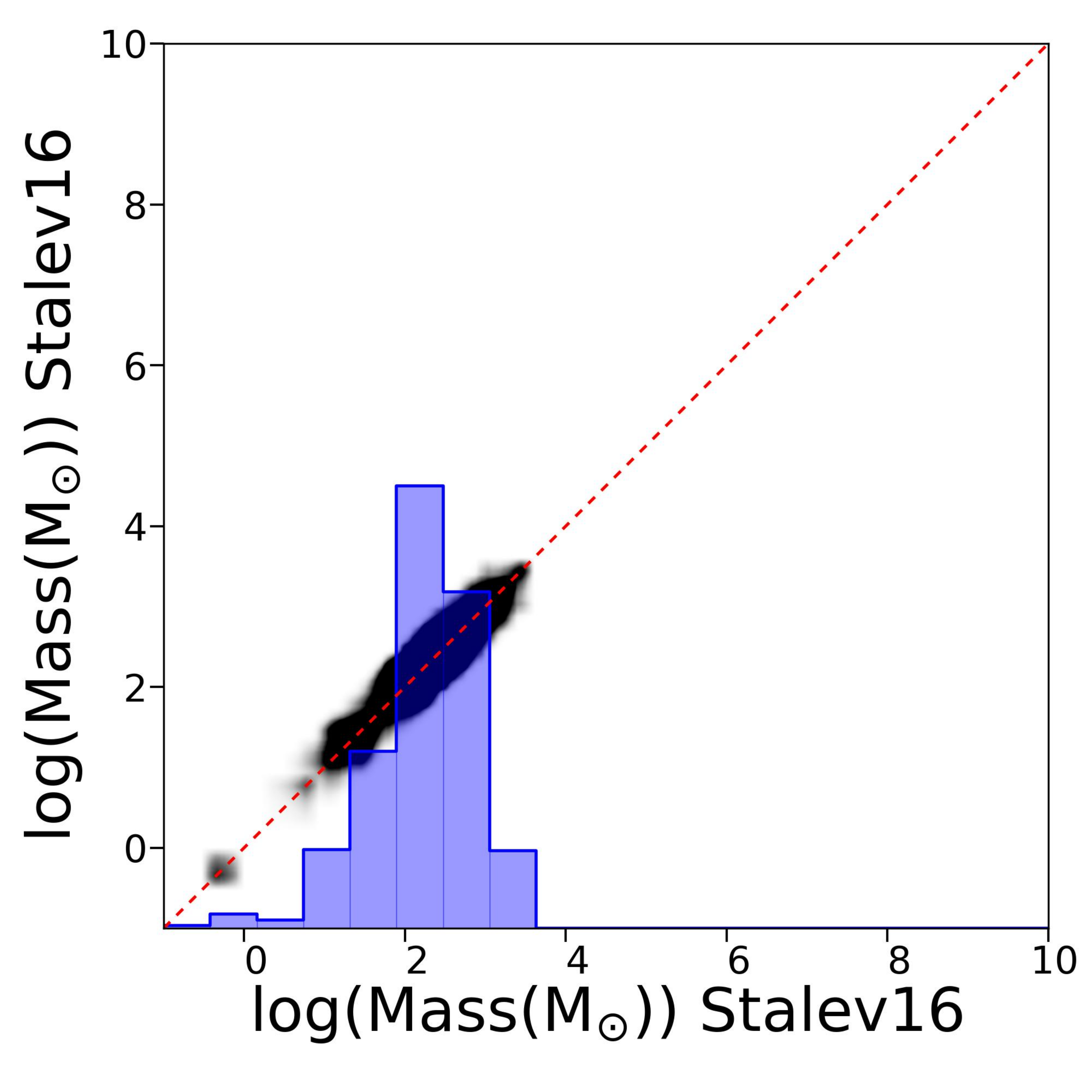}
\includegraphics[width=0.68\columnwidth]{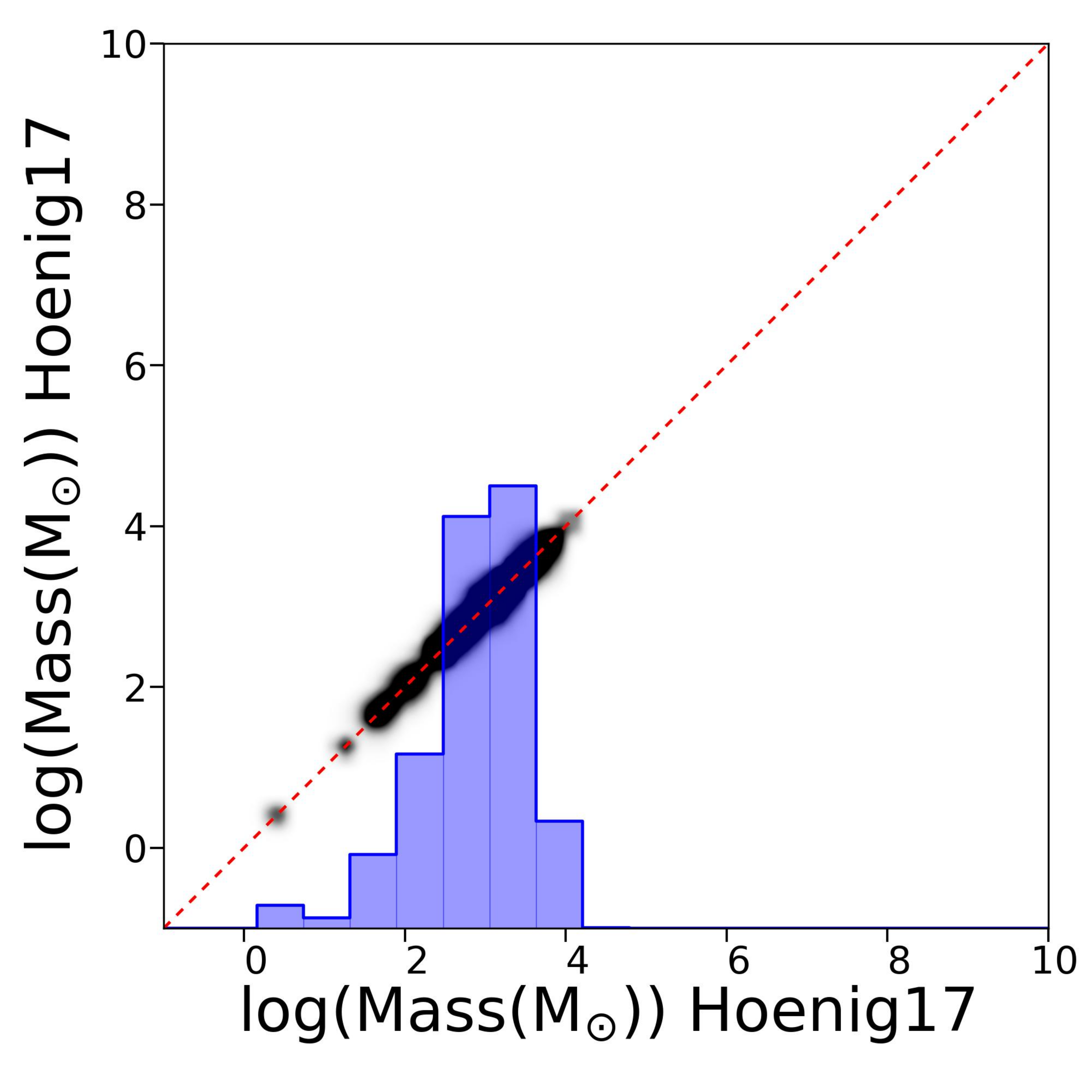}
\caption{Parameter versus parameter plot using the sample PDF distributions for the total dust mass. The blue-filled histograms show the total dust mass distribution per model (maximum of the distribution arbitrarily scaled to $\rm{log(M_{dust})=4.5}$ for clarity of the plot). The dotted line shows the one-to-one relationship expected for the best accuracy determination between parameters.}
\label{fig:massselffit}
\end{flushleft}
\end{figure*}

\begin{figure*}[!ht]
\begin{flushright}
\includegraphics[width=0.68\columnwidth]{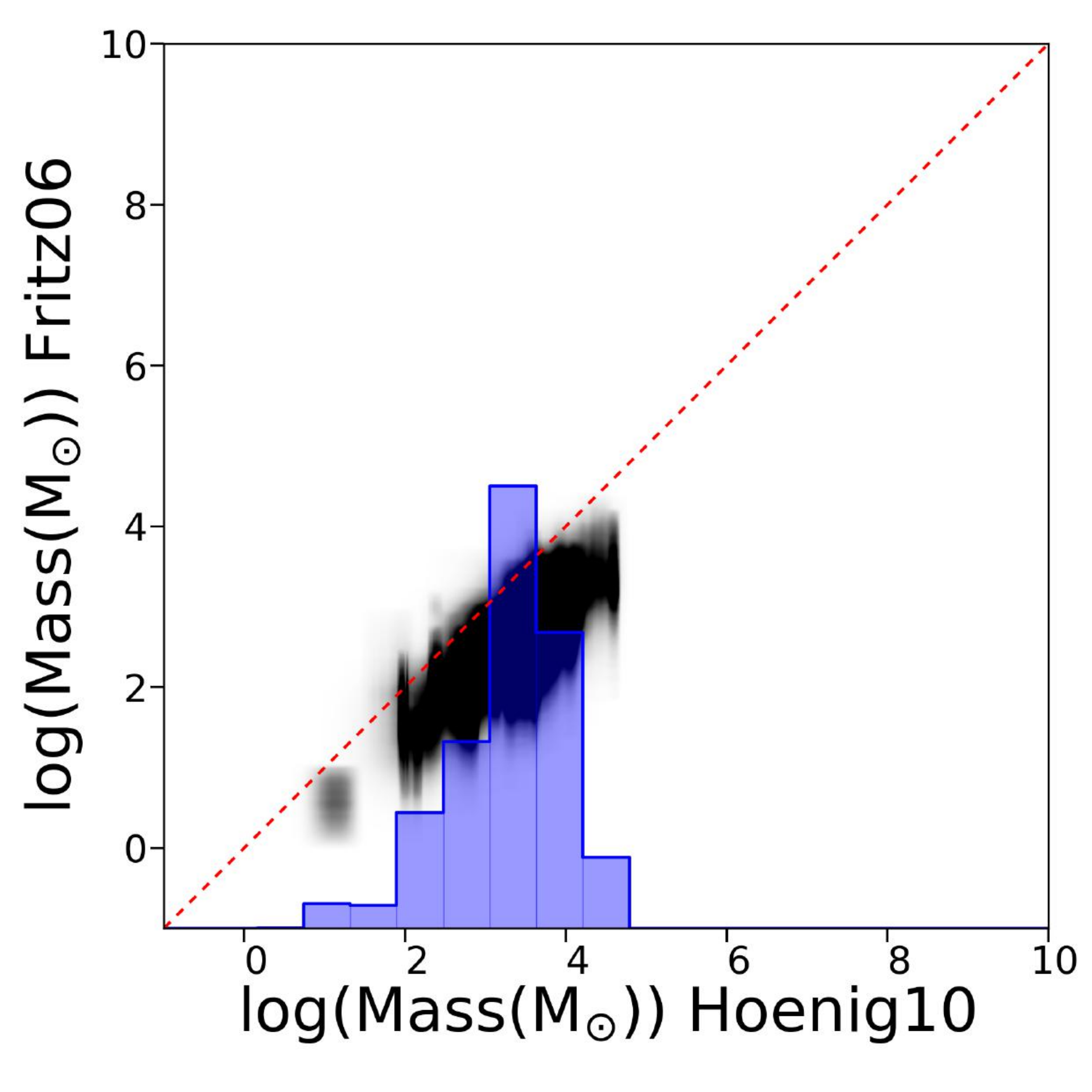}
\includegraphics[width=0.68\columnwidth]{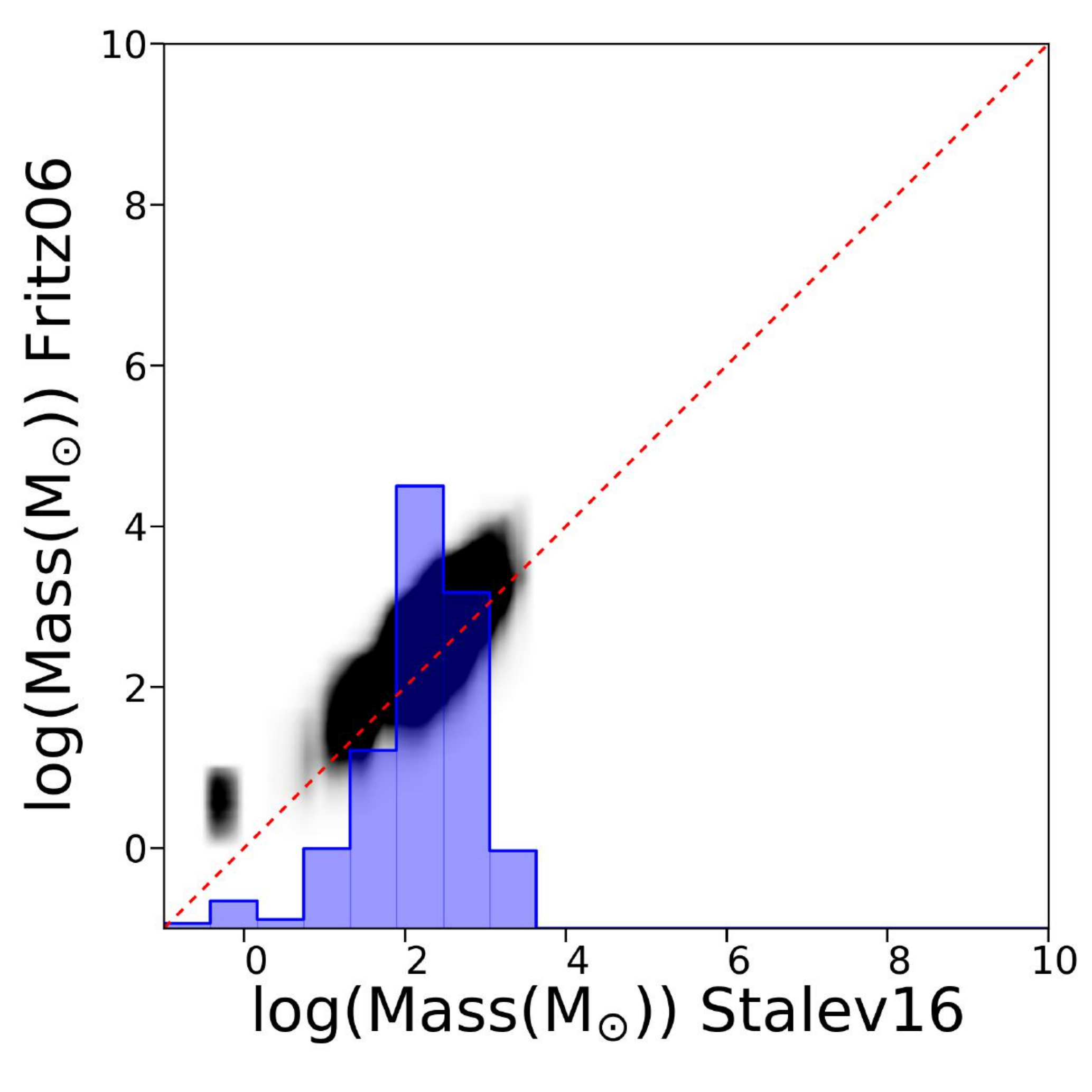}
\includegraphics[width=0.68\columnwidth]{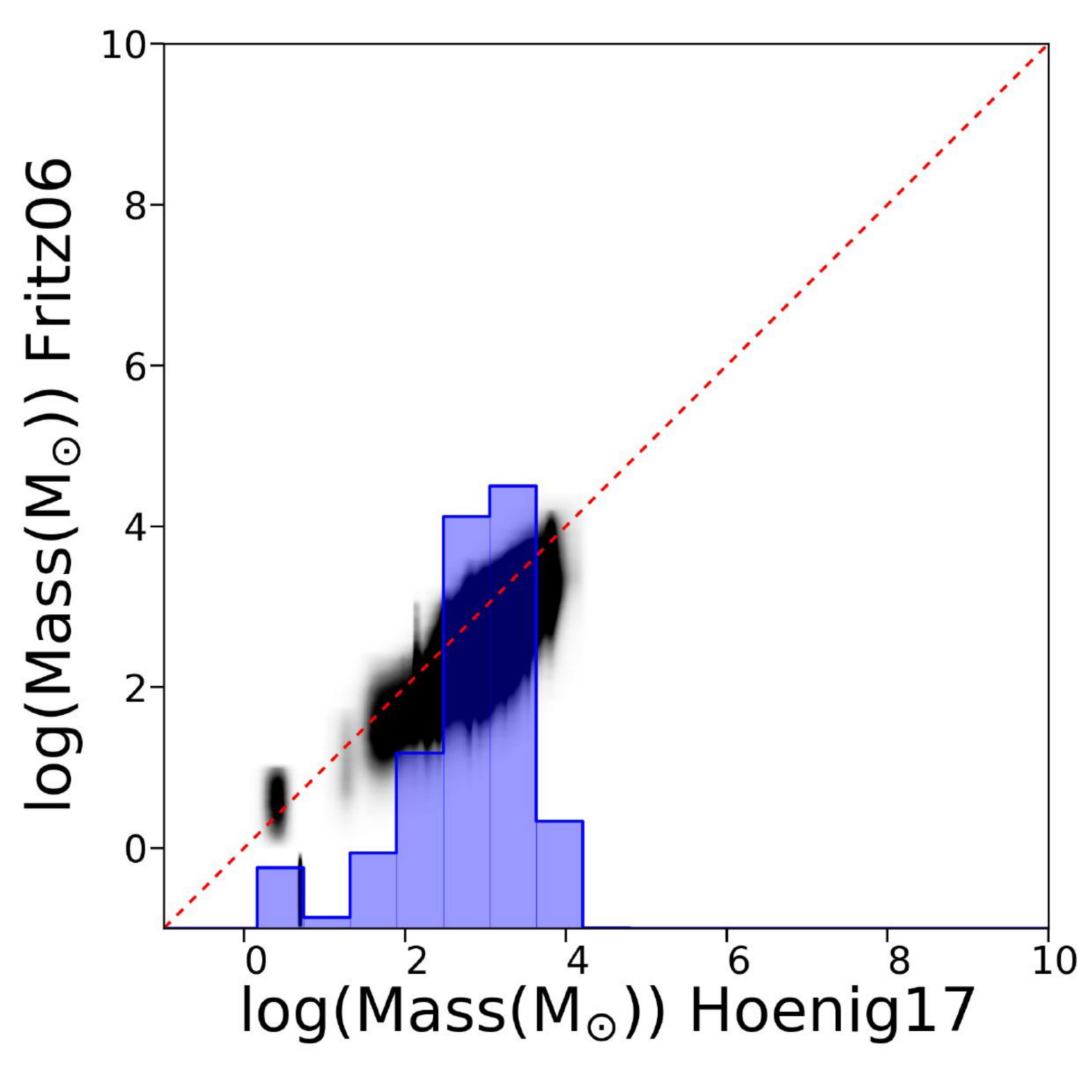} 
\includegraphics[width=0.68\columnwidth]{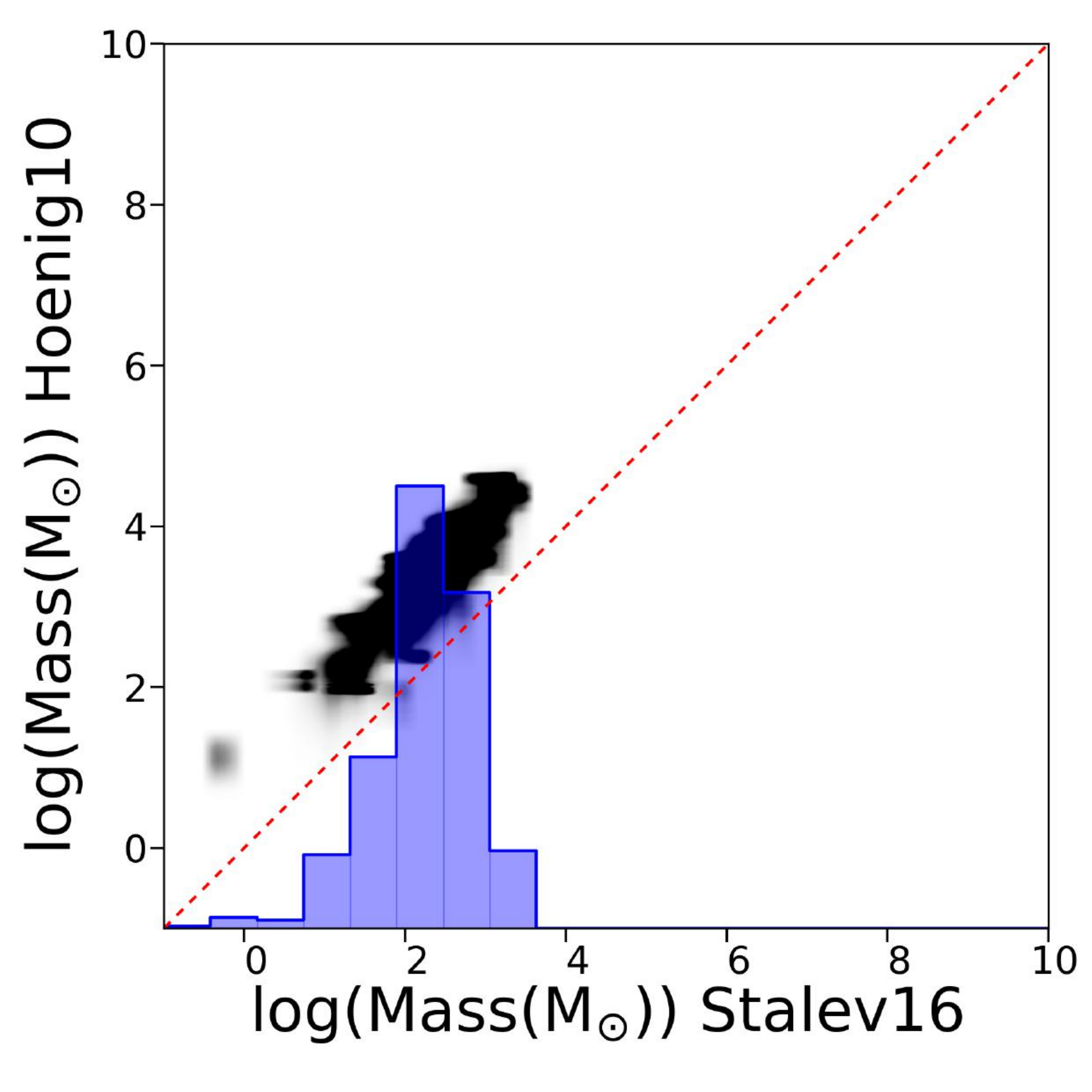} 
\includegraphics[width=0.68\columnwidth]{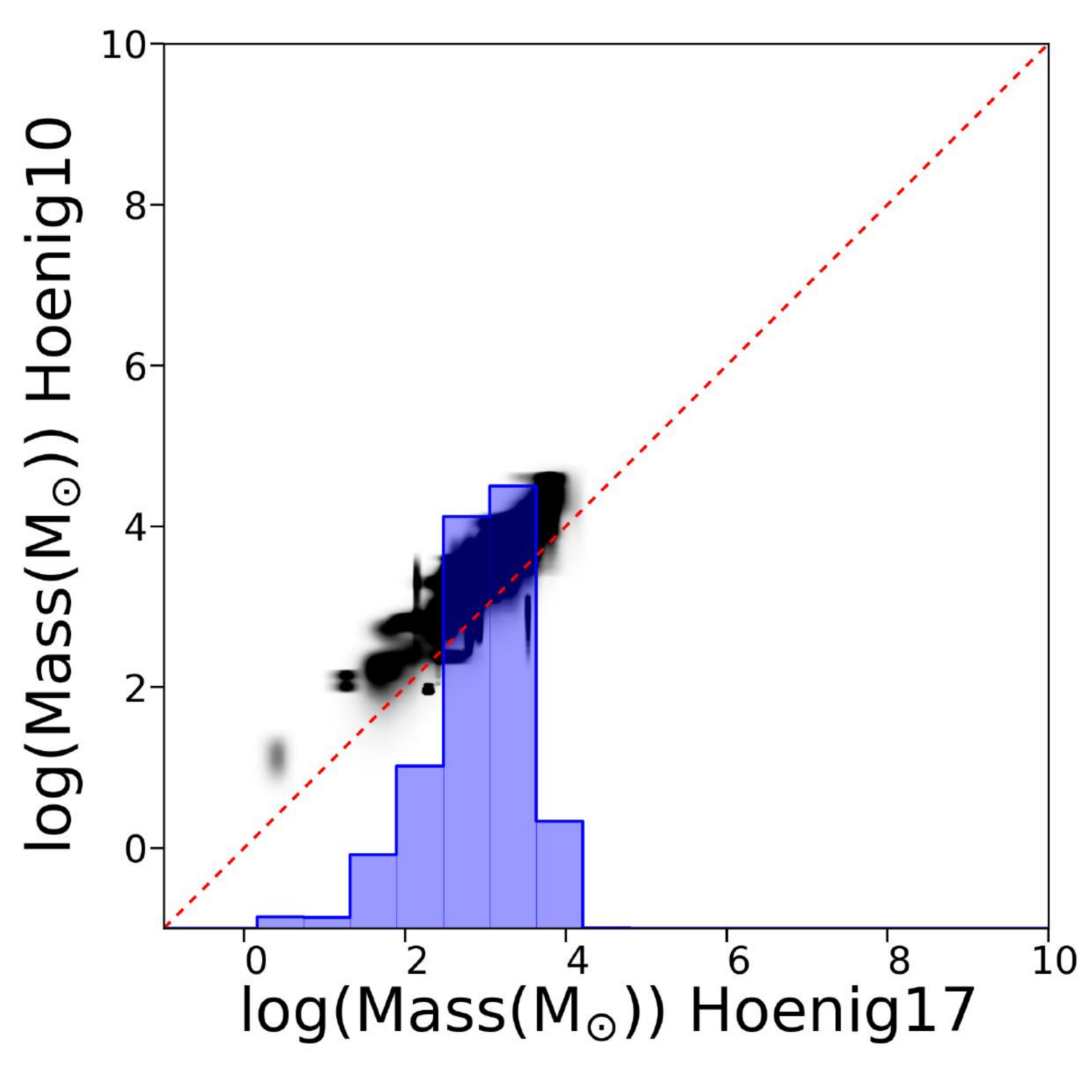} 
\includegraphics[width=0.68\columnwidth]{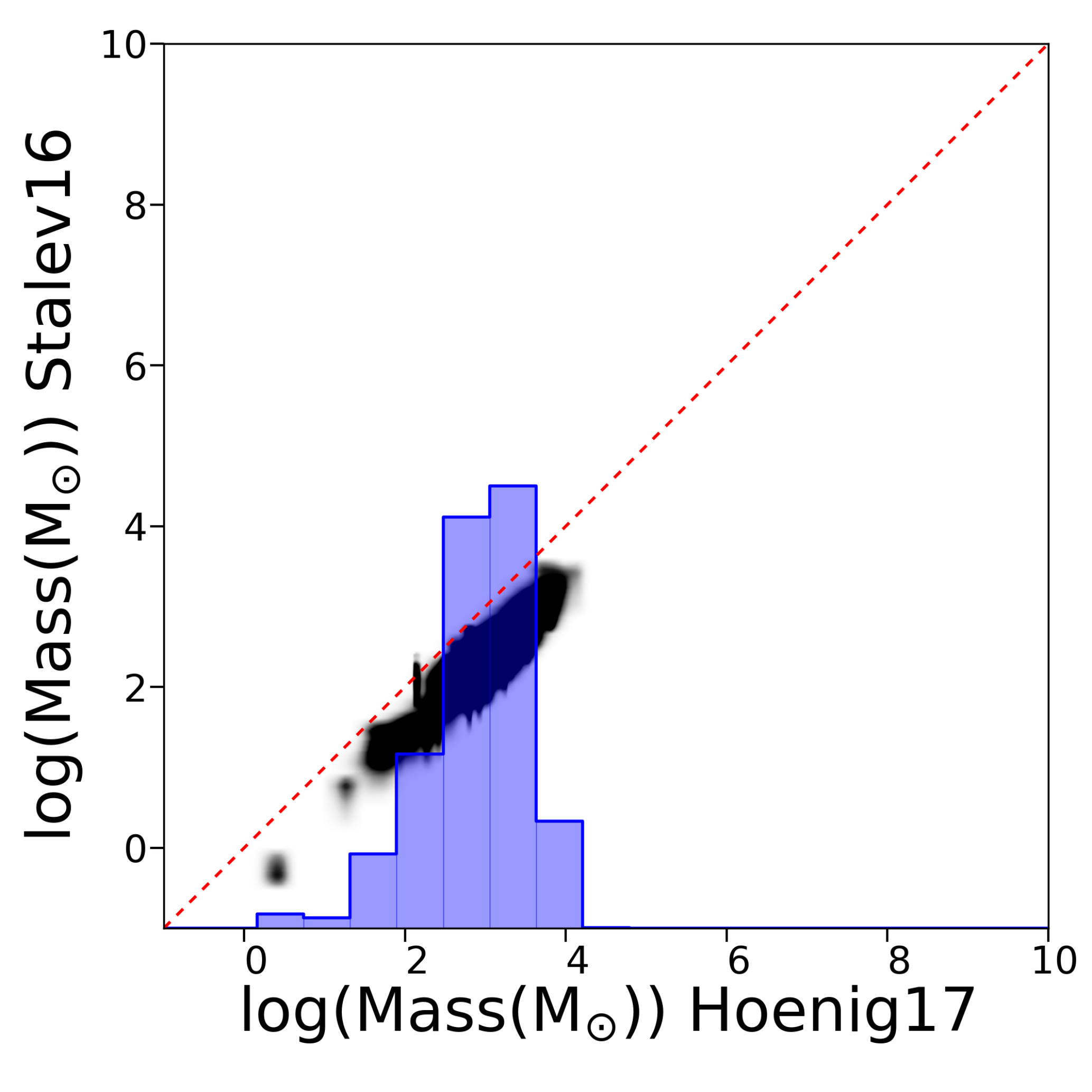}
\caption{Correlations found among models PDF distribution for the sample total dust mass. The blue-filled histograms show the total dust mass distribution per model (maximum of the distribution arbitrarily scaled to $\rm{log(M_{dust})=4.5}$ for clarity of the plot). The dotted line shows the one-to-one relationship expected for the best accuracy determination between parameters.}
\label{fig:massotherfit}
\end{flushright}
\end{figure*}

\subsection{Derived quantities: outer radius, covering factor, and total dust mass}

Although the values obtained for the individual parameters might not be well constrained (and, as shown above, the results depend on the model), some derived quantities could still be robust \citep[e.g. the covering factor as suggested by][]{Ramos-Almeida11}. We explore here three derived quantities with strong physical significance: (1) outer radius of the torus; (2) total AGN dust mass; and (3) covering factor. We produced the PDF for each quantity, model, and object deriving it from the PDFs of the individual parameters involved.

We provide below the results for these derived quantities. As a general note, the following analysis was performed using the entire sample but we found the same using only good spectral-fit results (i.e. $\rm{\chi^2/dof<2}$) or using AGN-dominated objects only. Furthermore, no significat differences were found on the distribution of the derived quantities by splitting the sample into AGN types\footnote{We did not split the sample into bolometric luminosity bins because all these quantities depend on the luminosity itself.}. Finally, we have found awkward distributions for [Sieben15]. In particular, it shows a narrow range of covering factors and unrealistic dust masses ($\rm{log(M_{dust})>6}$) for a large fraction of sources. 

\subsubsection{Outer radius of the torus $\rm{R_{out}}$}

The $Y$ parameter is fixed to a large value for all the SEDs in [Hoenig10], [Sieben15], and [Hoenig17]. Therefore, deriving $\rm{R_{out}}$ makes no sense for these three models. We derived $\rm{R_{out}}$ using the $Y$ parameter PDF for [Fritz06], [Nenkova08], and [Stalev16]. Fig.\,\ref{fig:routselffit} shows the parameter versus parameter plots for the $\rm{R_{out}}$ PDFs of the full sample. The correlation coefficient is $r>0.8$ in all cases, except for [Fritz06] versus [Nenkova08] (r=0.5). The three models rely on the assumption that their inner radius is linked to the dust sublimation radius, and therefore linked to the AGN bolometric luminosity\footnote{Note that the bolometric luminosity is derived from the 2-10 keV X-ray luminosity, using the relation proposed by \citet{Marconi04}.}. Therefore, the linear relation might be the result of the bolometric luminosity being used in all the models to compute $R_{out}$. More interestingly, [Nenkova08] shows a large range of values $\rm{R_{out}<12\,pc}$, while [Fritz06] (except for a small fraction of the sample with $\rm{8\,pc<R_{out}<10\,pc}$) and [Stalev16] show a small range of values $\rm{R_{out}<3\,pc}$. In general, $\rm{R_{out}}$ is consistent among the three models for small values (i.e. when $\rm{R_{out}<1\,pc}$).

\subsubsection{Covering factor $\rm{F_{cov}}$} 

We computed $\rm{F_{cov}}$ as the unity minus the escape probability, i.e. $\rm{\int e^{-\tau_{\nu}(los)}cos(\theta)d\theta}$ where $\rm{\theta}$ is the azimuthal angle and $\rm{\tau_{\nu}(los)}$ is the line of sight opacity. The latter is computed from the equatorial opacity and the density distribution for smooth models, and from the distribution of clouds for clumpy models. The integration was made with the {\sc quad} function within {\sc scipy} (python 2.7). Fig.\,\ref{fig:fcovselffit} shows the full sample $\rm{F_{cov}}$ distributions (through the parameter versus parameter plot) for each model, respectively. $\rm{F_{cov}}$ is poorly constrained for all the models (a non-negligible probability above 1\% of $\rm{F_{cov}}$ is shown at any value) except for [Hoenig10] and [Hoenig17] (see Fig.\,\ref{fig:fcovselffit}). Indeed, only [Hoenig10] and [Hoenig17] show a marginal correlation ($\rm{r\sim 0.6}$), suggesting that the parameter is constrained. Furthermore, most models show a maximum of $\rm{F_{cov}}$ distribution at large values with a tail reaching $\rm{F_{cov}\sim 0.1}$. The exceptions are [Sieben15] ($\rm{0.1<F_{cov}<0.2}$) and [Hoenig17] ($\rm{0.1<F_{cov}<0.6}$).

\subsubsection{Total dust mass $\rm{M_{dust}}$} 

We computed $\rm{M_{dust}}$ integrating the density distribution function for each model within the dusty structure volume. 
We compute the parameter versus parameter plot for $\rm{M_{dust}}$ to evaluate if the mass is restricted (Fig.\,\ref{fig:massselffit}) and if there is a linear relation among $\rm{M_{dust}}$ derived from different models (Fig.\,\ref{fig:massotherfit}). [Hoenig10], [Stalev16], and [Hoenig17] show a total mass that is roughly twice better restricted than the other three models. In general, [Fritz06], [Hoenig10], [Stalev16], and [Hoenig17] overlap in their estimates of $\rm{M_{dust}}$ with a range $\rm{0<log(M_{dust})<4}$. [Nenkova08] tends to show larger $\rm{M_{dust}}$ ($\rm{2<log(M_{dust})<8}$) compared to other models. When using [Sieben15], we recover a bimodal distribution of $\rm{M_{dust}}$ with two ranges, one overlapping with other models ($\rm{1<log(M_{dust})<4}$) and another one showing large $\rm{M_{dust}}$ ($\rm{6<log(M_{dust})<9}$). $\rm{M_{dust}}$ correlates among models ($\rm{r>0.8}$) for [Fritz06], [Hoenig10], [Stalev16], and [Hoenig17]. These estimates show a shift (within a factor of 10) with the lowest values for [Stalev16], and increasing for [Fritz06], [Hoenig17], and [Hoenig10].

\section{Discussion} \label{sec:discussion}

\subsection{Best model}\label{sec:discussion:bestmodel}

The model that better reproduces the \emph{Spitzer} spectra among the six models analysed is [Hoenig17]. This model is built on the idea that the dust around the AGN is distributed in an inflowing disk and an outflowing wind. This model is claimed to better reproduce the 3-5\,$\rm{\mu m}$ bump observed in many Sy1, which clumpy models do not \citep{Garcia-Gonzalez17}. Indeed, the disk-wind model [Hoenig17] is particularly good at describing Sy1 in our sample, while Sy2 (and, despite their low number, beamed AGN) are equally fitted to the clumpy torus model [Nenkova08] (see Fig.\,\ref{fig:data:goodfitstat2}). As shown in Figs.\,\ref{fig:genfit1} and \ref{fig:genfit2}, [Hoenig17] is good at describing near- and mid-infrared slopes although fails to describe the far-infrared slopes in our sample. Far-infrared slopes are better reproduced by [Fritz06], [Nenkova08], and [Stalev16].

The disk-wind model is consistent with an old interpretation of the dusty region in which the torus might be the dusty portion of an outflowing wind coming off the accretion disk \citep{Efstathiou95,Elitzur06,Suganuma06,Nenkova08B,Netzer15,Ramos-Almeida17,Lyu18}, which naturally extends the clumpy behavior of the BLR to further distances \citep{Risaliti02}. Indeed, a significant part of the mid-infrared emission appears to come from the ionization cones \citep{Braatz93,Cameron93,Hoenig13,Asmus16}. 

We also found that the best fitting model could be different for high and low luminosity AGN, with a larger percentage of best fits for [Hoenig17] for high luminous objects, while a larger number of low luminous AGN are better fitted to the clumpy torus model [Nenkova08] (see Fig.\,\ref{fig:data:goodfitstat}), also consistent with our findings when comparing best models describing Sy1 and Sy2. This result is expected if outflows dominate the high-luminosity end of AGN, while inefficient accretion flows are not able to produce these outflows at the low-luminosity end \citep[also predicted by][]{Elitzur06}. This is also consistent with outflows producing the short-term time variations associated to changes on the obscuration measured at X-rays in Sy1 \citep{Gonzalez-Martin18} and with the dependence on the covering factor on the Eddington rate shown by \citet{Ricci17}. Unfortunately, we were not able to explore the dependence on our results on the Eddington rate because we found Eddington rates (or BH masses) only for $\rm{\sim}$50\% of the sample (and they are biased toward the high accreting sources). Interestingly, our small sample of beamed AGN are reproduced equally well by [Nenkova08] and [Hoenig17] as it is the case for Sy2, indicating that some face-on AGN lack of the dusty wind signatures. 

\subsection{Are the models good enough?}\label{sec:discussion:goodenough}

Although able to statistically describe a good fraction of the spectra (up to 50\% for [Hoenig17]) all the spectral fits show relatively large residuals (see Fig.\,\ref{fig:data:residuals}), indicating that the scenario is more complex than shown by current models. Interestingly, the two models better describing a large portion of the sample (the clumpy torus model [Nenkova08] and clumpy disk-wind model [Hoenig17]) are the ones with the largest produced SED libraries, (1,250,000 and 132,000 SEDs, respectively). Another extreme case is [Sieben15] which manages to provide good fits for less than 5\% of the spectra. This might be due to the small number of SEDs provided (3,600) and a fixed density distribution used for this model (which is also visible in the unrealistic large dust masses and narrow covering factor obtained using this model). Indeed, synthetic spectra showed that the model [Hoenig17] cannot be distinguished (using current or future facilities) from the clumpy model [Hoenig10] based solely on SED fitting (see Paper I). However, data do seem to prefer the latest wind version of the model [Hoenig17]. This might be due to a better coverage of the parameter space, including over 132,000 SEDs compared to 1,700 for [Hoenig10]. This embosses another important aspect when comparing models: modelers might want to provide well-sampled SED libraries to better compare them with data. It is also worth to mention that [Hoenig17] is better describing at the same time the near- and mid-infrared slopes (see Fig.\,\ref{fig:genfit1}), which might explain partially why this model is preferred against its partner model [Hoenig10].

The number of SEDs is not the only variable to be revisited when providing new SED libraries. Another important aspect is the parameter space, which in most cases is not appropriate for reproducing the SED diversity (see Figs.\,\ref{fig:genfit1}-\ref{fig:genfit3}). We show that many parameters tend to cluster towards their upper or lower limits provided by the SED library (see Table\,\ref{tab:parameterresults}). For instance, the parameter governing the dust radial distribution (either in clumps or smooth) could be expanded to find a better match to the data. Moreover, other parameters seem to be over sampled; e.g. the $Y$ parameter in the clumpy torus model [Nenkova08B] does not seem to require values larger than $Y=50$. This can be easily spotted in Figs.\,\ref{fig:genfit1}-\ref{fig:genfit3}, where many SEDs do not describe any AGN spectrum. Perhaps our conclusions are partially biased to the models with better sampled SED libraries.

\subsection{Goodness of parameter determination}\label{sec:discussion:parameter}

To decide which is the best model solution, attention should be paid to the actual parameters found and to the meaningful picture they should provide in terms of plausible physical description (see below). Furthermore, we have found misconceptions about the models. For example, the clumpy torus model [Nenkova08] is claimed to reproduce the spectra with smaller torus radii, consistent with the interferometric data that have constrained the radial extent of the dusty structure to 10 pc \citep[e.g. for Circinus galaxy, and Centaurus\,A,][respectively]{Tristram07,Meisenheimer07}. However, this claim is actually inconsistent with our results. The smooth model [Fritz06] and the two phase model [Stalev16] provide twice smaller $\rm{R_{out}}$ compared to the clumpy model [Nenkova08] for the very same data analyzed here. This is in fact the expected behavior because smooth distributions are able to stock larger amount of dust than clumpy distributions of dust in the same volume. The historical reason behind the large tori found for smooth torus is that pioneer works try to fit at the same time mid- and far-infrared AGN SEDs \citep[e.g.][]{Pier92}, being the latter dominated by dust heated by star-formation rather than the AGN. This resulted in unrealistic large tori.

Many attempts have been made to estimate the torus parameters using SED fitting, with photometric data \citep[e.g.][]{Ramos-Almeida09,Ramos-Almeida11} or adding them to ground-based spectra \citep[e.g.][]{Alonso-Herrero11}. We demonstrate in Paper I that full-coverage 5-30\,$\rm{\mu m}$ spectra are able to constrain all the parameters for most of the models studied here, with the exception of the wind model [Hoenig17]. Consistent with that, we found that the spectral coverage of the low resolution \emph{Spitzer}/IRS spectra can restrict $\rm{\sim80\%}$ of the parameters within $\rm{\sim10-20\%}$ of the space parameter.

\citet{Ramos-Almeida14} used Bayesian techniques to constrain the parameters of the clumpy dusty model [Nenkova08]. They concluded that the torus width, viewing angle, and distribution of the clouds could be constrained with near- and mid-infrared photometry only, but mid-infrared spectroscopy is necessary to restrict the posterior distributions of the number of clouds and their optical depth. Furthermore, they show that 8-13\,$\rm{\mu m}$ ground-based spectra alone reliably restrict some of these parameters \citep[see also][]{Alonso-Herrero11}. Using the full coverage of the low-resolution IRS/\emph{Spitzer} spectra, we show that all the parameters of [Nenkova08] can be restricted, being $q$, $Y$, and $\tau_{\nu}$ the best constrained. The range $\rm{R_{out}}$ found here when using the clumpy [Nenkova08] is fully consistent with that derived using Bayesian techniques for low-luminous AGN \citep{Gonzalez-Martin17}, Seyferts \citep{Ramos-Almeida09,Alonso-Herrero11}, and QSOs \citep{Martinez-Paredes17}.

\citet{Fritz06} also studied the parameter restriction using infrared data by considering all the solutions with $\rm{\chi^2/dof < \chi^2/dof (best) + 3}$ (being $\rm{\chi^2/dof (best)}$ the minimum $\rm{\chi^2}$ statistics). They found that $\rm{R_{out}}$ differs up to a factor of 3 while the parameters associated to the dust density distribution are better constrained. This is consistent with our finding where $\rm{R_{out}}$ and the covering factor strongly depend on the selection of the model while $\rm{M_{dust}}$ is better constrained irrespective of the model. We further explore the degeneracy of the parameters within a model, finding no indication of linear relationships among them.

Note, however, that the resulting parameters are themselves dependent on the model assumed. In fact, we did not find any equivalence between parameters of different models, although expected in some cases (e.g. the viewing angle toward the dusty structure, see also Paper I). Therefore, we strongly suggest to evaluate the goodness of the model using a set of models before drawing conclusions from the resulting parameters. 

However, there are two major concerns to derive the model parameters using \emph{Spitzer}/IRS spectra: suitability of the models at describing the data and host galaxy dilution due to low spatial resolution data. The \emph{Spitzer}/IRS data (with a spatial resolution of $\rm{\sim}$3 arcsec at 10\,$\rm{\mu m}$) are prone to contamination from the host galaxy or dust heated by circumnuclear star-formation \citep{Fritz06,Netzer07,Siebenmorgen15}. Indeed, this is the result presented in this paper, where 30-50\% of the sample (depending on the model used) is contaminated by more than 50\% continuum emission not associated with the AGN (see Fig.\,\ref{fig:data:goodfitstat}). These results are consistent with previous results \citep{Rodriguez-Espinosa87,Dultzin88,Dultzin94,Dultzin96}. Decomposition methods as those used in this paper, and previously in \citet{Netzer07} \citep[see also][]{Hao07,Hernan-Caballero15} are then needed to decontaminate the torus component from circumnuclear contributors. However, as shown in this paper, when this contribution is large (stellar component contributing at least 50\% compared to the torus component at 5\,$\rm{\mu m}$, or dust heated by star-formation contributing at least 50\% compared to the torus component at 30\,$\rm{\mu m}$) dilution strongly influences the parameter determination. High spatial resolution observations are needed for the best isolation of the dust associated the AGN from other contributors, as those provided with ground-based facilities \citep{Ramos-Almeida09,Alonso-Herrero11,Fuller16,Martinez-Paredes17,Garcia-Bernete19}, the novel combination of VLTI+GRAVITY and the future \emph{JWST}. Note however that both N- and Q-bands ground-based spectra are available only for NGC\,1068. Interestingly, both bands are not easily fitted with a single model (Victoria-Ceballos in prep.) 

One way to solve the parameter space determination is the use of interferometric observations together with SED fitting \citep{Hoenig06,Hoenig10B}. While interferometry provides the most direct access to the dust AGN structure, observations are limited to the brightest AGN with current facilities. Indeed, ALMA is providing for the first time the kinematics of the cold dust, finding large dust structures kinematically decoupled from the host galaxy \citep{Garcia-Burillo16,Gallimore16,Alonso-Herrero18,Combes19}. Alternatively, we propose the use of X-ray reflected component (associated to the AGN dust structure) together with the mid-infrared spectrum to infer the torus properties (Esparza-Arredondo et al. in prep.). 

\section{Summary} \label{sec:summary}

The main results from the fitting of \emph{Spitzer}/IRS spectra of a sample of 110 AGN using six dust models:

\begin{enumerate}

    \item When mid-infrared AGN-dominated spectra are selected, all but [Sieben15] show satisfactory residuals. However, residuals are larger than expected if we select the right model (i.e. self-fitting synthetic spectra to the same model). The main discrepancies are the same than those reported by our synthetic spectra (Paper I); i.e. slopes below $\rm{\sim}$7\,$\rm{\mu m}$ and above $\rm{\sim}$25\,$\rm{\mu m}$ and around the 10 and 18\,$\rm{\mu m}$ silicate features. 
    
    \item The fraction of objects requiring a prominent circumnuclear component is small (20\%) when using [Fritz06]; although up to 30-50\% of the spectra require large circumnuclear components contribution for the other models.
    
    \item The largest percentage of good fits (40\%) is obtained when using [Hoenig17]; [Nenkova08] is the next one showing the largest percentage (30\%); roughly 10\% can be fitted with either [Fritz06] or [Hoenig10]; and less than 5\% can be fitted with [Sieben15]. [Hoenig17] seems to be particularly good for Sy1 and [Nenkova08] for Sy2. [Hoenig17] is better suited for high and [Nenkova08] for low AGN luminosities.
    
    \item When the model is good enough (i.e. $\rm{\chi^2/dof <2}$) to reproduce the data, we manage to constrain 80\% of the parameters, irrespective of the optical class or the X-ray luminosity. The parameters are restricted to 10-20\% of its parameter space and this percentage increases to the low and/or high limit for most of the parameters. Many of them tend to be clustered at the low and/or high limit. This might further suggest that a strong revision of the SED libraries could result on a better match of the mid-infrared spectra of AGN.
    
    \item We did not find equivalencies between parameters of different models. Therefore, the parameter results strongly depend on the chosen model. We also computed derived quantities and we compared the results using different models: 
    	\begin{itemize}
		\item The derived dust masses are equivalent for most of the models. 
		\item The outer radius of the dusty structure in physical units is roughly twice larger for [Nenkova08] compared to [Fritz06] and [Stalev16]. 
		\item The covering factor strongly depends on the model used, showing a wide range of values for [Fritz06], [Nenkova08], [Hoenig10], and [Stalev16] and a narrow distribution for [Sieben15] and [Hoenig17]. 
    	\end{itemize}
\end{enumerate}

\citet{vanBemmel03} said more than 15 years ago that any ``improvement of the models is only relevant when better observations are available to constrain the models, and when we have a better understanding of the contribution of star-formation to the infrared SED". New data that \emph{JWST} will provide soon are ideal for those purposes. Therefore, observers and modelers need to work together to obtain a better sample of well isolated mid-infrared spectra and on new SED libraries with a careful selection of the parameter space to be confronted against data.

\acknowledgments

We thank to the anonymous referee for his/her comments and suggestions which have improved significantly the results of this research. This research is mainly funded by the UNAM PAPIIT project IA103118 (PI OG-M). MM-P acknowledges support by KASI postdoctoral fellowships. IM and JM acknowledge financial support from the research project AYA2016-76682-C3-1-P (AEI/FEDER, UE). JMR-E acknowledge support from the Spanish Ministry of Science under grant AYA2015-70498-C2-1, and AYA2017-84061-P.  IG-B acknowledges financial support from the Spanish Ministry of Science and Innovation (MICINN) through projects PN AYA2015-64346-C2-1-P and AYA2016-76682-C3-2-P. I.M. and J.M. acknowledge financial support from the State Agency for Research of the Spanish MCIU through the ``Center of Excellence Severo Ochoa" award for the Instituto de Astrof\'isica de Andaluc\'ia (SEV-2017-0709). D. E.-A. acknowledges support from a CONACYT scholarship. D-D acknowledges PAPIIT UNAM support from grant IN113719. This research has made use of dedicated servers maintained by Jaime Perea (HyperCat at IAA-CSIC), Alfonso Ginori Gonz\'alez, Gilberto Zavala, and Miguel Espejel (Galaxias, Posgrado04, and Arambolas at IRyA-UNAM) and Daniel D\'iaz-Gonz\'alez (IRyAGN1 and IRyAGN2). All of them are gratefully acknowledged.

\appendix

\section{Spectral fits for the AGN sample}\label{appendix:tableresults}

Table \ref{tab:samplefit} shows the results on the model fitting of 110 AGN selected from the \emph{Swift}/BAT sample with available \emph{Spitzer} spectra. 

\addtolength{\oddsidemargin}{-.3in}
\addtolength{\evensidemargin}{-.3in}
\renewcommand{\baselinestretch}{0.6}
\begin{table*}
\scriptsize
\begin{center}
\begin{tabular}{llcccccccccccc}
\hline \hline
Obj.	& Mod	&  A / S / I   &   $\rm{\chi^{2}/dof}$ & $E_{(B-V)}$ &\multicolumn{8}{c}{Parameters}       \\
	                  &          & \%     &                                       &      &   &     &     &     &     &      &  &   \\ \hline
				&	F06&			&			&							& $i$ & $\sigma$ & $\Gamma$ & $\beta$ & $Y$ & $\tau_{9.7\mu m}$ &  & \\
				&	N08&			&			&							& $i$ & $N_{0}$  & $\sigma$ & $Y$ & $q$ &$\tau_{v}$ &  & \\
				&	H10&			&			&							& $i$ & $N_{0}$  & $\theta$ & $a$ & $\tau_{cl}$ &  &  &  \\
				&	S15&			&			&							& $i$ & $R_{in}$ & $\eta$ & $\tau_{cl}$ & $\tau_{disk}$ &  & & \\
				&	S16&			&			&							& $i$ & $\sigma$ & $p$ & $q$ & $Y$ & $\tau_{9.7\mu m}$ &  & \\
				&	H17&			&			&							& $i$ & $N_{0}$ & $a$  & $\sigma$ & $\theta$ & $a_{w}$ & $h$ & $f_{w}$ \\ \hline \hline
				&		&			&			&							&		&		&		&			&			&			&		&		\\
NGC235A                & F06              &  49.8/  0.0/ 50.2 &   1.83 & 0.9$_{0.8}^{1.0}$ & $<$15.3 & $<$23.9 & $<$0.1 & $>$-0.1 & 27.4$_{23.4}^{29.1}$ & 2.0$_{1.9}^{2.3}$ &  &  &  \\ 
                       & N08              &  36.1/  6.7/ 57.2 &   1.98 & 1.5$_{1.4}^{1.6}$ & 52.4$_{49.1}^{55.9}$ & $>$14.8 & $>$69.2 & 19.9$_{19.6}^{20.1}$ & 1.0$_{0.98}^{1.02}$ & 20.0$_{19.9}^{20.1}$ &  &  &  \\ 
                       & H10              &  23.9/  8.0/ 68.1 &   2.35 & 2.0$_{1.9}^{2.1}$ & $>$83.7 & 3.1$_{2.8}^{3.5}$ & $>$5.0 & $>$-0.1 & $>$78.4 &  &  &  &  \\ 
                       & S15              &  27.8/  6.8/ 65.4 &   2.43 & 2.2$_{2.1}^{2.3}$ & $<$21.3 & $<$3.0 & $>$77.3 & $<$0.1 & 13.5$_{13.5}^{13.53}$ &  &  &  &  \\ 
                       & S16$\circ\bullet$ &  56.1/  0.0/ 43.9 &   1.79 & 0.8$_{0.7}^{1.0}$ & 86.1$_{78.6}^{89.1}$ & 24.6$_{21.1}^{31.5}$ & 1.3$_{1.2}^{1.4}$ & $>$0.0 & $>$21.5 & $<$4.1 &   &  &  \\ 
                       & H17              &  30.5/  0.0/ 69.5 &   2.07 & 2.1$_{2.1}^{2.2}$ & 21.3$_{21.1}^{22.0}$ & $>$9.9 & $<$-3.0 & $>$14.7 & $>$44.5 & $>$-0.5 & $<$0.1 & $>$0.7 &  \\ 
Mrk348                 & F06$\circ$       &  94.9/  5.1/  0.0 &   1.79 & $<$0.5 & $<$11.0 & $<$20.3 & 3.4$_{3.3}^{3.6}$ & $>$-0.0 & $<$10.0 & 6.0$_{5.8}^{6.1}$ &  &  &  \\ 
                       & N08$\circ$       &  90.4/  9.6/  0.0 &   1.59 & 0.4$_{0.3}^{0.4}$ & $>$87.5 & $>$14.9 & $>$68.6 & 14.8$_{12.8}^{17.7}$ & 1.9$_{1.8}^{2.0}$ & $<$10.3 &  &  &  \\ 
                       & H10$\circ\bullet$ &  79.9/  9.3/ 10.8 &   1.44 & 0.2$_{0.1}^{0.2}$ & 75.0$_{72.9}^{78.7}$ & 7.1$_{6.1}^{7.8}$ & $<$8.8 & -1.05$_{-1.09}^{-0.95}$ & 44.8$_{41.5}^{47.6}$ &  &  &  &  \\ 
                       & S15$\circ$       &  81.9/ 11.1/  7.0 &  12.06 & 1.2$_{1.1}^{1.2}$ & $>$85.9 & $<$3.0 & 7.7$_{7.45}^{7.71}$ & $<$0.0 & 13.5$_{13.5}^{13.51}$ &  &  &  &  \\ 
                       & S16$\circ\bullet$ &  91.2/  5.7/  3.2 &   1.53 & $<$0.5 & 79.8$_{74.1}^{83.1}$ & 73.4$_{66.0}^{78.6}$ & $>$1.5 & $>$1.1 & $<$10.2 & 7.0$_{6.6}^{7.7}$ &   &  &  \\ 
                       & H17$\circ\bullet$ &  86.3/  6.9/  6.9 &   1.09 & 0.3$_{0.3}^{0.4}$ & 59.9$_{59.3}^{60.6}$ & $>$9.8 & -2.5$_{-2.52}^{-2.48}$ & $<$7.1 & $>$44.6 & $>$-0.5 & $>$0.5 & 0.45$_{0.43}^{0.47}$ &  \\ 
Mrk352                 & F06$\circ$       &  92.4/  7.6/  0.0 &   1.68 & $<$0.5 & 62.3$_{59.4}^{65.5}$ & $<$20.3 & $>$6.0 & $<$-1.0 & $<$10.0 & $>$9.0 &  &  &  \\ 
                       & N08$\circ\bullet$ &  94.2/  5.8/  0.0 &   1.14 & $<$0.5 & $>$72.4 & $<$3.4 & $>$15.0 & $<$7.7 & $>$2.2 & 33.1$_{23.3}^{61.1}$ &  &  &  \\ 
                       & H10$\circ\bullet$ &  93.5/  4.0/  2.5 &   1.14 & $<$0.5 & $<$59.6 & $<$4.5 & $>$44.5 & -1.32$_{-1.37}^{-1.27}$ & $>$71.5 &  &  &  &  \\ 
                       & S15$\circ$       &  85.2/ 14.8/  0.0 &   5.42 & $<$0.5 & 33.1$_{32.8}^{33.5}$ & $<$3.0 & $<$1.5 & $<$0.0 & 13.5$_{13.5}^{13.51}$ &  &  &  &  \\ 
                       & S16$\circ$       &  93.3/  6.7/  0.0 &   1.87 & $<$0.5 & 30.9$_{22.3}^{34.0}$ & $>$78.8 & $>$1.5 & $>$1.5 & $<$10.0 & $<$3.0 &   &  &  \\ 
                       & H17$\circ\bullet$ & 100.0/  0.0/  0.0 &   1.20 & $<$0.5 & $<$15.0 & $>$5.2 & -2.81$_{-2.86}^{-2.76}$ & $>$7.0 & 34.2$_{31.7}^{35.5}$ & $>$-0.9 & $<$0.4 & $>$0.2 &  \\ 
NGC454E                & F06$\circ\bullet$ &  74.4/  5.0/ 20.6 &   1.26 & $<$0.5 & 48.0$_{45.2}^{54.0}$ & 34.2$_{31.9}^{36.9}$ & $<$0.0 & -0.5$_{-0.8}^{-0.2}$ & $<$10.9 & 3.9$_{3.6}^{4.2}$ &  &  &  \\ 
                       & N08$\circ\bullet$ &  66.7/ 10.5/ 22.8 &   1.08 & $<$0.5 & $>$75.3 & 10.8$_{10.1}^{13.9}$ & 25.2$_{15.7}^{58.0}$ & 19.8$_{15.5}^{21.3}$ & 1.5$_{0.9}^{1.7}$ & 22.8$_{19.9}^{27.2}$ &  &  &  \\ 
                       & H10              &  56.5/ 11.6/ 31.9 &   1.09 & $<$0.5 & $>$77.5 & 3.6$_{2.9}^{4.0}$ & $>$45.7 & $>$-0.1 & 70.7$_{67.7}^{79.6}$ &  &  &  &  \\ 
                       & S15              &  65.2/  7.8/ 27.0 &   2.39 & $<$0.5 & $>$85.9 & $<$3.0 & 7.7$_{7.5}^{7.9}$ & 30.0$_{28.1}^{30.9}$ & 13.6$_{13.5}^{14.2}$ &  &  &  &  \\ 
                       & S16              &  66.4/  9.0/ 24.6 &   1.08 & $<$0.5 & 17.2$_{14.8}^{32.2}$ & $>$73.6 & $<$0.0 & $<$0.1 & 13.0$_{12.8}^{14.0}$ & 4.9$_{4.1}^{5.3}$ &   &  &  \\ 
                       & H17              &  58.7/  9.4/ 31.9 &   0.96 & $<$0.5 & 61.5$_{53.0}^{65.0}$ & $>$8.4 & -2.3$_{-2.4}^{-2.2}$ & $>$13.3 & $>$42.7 & $>$-0.9 & $>$0.2 & $>$0.7 &  \\ 
NGC526A                & F06$\circ$       &  94.8/  5.2/  0.0 &  12.76 & $<$0.5 & 20.0$_{19.9}^{20.2}$ & $<$20.0 & $>$6.0 & $>$-0.0 & $<$10.0 & $>$10.0 &  &  &  \\ 
                       & N08$\circ$       &  94.6/  5.4/  0.0 &   8.70 & $<$0.5 & $>$86.5 & 9.0$_{8.8}^{9.1}$ & $>$66.8 & $<$5.0 & 0.3$_{0.1}^{0.5}$ & $<$10.0 &  &  &  \\ 
                       & H10$\circ$       &  93.9/  6.1/  0.0 &   1.29 & $<$0.5 & 78.1$_{73.7}^{88.9}$ & $>$8.7 & $>$33.2 & -1.43$_{-1.49}^{-1.38}$ & 53.9$_{50.7}^{61.0}$ &  &  &  &  \\ 
                       & S15$\circ$       &  91.8/  8.2/  0.0 &  34.65 & 0.4$_{0.4}^{0.4}$ & $>$86.0 & $<$3.0 & 1.62$_{1.61}^{1.63}$ & $<$0.0 & 13.5$_{13.5}^{13.51}$ &  &  &  &  \\ 
                       & S16$\circ$       &  96.7/  3.3/  0.0 &  10.43 & $<$0.5 & $>$79.5 & $>$79.9 & $>$1.5 & 0.67$_{0.64}^{0.68}$ & $<$10.0 & $<$3.0 &   &  &  \\ 
                       & H17$\circ\bullet$ &  96.8/  3.2/  0.0 &   0.75 & 0.3$_{0.2}^{0.3}$ & 60.0$_{58.3}^{60.5}$ & $>$7.3 & -2.66$_{-2.69}^{-2.59}$ & $>$14.8 & $>$44.5 & -0.82$_{-1.45}^{-0.77}$ & 0.41$_{0.39}^{0.44}$ & $<$0.3 &  \\ 
E297-018             & F06$\circ$       &  93.6/  0.0/  6.4 &   1.70 & 0.6$_{0.6}^{0.7}$ & $<$0.0 & $<$20.2 & 2.15$_{2.13}^{2.18}$ & $<$-1.0 & $<$10.0 & 3.0$_{2.99}^{4.03}$ &  &  &  \\ 
                       & N08$\circ\bullet$ &  87.6/  4.0/  8.4 &   0.93 & 0.5$_{0.5}^{0.6}$ & $>$74.3 & $>$14.7 & $>$66.9 & $<$5.0 & 1.13$_{1.06}^{1.24}$ & $<$10.4 &  &  &  \\ 
                       & H10              &  82.7/  2.6/ 14.6 &   0.99 & 0.6$_{0.6}^{0.6}$ & 60.0$_{29.7}^{67.5}$ & $>$9.6 & $<$23.7 & -1.91$_{-1.93}^{-1.88}$ & $>$78.4 &  &  &  &  \\ 
                       & S15$\circ$       &  84.3/  7.9/  7.8 &  15.11 & 0.8$_{0.8}^{0.9}$ & $>$85.9 & $<$3.0 & $<$1.5 & $<$0.0 & 17.0$_{16.7}^{17.3}$ &  &  &  &  \\ 
                       & S16$\circ$       &  90.7/  1.2/  8.1 &   2.32 & 0.7$_{0.6}^{0.7}$ & $>$88.8 & 70.5$_{69.8}^{71.6}$ & $>$1.5 & $>$1.5 & $<$10.0 & 5.0$_{4.98}^{5.04}$ &   &  &  \\ 
                       & H17              &  88.7/  0.0/ 11.3 &   0.87 & 0.6$_{0.6}^{0.7}$ & $>$80.1 & $>$7.7 & $<$-2.9 & $<$11.4 & 36.1$_{34.0}^{41.7}$ & -1.2$_{-1.4}^{-1.1}$ & 0.25$_{0.17}^{0.33}$ & $>$0.6 &  \\ 
NGC788                 & F06$\circ$       & 100.0/  0.0/  0.0 &   3.34 & $<$0.5 & 0.3$_{0.1}^{0.7}$ & $<$20.3 & 4.81$_{4.78}^{4.83}$ & $>$-0.0 & $<$10.0 & $>$9.8 &  &  &  \\ 
                       & N08$\circ\bullet$ &  94.2/  4.0/  1.8 &   1.30 & $<$0.5 & $>$77.3 & 11.3$_{10.4}^{13.5}$ & $>$47.2 & 10.0$_{9.6}^{11.8}$ & 2.1$_{2.0}^{2.3}$ & 20.0$_{17.9}^{21.0}$ &  &  &  \\ 
                       & H10$\circ$       &  83.8/  3.3/ 12.9 &   2.16 & $<$0.5 & 84.0$_{80.0}^{88.7}$ & $>$9.8 & $>$56.8 & -0.9$_{-1.0}^{-0.8}$ & 37.0$_{34.2}^{40.6}$ &  &  &  &  \\ 
                       & S15$\circ$       &  91.7/  3.9/  4.4 &  20.71 & 0.8$_{0.8}^{0.8}$ & $>$86.0 & $<$3.0 & 7.4$_{7.1}^{7.7}$ & $<$0.0 & 13.5$_{13.5}^{13.51}$ &  &  &  &  \\ 
                       & S16$\circ$       &  97.0/  0.0/  3.0 &   2.54 & $<$0.5 & 19.9$_{16.5}^{21.0}$ & $>$79.0 & $>$1.5 & 0.65$_{0.58}^{0.7}$ & $<$10.0 & 8.9$_{8.4}^{9.3}$ &   &  &  \\ 
                       & H17$\circ\bullet$ &  91.1/  0.0/  8.9 &   1.25 & 0.2$_{0.2}^{0.3}$ & 50.5$_{49.1}^{51.7}$ & 7.0$_{6.8}^{7.2}$ & $<$-3.0 & 13.7$_{11.8}^{14.8}$ & $>$44.8 & -1.5$_{-1.51}^{-1.49}$ & $>$0.5 & $>$0.7 &  \\ 
Mrk590                 & F06$\circ$       &  99.4/  0.6/  0.0 &   8.91 & $<$0.5 & $<$0.0 & $<$20.0 & $>$6.0 & -0.25$_{-0.25}^{-0.22}$ & 10.44$_{10.39}^{10.53}$ & 9.3$_{9.1}^{9.6}$ &  &  &  \\ 
                       & N08$\circ$       &  96.0/  4.0/  0.0 &   2.96 & $<$0.5 & $>$83.2 & 5.7$_{5.3}^{6.0}$ & $>$62.6 & 7.2$_{6.8}^{7.9}$ & $>$2.5 & 103.6$_{99.3}^{106.5}$ &  &  &  \\ 
                       & H10$\circ$       &  91.6/  2.8/  5.6 &   4.04 & $<$0.5 & 51.7$_{49.9}^{52.3}$ & 9.2$_{8.6}^{9.9}$ & $>$57.9 & $>$-0.0 & 40.7$_{39.2}^{42.1}$ &  &  &  &  \\ 
                       & S15$\circ$       &  96.3/  3.7/  0.0 &  24.09 & 0.4$_{0.4}^{0.4}$ & $>$86.0 & $<$3.0 & $<$1.5 & $<$0.0 & 42.0$_{41.7}^{42.4}$ &  &  &  &  \\ 
                       & S16$\circ$       &  99.3/  0.7/  0.0 &   6.64 & $<$0.5 & 60.3$_{59.4}^{61.3}$ & $>$79.8 & $<$0.0 & $>$1.5 & $<$10.0 & 6.5$_{6.4}^{6.6}$ &   &  &  \\ 
                       & H17$\circ\bullet$ &  95.4/  0.0/  4.6 &   1.15 & 0.1$_{0.1}^{0.1}$ & 30.0$_{28.3}^{31.3}$ & $>$9.5 & $<$-3.0 & 11.4$_{11.2}^{11.6}$ & 36.1$_{33.3}^{37.3}$ & $>$-0.5 & $<$0.1 & 0.4$_{0.37}^{0.42}$ &  \\ 
IC1816                 & F06$\circ$       &  78.9/  0.0/ 21.1 &   2.23 & $<$0.5 & 62.0$_{60.8}^{63.1}$ & $<$20.2 & 0.08$_{0.06}^{0.1}$ & $>$-0.0 & $<$10.0 & $>$9.0 &  &  &  \\ 
                       & N08              &  66.5/  5.0/ 28.5 &   1.18 & 0.6$_{0.5}^{0.6}$ & $>$72.0 & $>$13.2 & $>$60.6 & 7.2$_{6.2}^{8.9}$ & $<$2.1 & 58.2$_{52.5}^{63.2}$ &  &  &  \\ 
                       & H10              &  68.9/  2.5/ 28.6 &   2.56 & 0.4$_{0.2}^{0.5}$ & 65.3$_{60.0}^{67.7}$ & $>$8.1 & $>$58.9 & $>$-0.1 & 62.9$_{57.4}^{66.2}$ &  &  &  &  \\ 
                       & S15$\circ$       &  73.8/  2.3/ 24.0 &   5.08 & 0.8$_{0.8}^{0.8}$ & $>$85.8 & $<$3.0 & 4.0$_{3.9}^{4.1}$ & $<$0.0 & $>$44.6 &  &  &  &  \\ 
                       & S16              &  69.2/  1.5/ 29.3 &   1.39 & $<$0.0 & 29.6$_{27.9}^{31.5}$ & $>$77.9 & $<$0.0 & $<$0.0 & 10.3$_{10.2}^{10.4}$ & $>$10.7 &   &  &  \\ 
                       & H17              &  63.6/  4.6/ 31.8 &   1.36 & $<$0.0 & $>$89.9 & $>$9.8 & $>$-0.5 & $>$14.3 & $<$30.1 & $>$-0.6 & $<$0.1 & 0.58$_{0.49}^{0.64}$ &  \\ 
NGC973                 & F06              &  17.6/ 18.7/ 63.7 &   2.76 & 2.9$_{2.8}^{3.2}$ & $<$0.0 & $>$59.7 & 0.09$_{0.07}^{0.13}$ & $<$-1.0 & $<$10.1 & 3.0$_{2.98}^{3.03}$ &  &  &  \\ 
                       & N08              &  17.2/ 20.2/ 62.6 &   2.01 & 3.8$_{3.6}^{4.3}$ & 35.9$_{31.0}^{41.3}$ & $>$14.9 & $>$69.5 & 9.3$_{9.0}^{9.7}$ & $<$0.0 & 20.0$_{19.8}^{20.3}$ &  &  &  \\ 
                       & H10              &   5.8/ 16.5/ 77.7 &   2.59 & 5.5$_{5.3}^{5.7}$ & $>$80.8 & $>$9.0 & 44.4$_{38.1}^{49.7}$ & $<$-2.0 & $>$75.5 &  &  &  &  \\ 
                       & S15              &  24.9/ 23.0/ 52.1 &   7.79 & 0.6$_{0.5}^{0.7}$ & $<$19.5 & $<$3.0 & 7.7$_{7.5}^{8.0}$ & $>$956.2 & $>$44.4 &  &  &  &  \\ 
                       & S16              &  13.4/ 19.8/ 66.8 &   3.03 & 3.6$_{3.5}^{3.7}$ & $>$79.0 & $>$75.6 & $>$1.5 & $>$1.4 & $<$10.1 & 5.0$_{4.5}^{5.4}$ &   &  &  \\ 
                       & H17              &   8.8/ 16.6/ 74.5 &   2.09 & 4.7$_{4.6}^{4.9}$ & 60.1$_{54.3}^{63.3}$ & $<$5.2 & $<$-3.0 & $>$13.4 & $>$42.8 & -1.7$_{-1.74}^{-1.62}$ & $<$0.1 & 0.6$_{0.5}^{0.7}$ &  \\ 
NGC1052                & F06$\circ$       &  95.6/  4.4/  0.0 &   4.38 & $<$0.5 & $<$0.1 & $<$20.1 & $>$6.0 & $>$-0.0 & 13.7$_{13.6}^{14.6}$ & 8.5$_{8.3}^{8.6}$ &  &  &  \\ 
                       & N08$\circ$       &  92.6/  5.1/  2.4 &   1.79 & $<$0.5 & $>$87.8 & 3.8$_{3.6}^{4.1}$ & $<$16.1 & 10.0$_{9.9}^{10.3}$ & 0.9$_{0.8}^{1.1}$ & 57.8$_{52.9}^{62.9}$ &  &  &  \\ 
                       & H10$\circ$       &  87.4/  5.2/  7.5 &   1.75 & 0.2$_{0.1}^{0.2}$ & 49.7$_{44.3}^{52.6}$ & 5.7$_{5.3}^{6.2}$ & $>$57.8 & $>$-0.0 & 72.7$_{67.9}^{77.3}$ &  &  &  &  \\ 
                       & S15$\circ$       &  88.9/  6.4/  4.7 &  11.57 & $<$0.5 & $>$86.0 & $<$3.0 & $<$1.5 & $<$0.0 & $>$44.9 &  &  &  &  \\ 
                       & S16$\circ$       &  93.9/  3.8/  2.2 &   4.14 & $<$0.5 & 57.1$_{54.1}^{59.4}$ & $>$76.5 & $<$0.0 & $>$1.5 & 11.7$_{11.6}^{12.0}$ & 5.8$_{5.6}^{6.2}$ &   &  &  \\ 
                       & H17$\circ\bullet$ &  92.4/  0.0/  7.6 &   0.95 & 0.2$_{0.2}^{0.3}$ & 29.9$_{27.1}^{30.4}$ & 8.3$_{7.6}^{9.3}$ & -2.51$_{-2.62}^{-2.48}$ & $>$13.7 & 35.1$_{33.8}^{36.6}$ & $>$-0.5 & $<$0.1 & $>$0.7 &  \\ 
\hline \hline
\end{tabular}
\caption{Spectral fit results. Best-fit results per object and model. Models are quoted in Col.\,2 as follows. F06: [Fritz06]; N08: [Nenkova08]; H10: [Hoenig10]; S15: [Sieben15]; S16: [Stalev16]; and H17: [Hoenig17]. Col.\,3 includes the percentage contribution to the 5-30\,$\rm{\mu m}$ waveband per component (A: AGN; S: Stellar; and I: ISM), the reduced $\rm{\chi^2}$ ($\rm{\chi^2/dof}$) is included in Col.\,4, color excess for the foreground extinction $\rm{E(B-V)}$ is included in Col.\,5, and the final parameters per model. AGN-dominated spectra are marked with empty circles and good fits ($\rm{\chi^2/dof< min(\chi^2/dof)+0.5}$) are marked with filled circles next to the model name. 
}
\label{tab:samplefit}
\end{center}
\end{table*}

\begin{table*}
\addtocounter{table}{-1}
\scriptsize
\begin{center}
\begin{tabular}{llcccccccccccc}
\hline \hline
Obj.	& Mod	&  A / S / I   &   $\rm{\chi^{2}/dof}$ & $E_{(B-V)}$ &\multicolumn{8}{c}{Parameters}       \\
	                  &          & \%     &                                       &      &   &     &     &     &     &      &  &   \\ \hline
				&	F06&			&			&							& $i$ & $\sigma$ & $\Gamma$ & $\beta$ & $Y$ & $\tau_{9.7\mu m}$ &  & \\
				&	N08&			&			&							& $i$ & $N_{0}$  & $\sigma$ & $Y$ & $q$ &$\tau_{v}$ &  & \\
				&	H10&			&			&							& $i$ & $N_{0}$  & $\theta$ & $a$ & $\tau_{cl}$ &  &  &  \\
				&	S15&			&			&							& $i$ & $R_{in}$ & $\eta$ & $\tau_{cl}$ & $\tau_{disk}$ &  & & \\
				&	S16&			&			&							& $i$ & $\sigma$ & $p$ & $q$ & $Y$ & $\tau_{9.7\mu m}$ &  & \\
				&	H17&			&			&							& $i$ & $N_{0}$ & $a$  & $\sigma$ & $\theta$ & $a_{w}$ & $h$ & $f_{w}$ \\
\hline \hline
\end{tabular}
\caption{Spectral fit results (continuation). 
}
\label{tab:samplefit}
\end{center}
\end{table*}

\begin{table*}
\addtocounter{table}{-1}
\scriptsize
\begin{center}
\begin{tabular}{llcccccccccccc}
\hline \hline
Obj.	& Mod	&  A / S / I   &   $\rm{\chi^{2}/dof}$ & $E_{(B-V)}$ &\multicolumn{8}{c}{Parameters}       \\
	                  &          & \%     &                                       &      &   &     &     &     &     &      &  &   \\ \hline
				&	F06&			&			&							& $i$ & $\sigma$ & $\Gamma$ & $\beta$ & $Y$ & $\tau_{9.7\mu m}$ &  & \\
				&	N08&			&			&							& $i$ & $N_{0}$  & $\sigma$ & $Y$ & $q$ &$\tau_{v}$ &  & \\
				&	H10&			&			&							& $i$ & $N_{0}$  & $\theta$ & $a$ & $\tau_{cl}$ &  &  &  \\
				&	S15&			&			&							& $i$ & $R_{in}$ & $\eta$ & $\tau_{cl}$ & $\tau_{disk}$ &  & & \\
				&	S16&			&			&							& $i$ & $\sigma$ & $p$ & $q$ & $Y$ & $\tau_{9.7\mu m}$ &  & \\
				&	H17&			&			&							& $i$ & $N_{0}$ & $a$  & $\sigma$ & $\theta$ & $a_{w}$ & $h$ & $f_{w}$ \\
\hline \hline
\end{tabular}
\caption{Spectral fit results (continuation). 
}
\label{tab:samplefit}
\end{center}
\end{table*}

\begin{table*}
\addtocounter{table}{-1}
\scriptsize
\begin{center}
\begin{tabular}{llcccccccccccc}
\hline \hline
Obj.	& Mod	&  A / S / I   &   $\rm{\chi^{2}/dof}$ & $E_{(B-V)}$ &\multicolumn{8}{c}{Parameters}       \\
	                  &          & \%     &                                       &      &   &     &     &     &     &      &  &   \\ \hline
				&	F06&			&			&							& $i$ & $\sigma$ & $\Gamma$ & $\beta$ & $Y$ & $\tau_{9.7\mu m}$ &  & \\
				&	N08&			&			&							& $i$ & $N_{0}$  & $\sigma$ & $Y$ & $q$ &$\tau_{v}$ &  & \\
				&	H10&			&			&							& $i$ & $N_{0}$  & $\theta$ & $a$ & $\tau_{cl}$ &  &  &  \\
				&	S15&			&			&							& $i$ & $R_{in}$ & $\eta$ & $\tau_{cl}$ & $\tau_{disk}$ &  & & \\
				&	S16&			&			&							& $i$ & $\sigma$ & $p$ & $q$ & $Y$ & $\tau_{9.7\mu m}$ &  & \\
				&	H17&			&			&							& $i$ & $N_{0}$ & $a$  & $\sigma$ & $\theta$ & $a_{w}$ & $h$ & $f_{w}$ \\
\hline \hline
\end{tabular}
\caption{Spectral fit results (continuation). 
}
\label{tab:samplefit}
\end{center}
\end{table*}

\begin{table*}
\addtocounter{table}{-1}
\scriptsize
\begin{center}
\begin{tabular}{llcccccccccccc}
\hline \hline
Obj.	& Mod	&  A / S / I   &   $\rm{\chi^{2}/dof}$ & $E_{(B-V)}$ &\multicolumn{8}{c}{Parameters}       \\
	                  &          & \%     &                                       &      &   &     &     &     &     &      &  &   \\ \hline
				&	F06&			&			&							& $i$ & $\sigma$ & $\Gamma$ & $\beta$ & $Y$ & $\tau_{9.7\mu m}$ &  & \\
				&	N08&			&			&							& $i$ & $N_{0}$  & $\sigma$ & $Y$ & $q$ &$\tau_{v}$ &  & \\
				&	H10&			&			&							& $i$ & $N_{0}$  & $\theta$ & $a$ & $\tau_{cl}$ &  &  &  \\
				&	S15&			&			&							& $i$ & $R_{in}$ & $\eta$ & $\tau_{cl}$ & $\tau_{disk}$ &  & & \\
				&	S16&			&			&							& $i$ & $\sigma$ & $p$ & $q$ & $Y$ & $\tau_{9.7\mu m}$ &  & \\
				&	H17&			&			&							& $i$ & $N_{0}$ & $a$  & $\sigma$ & $\theta$ & $a_{w}$ & $h$ & $f_{w}$ \\
\hline \hline
\end{tabular}
\caption{Spectral fit results (continuation). 
}
\label{tab:samplefit}
\end{center}
\end{table*}

\begin{table*}
\addtocounter{table}{-1}
\scriptsize
\begin{center}
\begin{tabular}{llcccccccccccc}
\hline \hline
Obj.	& Mod	&  A / S / I   &   $\rm{\chi^{2}/dof}$ & $E_{(B-V)}$ &\multicolumn{8}{c}{Parameters}       \\
	                  &          & \%     &                                       &      &   &     &     &     &     &      &  &   \\ \hline
				&	F06&			&			&							& $i$ & $\sigma$ & $\Gamma$ & $\beta$ & $Y$ & $\tau_{9.7\mu m}$ &  & \\
				&	N08&			&			&							& $i$ & $N_{0}$  & $\sigma$ & $Y$ & $q$ &$\tau_{v}$ &  & \\
				&	H10&			&			&							& $i$ & $N_{0}$  & $\theta$ & $a$ & $\tau_{cl}$ &  &  &  \\
				&	S15&			&			&							& $i$ & $R_{in}$ & $\eta$ & $\tau_{cl}$ & $\tau_{disk}$ &  & & \\
				&	S16&			&			&							& $i$ & $\sigma$ & $p$ & $q$ & $Y$ & $\tau_{9.7\mu m}$ &  & \\
				&	H17&			&			&							& $i$ & $N_{0}$ & $a$  & $\sigma$ & $\theta$ & $a_{w}$ & $h$ & $f_{w}$ \\
\hline \hline
\end{tabular}
\caption{Spectral fit results (continuation). 
}
\label{tab:samplefit}
\end{center}
\end{table*}

\begin{table*}
\addtocounter{table}{-1}
\scriptsize
\begin{center}
\begin{tabular}{llcccccccccccc}
\hline \hline
Obj.	& Mod	&  A / S / I   &   $\rm{\chi^{2}/dof}$ & $E_{(B-V)}$ &\multicolumn{8}{c}{Parameters}       \\
	                  &          & \%     &                                       &      &   &     &     &     &     &      &  &   \\ \hline
				&	F06&			&			&							& $i$ & $\sigma$ & $\Gamma$ & $\beta$ & $Y$ & $\tau_{9.7\mu m}$ &  & \\
				&	N08&			&			&							& $i$ & $N_{0}$  & $\sigma$ & $Y$ & $q$ &$\tau_{v}$ &  & \\
				&	H10&			&			&							& $i$ & $N_{0}$  & $\theta$ & $a$ & $\tau_{cl}$ &  &  &  \\
				&	S15&			&			&							& $i$ & $R_{in}$ & $\eta$ & $\tau_{cl}$ & $\tau_{disk}$ &  & & \\
				&	S16&			&			&							& $i$ & $\sigma$ & $p$ & $q$ & $Y$ & $\tau_{9.7\mu m}$ &  & \\
				&	H17&			&			&							& $i$ & $N_{0}$ & $a$  & $\sigma$ & $\theta$ & $a_{w}$ & $h$ & $f_{w}$ \\
\hline \hline
\end{tabular}
\caption{Spectral fit results (continuation). 
}
\label{tab:samplefit}
\end{center}
\end{table*}

\begin{table*}
\addtocounter{table}{-1}
\scriptsize
\begin{center}
\begin{tabular}{llcccccccccccc}
\hline \hline
Obj.	& Mod	&  A / S / I   &   $\rm{\chi^{2}/dof}$ & $E_{(B-V)}$ &\multicolumn{8}{c}{Parameters}       \\
	                  &          & \%     &                                       &      &   &     &     &     &     &      &  &   \\ \hline
				&	F06&			&			&							& $i$ & $\sigma$ & $\Gamma$ & $\beta$ & $Y$ & $\tau_{9.7\mu m}$ &  & \\
				&	N08&			&			&							& $i$ & $N_{0}$  & $\sigma$ & $Y$ & $q$ &$\tau_{v}$ &  & \\
				&	H10&			&			&							& $i$ & $N_{0}$  & $\theta$ & $a$ & $\tau_{cl}$ &  &  &  \\
				&	S15&			&			&							& $i$ & $R_{in}$ & $\eta$ & $\tau_{cl}$ & $\tau_{disk}$ &  & & \\
				&	S16&			&			&							& $i$ & $\sigma$ & $p$ & $q$ & $Y$ & $\tau_{9.7\mu m}$ &  & \\
				&	H17&			&			&							& $i$ & $N_{0}$ & $a$  & $\sigma$ & $\theta$ & $a_{w}$ & $h$ & $f_{w}$ \\
\hline \hline
\end{tabular}
\caption{Spectral fit results (continuation). 
}
\label{tab:samplefit}
\end{center}
\end{table*}

\begin{table*}
\addtocounter{table}{-1}
\scriptsize
\begin{center}
\begin{tabular}{llcccccccccccc}
\hline \hline
Obj.	& Mod	&  A / S / I   &   $\rm{\chi^{2}/dof}$ & $E_{(B-V)}$ &\multicolumn{8}{c}{Parameters}       \\
	                  &          & \%     &                                       &      &   &     &     &     &     &      &  &   \\ \hline
				&	F06&			&			&							& $i$ & $\sigma$ & $\Gamma$ & $\beta$ & $Y$ & $\tau_{9.7\mu m}$ &  & \\
				&	N08&			&			&							& $i$ & $N_{0}$  & $\sigma$ & $Y$ & $q$ &$\tau_{v}$ &  & \\
				&	H10&			&			&							& $i$ & $N_{0}$  & $\theta$ & $a$ & $\tau_{cl}$ &  &  &  \\
				&	S15&			&			&							& $i$ & $R_{in}$ & $\eta$ & $\tau_{cl}$ & $\tau_{disk}$ &  & & \\
				&	S16&			&			&							& $i$ & $\sigma$ & $p$ & $q$ & $Y$ & $\tau_{9.7\mu m}$ &  & \\
				&	H17&			&			&							& $i$ & $N_{0}$ & $a$  & $\sigma$ & $\theta$ & $a_{w}$ & $h$ & $f_{w}$ \\
\hline \hline
\end{tabular}
\caption{Spectral fit results (continuation). 
}
\label{tab:samplefit}
\end{center}
\end{table*}

\begin{table*}
\addtocounter{table}{-1}
\scriptsize
\begin{center}
\begin{tabular}{llcccccccccccc}
\hline \hline
Obj.	& Mod	&  A / S / I   &   $\rm{\chi^{2}/dof}$ & $E_{(B-V)}$ &\multicolumn{8}{c}{Parameters}       \\
	                  &          & \%     &                                       &      &   &     &     &     &     &      &  &   \\ \hline
				&	F06&			&			&							& $i$ & $\sigma$ & $\Gamma$ & $\beta$ & $Y$ & $\tau_{9.7\mu m}$ &  & \\
				&	N08&			&			&							& $i$ & $N_{0}$  & $\sigma$ & $Y$ & $q$ &$\tau_{v}$ &  & \\
				&	H10&			&			&							& $i$ & $N_{0}$  & $\theta$ & $a$ & $\tau_{cl}$ &  &  &  \\
				&	S15&			&			&							& $i$ & $R_{in}$ & $\eta$ & $\tau_{cl}$ & $\tau_{disk}$ &  & & \\
				&	S16&			&			&							& $i$ & $\sigma$ & $p$ & $q$ & $Y$ & $\tau_{9.7\mu m}$ &  & \\
				&	H17&			&			&							& $i$ & $N_{0}$ & $a$  & $\sigma$ & $\theta$ & $a_{w}$ & $h$ & $f_{w}$ \\
\hline \hline
\end{tabular}
\caption{Spectral fit results (continuation). 
}
\label{tab:samplefit}
\end{center}
\end{table*}

\newpage

\section{Parameters for the full sample}\label{appendix:dataspecfitPDF}

\begin{figure*}[!ht]
\begin{center}
\includegraphics[width=0.32\columnwidth]{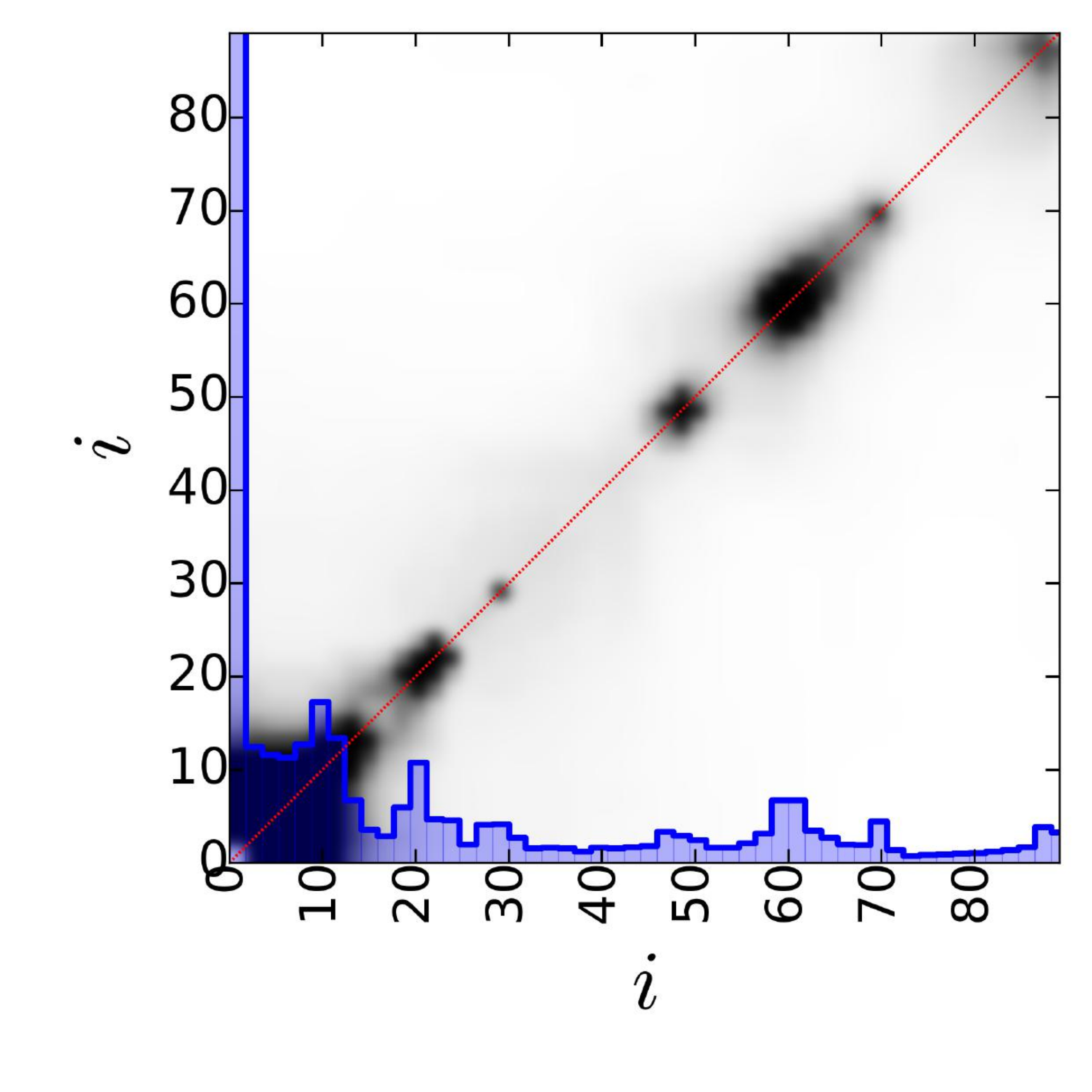}
\includegraphics[width=0.32\columnwidth]{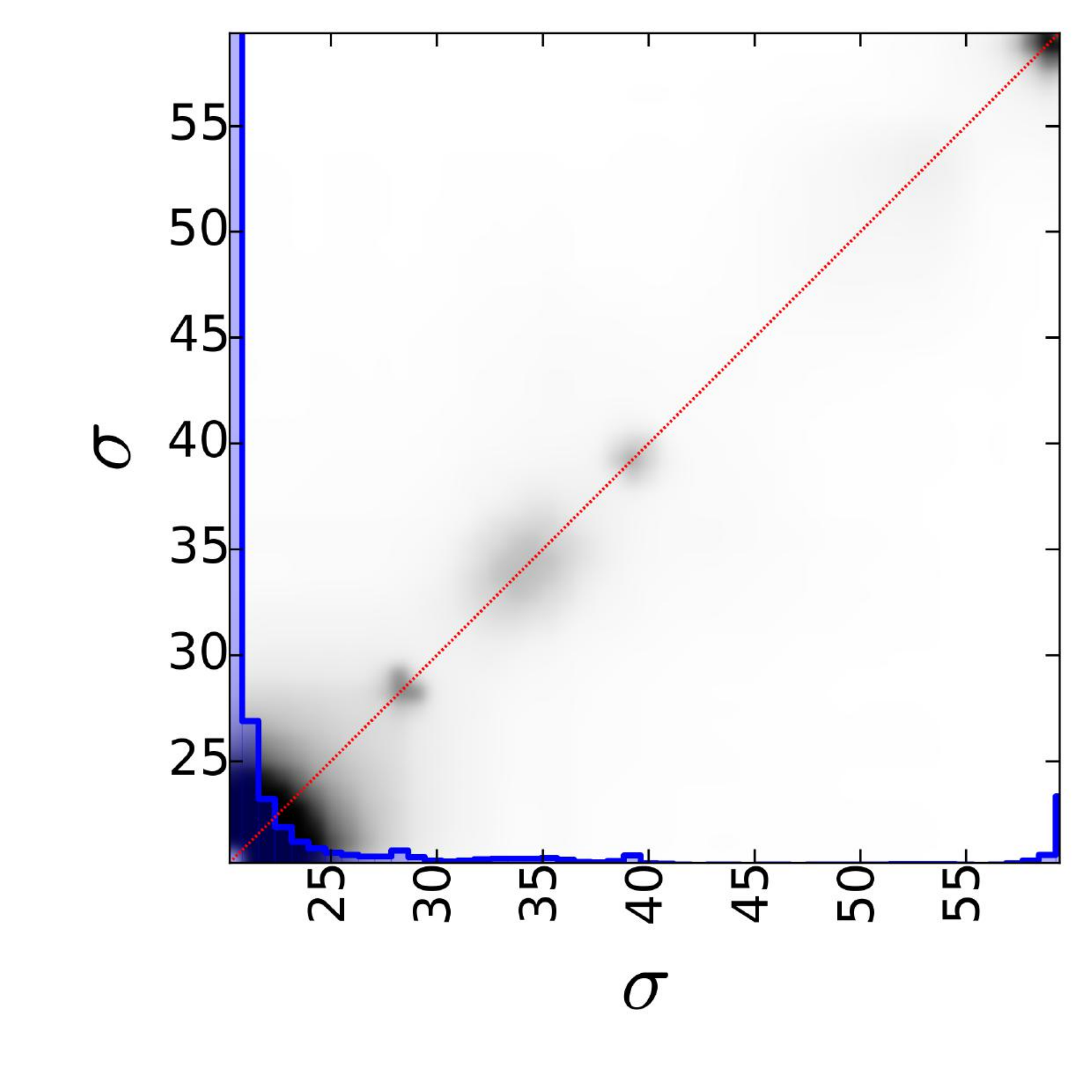}
\includegraphics[width=0.32\columnwidth]{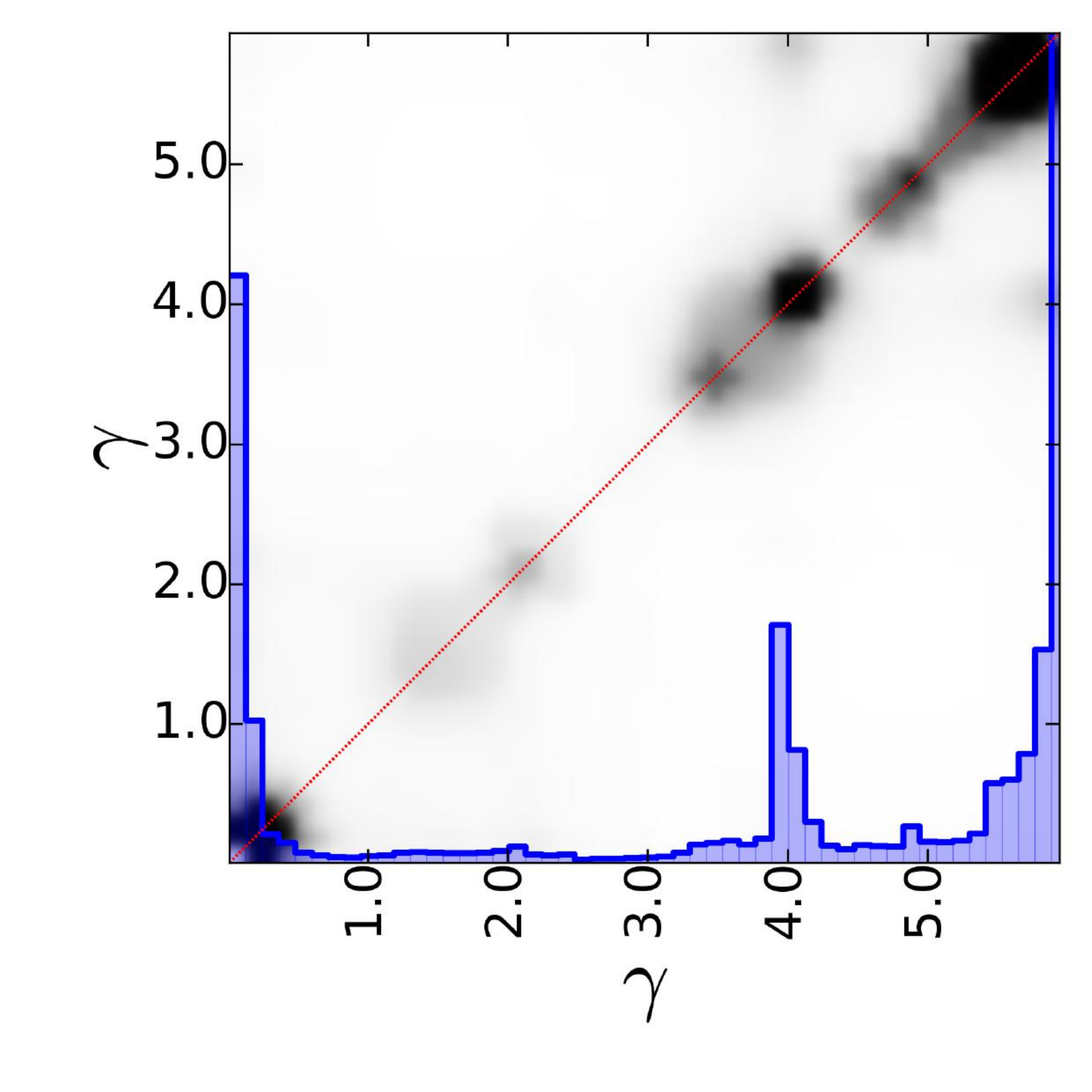}
\includegraphics[width=0.32\columnwidth]{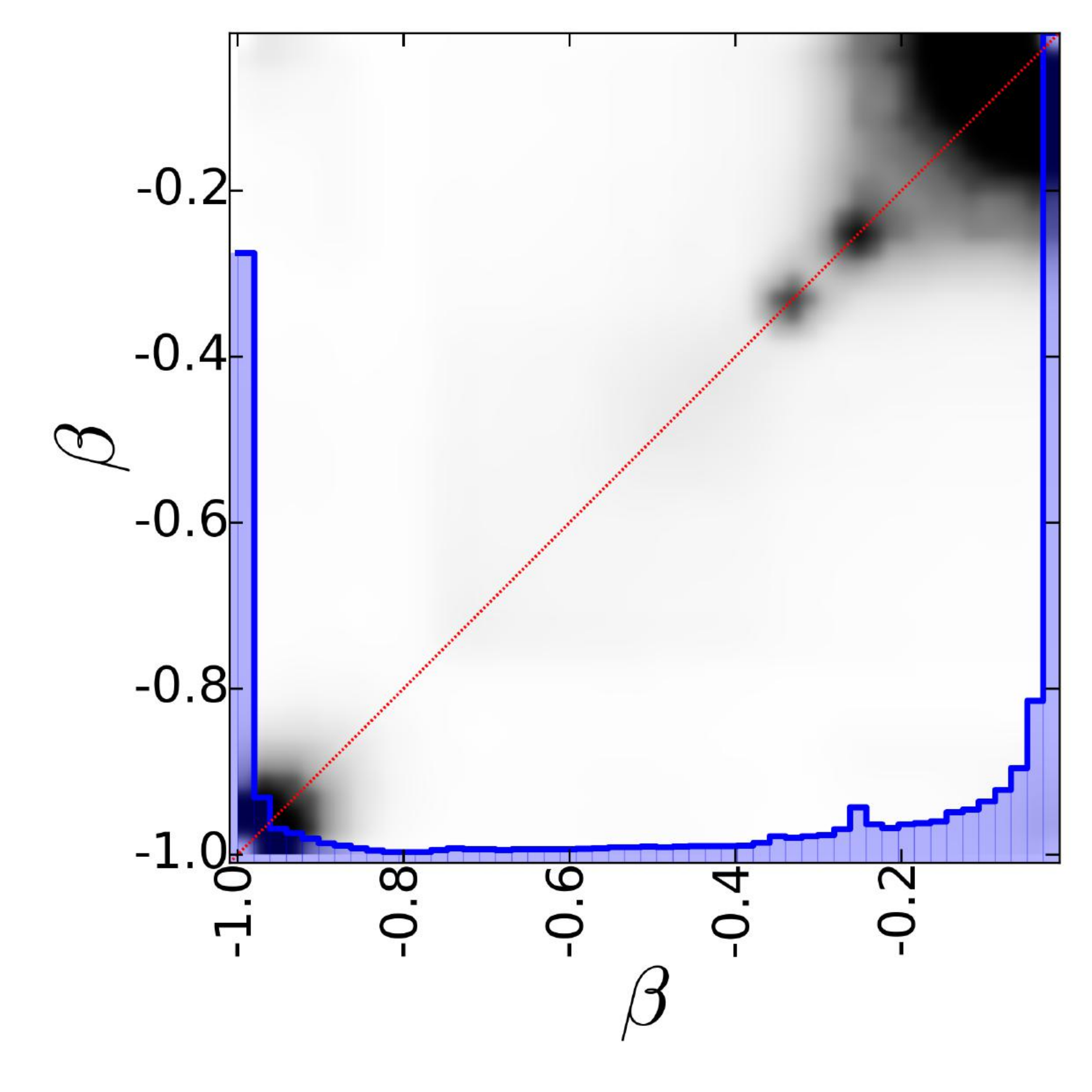}
\includegraphics[width=0.32\columnwidth]{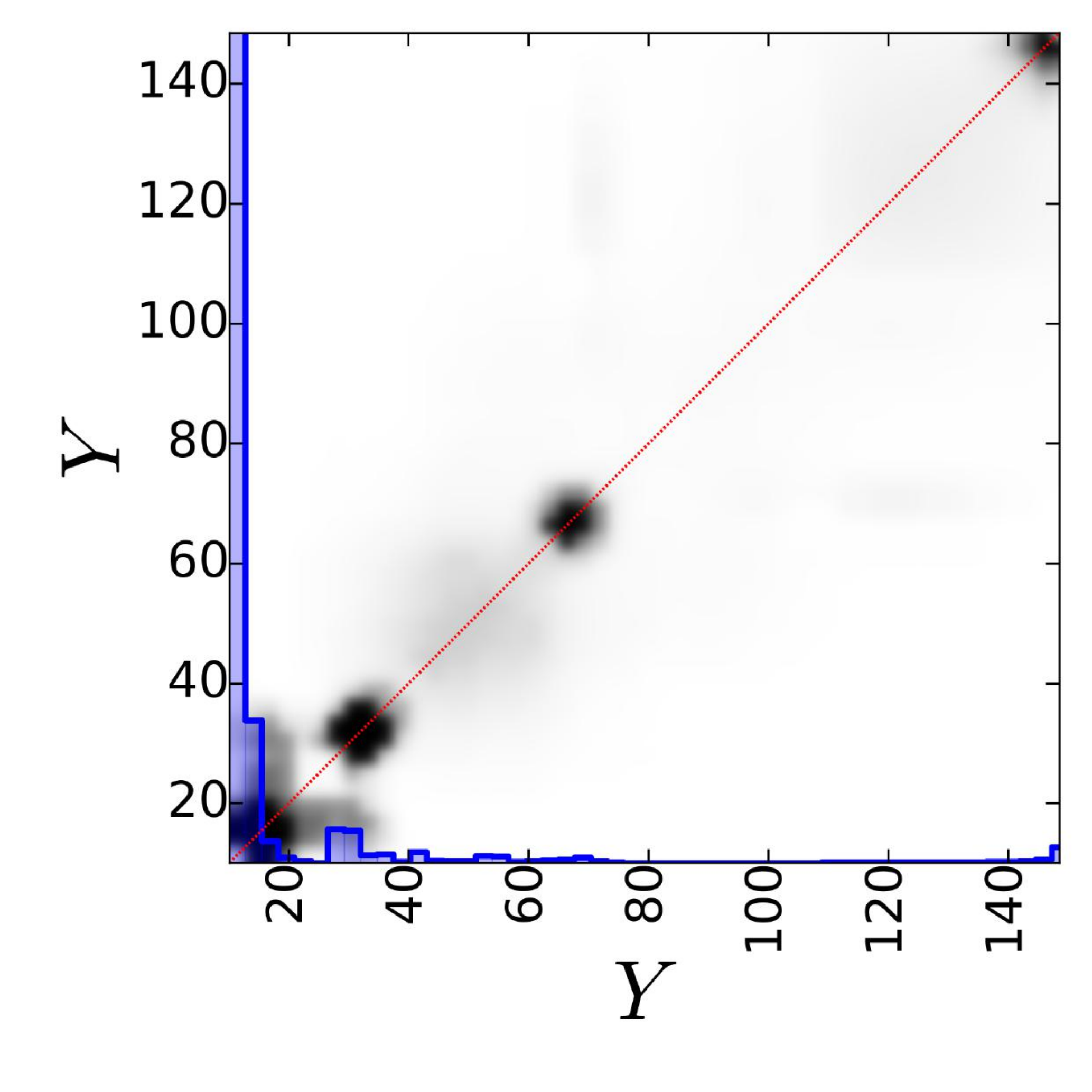}
\includegraphics[width=0.32\columnwidth]{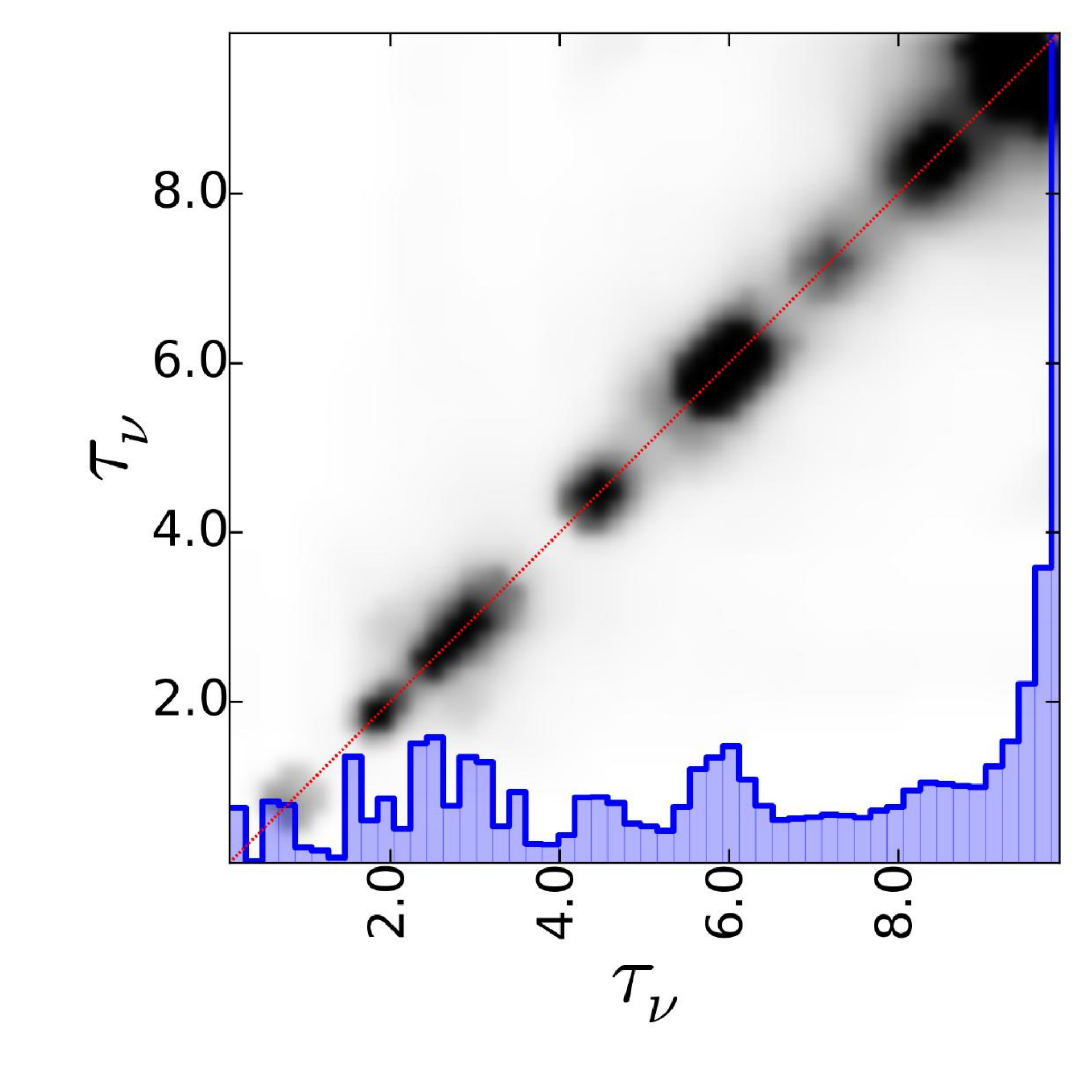}
\caption{Parameter versus parameter estimated from the sample drawn from the PDF (see text) for the model [Fritz06]. The blue histogram shows the total distribution for each parameter.}
\label{fig:dataspecfitPDF}
\end{center}
\end{figure*}

\begin{figure*}[!ht]
\begin{center}
\includegraphics[width=0.32\columnwidth]{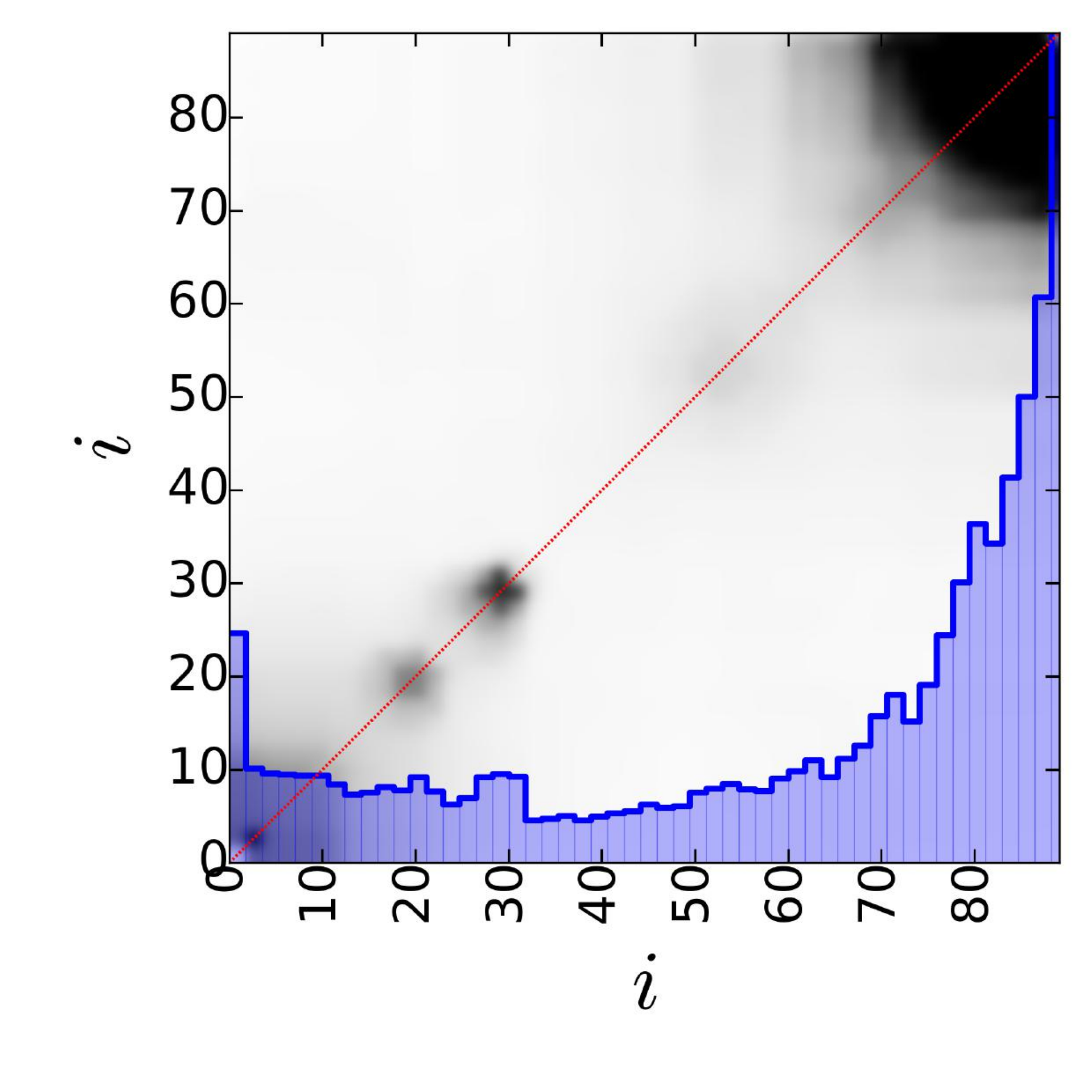}
\includegraphics[width=0.32\columnwidth]{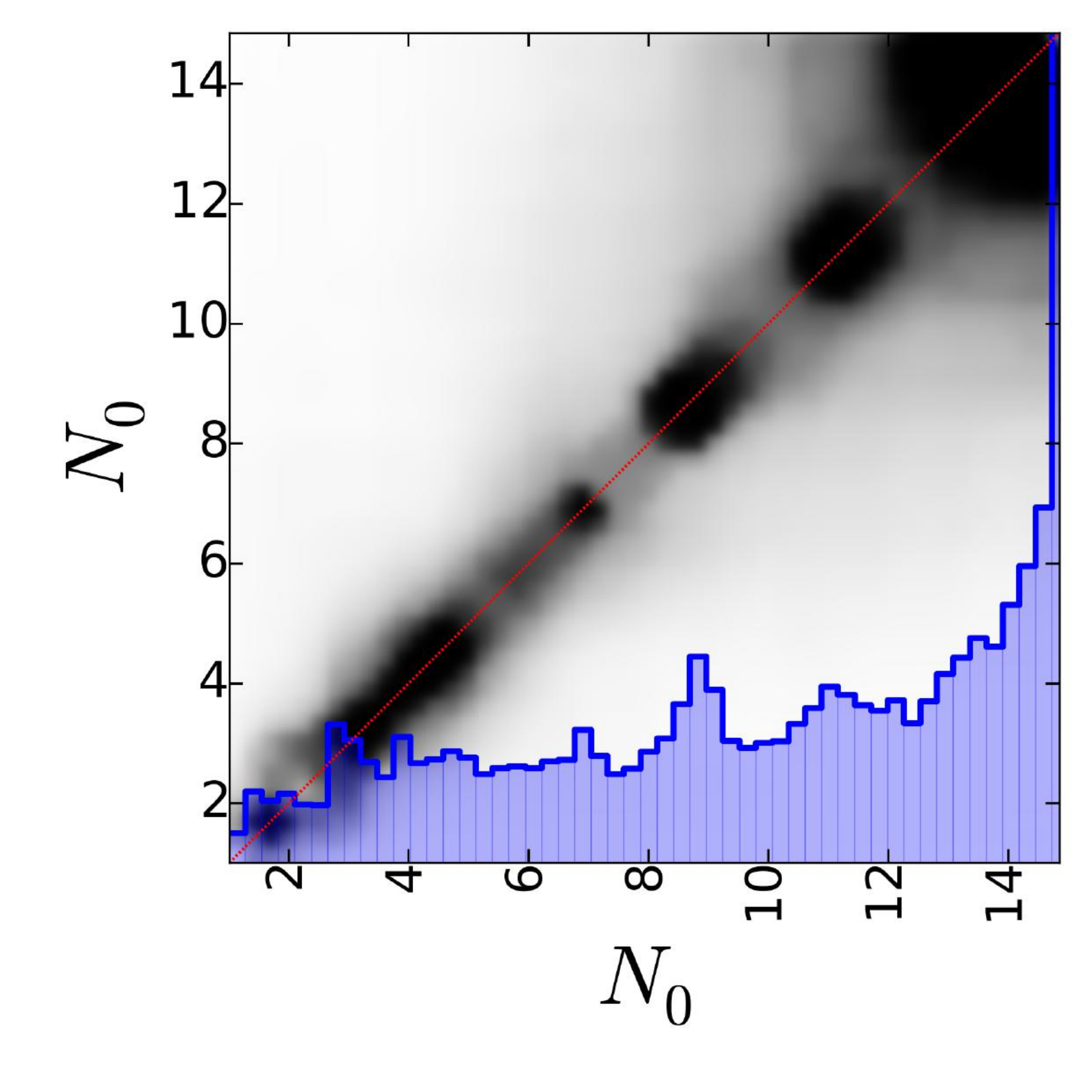}
\includegraphics[width=0.32\columnwidth]{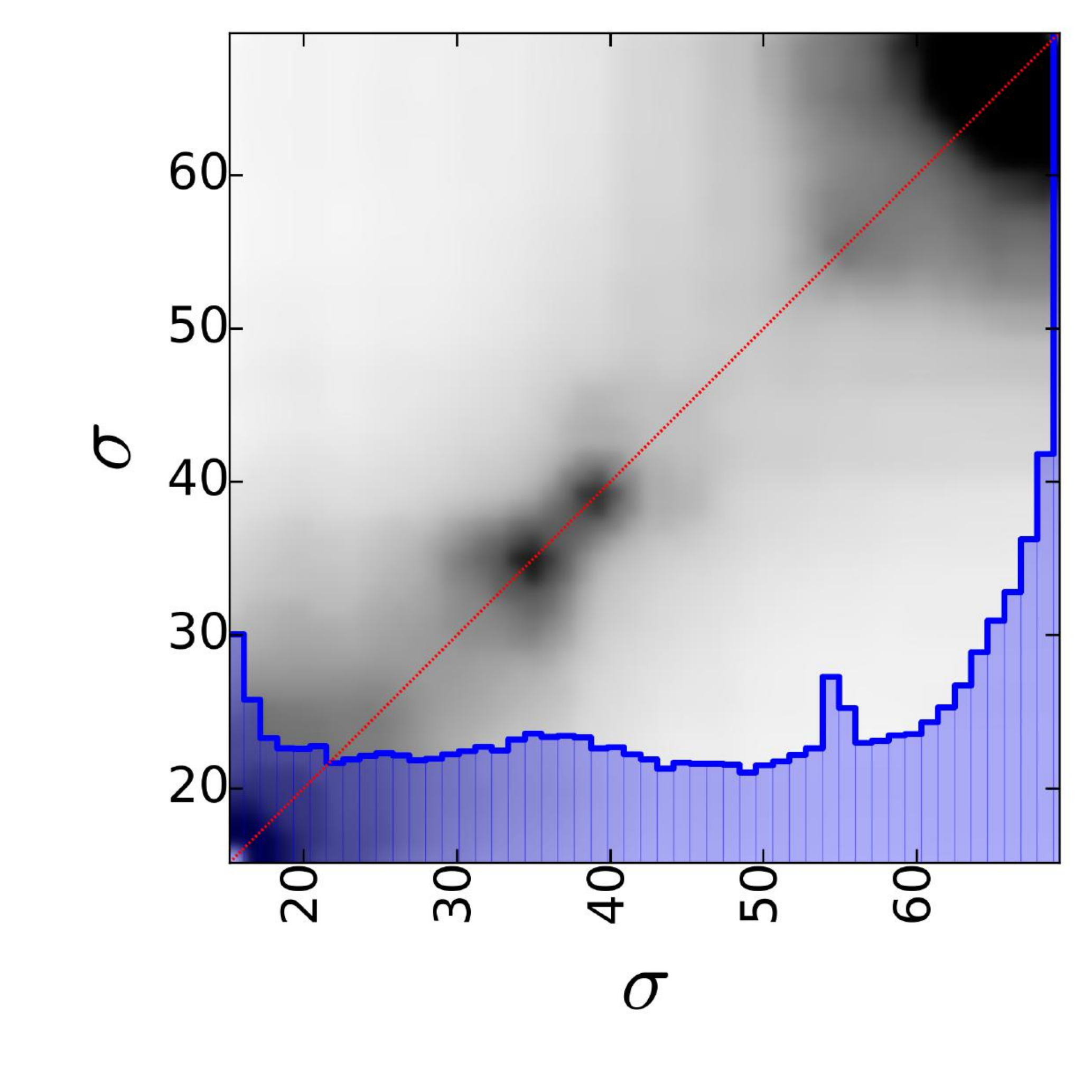}
\includegraphics[width=0.32\columnwidth]{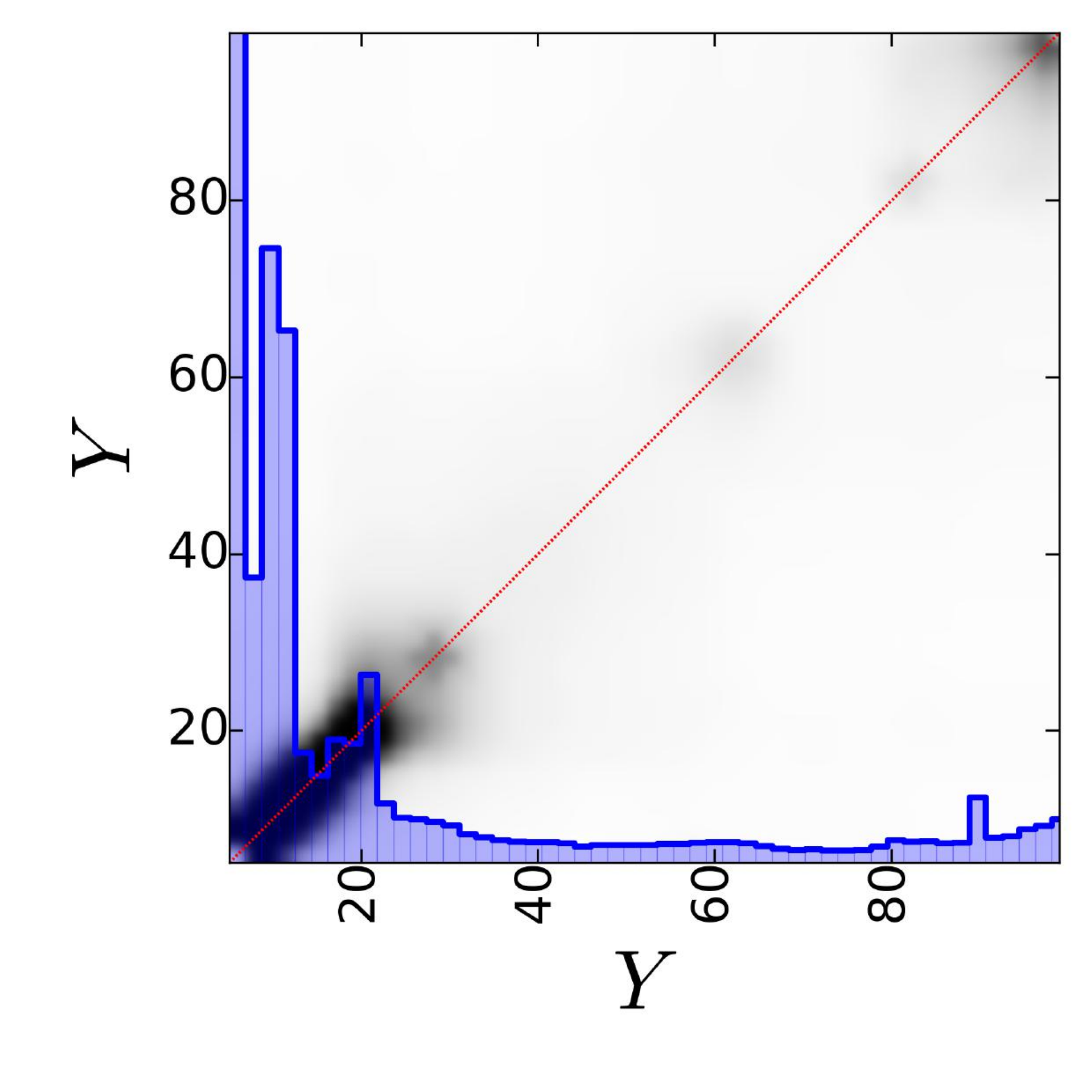}
\includegraphics[width=0.32\columnwidth]{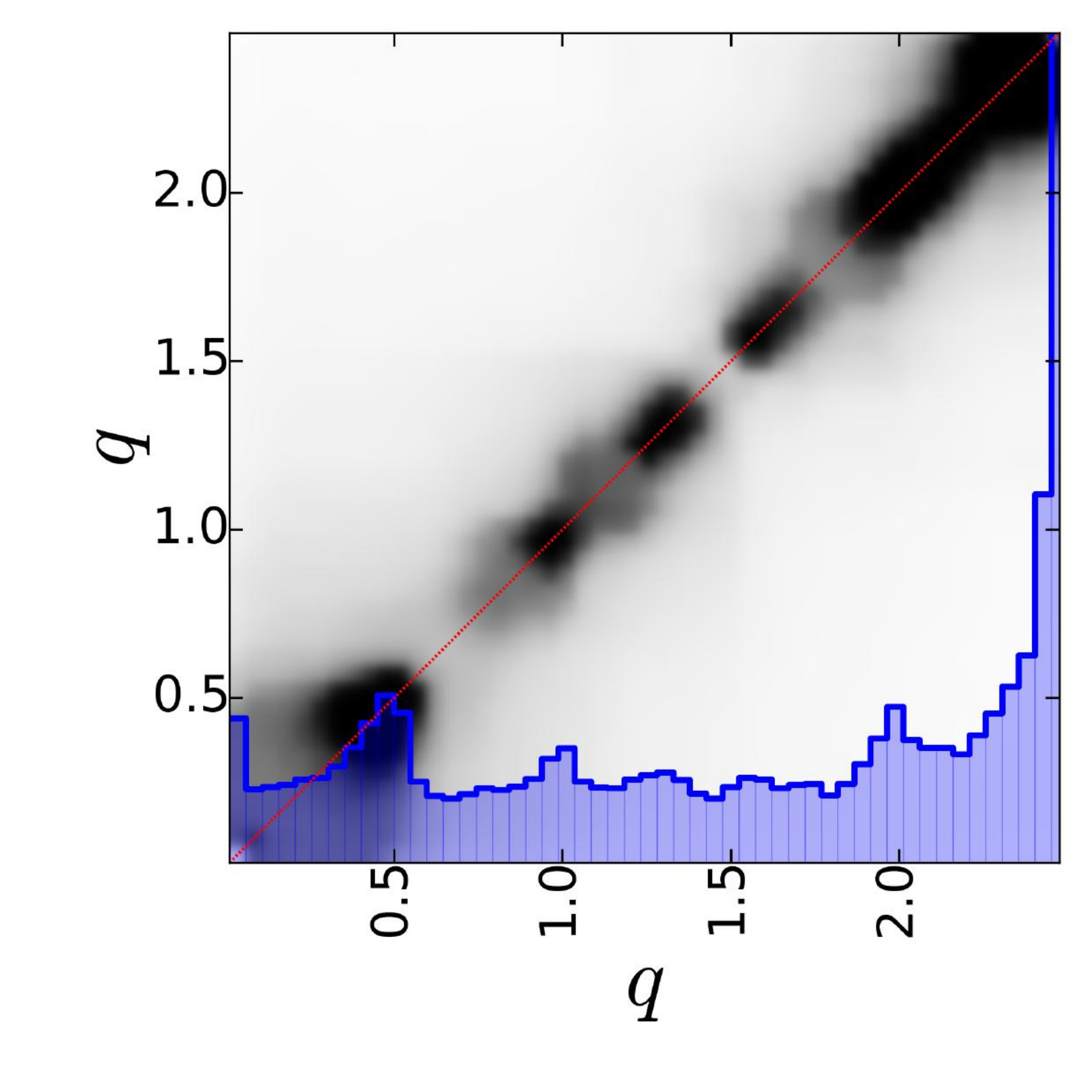}
\includegraphics[width=0.32\columnwidth]{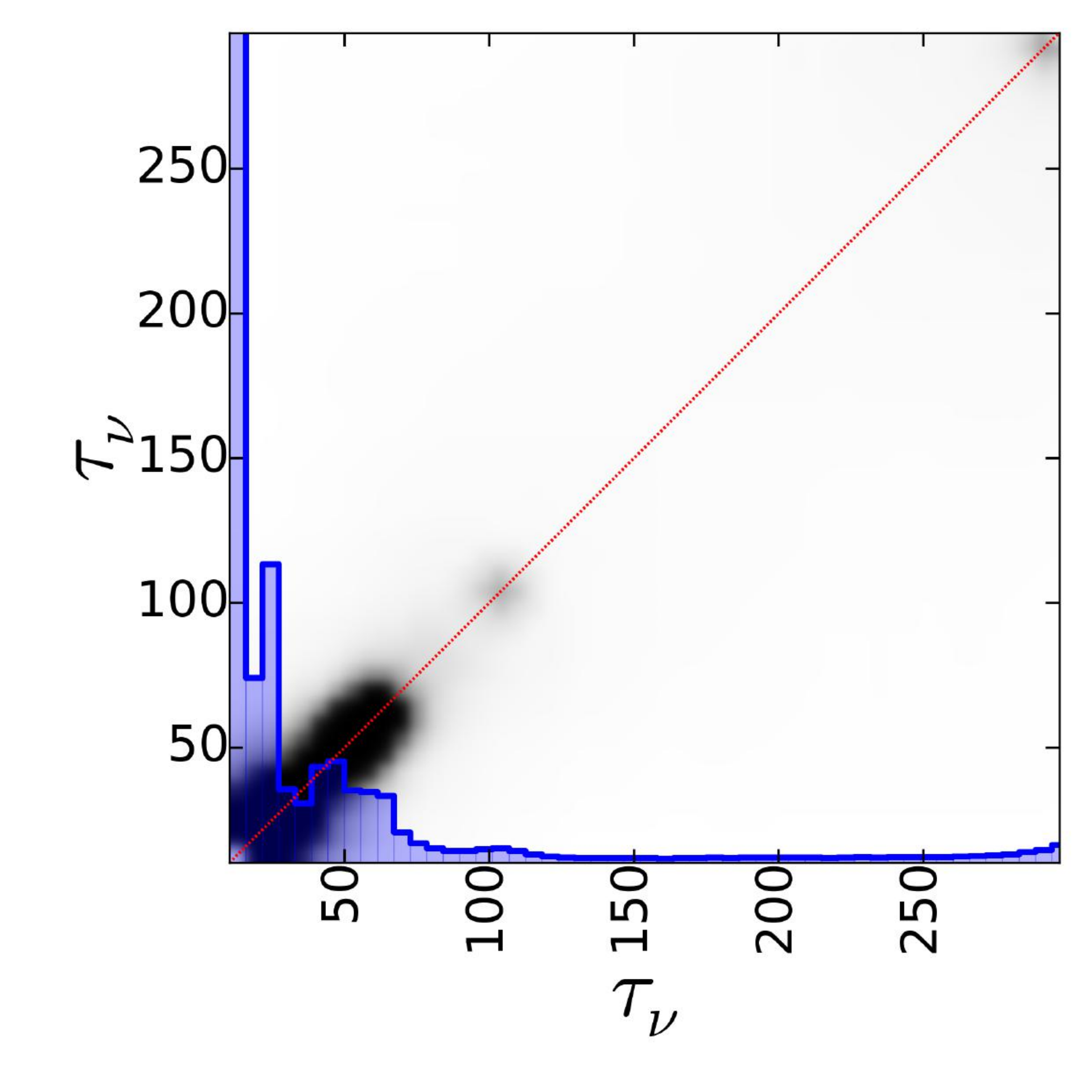}
\caption{Parameter versus parameter estimated from the sample drawn from the PDF (see text) for the model [Nenkova08]. The blue histogram shows the total distribution for each parameter.}
\label{fig:dataspecfitPDF}
\end{center}
\end{figure*}

\begin{figure*}[!ht]
\begin{flushleft}
\includegraphics[width=0.32\columnwidth]{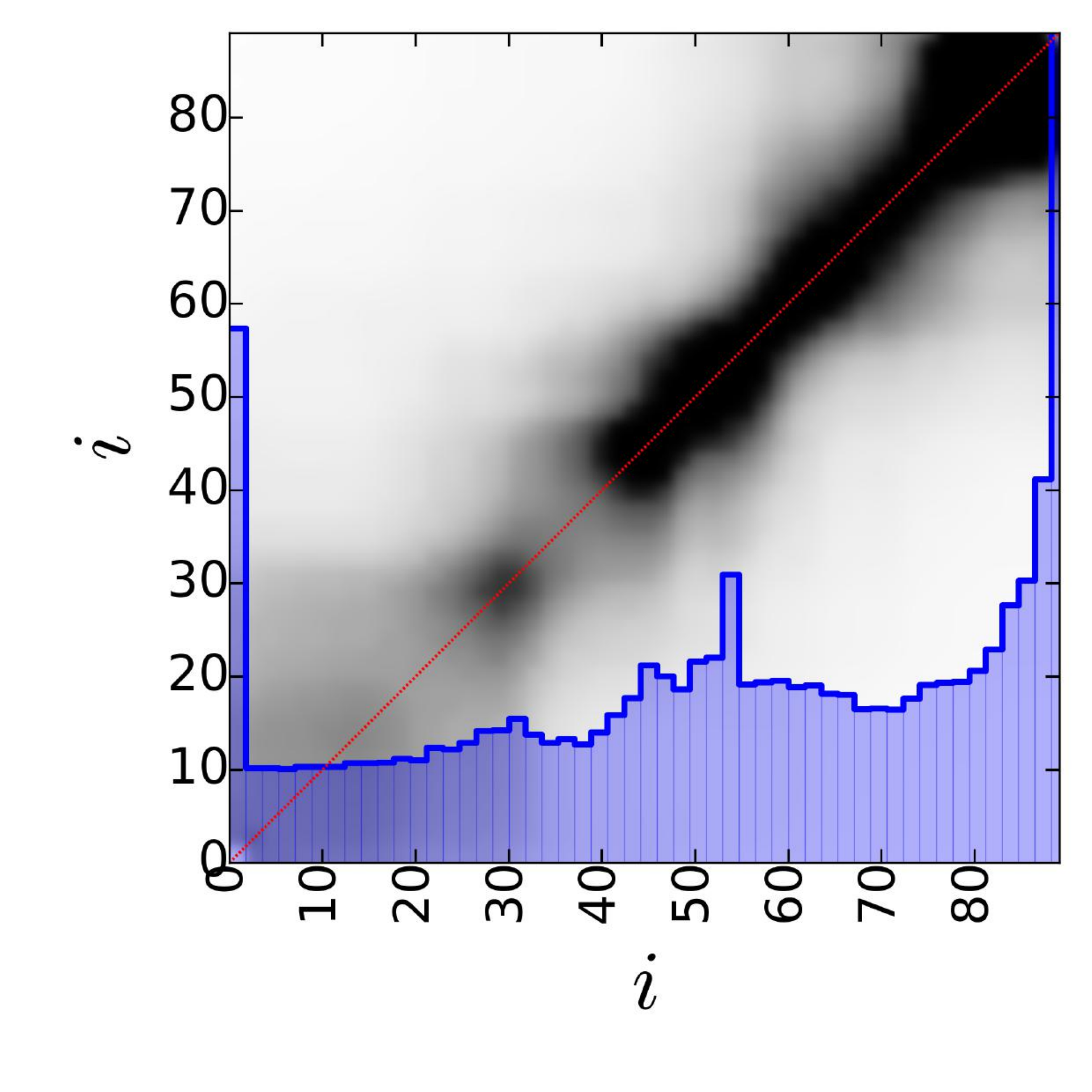}
\includegraphics[width=0.32\columnwidth]{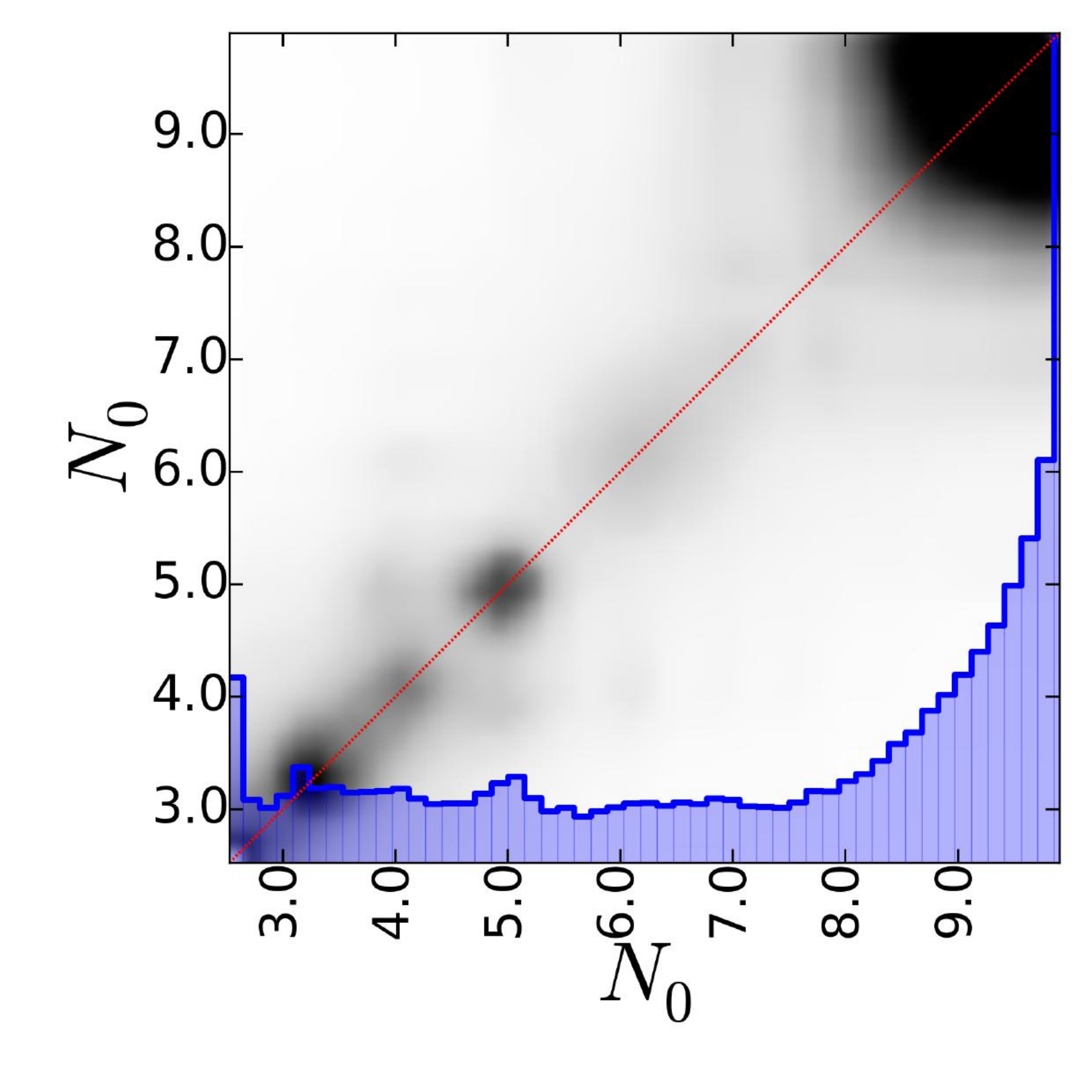}
\includegraphics[width=0.32\columnwidth]{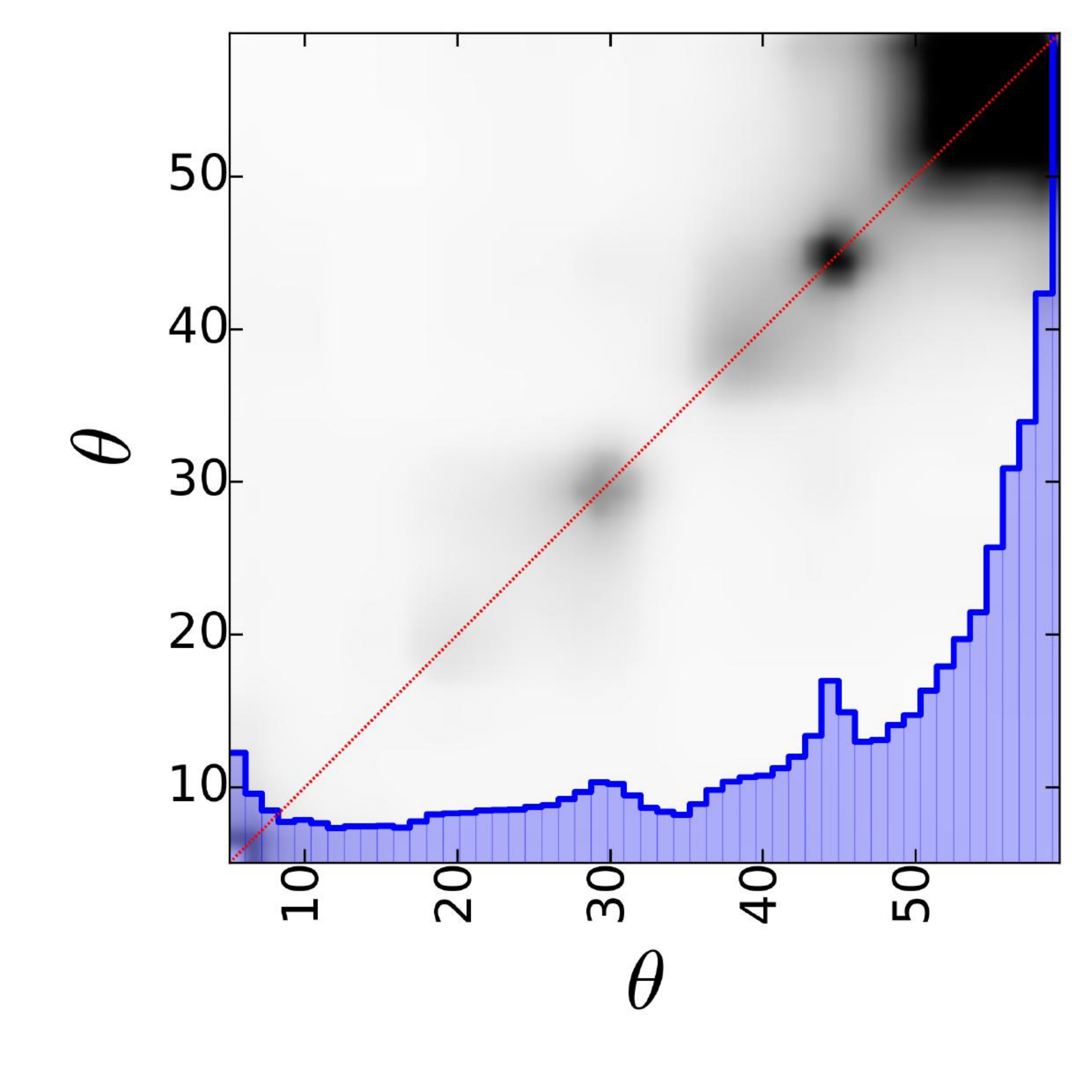}
\includegraphics[width=0.32\columnwidth]{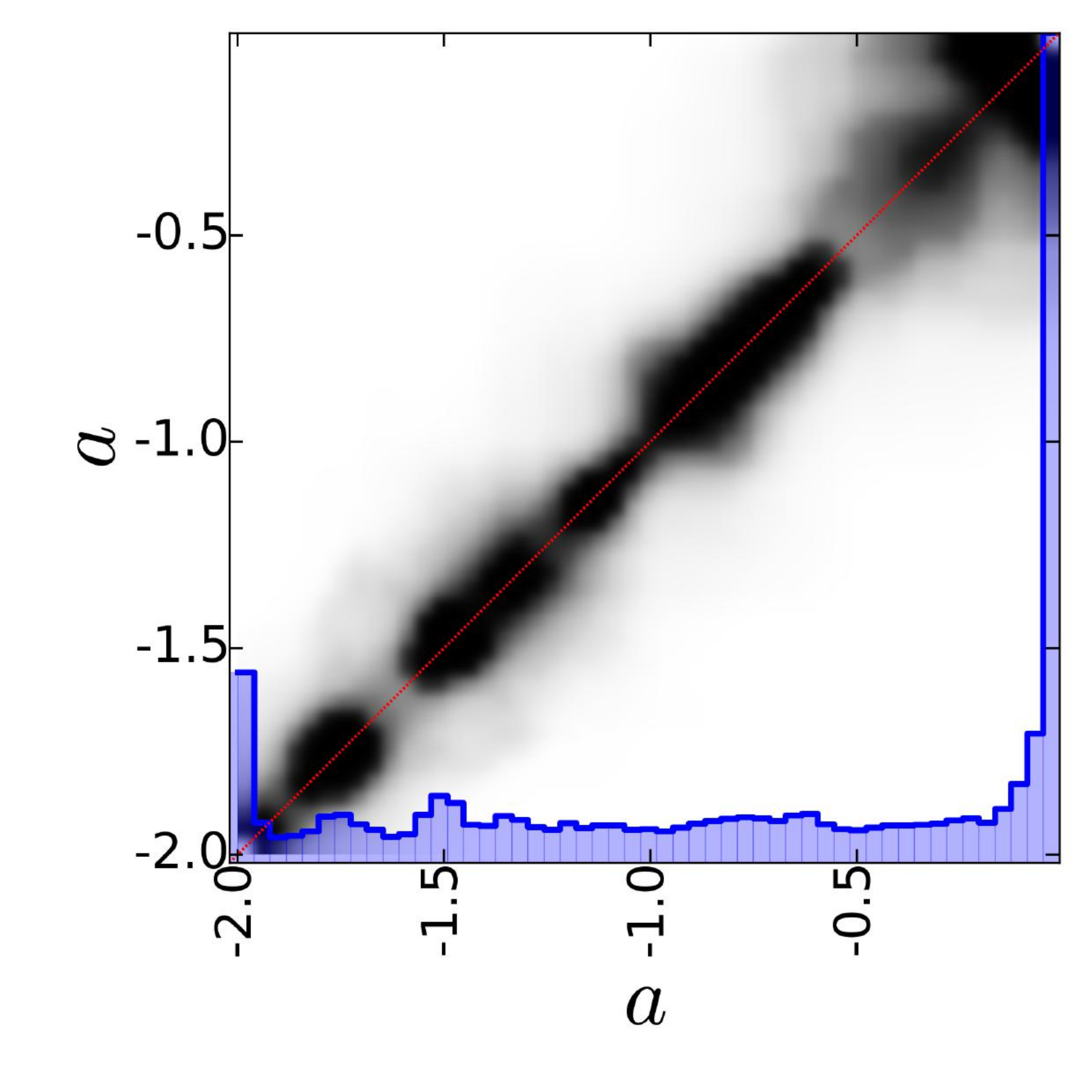}
\includegraphics[width=0.32\columnwidth]{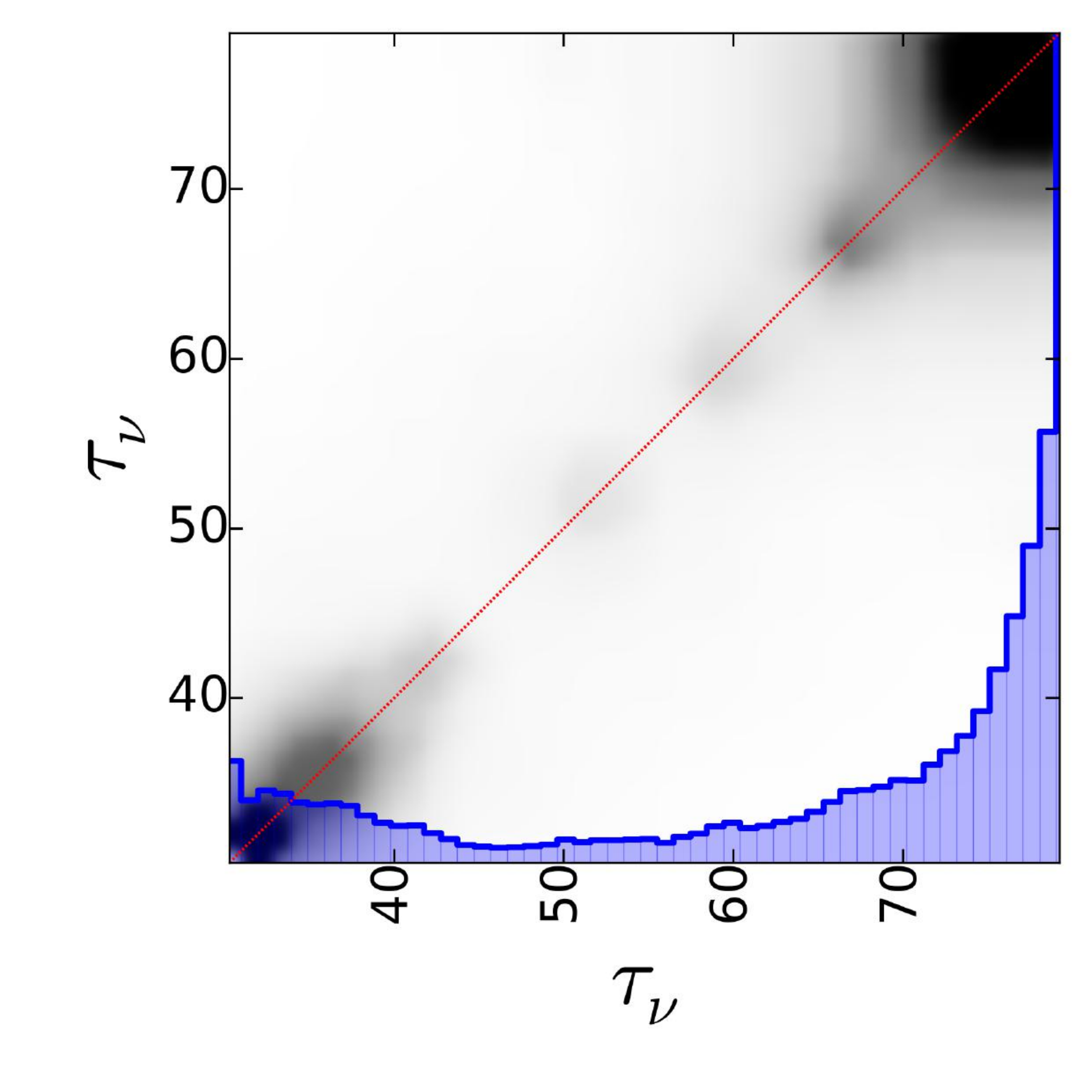}
\caption{Parameter versus parameter estimated from the sample drawn from the PDF (see text) for the model [Hoenig10]. The blue histogram shows the total distribution for each parameter.}
\label{fig:dataspecfitPDF}
\end{flushleft}
\end{figure*}

\begin{figure*}[!ht]
\begin{flushleft}
\includegraphics[width=0.32\columnwidth]{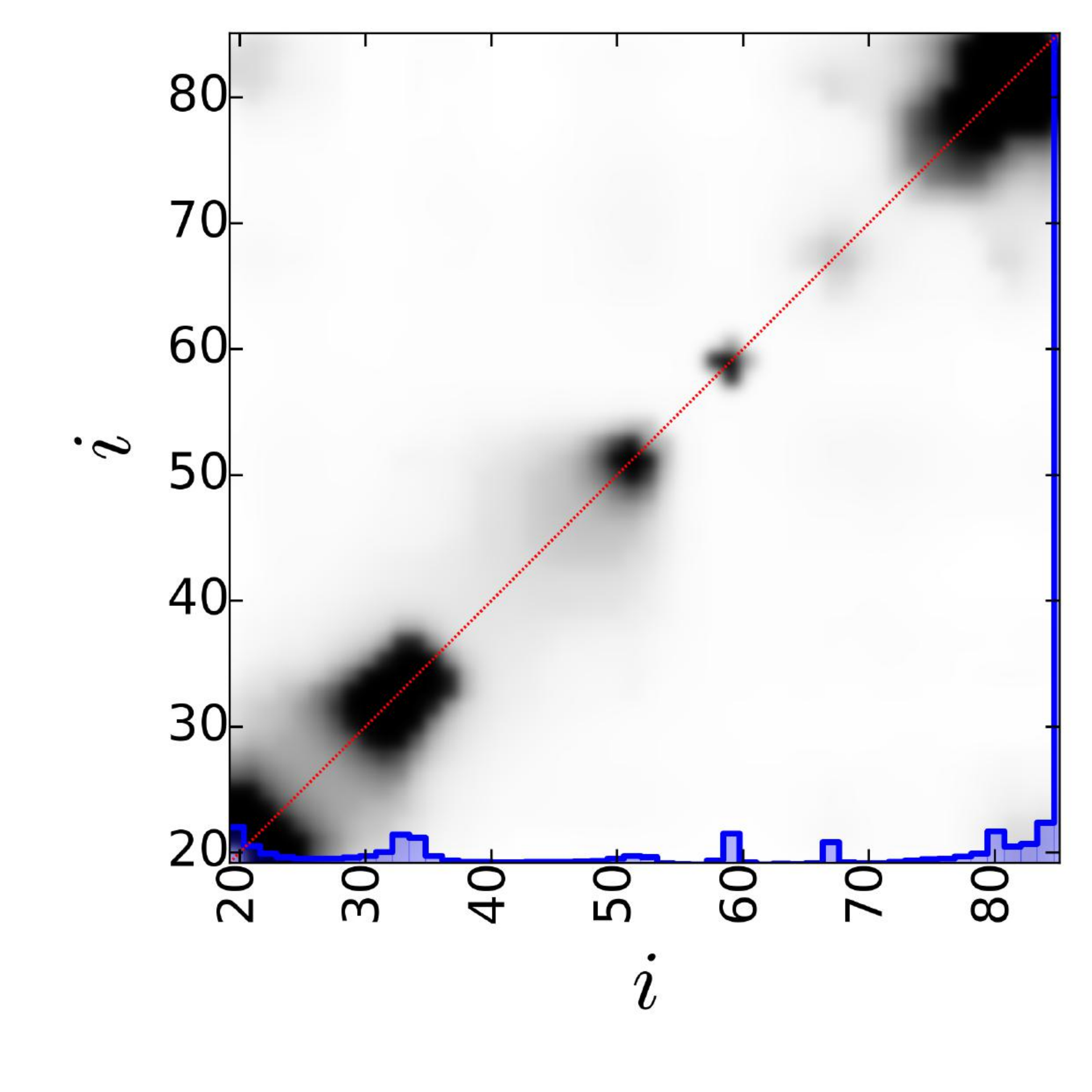}
\includegraphics[width=0.32\columnwidth]{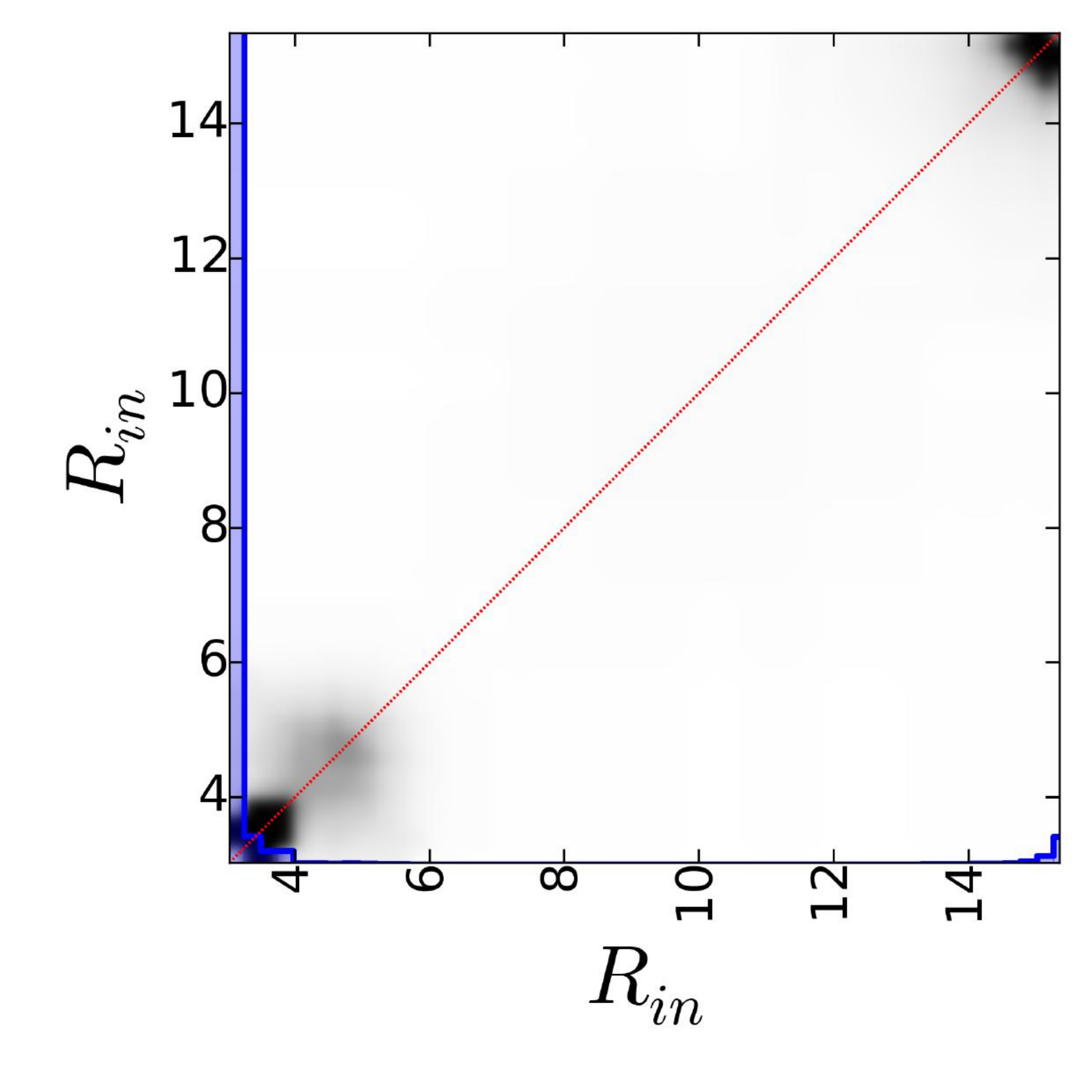}
\includegraphics[width=0.32\columnwidth]{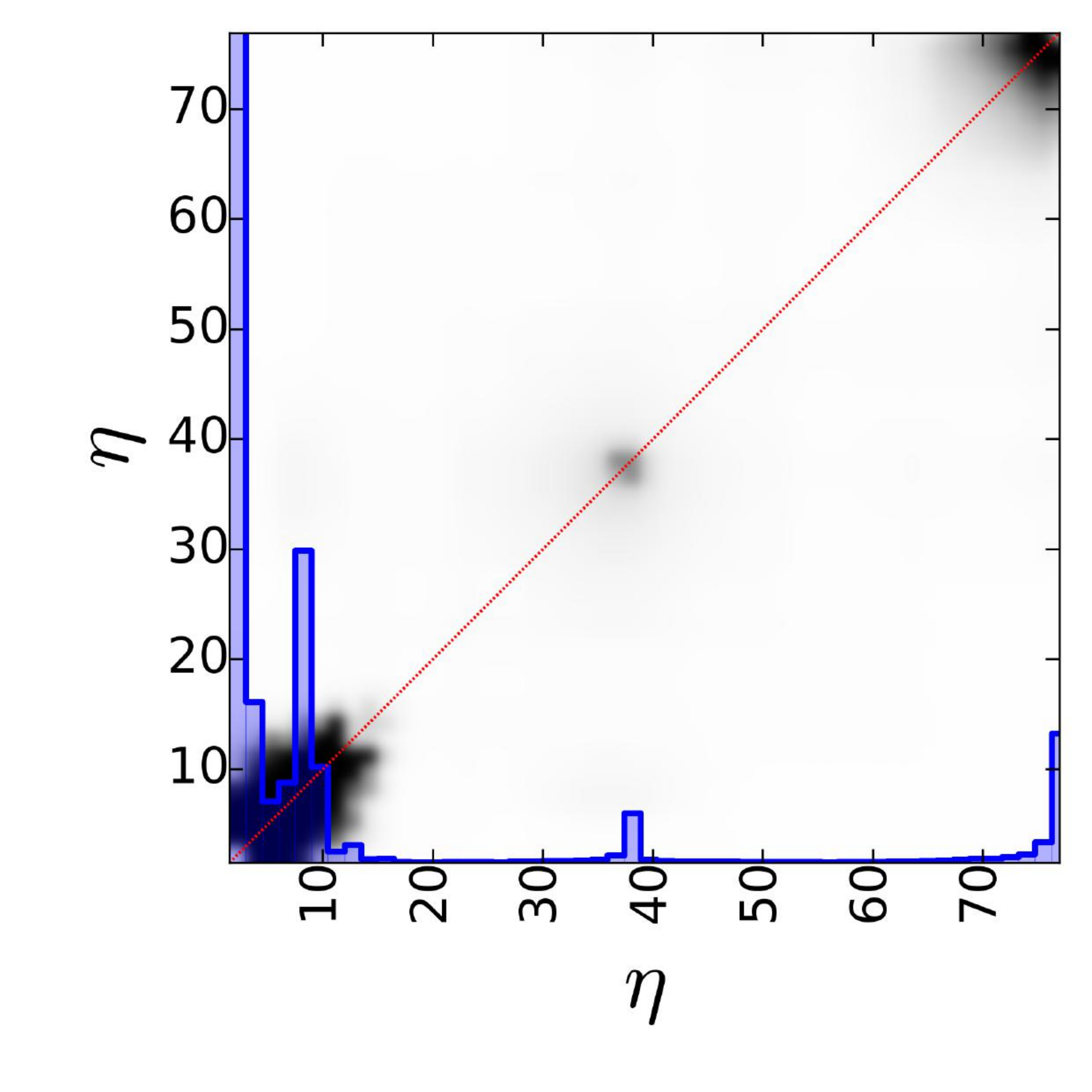}
\includegraphics[width=0.32\columnwidth]{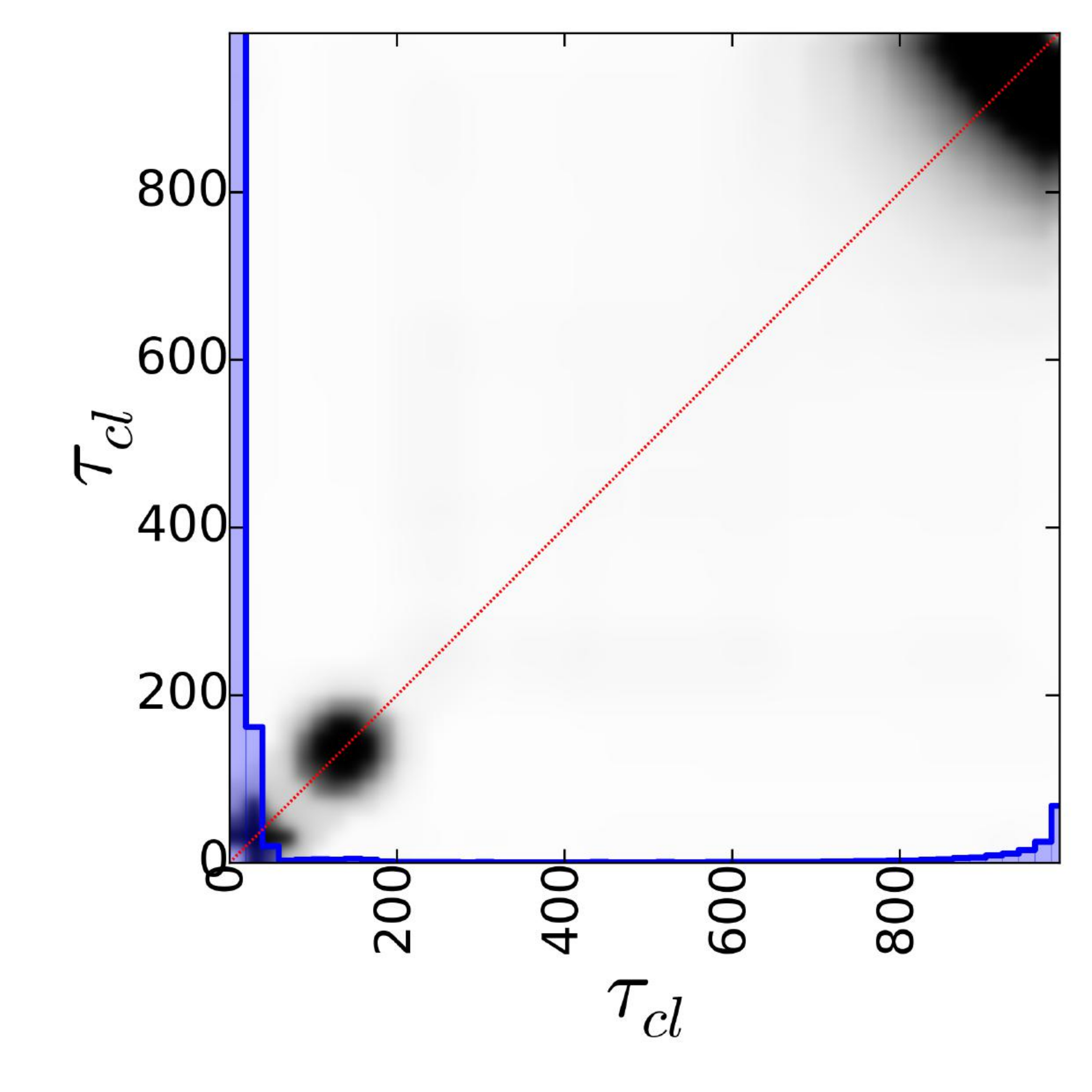}
\includegraphics[width=0.32\columnwidth]{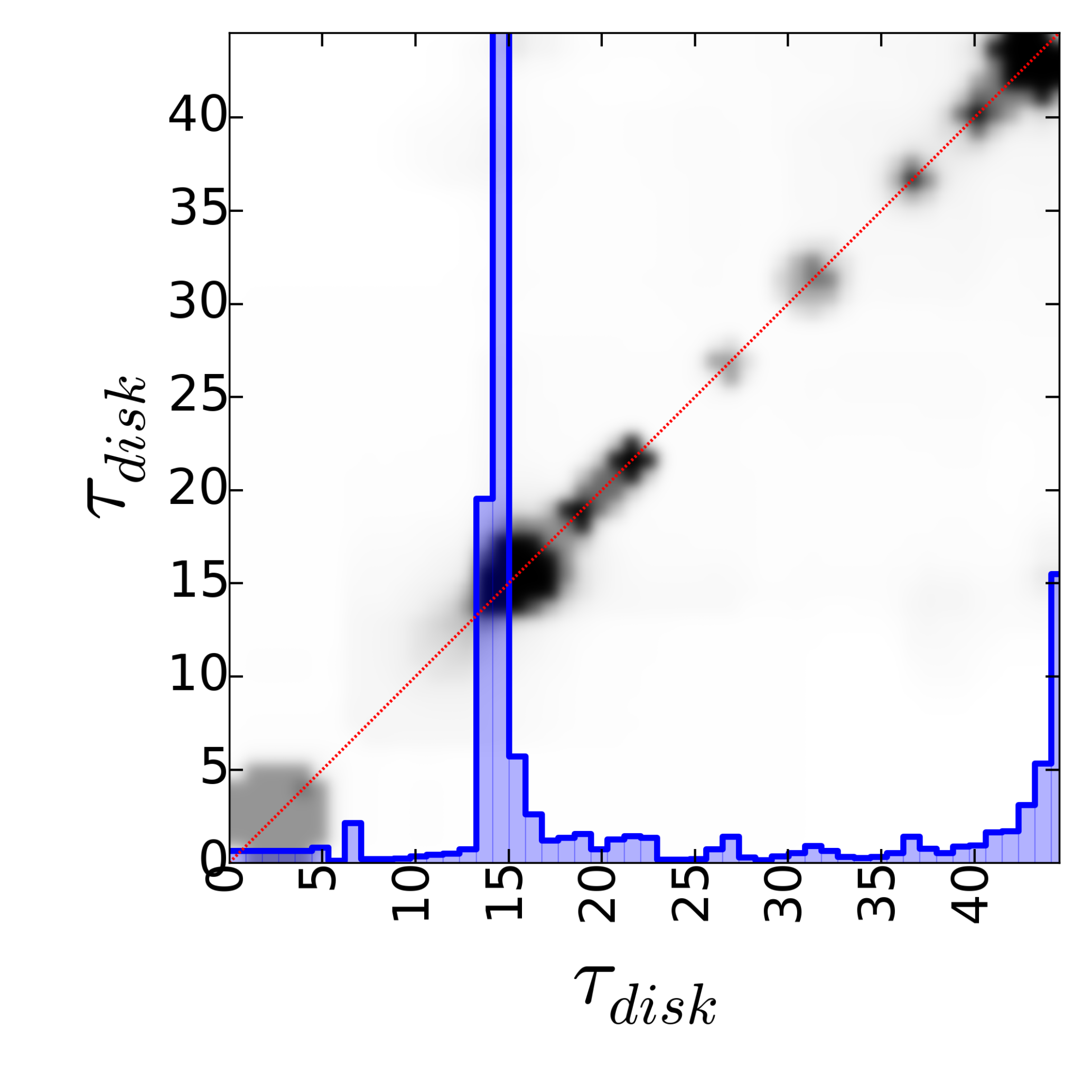}
\caption{Parameter versus parameter estimated from the sample drawn from the PDF (see text) for the model [Sieben15]. The blue histogram shows the total distribution for each parameter.}
\label{fig:dataspecfitPDF}
\end{flushleft}
\end{figure*}

\begin{figure*}[!ht]
\begin{center}
\includegraphics[width=0.32\columnwidth]{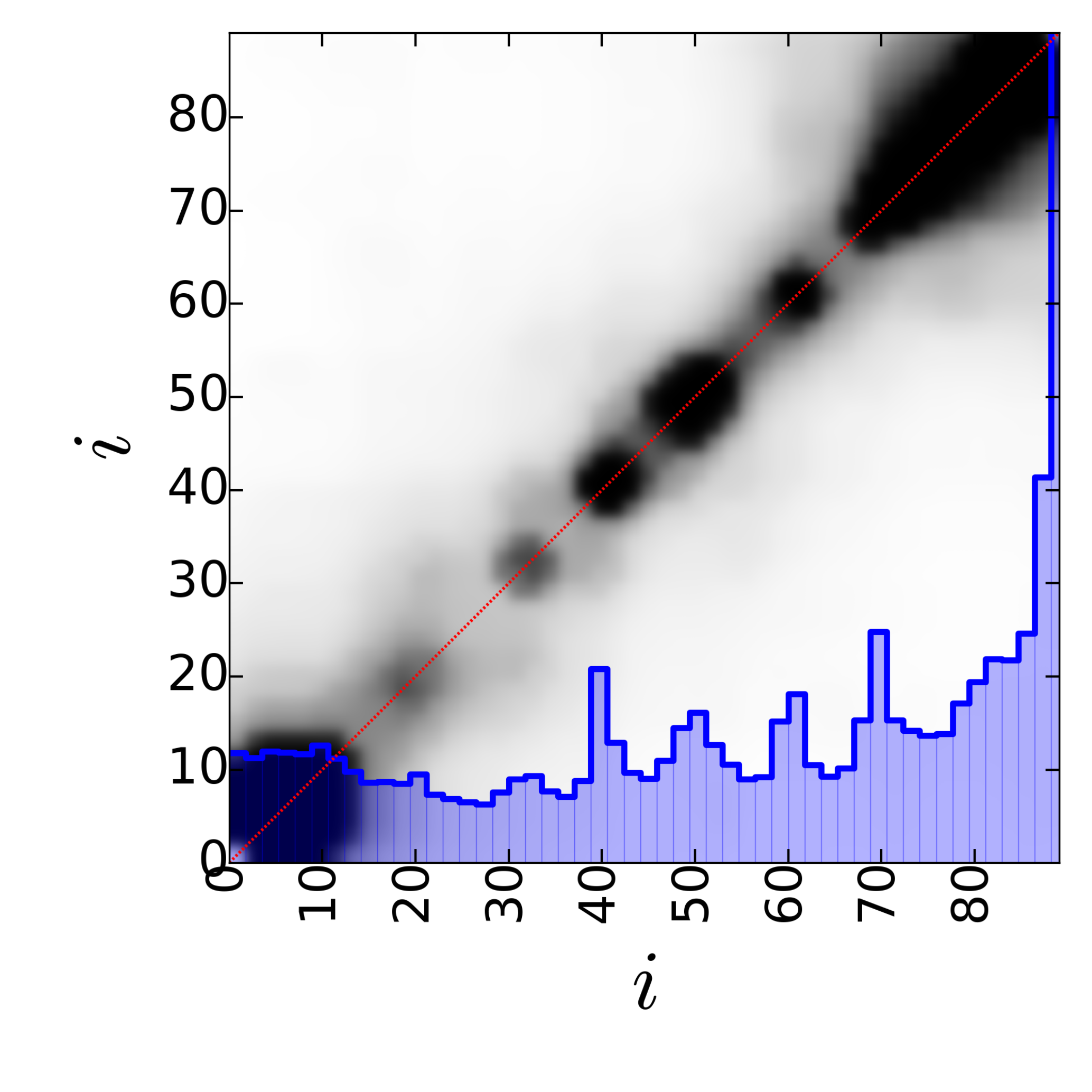}
\includegraphics[width=0.32\columnwidth]{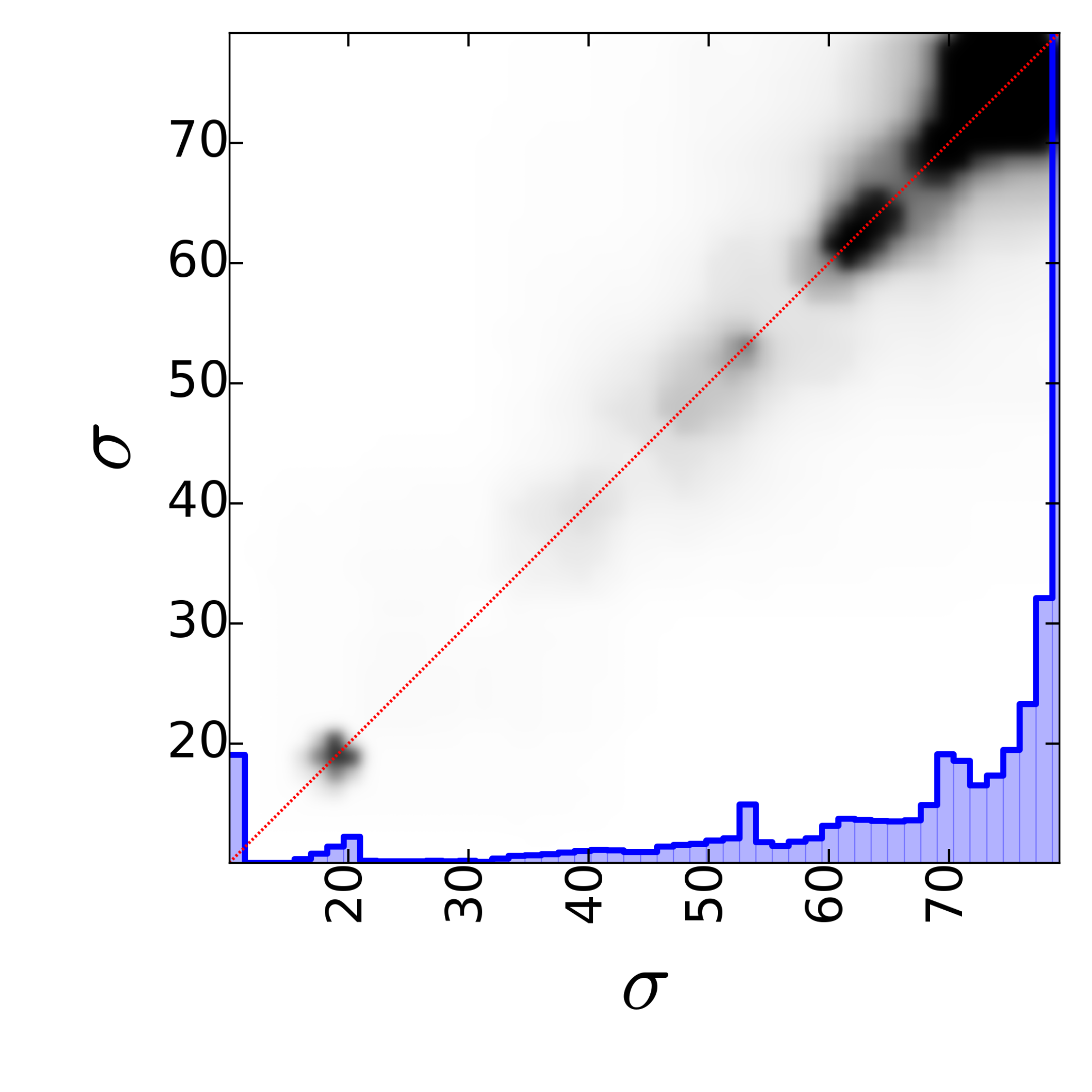}
\includegraphics[width=0.32\columnwidth]{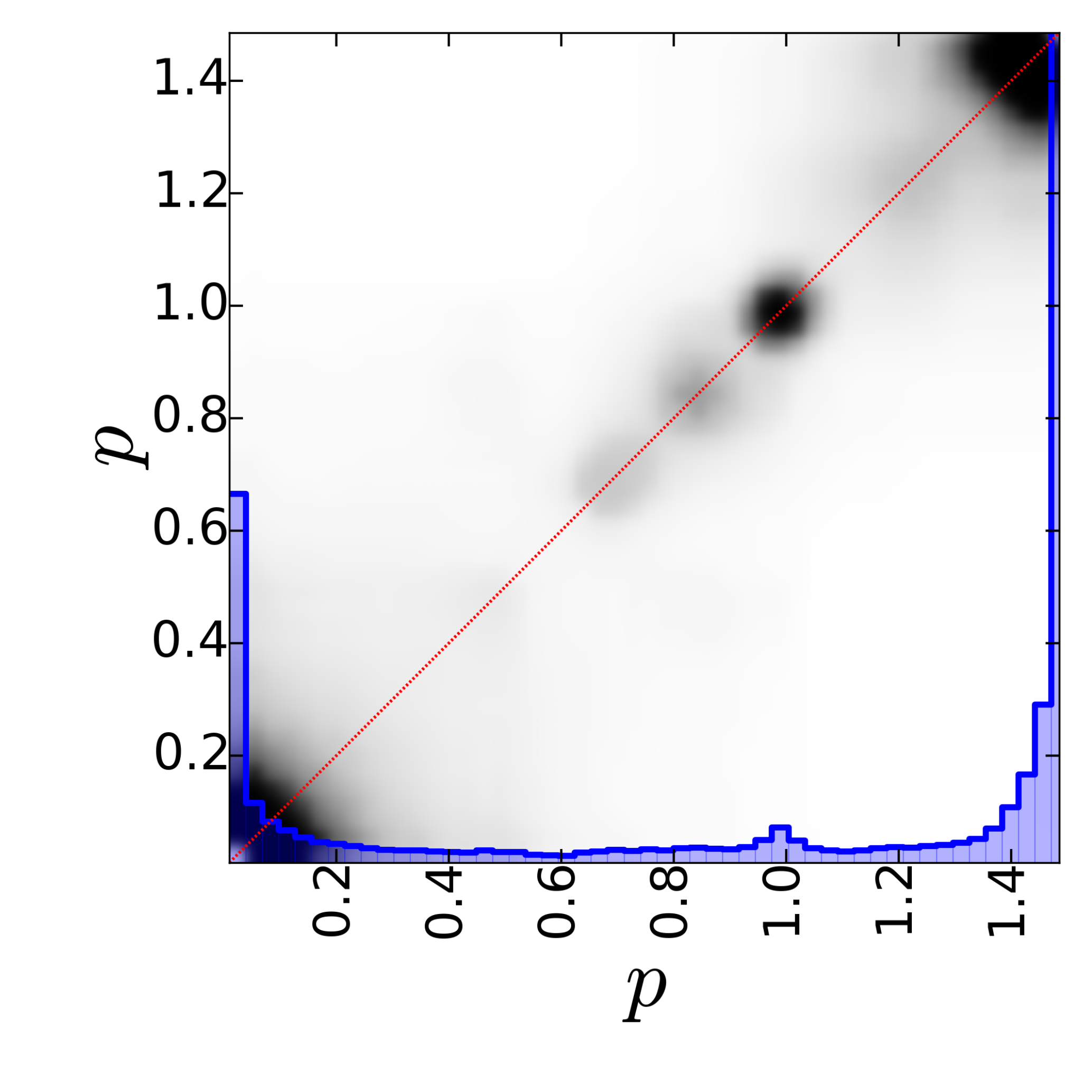}
\includegraphics[width=0.32\columnwidth]{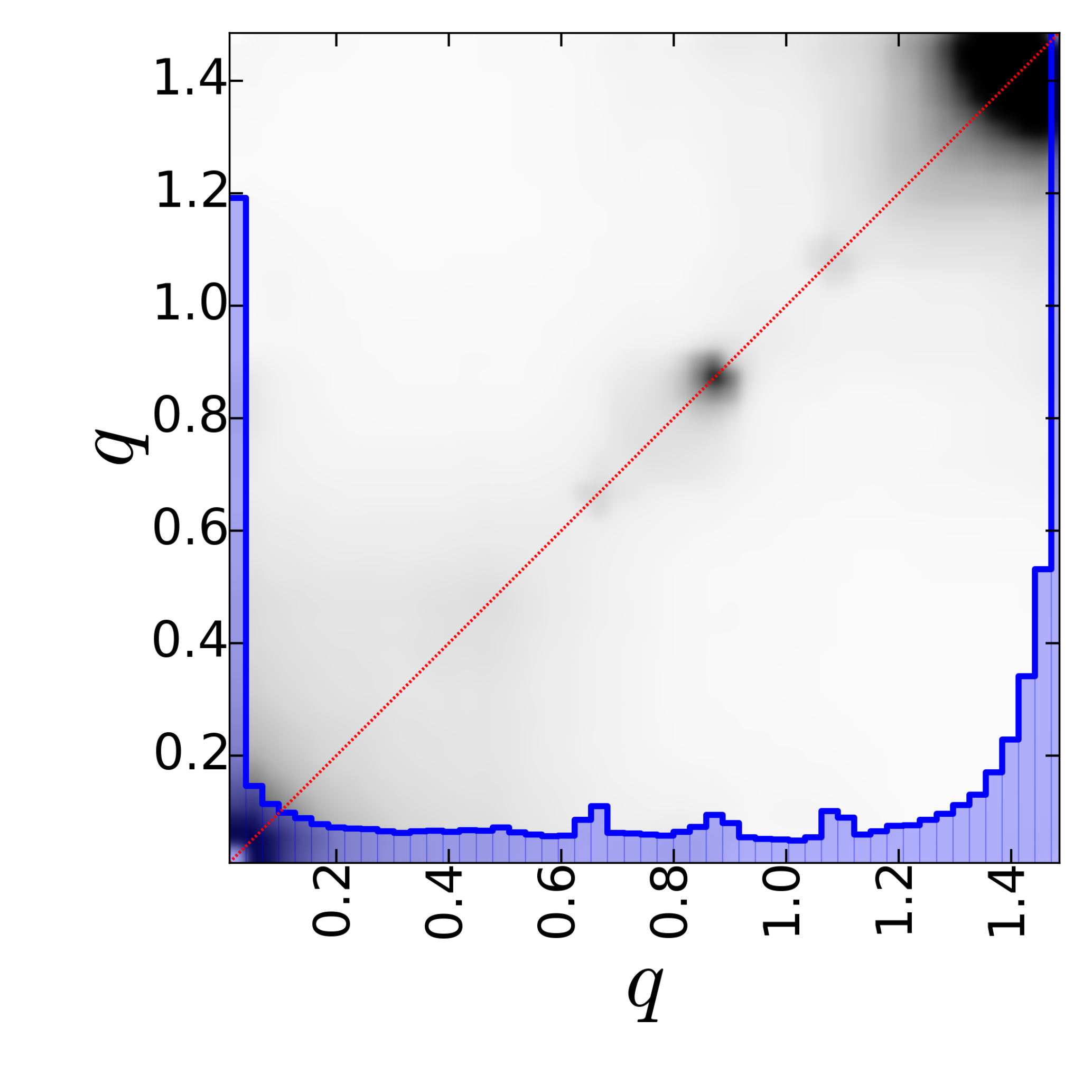}
\includegraphics[width=0.32\columnwidth]{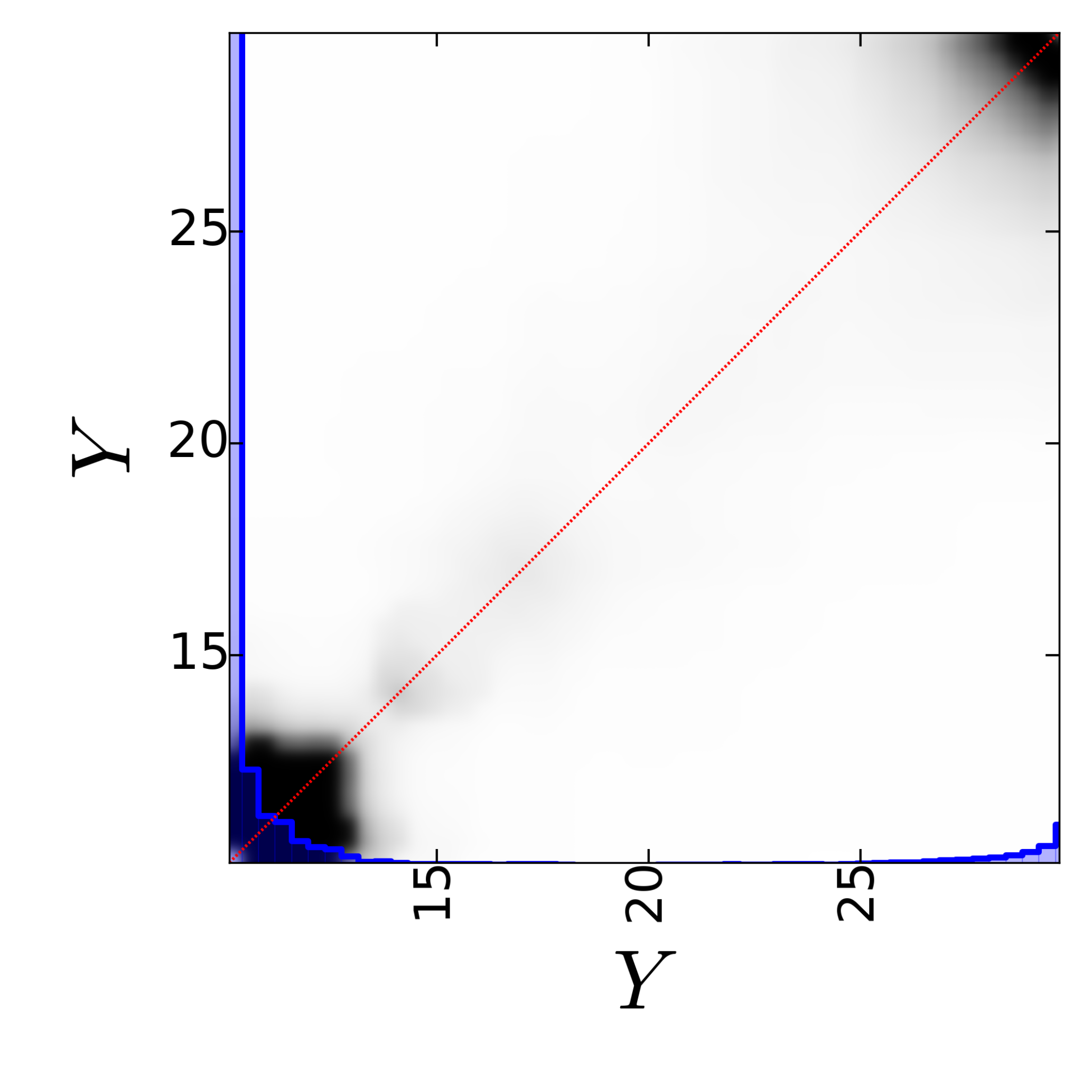}
\includegraphics[width=0.32\columnwidth]{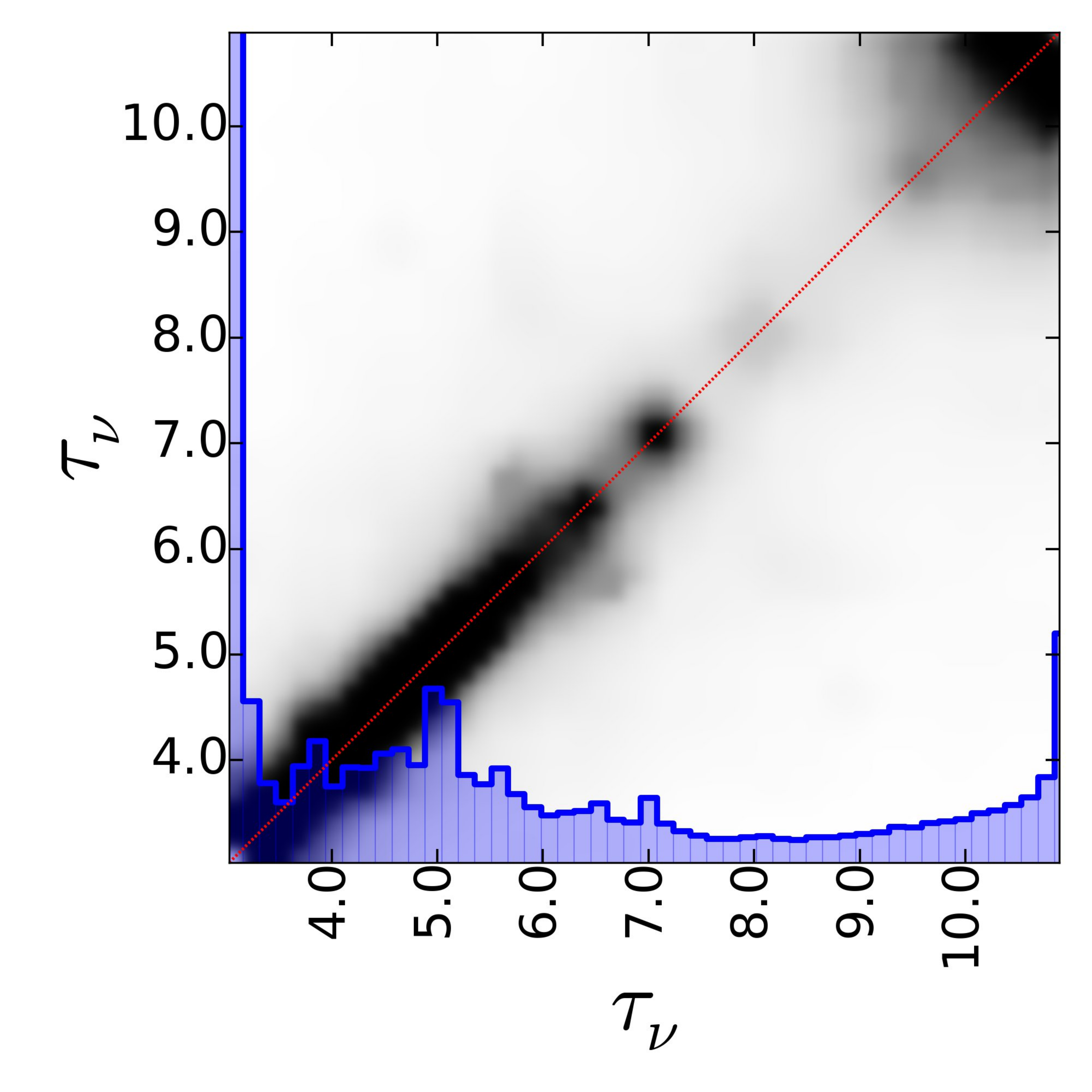}
\caption{Parameter versus parameter estimated from the sample drawn from the PDF (see text) for the model [Stalev16]. The blue histogram shows the total distribution for each parameter.}
\label{fig:dataspecfitPDF}
\end{center}
\end{figure*}

\begin{figure*}[!ht]
\begin{flushleft}
\includegraphics[width=0.32\columnwidth]{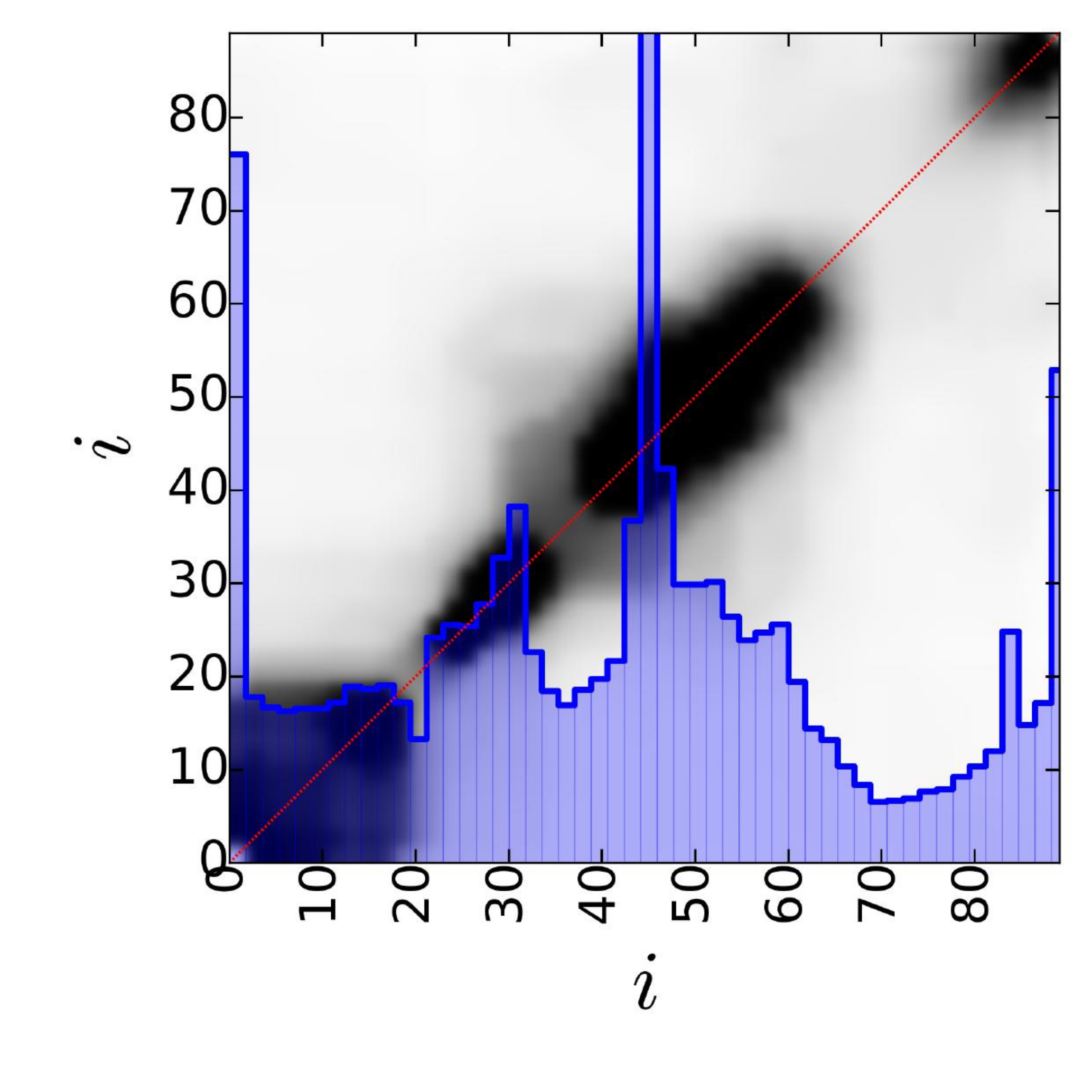}
\includegraphics[width=0.32\columnwidth]{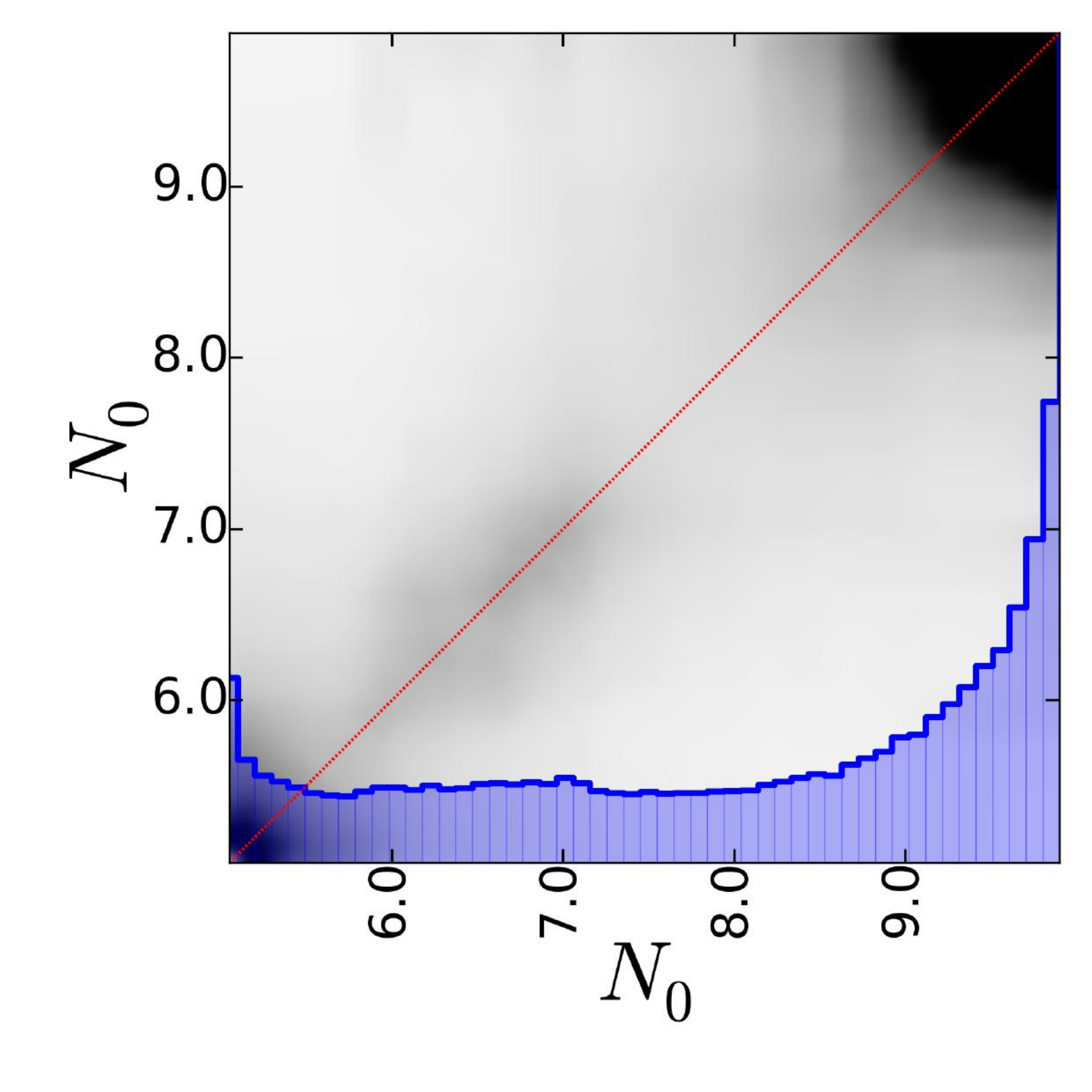}
\includegraphics[width=0.32\columnwidth]{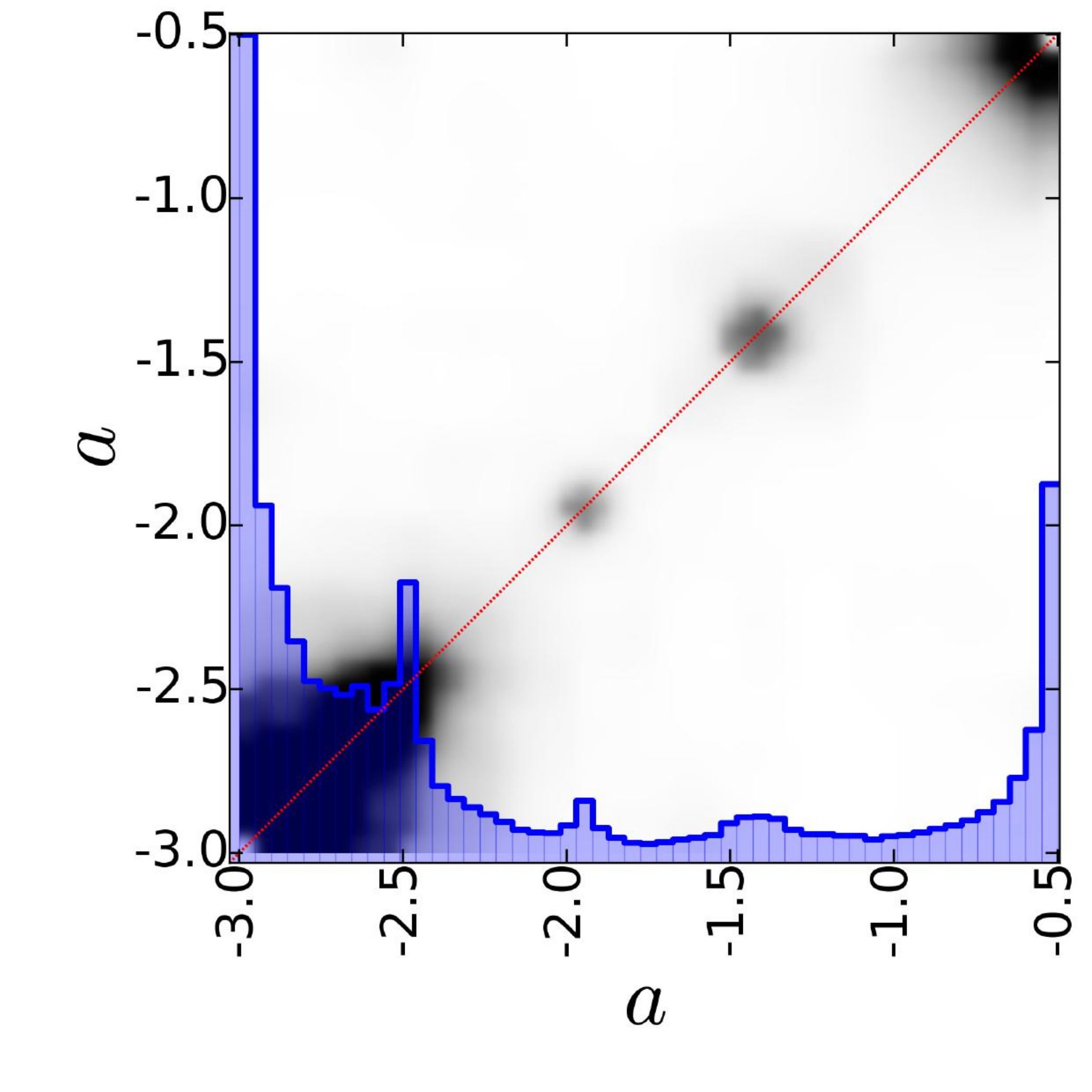}
\includegraphics[width=0.32\columnwidth]{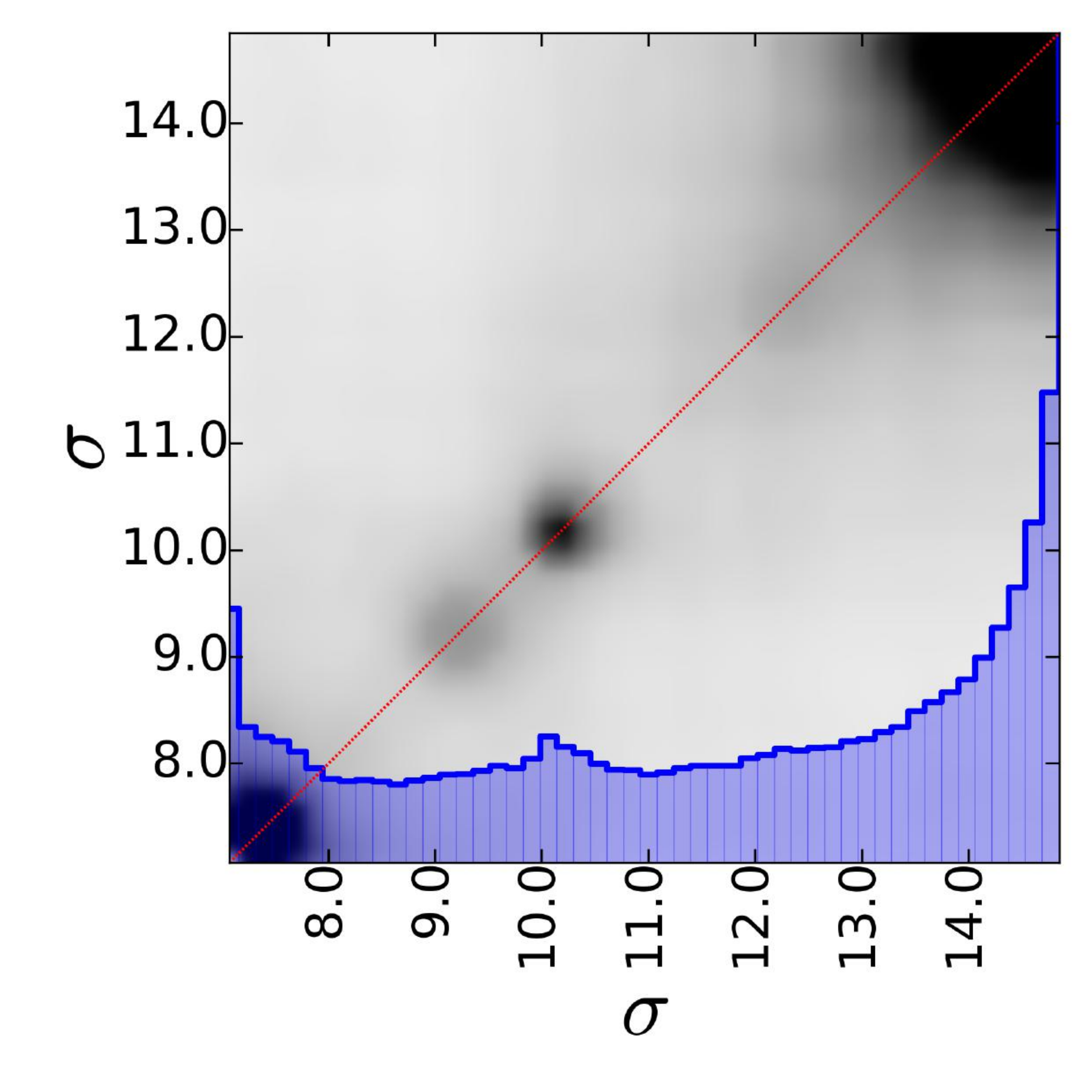}
\includegraphics[width=0.32\columnwidth]{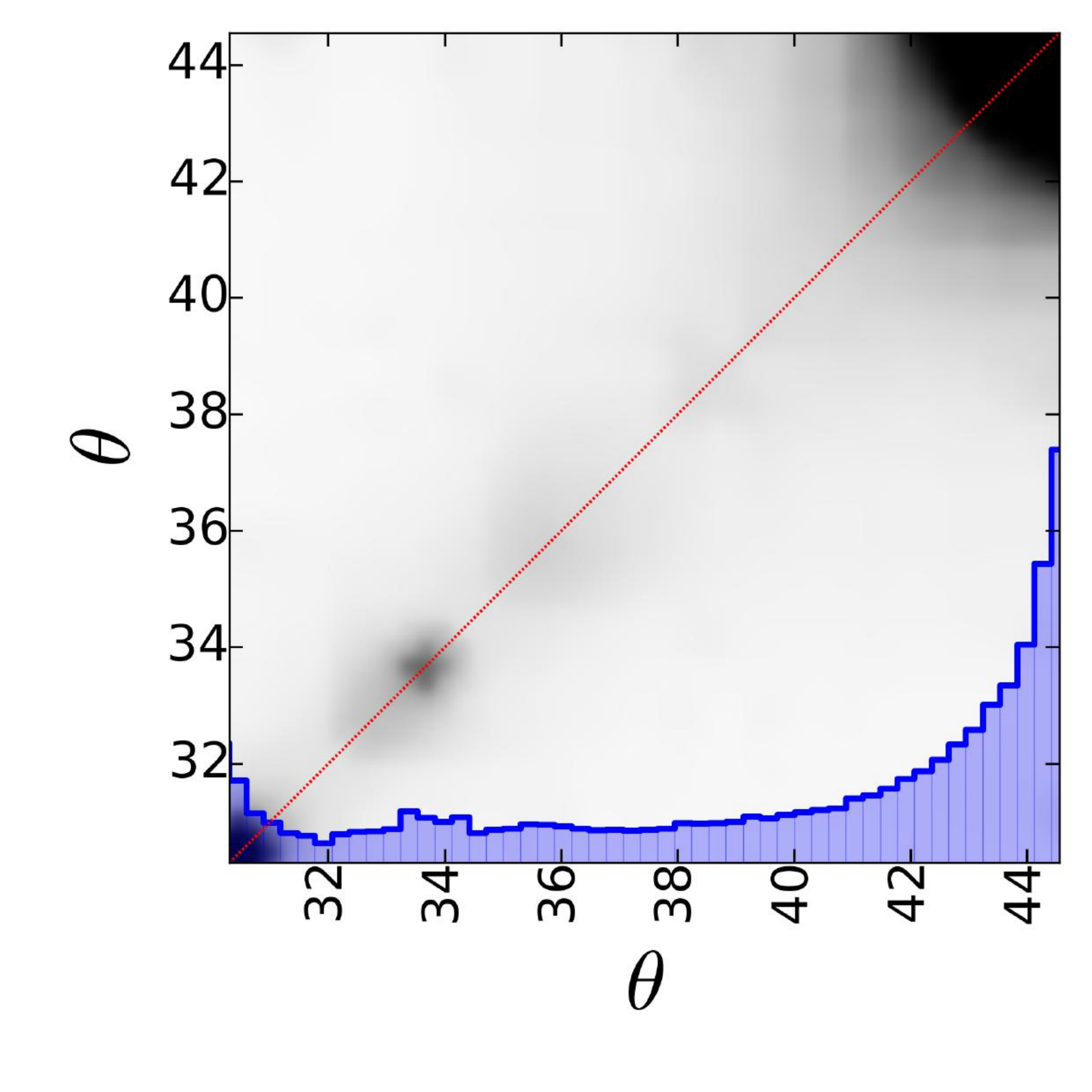}
\includegraphics[width=0.32\columnwidth]{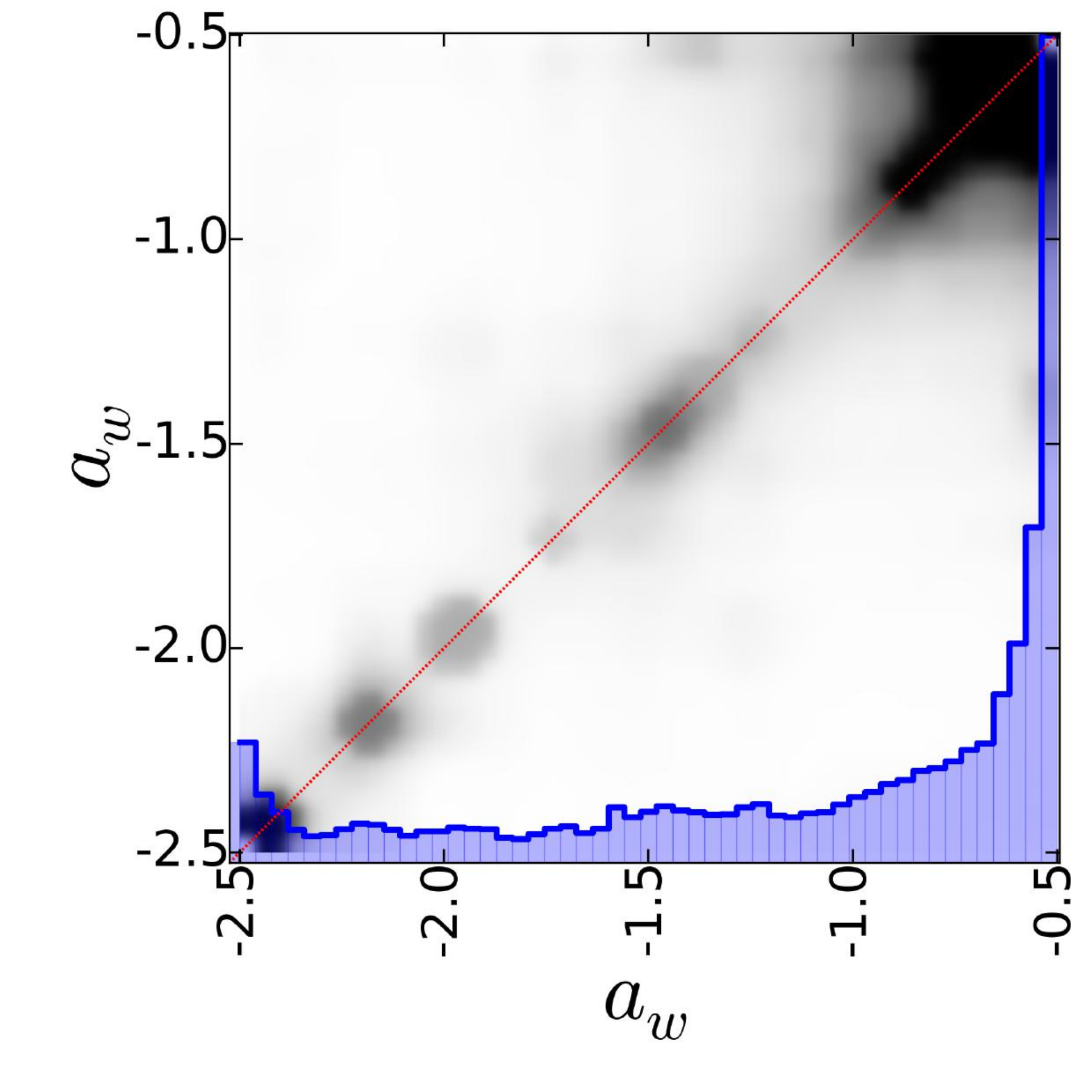}
\includegraphics[width=0.32\columnwidth]{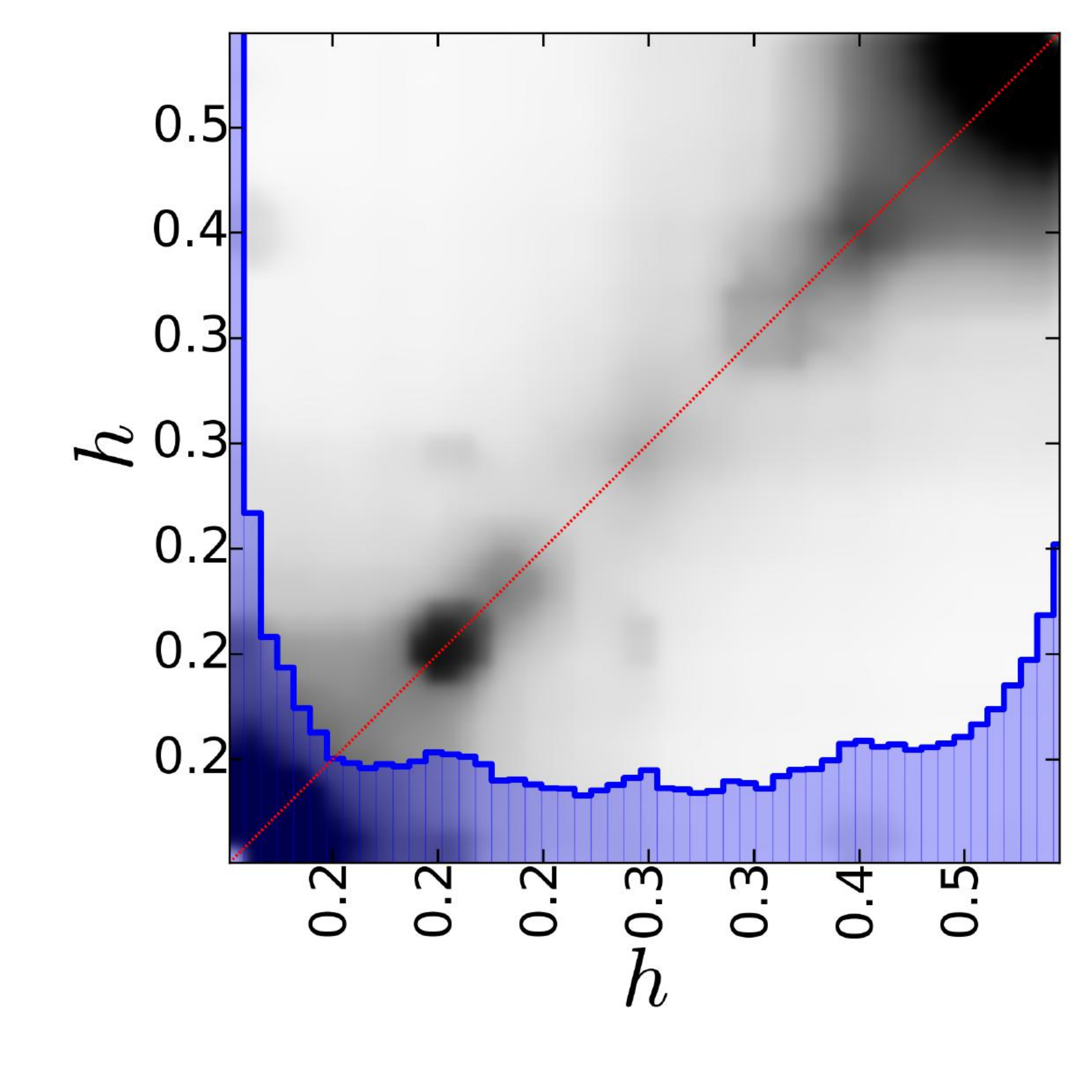}
\includegraphics[width=0.32\columnwidth]{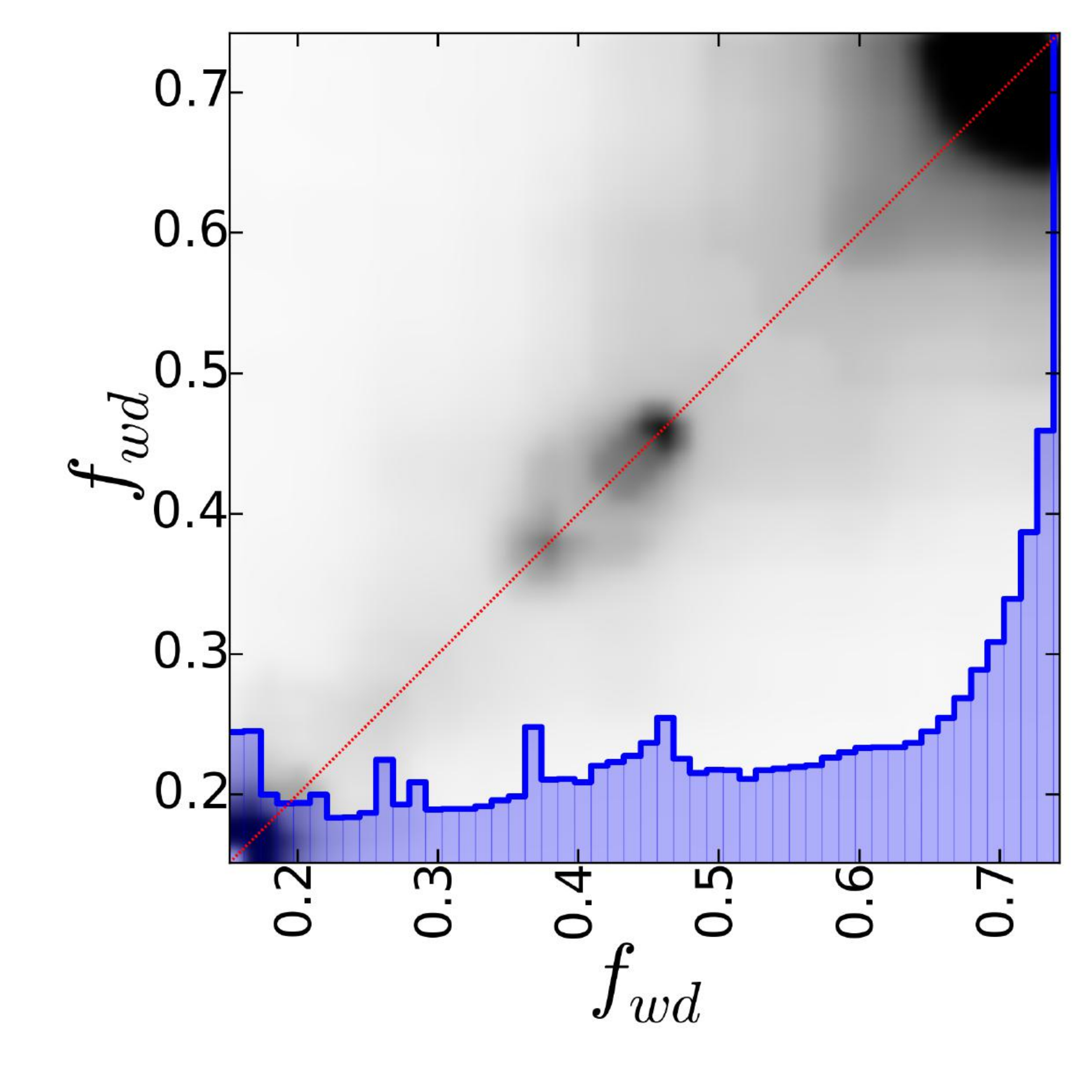}
\caption{Parameter versus parameter estimated from the sample drawn from the PDF (see text) for the model [Hoenig17]. The blue histogram shows the total distribution for each parameter.}
\label{fig:dataspecfitPDF}
\end{flushleft}
\end{figure*}


\begin{thebibliography}{}
\bibitem[Antonucci, \& Miller(1985)]{Antonucci85} Antonucci, R.~R.~J., \& Miller, J.~S.\ 1985, \apj, 297, 621.
\bibitem[Alonso-Herrero et al.(2011)]{Alonso-Herrero11} Alonso-Herrero, A., Ramos Almeida, C., Mason, R., et al.\ 2011, \apj, 736, 82.
\bibitem[Alonso-Herrero et al.(2018)]{Alonso-Herrero18} Alonso-Herrero, A., Pereira-Santaella, M., Garc{\'\i}a-Burillo, S., et al.\ 2018, \apj, 859, 144.
\bibitem[Arnaud(1996)]{Arnaud96} Arnaud, K.~A.\ 1996, Astronomical Data Analysis Software and Systems V, 101, 17.
\bibitem[Asmus et al.(2016)]{Asmus16} Asmus, D., H{\"o}nig, S.~F., \& Gandhi, P.\ 2016, \apj, 822, 109
\bibitem[Braatz et al.(1993)]{Braatz93} Braatz, J.~A., Wilson, A.~S., Gezari, D.~Y., et al.\ 1993, \apj, 409, L5.
\bibitem[Bruzual \& Charlot(2003)]{Bruzual03} Bruzual, G., \& Charlot, S.\ 2003, \mnras, 344, 1000. 
\bibitem[Burtscher et al.(2013)]{Burtscher13} Burtscher, L., Meisenheimer, K., Tristram, K.~R.~W., et al.\ 2013, \aap, 558, A149.
\bibitem[Cameron et al.(1993)]{Cameron93} Cameron, M., Storey, J.~W.~V., Rotaciuc, V., et al.\ 1993, \apj, 419, 136.
\bibitem[Combes et al.(2019)]{Combes19} Combes, F., Garc{\'\i}a-Burillo, S., Audibert, A., et al.\ 2019, \aap, 623, A79.
\bibitem[Dullemond, \& van Bemmel(2005)]{Dullemond05} Dullemond, C.~P., \& van Bemmel, I.~M.\ 2005, \aap, 436, 47.
\bibitem[Efstathiou, \& Rowan-Robinson(1995)]{Efstathiou95} Efstathiou, A., \& Rowan-Robinson, M.\ 1995, \mnras, 273, 649.
\bibitem[Dultzin-Hacyan et al.(1988)]{Dultzin88} Dultzin-Hacyan, D., Moles, M., \& Masegosa, J.\ 1988, \aap, 206, 95.
\bibitem[Dultzin-Hacyan, \& Benitez(1994)]{Dultzin94} Dultzin-Hacyan, D., \& Benitez, E.\ 1994, \aap, 291, 720.
\bibitem[Dultzin-Hacyan, \& Ruano(1996)]{Dultzin96} Dultzin-Hacyan, D., \& Ruano, C.\ 1996, \aap, 305, 719.
\bibitem[Elitzur, \& Shlosman(2006)]{Elitzur06} Elitzur, M., \& Shlosman, I.\ 2006, \apj, 648, L101.
\bibitem[Feltre et al.(2012)]{Feltre12} Feltre, A., Hatziminaoglou, E., Fritz, J., \& Franceschini, A.\ 2012, \mnras, 426, 120.
\bibitem[Fritz et al.(2006)]{Fritz06} Fritz, J., Franceschini, A., \& Hatziminaoglou, E.\ 2006, \mnras, 366, 767 [Fritz06].
\bibitem[Fuller et al.(2016)]{Fuller16} Fuller, L., Lopez-Rodriguez, E., Packham, C., et al.\ 2016, \mnras, 462, 2618. 
\bibitem[Gallimore et al.(2016)]{Gallimore16} Gallimore, J.~F., Elitzur, M., Maiolino, R., et al.\ 2016, \apjl, 829, L7
\bibitem[Garc{\'\i}a-Bernete et al.(2019)]{Garcia-Bernete19} Garc{\'\i}a-Bernete, I., Ramos Almeida, C., Alonso-Herrero, A., et al.\ 2019, \mnras, 486, 4917
\bibitem[Garc{\'{\i}}a-Burillo et al.(2016)]{Garcia-Burillo16} Garc{\'{\i}}a-Burillo, S., Combes, F., Ramos Almeida, C., et al.\ 2016, \apjl, 823, L12.
\bibitem[Garc{\'{\i}}a-Gonz{\'a}lez et al.(2017)]{Garcia-Gonzalez17} Garc{\'{\i}}a-Gonz{\'a}lez, J., Alonso-Herrero, A., H{\"o}nig, S.~F., et al.\ 2017, \mnras, 470, 2578.
\bibitem[Gonz{\'a}lez-Mart{\'{\i}}n et al.(2019)]{Gonzalez-Martin19A} Gonz{\'a}lez-Mart{\'{\i}}n, O., Masegosa, J., Garc\'ia-Bernete, I., et al.\ 2019, submitted (Paper I).
\bibitem[Gonz{\'a}lez-Mart{\'{\i}}n(2018)]{Gonzalez-Martin18} Gonz{\'a}lez-Mart{\'{\i}}n, O.\ 2018, \apj, 858, 2.
\bibitem[Gonz{\'a}lez-Mart{\'{\i}}n et al.(2017)]{Gonzalez-Martin17} Gonz{\'a}lez-Mart{\'{\i}}n, O., Masegosa, J., Hern{\'a}n-Caballero, A., et al.\ 2017, \apj, 841, 37.
\bibitem[Gonz{\'a}lez-Mart{\'{\i}}n et al.(2015)]{Gonzalez-Martin15} Gonz{\'a}lez-Mart{\'{\i}}n, O., Masegosa, J., M{\'a}rquez, I., et al.\ 2015, \aap, 578, A74.
\bibitem[Gonz{\'a}lez-Mart{\'{\i}}n et al.(2013)]{Gonzalez-Martin13} Gonz{\'a}lez-Mart{\'{\i}}n, O., Rodr{\'{\i}}guez-Espinosa, J.~M., D{\'{\i}}az-Santos, T., et al.\ 2013, \aap, 553, A35.
\bibitem[Granato et al.(1997)]{Granato97} Granato, G.~L., Danese, L., \& Franceschini, A.\ 1997, \apj, 486, 147.
\bibitem[Hao et al.(2007)]{Hao07} Hao, L., Weedman, D.~W., Spoon, H.~W.~W., et al.\ 2007, \apjl, 655, L77 
\bibitem[Hern{\'a}n-Caballero et al.(2015)]{Hernan-Caballero15} Hern{\'a}n-Caballero, A., Alonso-Herrero, A., Hatziminaoglou, E., et al.\ 2015, \apj, 803, 109 
\bibitem[H{\"o}nig et al.(2006)]{Hoenig06} H{\"o}nig, S.~F., Beckert, T., Ohnaka, K., \& Weigelt, G.\ 2006, \aap, 452, 459 
\bibitem[H{\"o}nig et al.(2010)]{Hoenig10A} H{\"o}nig, S.~F., Kishimoto, M., Gandhi, P., et al.\ 2010, \aap, 515, A23
\bibitem[H{\"o}nig \& Kishimoto(2010)]{Hoenig10B} H{\"o}nig, S.~F., \& Kishimoto, M.\ 2010, \aap, 523, A27 [Hoenig10].
\bibitem[H{\"o}nig et al.(2013)]{Hoenig13} H{\"o}nig, S.~F., Kishimoto, M., Tristram, K.~R.~W., et al.\ 2013, \apj, 771, 87.
\bibitem[H{\"o}nig \& Kishimoto(2017)]{Hoenig17} H{\"o}nig, S.~F., \& Kishimoto, M.\ 2017, \apjl, 838, L20 [Hoenig17].
\bibitem[Lebouteiller et al.(2011)]{Lebouteiller11} Lebouteiller, V., Barry, D.~J., Spoon, H.~W.~W., et al.\ 2011, \apjs, 196, 8
\bibitem[Lyu, \& Rieke(2018)]{Lyu18} Lyu, J., \& Rieke, G.~H.\ 2018, \apj, 866, 92.
\bibitem[L{\'o}pez-Gonzaga et al.(2016)]{Lopez-Gonzaga16} L{\'o}pez-Gonzaga, N., Burtscher, L., Tristram, K.~R.~W., Meisenheimer, K., \& Schartmann, M.\ 2016, \aap, 591, A47.
\bibitem[Marconi et al.(2004)]{Marconi04} Marconi, A., Risaliti, G., Gilli, R., et al.\ 2004, \mnras, 351, 169.
\bibitem[Mart{\'{\i}}nez-Paredes et al.(2015)]{Martinez-Paredes15} Mart{\'{\i}}nez-Paredes, M., Alonso-Herrero, A., Aretxaga, I., et al.\ 2015, \mnras, 454, 3577.
\bibitem[Mart{\'{\i}}nez-Paredes et al.(2017)]{Martinez-Paredes17} Mart{\'{\i}}nez-Paredes, M., Aretxaga, I., Alonso-Herrero, A., et al.\ 2017, \mnras, 468, 2.
\bibitem[Meisenheimer et al.(2007)]{Meisenheimer07} Meisenheimer, K., Tristram, K.~R.~W., Jaffe, W., et al.\ 2007, \aap, 471, 453 
\bibitem[Mor, \& Netzer(2012)]{Mor12} Mor, R., \& Netzer, H.\ 2012, \mnras, 420, 526.
\bibitem[Nenkova et al.(2008A)]{Nenkova08A} Nenkova, M., Sirocky, M.~M., Ivezi{\'c}, {\v{Z}}., et al.\ 2008, \apj, 685, 147.
\bibitem[Nenkova et al.(2008B)]{Nenkova08B} Nenkova, M., Sirocky, M.~M., Nikutta, R., Ivezi{\'c}, {\v Z}., \& Elitzur, M.\ 2008, \apj, 685, 160-180 [Nenkova08].
\bibitem[Netzer et al.(2007)]{Netzer07} Netzer, H., Lutz, D., Schweitzer, M., et al.\ 2007, \apj, 666, 806 
\bibitem[Netzer(2015)]{Netzer15} Netzer, H.\ 2015, Annual Review of Astronomy and Astrophysics, 53, 365.
\bibitem[Pier, \& Krolik(1992)]{Pier92} Pier, E.~A., \& Krolik, J.~H.\ 1992, \apj, 401, 99.
\bibitem[Ramos Almeida et al.(2009)]{Ramos-Almeida09} Ramos Almeida, C., Levenson, N.~A., Rodr{\'{\i}}guez Espinosa, J.~M., et al.\ 2009, \apj, 702, 1127 
\bibitem[Ramos Almeida et al.(2011)]{Ramos-Almeida11} Ramos Almeida, C., Levenson, N.~A., Alonso-Herrero, A., et al.\ 2011, \apj, 731, 92 
\bibitem[Ramos Almeida et al.(2014)]{Ramos-Almeida14} Ramos Almeida, C., Alonso-Herrero, A., Levenson, N.~A., et al.\ 2014, \mnras, 439, 3847 
\bibitem[Ramos Almeida \& Ricci(2017)]{Ramos-Almeida17} Ramos Almeida, C., \& Ricci, C.\ 2017, Nature Astronomy, 1, 679 
\bibitem[Ricci et al.(2017)]{Ricci17} Ricci, C., Trakhtenbrot, B., Koss, M.~J., et al.\ 2017, \nat, 549, 488
\bibitem[Risaliti et al.(2002)]{Risaliti02} Risaliti, G., Elvis, M., \& Nicastro, F.\ 2002, \apj, 571, 234.
\bibitem[Rodriguez Espinosa et al.(1987)]{Rodriguez-Espinosa87} Rodriguez Espinosa, J.~M., Rudy, R.~J., \& Jones, B.\ 1987, \apj, 312, 555.
\bibitem[Siebenmorgen et al.(2015)]{Siebenmorgen15} Siebenmorgen, R., Heymann, F., \& Efstathiou, A.\ 2015, \aap, 583, A120 [Sieben15].
\bibitem[Smith et al.(2007)]{Smith07} Smith, J.~D.~T., Draine, B.~T., Dale, D.~A., et al.\ 2007, \apj, 656, 770. 
\bibitem[Stalevski et al.(2012)]{Stalevski12} Stalevski, M., Fritz, J., Baes, M., et al.\ 2012, \mnras, 420, 2756.
\bibitem[Stalevski et al.(2016)]{Stalevski16} Stalevski, M., Ricci, C., Ueda, Y., et al.\ 2016, \mnras, 458, 2288 [Stalev16].
\bibitem[Oh et al.(2018)]{Oh18} Oh, K., Koss, M., Markwardt, C.~B., et al.\ 2018, \apjs, 235, 4 
\bibitem[Pasetto et al.(2019)]{Pasetto19} Pasetto, A., Gonz{\'a}lez-Mart{\'\i}n, O., Esparza-Arredondo, D., et al.\ 2019, \apj, 872, 69.
\bibitem[Pei(1992)]{Pei92} Pei, Y.~C.\ 1992, \apj, 395, 130
\bibitem[Schartmann et al.(2008)]{Schartmann08} Schartmann, M., Meisenheimer, K., Camenzind, M., et al.\ 2008, \aap, 482, 67
\bibitem[Suganuma et al.(2006)]{Suganuma06} Suganuma, M., Yoshii, Y., Kobayashi, Y., et al.\ 2006, \apj, 639, 46.
\bibitem[Tristram et al.(2007)]{Tristram07} Tristram, K.~R.~W., Meisenheimer, K., Jaffe, W., et al.\ 2007, \aap, 474, 837 
\bibitem[van Bemmel, \& Dullemond(2003)]{vanBemmel03} van Bemmel, I.~M., \& Dullemond, C.~P.\ 2003, \aap, 404, 1
\end{thebibliography}
\end{document}